\documentclass[10pt]{elsarticle}

\usepackage[english]{babel}
\usepackage[utf8x]{inputenc}
\usepackage[T1]{fontenc}

\usepackage[a4paper,top=3cm,bottom=2cm,left=3cm,right=3cm,marginparwidth=1.75cm]{geometry}

\usepackage{amsmath}
\usepackage{graphicx}
\usepackage{algorithm}
\usepackage{algpseudocode}
\usepackage[colorlinks=true, allcolors=blue]{hyperref}
\usepackage{natbib}
\usepackage{caption}
\usepackage{subcaption}
\usepackage{multirow}
\usepackage{array}
\usepackage{rotating}
\usepackage{hyperref}
\usepackage{amssymb}
\usepackage{comment}
\usepackage{amsmath}
\usepackage{multirow}
\usepackage{colortbl}
\usepackage{textcomp}
\usepackage{gensymb}
\usepackage{lscape}
\usepackage{booktabs}
\usepackage{arydshln}
\usepackage{dcolumn}
\usepackage{bm}
\usepackage{longtable}
\usepackage{multirow}
\usepackage{array}
\usepackage{lscape}
\usepackage{comment}
\usepackage{enumitem}  
\usepackage{float} 

\usepackage{nomencl}
\makenomenclature

\usepackage{etoolbox}
\renewcommand\nomgroup[1]{%
  \item[\bfseries
  \ifstrequal{#1}{A}{Geometric}{%
  \ifstrequal{#1}{B}{Operational}{%
  \ifstrequal{#1}{C}{Environmental}{%
  \ifstrequal{#1}{D}{Nondimensional}{}}}}%
]}

\DeclareMathOperator*{\argmin}{argmin}
\DeclareMathOperator*{\argmax}{argmax}

\begin{document}
\begin{frontmatter}

\title{Evaluation and Modeling of Pneumatic Percussive Drill for Martian Subsurface Access}
\selectlanguage{english}

\author[1]{Luis Phillipe Tosi}
\ead{phillipe@kilnenergy.com}

\author[1]{Marcel Veismann}

\author[1]{Kristopher Sherrill}

\author[1]{Marcello Gori}

\author[1]{Scott Perl}

\cortext[cor1]{Corresponding author}

\address[1]{{NASA Jet Propulsion Laboratory, California Institute of Technology\\
4800 Oak Grove Dr., Pasadena, CA, 91109, USA}}


\begin{abstract}
Deep subsurface access on Mars would enable investigation of ancient lacustrine deposits, volatile-rich horizons, and other geologic targets that are inaccessible to present shallow drilling systems. This study evaluates a wireline pneumatic (WiP) rotary-percussive drill concept intended for deep Martian subsurface access using compressed atmospheric CO$_2$ as both the actuation and transport fluid. The concept combines a pneumatically driven hammer, a magnetic flapper-valve, and incremental bit indexing within a compact bottom-hole assembly intended for low-power wireline deployment.

To analyze the system, we developed a reduced-order model of the hammer and chamber dynamics that captures the coupled pressure, flow, and impact behavior during a strike. The model was compared against benchtop percussion experiments and used to interpret key performance quantities including hammer velocity, displacement, and impact energy. A modified testbed was then used to drill Martian rock simulants spanning weaker sandstone and stronger Saddleback basalt cases, allowing the drilling response to be related to both operating pressure and material properties.

The experiments showed that the WiP concept can deliver repeatable percussive impacts and achieve mechanical specific energy (MSE) values from 74 to 360~MJ/m$^3$, with lower MSE in the weaker simulant and higher MSE in the stronger basalt, as expected. The results also indicate that the system is most effective when operated in a percussion-dominant mode with bit geometry matched to the available impact energy. Taken together, the architecture study, validated model, and experimental results support the WiP drill as a credible candidate for low-power deep drilling on Mars while also identifying the remaining engineering work needed to improve robustness, cuttings removal, and full-system integration.

\end{abstract}


\end{frontmatter}

\section{Introduction} 
\label{sec:introduction}

Planetary deep drilling presents major opportunities for space exploration by providing direct access to subsurface environments. Such access is essential for studying geological history, searching for evidence of past life, and mining resources such as water for in-situ resource utilization (ISRU). It is also relevant to future exploration of ocean worlds such as Europa, Enceladus, and Titan, where access to the shallow subsurface may help constrain habitability and geologic history~\cite{zacny2008drilling, knez2021review}. High-value drilling targets therefore include Mars, the Moon, asteroids, comets, and ocean worlds. Near-surface drilling, which typically collects samples from depths of up to a few centimeters, has already been demonstrated in several missions, including the rotary-percussive drills on NASA's Curiosity and Perseverance rovers for rock core sampling and the rasping drill on the Phoenix lander for sampling ice-rich surface layers~\cite{glass2020future, zacny2009drilling,zacny2009extraterrestrial,zacny2013reaching}. These near-surface drills rely on surface-mounted mechanisms and are relatively straightforward and effective for shallow penetration into regolith or rock. In contrast, drilling beyond about 1~m rapidly increases system complexity and size while also requiring a comprehensive strategy for cuttings transport out of the borehole. A well-known example is NASA's \textit{InSight} mission, whose self-hammering probe was intended to reach depths of up to 5~m but did not achieve that objective.

\begin{landscape}
 \begin{table}[p]
   \centering
   \resizebox{0.97\linewidth}{!}{%
    \begin{tabular}{|p{4.7cm}|p{1.0cm}p{3.5cm}p{1.5cm}p{1.8cm}p{1.8cm}p{1.8cm}p{1.8cm}p{0.8cm}p{0.8cm}p{1.5cm}p{1.5cm}p{3cm}p{1.25cm}p{1cm}p{4.5cm}p{1.5cm}p{0.8cm}p{0.8cm}p{1.5cm}p{1.5cm}p{1.5cm}p{1.5cm}p{1.5cm}}       

      \textbf{Drill Name/Company} & \textbf{Design Depth (m)} & \textbf{Drill Action} & \textbf{Drill Actuation} & \textbf{Drive Location} & \textbf{Cuttings Removal} & \textbf{Cutting Extraction} & \textbf{Wellbore Stability } & \textbf{Size (m)} & \textbf{Mass (kg)} & \textbf{Ref.} & \textbf{Drilled Depth (m)} & \textbf{Test Site} & \textbf{T (deg)} & \textbf{p (Torr)} & \textbf{Substrate} & \textbf{D$_{\text{Wellbore}}$ (mm)} & \textbf{D$_{\text{Core}}$ (mm)} & \textbf{L$_{\text{Core}}$ (mm)} & \textbf{WOB (N)} & \textbf{ROP (cm/h)} & \textbf{Power (W)} & \textbf{Energy  (Wh/m)} & \textbf{MSE (MJ/m$^3$)} \\ \hline
            \textbf{CRUX (HB)} & 2 & Rotary-Percussive & E & TS & Auger & ~ & ~ & ~ & ~ & \cite{paulsen2006robotic} & 8.2 (6 holes) & Haughton Crater & ~ & 760 & Breccia and ice & 38 & 25 & ~ & ~ & ~ & ~ & ~ & \\ 
        \textbf{MARTE (HB)} & 10 & Rotary  & E & TS & Auger & bailing + pneumatics & None & 1.5+10x1 & ~ & \cite{cannon2007marte, stoker20082005, paulsen2006robotic} & 6.1 & Rio Tinto, Spain & ~ & 760 & Clay+Gossan & 48 & 27 & 200 & ~ & 14 & 50 & 357 & 1039 \\ 
        \textbf{DAME (HB)} & 10 & Rotary  & E & TS & Auger & Auger & None & ~ & (20) & \cite{zacny2009extraterrestrial,paulsen2006robotic} & 3.2 & Haughton Crater & <0 & 760 & Breccia and ice & 44.45 & N/A & N/A & 2700 & 10.32 & 58 & 562 & 1304 \\ 
        \textbf{Trident (HB)} & 1 & Rotary-Percussive & E & TS & Auger & 10cm bits & ~ & 1m & (20) & \cite{glass2024trident,stoker2023mission} & 7.8 (8 holes) & Atacama & ~ & 760 & sand/clay & 24.5 & N/A & N/A & 100 & ~ & 200 & ~ & ~ \\ 
        \textbf{DeeDri (ASI)} & 2.5 & Rotary & E & TS & Auger & Auger & None & 10x0.23 & (8.3) & \cite{magnani2004deep} & 0.25 & lab & ~ & 760 & Sand & 35 & 14 & 25 & 3 & 60 & 7.5 & 13 & 56 \\ 
        \textbf{} & ~ & ~ & ~ & ~ & ~ & ~ & ~ & ~ & ~ & ~ & ~ & lab & ~ & 760 & 1-2MPa gas concrete & 35 & ~ & ~ & 15 & 24 & 14.5 & 60 & 269 \\ 
        \textbf{} & ~ & ~ & ~ & ~ & ~ & ~ & ~ & ~ & ~ & ~ & ~ & lab & ~ & 760 & 10-50MPa Tuff & 35 & ~ & ~ & 6 & 8.4 & 7.5 & 89 & 398 \\ 
        \textbf{} & ~ & ~ & ~ & ~ & ~ & ~ & ~ & ~ & ~ & ~ & ~ & lab & ~ & 760 & 40-60MPa Travertine & 35 & ~ & ~ & 200 & 7.8 & 29 & 372 & 1656 \\ 
        \textbf{ExoMars} & 2 & Rotary & E & TS & Auger & Auger & None & 0.7+3x0.5 & 7.25 (17) & \cite{van2005development, magnani2011testing} & ~ & ~ & ~ & ~ & ~ & 25 & 10 & 30 & ~ & ~ & ~ & ~ & ~ \\ 
        \textbf{ExoMars (MDR)} & 2 & Rotary & E & TS & Auger & ~ & ~ & 10x0.2 & ~ & \cite{van2005development} & ~ & ~ & ~ & ~ & ~ & 20 & ~ & ~ & ~ & ~ & ~ & ~ & ~ \\ 
        \textbf{} & ~ & ~ & ~ & ~ & ~ & ~ & ~ & ~ & ~ & ~ & ~ & ~ & ~ & ~ & ~ & ~ & ~ & ~ & ~ & ~ & ~ & ~ & ~ \\ 
        \textbf{} & ~ & ~ & ~ & ~ & ~ & ~ & ~ & ~ & ~ & ~ & ~ & ~ & ~ & ~ & ~ & ~ & ~ & ~ & ~ & ~ & ~ & ~ & ~ \\ 
        \textbf{} & ~ & ~ & ~ & ~ & ~ & ~ & ~ & ~ & ~ & ~ & ~ & ~ & ~ & ~ & ~ & ~ & ~ & ~ & ~ & ~ & ~ & ~ & ~ \\ 
        \textbf{MPDS (ATK Space)} & 10-20 & Rotary & E & TS & Auger & Bailing & None & xxxx1 & 25 & \cite{guerrero2008final} & 0.25 & Idaho Falls & -35 - -18 & 760 & Ice & 20.32 & 15 & 10-62 & 25-200 & 16.8 & 45 & 268 & 6534 \\ 
        \textbf{} & ~ & ~ & ~ & ~ & ~ & ~ & ~ & ~ & ~ & ~ & 0.41 & Idaho Falls & -35 - -7 & 760 & Frozen soil & 20.32 & 15 & 10-62 & 80-260 & 12 & 67.5 & 563 & 13722 \\ 
        \textbf{} & ~ & ~ & ~ & ~ & ~ & ~ & ~ & ~ & ~ & ~ & 2.1 & Idaho Falls & -18 - -3 & 760 & Basalt & 20.32 & 15 & 10-62 & 200-450 & 10.5 & 85 & 810 & 34558 \\ 
        \textbf{} & ~ & ~ & ~ & ~ & ~ & ~ & ~ & ~ & ~ & ~ & 1.5 & lab & 20 & 760 & Limestone & 20.32 & 15 & 10-62 & ~ & 12 & 80 & 667 & 16263 \\ 
        \textbf{Ice Breaker (HB/ARC)} & 1 & Rotary-Percussive & E & TS & Auger & 2-stage auger, bite sampling & Auger & 1.2 & 40 & \cite{zacny2011testing} & 1 & Antarctica & -19 & 760 & 40MPa ice-cemented ground  & 25 & N/A & ~ & 70 & 120 & \~100 (<120) & 100 & 733 \\ 
        \textbf{} & ~ & ~ & ~ & ~ & ~ & ~ & ~ & ~ & ~ & ~ & 2.5 & Antarctica & -25 & 760 & Ice & 25 & N/A & ~ & \~70 & 100 & \~100 (<150) & 100 & 733 \\ 
        \textbf{} & ~ & ~ & ~ & ~ & ~ & ~ & ~ & ~ & ~ & ~ & 1 & lab & -20 & 6.4 & Ice & 25 & N/A & ~ & <100N & 100 & 60 (<80) & 60 & 440 \\ 
        \textbf{} & ~ & ~ & ~ & ~ & ~ & ~ & ~ & ~ & ~ & ~ & 1 & lab & -20 & 6.4 & Ice (2\% perchlorate)  & 25 & N/A & ~ & <100N & 50 & 60 (<80) & 120 & 880 \\ 
        \textbf{} & ~ & ~ & ~ & ~ & ~ & ~ & ~ & ~ & ~ & \cite{zacny2013reaching} & 1 & lab & 26 & 3 & 45MPa Limestone & 25.4 & N/A & bits & 80N  & ~ & \~50 (<70) & 60 & 426 \\ 
        \textbf{} & ~ & ~ & ~ & ~ & ~ & ~ & ~ & ~ & ~ & ~ & ~ & lab & 25 & 6.4 & Indiana Limestone & 28.6 & N/A & ~ & <100N & 40 & 55 & 150 & 841 \\ 
        \textbf{} & ~ & ~ & ~ & ~ & ~ & ~ & ~ & ~ & ~ & ~ & ~ & lab & -200 & 6.4 & Mars Mojave Simulant & 28.6 & N/A & ~ & <100N & 45 & 80 & 210 & 1177 \\ 
        \textbf{} & ~ & ~ & ~ & ~ & ~ & ~ & ~ & ~ & ~ & ~ & ~ & lab & -20 & 6.4 & Mars Mojave Simulant & 28.6 & N/A & ~ & <100N & 160 & 60 & 40 & 224 \\ 
        \textbf{} & ~ & ~ & ~ & ~ & ~ & ~ & ~ & ~ & ~ & ~ & ~ & lab & -20 & 6.4 & Ice & 28.6 & N/A & ~ & <100N & 200 & 45 & 25 & 140 \\ 
        \textbf{} & ~ & ~ & ~ & ~ & ~ & ~ & ~ & ~ & ~ & ~ & 1 & Antarctica & -19 & 760 & 40MPa Ice-cemented ground & 25 & N/A & bits & 80N  & 1.2 & \~80 (<100) & 100 & 733 \\ 
        \textbf{} & ~ & ~ & ~ & ~ & ~ & ~ & ~ & ~ & ~ & ~ & 0.8 & Antarctica & -19 & 760 & 40MPa Ice-cemented ground & 25 & N/A & cont. & 80N  & 60 & \~140 (<200) & 233 & 1711 \\ 
        \textbf{} & ~ & ~ & ~ & ~ & ~ & ~ & ~ & ~ & ~ & ~ & 2.5 & Antarctica & -25 & 760 & Massive ice & 25 & N/A & cont. &  80N  & 100 & 100 & 100 & 733 \\ 
        \textbf{JSC (NASA JSC/Baker)} & 10+ & Rotary & E & DH & Auger & Bailing & None & 2.1 & 8.5 & \cite{george2012planetary} & 2 & Arctic & ~ & 760 & Ice and Sandstone & 45 & 25 & 150 & 387 & 20 & 96 & 480 & 1572 \\ 
       \textbf{U/S Gopher  (JPL)} & 20-30m & Percussive & E+P & DH & Pneumatic & Pneumatic & None & ~ & 0.7 & \cite{badescu2006ultrasonic, barcohen2016autoASCE} & 1.76 & Antarctica & -16 - -5 & 760 & Ice & 64 & ~ & ~ & ~ & ~ & ~ & ~ & ~ \\ 
        \textbf{} & ~ & ~ & ~ & ~ & ~ & ~ & ~ & ~ & ~ & ~ & 1.25 & Mt. Hood Glacier & 0 & 760 & Ice & 64 & ~ & ~ & ~ & 25 & ~ & ~ & ~ \\ 
        \textbf{Auto-Gopher 1 (JPL/HB)} & 10+ & Rotary-Percussive & E & DH & Auger & Bailing & Auger & 1.5 & 22 & \cite{badescu2013auto, barcohen2016autoASCE,bar2017auto} & 0.014 & lab & ~ & 760 & Limestone & 71 & 50.8 & 100 & 43.7N  & 4.2 & 98 & 2333 & 4347 \\ 
        \textbf{} & ~ & ~ & ~ & ~ & ~ & ~ & ~ & ~ & ~ & ~ & 3.07 & US Gypsum quarry & ~ & 760 & 40MPa gypsum & 71 & ~ & ~ & ~ & 40 & 90-120 (0\%) & 350 & 652 \\ 
        \textbf{} & ~ & ~ & ~ & ~ & ~ & ~ & ~ & ~ & ~ & ~ & ~ & US Gypsum quarry & ~ & 760 & ~ & 71 & ~ & ~ & ~ & 80 & 220-250 (50\%, 5s) & 280 & 522 \\ 
        \textbf{} & ~ & ~ & ~ & ~ & ~ & ~ & ~ & ~ & ~ & ~ & ~ & US Gypsum quarry & ~ & 760 & ~ & 71 & ~ & ~ & ~ & 140 & 330-360 (50\%, 1s)  & 250 & 466 \\ 
        \textbf{} & ~ & ~ & ~ & ~ & ~ & ~ & ~ & ~ & ~ & ~ & ~ & US Gypsum quarry & ~ & 760 & ~ & 71 & ~ & ~ & ~ & 180 & 330-360 (100\%)  & 220 & 410 \\ 
        \textbf{Auto-Gopher 2 (JPL/HB)} & ~ & Rotary-Percussive & E & DH(wireline) & Auger & Bailing & ~ & 3.7 & 65 & \cite{badescu2019auto} & ~ & lab & ~ & 760 & 45MPa Limestone & ~ & ~ & ~ & ~ & 75.6 & <500W & 614 & 273 \\ 
        \textbf{} & ~ & ~ & ~ & ~ & ~ & ~ & ~ & ~ & ~ & ~ & 7.52 & US Gypsum quarry & -25 & 760 & 38MPa gypsum & ~ & ~ & ~ & ~ & ~ & ~ & ~ & ~ \\ 
        \textbf{ATC (Ratheon UTD)} & 200 & Rotary & E & DH & Auger & Core Barrel/Cuttings bucket & None & 2 & 7 (40) & \cite{shenhar2006autnomous} & 0.792 & Idaho Falls & ~ & 760 & Basalt & 37 & 25 & 110 & 450 & 0.69 & 69 & 10000 & 61608 \\ 
        \textbf{} & ~ & ~ & ~ & ~ & ~ & ~ & ~ & ~ & ~ & \cite{shenhar2005final} & 10 (7 holes) & lab & RT & 760 & 22MPa Limestone & ~ & ~ & ~ & 133 & 20 & 74 & 351 & 2162 \\ 
        \textbf{} & ~ & ~ & ~ & ~ & ~ & ~ & ~ & ~ & ~ & ~ & 0.075 & lab & -40 & 760 & Limestone & ~ & ~ & ~ & ~ & ~ & ~ & ~ & ~ \\ 
        \textbf{Redwater (HB)} & 10-20 & Rotary-Percussive & E+P & DH & Pneumatic  & Pneumatic & None & 0.84 & 6 & \cite{palmowski2022redwater} & 1.03 & lab & -60 & 7 & Ice & 70 & N/A & N/A & 197 & 194 & 131 & 68 & 63 \\ 
        \textbf{} & ~ & ~ & ~ & ~ & ~ & ~ & ~ & ~ & ~ & ~ & 1.4 & lab & -5 & 760 & Ice & 70 & N/A & N/A & 203 & 212.4 & 173 & 81 & 76 \\ 
        \textbf{Mole (Insight) DLR, ESA} & 3-5m & Percussive & E & ~ & ~ & ~ & ~ & ~ & ~ & \cite{wippermann2020penetration} & ~ & ~ & ~ & ~ & ~ & 27mm & N/A & N/A & 1.5 & ~ & ~ & ~ & ~ \\ 
    
    \end{tabular}
}
\caption{Overview of previously developed and tested planetary deep drill system}
\label{tab:review}
\end{table}
\end{landscape}

Off-Earth drilling systems face a common set of challenges, including reduced gravity, which limits the maximum weight on bit (WOB) that can be applied; extreme temperatures; thermal-management demands; borehole-stability concerns associated with low ambient pressure; strict power and mass constraints; the need for autonomous operation; and the effects of dust and other environmental conditions on system performance~\cite{knez2021review,khalilidermani2022survey}. Based on the current state of the art, planetary drilling to depths of 1--2~m is considered feasible on targets such as the Moon, Mars, Europa, and small bodies, whereas depths of 2-20~m have been projected on a longer development horizon~\cite{glass2020future}. Various extraterrestrial deep-drilling systems have been developed and tested for targets beyond 1~m depth, each tailored to specific mission objectives and environmental constraints. Table~\ref{tab:review} summarizes selected planetary deep-drilling systems and compares their architecture and relevant performance metrics. Key differentiators include drilling strategy, actuation method, drive location, cuttings-removal approach, and wellbore-stability strategy.

As shown in Table~\ref{tab:review}, the principal drill actuation methods considered for planetary drilling are rotary, percussive, and rotary-percussive. Rotary drilling removes rock by first embedding the bit into the formation through WOB and then shearing the material through continuous rotation, thereby breaking and removing rock fragments in a manner similar to conventional terrestrial drilling~\cite{hossain2015fundamentals,teale1965concept}. Rotary drills are particularly well suited for soft formations because the bit can be advanced easily into the material, enabling relatively high rates of penetration (ROP). Consequently, rotary drilling is often efficient in soft and porous rock, but its effectiveness decreases in harder formations where cutting and grinding become more difficult. In contrast, percussive drilling uses repeated axial impacts to fracture hard formations more efficiently, typically at lower WOB and often with higher ROP and reduced tool wear than purely rotary systems~\cite{bruno2005fundamental,han2005dynamically,sliwa2012drilling}. Compared with rotary drilling, in which the cutter is forced into the material and shears off chips, percussive drilling pulverizes the material through short-duration, high-force impacts~\cite{han2005dynamically}. Such systems commonly use flat-face, concave, or convex bits with sintered carbide inserts and internal flow passages for cooling and debris removal. While percussive systems generally outperform rotary systems in hard rock, they are less suitable for very soft or plastic formations~\cite{li2000analysis}; Table~\ref{tab:rockhardness} provides a rudimentary rock-hardness classification based on uniaxial compressive strength (UCS). Percussive performance is also generally less predictable and less well characterized than rotary drilling, which has been extensively used in the petroleum industry, and excessive hammer energy may introduce wellbore-stability concerns. Rotary-percussive drills combine impact fragmentation with rotary indexing and debris removal, making them particularly effective in hard and mixed formations and often enabling faster penetration, shorter drilling time, and reduced bit wear~\cite{liu2018rock, oparin2022evaluation}. 

Mechanical specific energy (MSE) is a standard metric for quantifying drilling efficiency because it measures the energy required to excavate a unit volume of rock~\cite{teale1965concept}. It may be written as~\cite{khalilidermani2022survey,tandanand1975drillability}:
\begin{equation}     \label{eq:MSE}
    \bar{E}=\frac{\text{Energy Input}}{\text{Volume of Rock Cut}}=\frac{E}{V_{\text{rock}}} = \frac{P}{A \cdot v_{\mathrm{drill}}}
\end{equation}
where $P$ is drill power, $v_{\mathrm{drill}}$ is the ROP, and $A$ is the borehole cross-sectional area. Lower MSE values indicate more efficient drilling. This metric is particularly relevant for planetary drill systems, which must operate within the limited power budgets of space missions. Rearranging Eq.~\ref{eq:MSE} also makes it possible to estimate ROP for a given power input if the specific energy of the formation is known. Consolidated results from the literature suggest generalized efficiency levels on the order of $\mathcal{O}(400~\mathrm{MJ/m^3})$ for percussive drilling and $\mathcal{O}(150~\mathrm{MJ/m^3})$ for rotary drilling across broad rock classes~\cite{zhang2022energy}, although the actual energy requirement can vary substantially with rock properties. MSE can therefore also be interpreted as a measure of a rock's resistance to penetration: softer formations consume less energy at high ROP, whereas harder or more compact rocks require more power to achieve similar ROP, or equivalently yield lower ROP at the same power input. Accordingly, MSE is strongly related to rock-strength metrics such as UCS, with higher-UCS rocks generally requiring more energy for drilling and therefore exhibiting higher MSE~\cite{tandanand1975drillability,li2000analysis}. For percussive drilling, ROP has also been reported to correlate with UCS~\cite{kahraman2003dominant}.

\begin{table}
    \centering
    \begin{tabular}{c|c}
        \textbf{UCS} & \textbf{Classification} \\
        \hline
        <20 MPa &  very soft \\
        20-50 MPa & soft \\
        50-120 MPa & medium hard \\
        120-200 MPa & hard \\
        >200 MPa & very hard \\
    \end{tabular}
    \caption{Rock hardness classification~\cite{li2000analysis}}
    \label{tab:rockhardness}
\end{table}

\begin{figure}[ht!]
    \centering
    \includegraphics[width=0.25\linewidth]{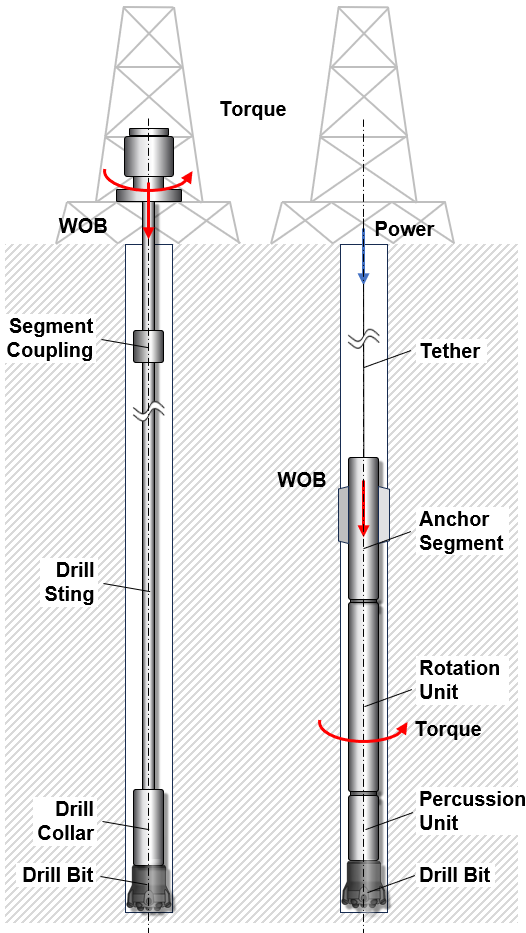}
    \caption{Comparison of different deep drill architectures: top-sided (left) and downhole (right)}
    \label{fig:TSvsDH}
\end{figure}

Drill-system architecture may also be classified by drive location into top-side and downhole (wireline) configurations~\cite{horne2015drilling,zacny2005strategies}. In top-side systems, also referred to as top-hammer systems, the drill is driven from the surface through a continuous drill rod that transmits force and torque to the bit, which generally makes such systems simpler and easier to operate for near-surface drilling (see Fig.~\ref{fig:TSvsDH}). Single-segment augers are suitable for shallow drilling, with achievable depth limited by the length and stiffness of the drill stem. Modular multi-segment systems can reach greater depths by progressively extending the drill string from a storage carousel as drilling advances~\cite{bar2012deep}. This approach, however, increases system and operational complexity and requires additional time for drill-string mating. Cuttings may be removed by augers in shallow systems or by circulation of a drilling fluid in deeper systems where auger friction becomes prohibitive. Under planetary mission mass and power constraints, top-sided systems are likely practical only to depths on the order of 0--10~m because of buckling concerns in the drill string and the large power required to transport cuttings to the surface from depth. For deeper planetary targets, wireline systems become more attractive. In a wireline architecture, the drill-bit and the full driving mechanism, that is, the borehole assembly (BHA), are deployed on a tether that provides power and data transmission and are lowered into the borehole~\cite{paulsen2016planetary}. WOB is then applied through an anchoring system, which also reacts cutting forces. This architecture enables drilling to substantial depth without a comparable increase in system mass or mechanical complexity, and is limited primarily by tether length. However, wireline systems must package the entire actuation system within the borehole diameter and remain vulnerable to borehole collapse. For planetary drilling, both top-side and downhole systems may use electric or pneumatic/hydraulic actuation, with hydraulic systems often projected to offer efficient low-power drilling at high ROP~\cite{knez2021review}. The trade space among these approaches includes mechanical complexity, robustness, drilling performance, tether mass, power throughput, and planetary-protection considerations.

To address the constraints of future Martian subsurface exploration, we propose a wireline pneumatic (WiP) rotary-percussive drill system capable of autonomously advancing to depths greater than 10~m. At such depths, a wireline architecture offers a compact and lightweight solution without the need for complex drill-string assembly. The drill, powered by compressed CO$_2$, is envisioned as part of the Martian lander mission DASER for deep subsurface access in the search for evidence of extinct microbial life~\cite{tosi2024deep}. Because bringing pressurized gas from Earth increases launch mass and introduces containment constraints, the mission concept instead anticipates high-pressure fluid production by ISRU compression of the native Martian atmosphere. The drill is also intended to operate within an approximately 100--200~W power budget consistent with a mid-class Martian lander mission similar to \textit{InSight}. The surface platform is assumed to rely on solar power, primarily for mission cost reasons.
In the proposed concept, the percussive mechanism performs the rock-fracture work, while the rotary feature is used primarily for tooth indexing between drilling cycles, that is, to rotate the bit slightly so that the teeth engage fresh material rather than repeatedly striking the same location. This indexing improves efficiency and reduces wear. A fully pneumatic actuation approach offers several advantages: (1) a simpler BHA without miniaturized electrical actuation components, which could otherwise introduce failure points or reduce efficiency; (2) improved thermal management, since a gas-driven BHA does not generate the same localized heating and temperature-sensitive power-system components can remain in the larger insulated surface station; (3) the potential to assist borehole stabilization through pressurization; and (4) pneumatic removal of drill cuttings without requiring repeated extraction of the drill assembly. To fully leverage this concept, we propose a full-face drill-bit so that the BHA can remain in the borehole throughout drilling, eliminating the need to extract drilled cores. Core extraction introduces a substantial jamming risk and, at large depths, the power required to pull out the full BHA together with a core becomes significant and leads to substantial downtime~\cite{shenhar2005final}. Similarly, mechanically transporting cuttings all the way to the surface becomes increasingly power intensive with depth because of friction. Instead, taking advantage of the low-gravity environment, cuttings can be conveyed pneumatically out of the borehole. If the drilled depth becomes too large for direct pneumatic transport to the surface, cuttings could instead be collected in a lightweight cuttings bucket of reduced diameter located above the BHA. Extracting only this smaller bucket, rather than the full BHA, reduces extraction power, operational downtime, and jamming risk.

Because planetary drilling systems are strongly constrained in mass and volume, the use of casing is limited, making collapse of an uncased borehole a significant risk. On Mars, however, reduced gravity should improve borehole stability, with prior analysis suggesting stable boreholes on the order of 100+~m under favorable conditions~\cite{edwards2020deep}. Sealing the borehole at the surface would also allow the CO$_2$ driving gas to be used for borehole pressurization after hammer actuation, helping to stabilize the wellbore and mitigate collapse. Specifically, maintaining a wellbore pressure above the formation pore pressure but below the fracture pressure can counter compressive stresses and reduce the risk of shear failure, which may occur when hoop stresses around the borehole exceed the strength of the surrounding formation~\cite{miles1949stresses,hossain2018drilling,westergaard1940plastic,aadnoy2022petroleum}. In terrestrial drilling, this stabilizing pressure is typically supplied by drilling fluid and is commonly discussed in terms of equivalent mud weight.
One potential challenge is whether permeability in the Martian megaregolith is sufficiently low to maintain the required wellbore pressure. This issue could be addressed in several ways, including deployment of a rolled-out sealing structure from the BHA. The intent of the present work, however, is not to resolve these mission-level details, but rather to demonstrate a pneumatic drilling mechanism that could enable such concepts.

This manuscript presents the development of the proposed planetary \textit{WiP} drill system for Martian subsurface access, with particular emphasis on the pneumatically driven percussion mechanism. This focus is motivated by the expectation of relatively hard Martian rock, for which impact-driven fracture is advantageous, and by the limited WOB that can realistically be applied in a mass-constrained planetary system. In other words, the concept is rotary-percussive at the system level, but the primary rock-breaking mechanism is percussive, while rotation is used mainly for indexing the bit between strikes rather than for sustained cutting. Given the uncertainty of planetary subsurface conditions on Mars, which may include hard volcanic rock, regolith, and ground ice, this percussion-dominant approach offers an attractive compromise between drilling effectiveness and system complexity.
We developed a comprehensive mathematical model of the hammer unit to improve understanding of system behavior and to support future design optimization. A full-scale benchtop BHA of the WiP percussion unit was then tested under laboratory conditions at JPL for performance characterization and model validation. During the test campaign, the percussion testbed successfully demonstrated the pneumatic hammering concept and drilled multiple holes with a combined depth of 0.5~m in soft sandstone and hard basalt.

\section{Drill System Architecture}
\subsection{The DASER mission}
The Martian DASER (Deep Access Subsurface Extraction \& Retrieval) mission has been proposed by JPL to access lacustrine deposits dating from the Noachian period of Mars history and to search for evidence of past microbial life~\cite{tosi2024deep}. Reaching these deposits requires deep-drilling technology capable of operating at depths greater than 10 m and potentially up to 100 m. The mission therefore features a pneumatically driven wireline rotary-percussive (\textit{WiP}) drill that operates on compressed CO$_2$.

The downhole assembly is fully pneumatically actuated, whereas the surface station contains the sensors and electronics, a drill-deployment system for the unconsolidated near-surface layer, and a compression unit that sources CO$_2$ from the native atmosphere, compresses it to liquefaction pressures, and delivers it to the drill through a pneumatic tether. To remain compatible with a mid-class mission concept, the surface station is assumed to be solar powered and to operate within a power budget of approximately 200 W. During drilling, the percussive mechanism and full-face drill-bit fragment the formation, and the resulting cuttings are pneumatically conveyed to the surface for analysis by the surface-station instruments. As the \textit{WiP} BHA advances into the borehole, the cuttings provide information about subsurface properties as a function of depth. Additional sensors could also be integrated into the BHA for in situ measurements or for borehole-wall reassessment during retrieval.
\begin{figure}[h!]
    \centering
    \includegraphics[width=0.5\linewidth]{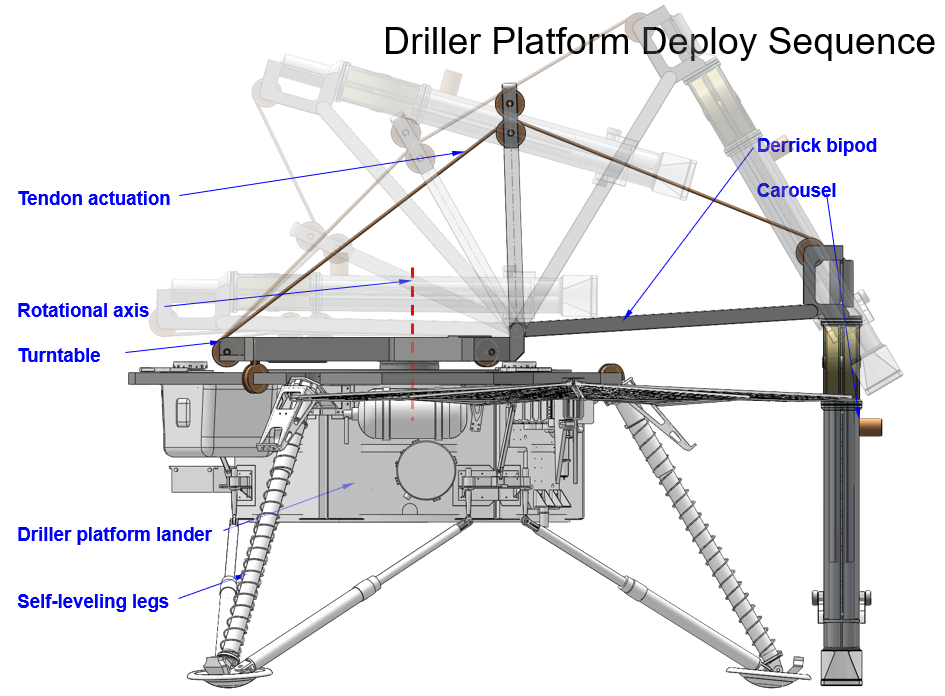}
    \caption{Drill deployment sequence of the DASER surface lander}
    \label{fig:DASER}
\end{figure}

\subsection{Design Features and Concept of Operations of the \textit{WiP} Bottom Hole Assembly (BHA)}
\begin{figure}[t!]
    \centering
    \includegraphics[width=1\linewidth]{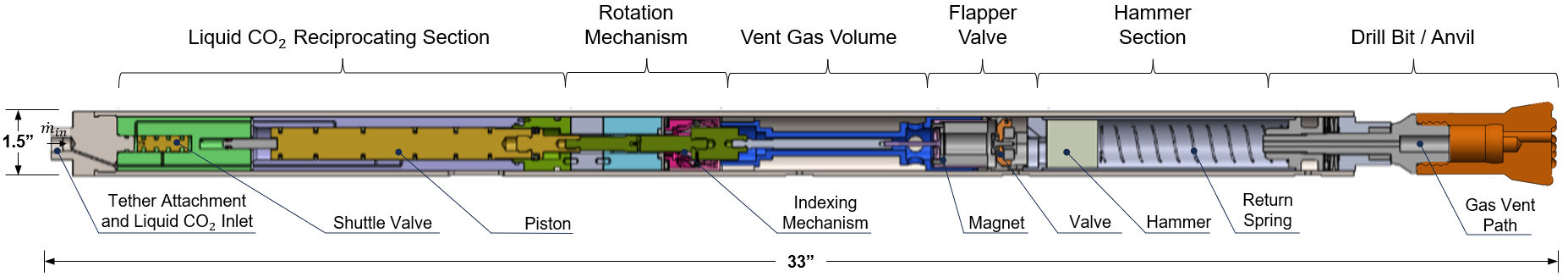}
    \caption{Cross-sectional view of the WiP BHA design}
    \label{fig:Drill_Architecture}
\end{figure}

Figure~\ref{fig:Drill_Architecture} shows a cross-sectional view of the \textit{WiP} drill BHA design. The BHA has a total length of 33", a casing diameter of 1.5", and a mass of approximately 3.8 kg. Its principal sections are the reciprocation section, rotation mechanism, vent-gas dead volume, flapper-valve, hammer section, and drill-bit.

A central feature of the concept is that the complete BHA, including both the percussion unit and the indexing mechanism, is actuated pneumatically using compressed CO$_2$ sourced directly from the Martian atmosphere. Because high-pressure fluid is already required for cuttings clearing, this approach avoids temperature-sensitive electrical actuation hardware in the borehole and instead relies on a comparatively simple and robust mechanical design. As a result, electrical power does not need to be transmitted through the tether, although a lightweight data line could still be included in future versions if downhole sensors are added. The umbilical used for fluid transmission and deployment is connected to the upper end of the BHA.

To reduce flow rate and tether size, the working fluid is supplied to the drill in liquid form. At Martian ambient temperatures, CO$_2$ can remain liquid at relatively modest pressures of approximately 8 bar. The pressurized liquid first enters the reciprocating section, where it drives a piston set that actuates the drill-bit rotation mechanism. This indexing mechanism rotates the bit incrementally during each drill stroke. Although the rotary motion is not intended to provide primary cutting action, it indexes the bit teeth into previously uncut material, improving cutting efficiency and ROP.

After driving the rotation mechanism, the fluid expands into gas in the vent volume, where the pressure gradually rises to approximately 100 psi. The main valve is held closed magnetically and opens automatically once the pressure force in the vent volume exceeds the magnetic closing force. When the valve opens, the pressurized gas accelerates the hammer into the drill-bit stem. Because the stem is free to move vertically relative to the rest of the BHA, the impact energy is transmitted to the rock below. The bit contains tungsten-carbide buttons or teeth and internal flow passages that vent high-pressure fluid at the cutting face. A sufficiently large exhaust area below the hammer is provided to clear the region beneath the bit and to avoid pressure build-up ahead of the hammer during the percussion stroke.

Modular drill-bits ranging from approximately 1.5" to 2" can be attached to the drill-bit stem. The current design does not include an anchoring system and therefore relies on its own weight to apply WOB. Future designs are expected to incorporate anchoring and anti-rotation features to improve impact-energy transfer to the formation and to increase ROP. The rotation mechanism and flapper-valve, which are the key actuation elements of the concept, are described in more detail below.

Note that the experimental campaign of this study did not test the full \textit{WiP} drill design shown in Fig.~\ref{fig:Drill_Architecture}, but rather a modified version of the BHA specifically developed to evaluate the hammer section and quantify percussive performance. The benchtop BHA used for testing is discussed in more detail in Sec.~\ref{sec:experiments}.

\subsection{Rotation Mechanism}
\begin{figure}[h!]
    \centering
    \includegraphics[width=1\linewidth]{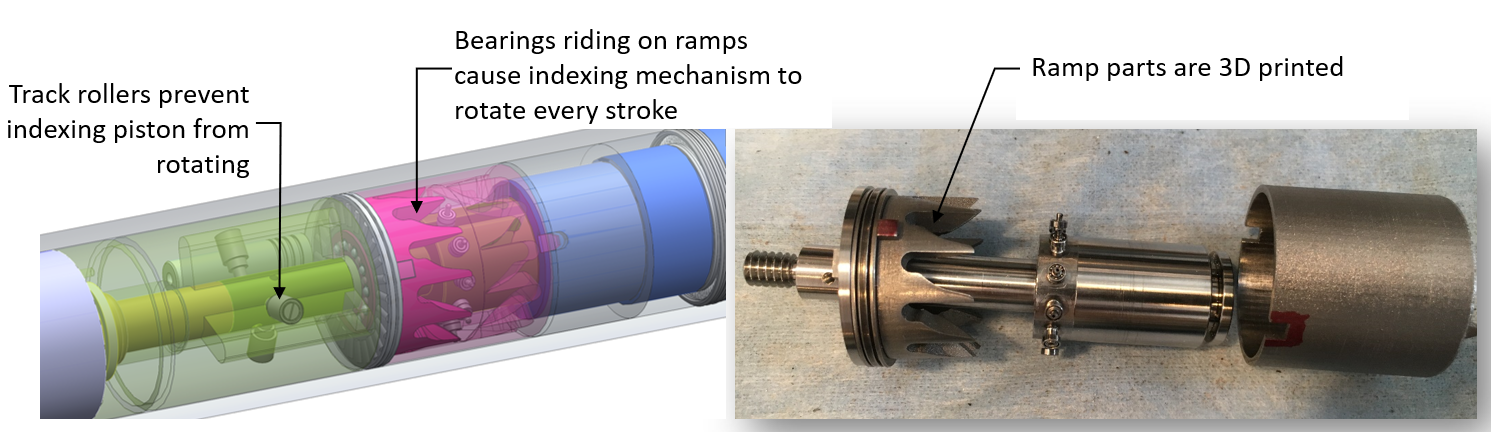}
    \caption{Design and prototype of the WiP rotation mechanism}
    \label{fig:RotationMechanism}
\end{figure}

In the fully liquid-CO$_2$-powered version, the high-pressure fluid generates reciprocating linear motion. The indexing mechanism shown in Fig.~\ref{fig:RotationMechanism} converts this reciprocating motion into partial rotation of the lower section of the drill, so that the bit indexes slightly on every stroke. Bearing or track rollers mounted on the upper section of the indexing shaft constrain its motion relative to the upper drill body. Offset nonlinear ramps between the upper and lower sections then convert the axial motion of the shaft into rotation. If flapper-valve actuation is synchronized with the indexing mechanism, the impact timing and the number of buttons on the drill-bit can be selected to ensure full coverage of the formation.

Future versions may alternatively incorporate an electrically driven rotation mechanism if electrical connections to the surface are available. In that case, a motor could be used to continuously rotate the lower section of the drill. Because the rotary components are used primarily for indexing rather than for primary cutting, the required torque is relatively low, especially under low WOB. The trade space among robustness, complexity, and drilling performance will be examined in future work.

\subsection{Flapper Valve}
\begin{figure}[b!]
     \centering
     \begin{subfigure}[b]{0.45\textwidth}
         \centering
         \includegraphics[width=\textwidth]{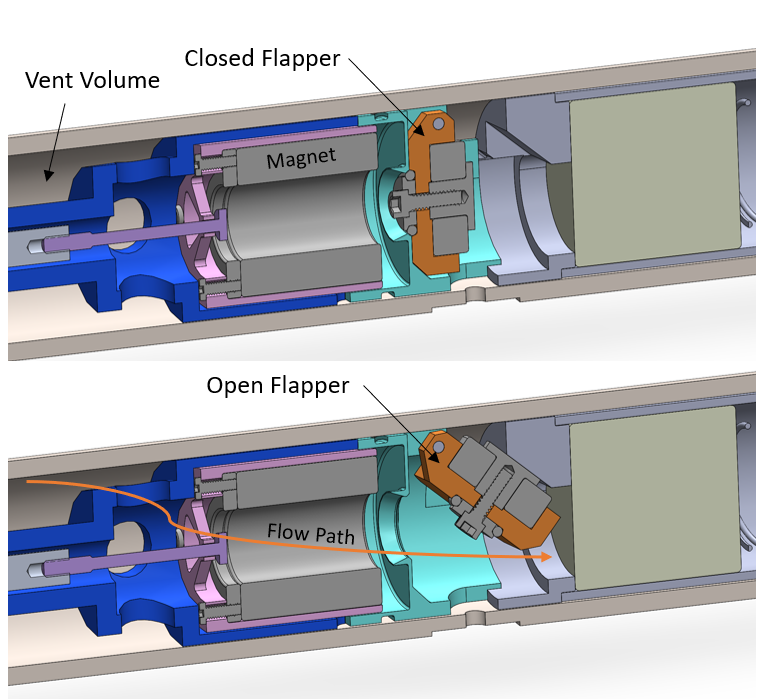}
         \caption{Schematic flapper-valve operation (closed and open valve)}
         \label{fig:flapperOpen}
     \end{subfigure}
     \hfill
     \begin{subfigure}[b]{0.47\textwidth}
         \centering
         \includegraphics[width=\textwidth]{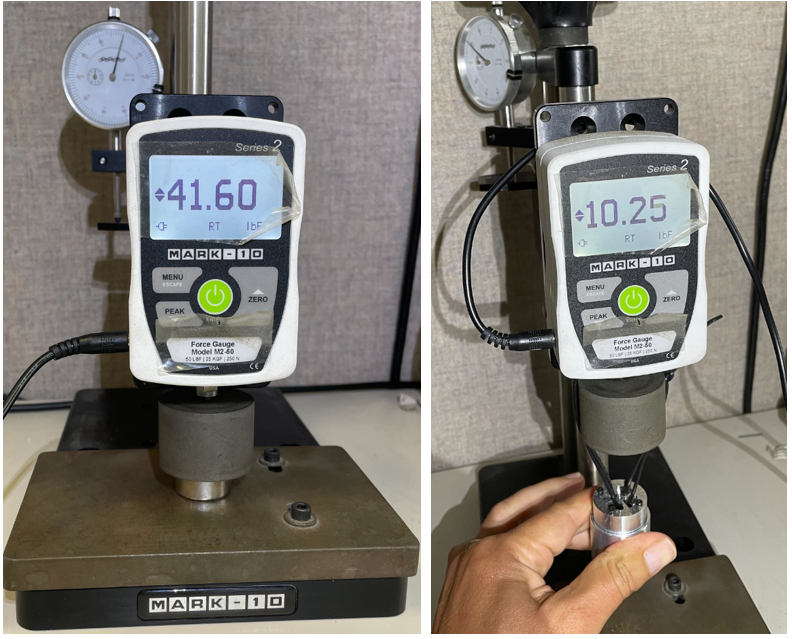}
         \caption{Force testing of the magnet and valve: maximum opening force at minimum stand-off distance (left), opening force at an increased stand-off distance of 1" (right)}
         \label{fig:MagnetTest}
     \end{subfigure}
     \caption{Flapper-valve design and testing}
\end{figure}

The \textit{WiP} drill uses a flapper-valve between the vent-gas volume and the hammer section so that pressure can build in the vent chamber before the hammer section is rapidly pressurized. The valve must pass the flow rates required to produce high hammer accelerations and impact energies, while also opening automatically at a prescribed pressure and resetting at the end of each stroke. The valve design is shown in Fig.~\ref{fig:flapperOpen}.

The valve consists of a cylindrical neodymium magnet, with a central passage for gas flow, and an alloy-steel flapper body that provides both magnetic attraction and impact toughness. In the closed configuration, a seal mounted on the flapper body presses against a thin magnetically permeable plate and seals the central orifice ($D_{v1} = 0.4$'') between the vent-volume side and the hammer section. Placing the O-ring on the flapper body helps prevent it from being pushed out by the expanding gas. Some leakage can be tolerated, so long-life seal materials may be preferable to softer but less durable alternatives.

As CO$_2$ gas enters the vent volume, the pressure rises until the pressure force on the flapper exceeds the magnetic closing force, at which point the valve opens automatically. The open valve then provides the flow path needed to pressurize the hammer section. Once the hammer reaches the end of its stroke and the pressures in the hammer and vent sections equilibrate, the magnetic force closes the flapper again, allowing the vent chamber to repressurize before the next cycle. The opening frequency is therefore governed by both the pressure drop during a stroke and the rate at which the vent chamber repressurizes.

The valve opening pressure is determined primarily by the stand-off distance between the magnet and the flapper body. Increasing this distance reduces the magnetic force and lowers the opening pressure. Figure~\ref{fig:MagnetTest} shows the force-gauge setup used to characterize both the bare magnets and the assembled flapper-valve. The measured opening force reached 41.6 lbf at the minimum stand-off distance, corresponding to a pressure capability of approximately 300 psi for an orifice area of 0.138 in$^2$. In practice, the percussion test campaign described in Sec.~\ref{sec:experiments} showed that the current design readily achieved the desired operating pressure of about 100 psi. If higher holding force is required in future designs, an additional magnet could be incorporated into the flapper body.

If the stand-off distance and magnetic force are set below the maximum available pressure force, the valve self-triggers whenever the vent pressure reaches the critical value. If the magnetic force is set above the maximum passive opening force, the valve could instead be triggered by actively retracting the magnet. That motion could be produced by the reciprocating piston, thereby preserving a fully pneumatic architecture while coupling the indexing and percussion functions.

Depending on the rock type and thermal environment, it may be advantageous to vary the valve opening pressure and therefore the delivered impact energy. The same adjustability may also be useful for managing CO$_2$ phase behavior under cold operating conditions. Although the current \textit{WiP} design uses a fixed magnet position, future versions could incorporate an adjustable magnet mount to vary the stand-off distance and thus the opening pressure. The modified percussion testbed described in Sec.~\ref{sec:experiments} already includes this capability for experimental evaluation.

\section{Pneumatic Drill Model} \label{sec:model}

We derive a reduced-order model in this section that maps the flow properties, thermodynamics, component geometry, and other design parameters to the hammer kinetic energy delivered to the drill-bit. The intent of the model is to provide enough fidelity to understand trends in parametric trades such that informed design decisions can be made. It is not intended to capture the full detailed physics, which would require fluid-structure-magnetic coupling and higher-fidelity numerical models. The goal is to provide a first-order representation of the observed system dynamics that can be semi-empirically calibrated with experimental data.
\begin{figure}[H]
    \centering
    \includegraphics[width=0.75\linewidth]{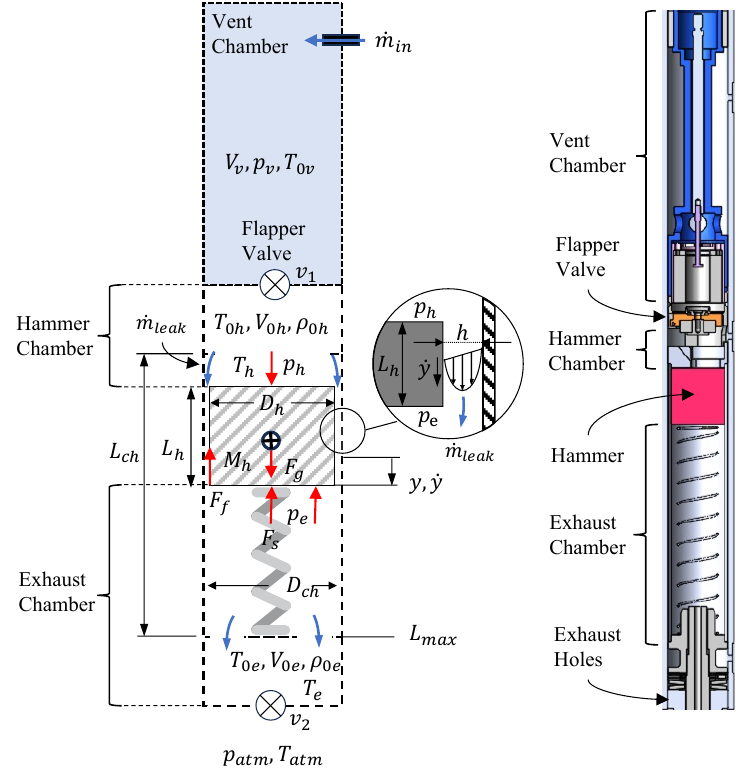}
    \caption{Diagram of coupled flow-hammer dynamics model in the lower drill section (\textit{left}), and the corresponding drill design segments (\textit{right}).  Flow path starts at the vent chamber, through valve 1, hammer chamber, exhaust chamber, and exits to the atmosphere through the exhaust holes (labelled as $v_2$ in schematic).}
    \label{fig:hammer_model}
\end{figure}
To capture these physics, the drill is segmented into multiple interconnected chambers in series that allow the working fluid to flow between them during distinct operating regimes. The thermodynamic coupling between each chamber is defined through the equations of motion of the fluid, and the resulting pressure is integrated over the hammer area to obtain the hammer driving force.
In this section, we first define the hammer dynamics and how they depend on the pressures in the hammer and exhaust chambers. We then define the flow and thermodynamic models for those chambers based on the drill geometry and the other parameters that characterize the drill.

We begin by defining the main operating segments of the drill. Figure~\ref{fig:hammer_model} shows a schematic of the lower drill segments together with the corresponding CAD geometry. Thermodynamic properties are defined in three chambers: the vent, hammer, and exhaust chambers, in addition to those of the surrounding environment. The flapper-valve separates the vent chamber from the hammer chamber, and the hammer itself separates the hammer chamber from the exhaust chamber. The hammer and exhaust chamber volumes change as the hammer moves. Both sections also have dead volumes that the hammer cannot reach. For example, the flapper-valve requires sufficient opening length for the flap to rotate fully around its hinge, as shown in Fig.~\ref{fig:flapperOpen}. Similarly, the exhaust chamber has a section at the very bottom that allows the retracting spring to approach its fully solid length. The dead volumes are $V_{0h}$ and $V_{0e}$ for the hammer and exhaust sections, respectively. The maximum displacement of the hammer within the hammer section is
\begin{equation}
    L_{max} = L_{ch} - L_h,
\end{equation}
where $L_{ch}$ is the total hammer section length, and $L_h$ is the hammer height.  

Before the drill is turned on, all chamber thermodynamic properties are in equilibrium with the environment (subscript $atm$). Once the drill is turned on, operation begins with the vent chamber filling at a constant mass flow rate $\dot{m}_{in}$ from the surface compression system, with the flapper-valve initially closed and the hammer at its maximum displaced position.
Once the vent-chamber pressure $p_v$ reaches the valve opening pressure $p_{o}$, which is set by the magnet position relative to the flapper seal surface, the flapper opens and the hammering regime begins.
As the flapper swings open about its hinge, the flow area $A_{v1}$ increases to a maximum value as a function of time, with flow moving from the vent chamber into the hammer chamber immediately after opening.
The detailed flapper dynamics are not part of the present model, but they could be implemented in future versions if the flapper inertia properties, magnetic force, and center of pressure can be reduced to a simple lumped-component model.
Even without the detailed flapper dynamics, $A_{v1}$ is still defined in time as an input to the model. Sensitivities to the opening time can therefore be investigated through its functional parameterization.
As the internal hammer-chamber pressure $p_h$ increases, it begins to accelerate the hammer downward, $\ddot{y}$, against the spring force $F_s$, the friction force on the chamber side walls $F_f$, and the exhaust pressure $p_e$, such that the total hammer-chamber volume $V_h$ increases. Because the hammer-section geometry has fixed total length, the total exhaust-chamber volume decreases by the same amount.
The hammer accelerates until it reaches the end of the hammer section by striking the drill-bit extension that protrudes slightly into the exhaust-chamber volume ($\sim 3$ mm).
The strike transfers energy to the drill-bit and the rock, which fractures. Flow then moves around the hammer through the gap $h$ and escapes through the exhaust holes. The holes themselves are modeled as a single orifice and denoted with the subscript $v_2$.

\subsection{Hammer Dynamics}
To describe the hammer acceleration, we use the free body diagram, Newton's second law in Fig.~\ref{fig:hammer_model}, and assume radially uniform $p_h$ and $p_e$, such that,
\begin{equation} \label{eq:FullHammerFBD}
    M_h \ddot{y}  = F_h =  A_h\left( p_h - p_e \right) - F_f - F_s + F_g,
\end{equation}
where 
\begin{equation} \label{eq:area}
    A_h = \frac{\pi D_h^2}{4},
\end{equation}
is the (equal) area of the top and bottom of the hammer.  Employing a viscous dissipation model for friction,  
\begin{equation} \label{eq:Ff}
    F_f = c_2 \dot{y} + c_1,
\end{equation}
where $c_1$ and $c_2$ are coefficients, and defining the gravitational force 
\begin{equation}
    F_g = M_h g,
\end{equation}
where $g$ is the gravitational acceleration, we can simplify Eq.~\ref{eq:FullHammerFBD}, 
\begin{equation} \label{eq:HammerFBDImplemented}
    \ddot{y} = \frac{1}{M_h} \left[ (p_h - p_e) A_h + F_{sh} - c_2 \dot{y} - k y + F_{BC} \right],
\end{equation}
with $k$ as the spring stiffness. We also define the constant term,
\begin{equation} \label{eq:Fsh}
    F_{sh} = -k \left[] L_s - \left(L_{ch} - L_h\right) \right] - c_1 + M_h g
\end{equation}
This term defines $y = 0$ at the uppermost position the hammer can reach, $L_{max}$. Embedded in this term is also the equilibrium hammer position, obtained by setting $\ddot{y}$ and $\dot{y}$ in Eq.~\ref{eq:FullHammerFBD} to 0 and solving for $y = y_{eq}$,
\begin{equation} \label{eq:yeq}
    y_{eq} = \frac{M_h g}{k}.
\end{equation}
To limit both the lower and upper bounds of the hammer displacement, we implement an extremely ``stiff and dissipative'' spring at those positions,
\begin{equation}
F_{BC} = \left\{
\begin{array}{ll}
      -k_{w} y - c_{w} \dot{y}  & \text{for } y \leq 0, \\
      -k_{w} \left( y - L_{max} \right) - c_{w} \dot{y}  & \text{for } y \geq L_{max}, \\
      0 & \text{otherwise};
\end{array} 
\right. 
\end{equation}
where $k_w$ and $c_w$ are the ``wall'' stiffness and damping coefficients. This term ensures that the hammer effectively hits a wall once it reaches those boundary conditions, and it does not affect the dynamics during the phase when pressure is accelerating the hammer. Finally, we can define the hammer and exhaust chamber volumes, respectively, as
\begin{equation} \label{eq:Vh}
    V_h = V_{0h} + A_h y,
\end{equation}
\begin{equation} \label{eq:Ve}
    V_e = V_{0e} - A_h y.
\end{equation}
These relationships allow the thermodynamic variables to be calculated as the volumes of the two regions evolve over time.
Finally, the kinetic energy in the hammer motion is the metric that quantifies drill performance. Its value at the end of the stroke,
\begin{equation} \label{eq:HammerKineticEnergy}
    E = \frac{1}{2} M_h \left. \dot{y} \right| _{L_{max}} ^2,
\end{equation}
is the value that we would like to maximize for any number of design and operational parametric constraints. 

\subsection{Flow and Thermodynamics} \label{sec:SubFlowThermoDynamics}
To calculate the forces on the hammer, we must define $p_h$ and $p_e$ while accounting for how each chamber is coupled through fluid flow and the associated thermodynamic variables.
In this section, we define each chamber as its own control volume (CV) and determine how the spatially uniform average pressure and temperature vary in time as material flows in and out.
The basis of this relation begins with the integral form of mass conservation,
\begin{equation} \label{eq:IntegralMassConservation}
    \frac{\partial}{\partial t} \left( \int_{V_h(t)} \rho \mathrm{d} V \right) + 
    \oint_{\partial V_h(t)} \rho {\bf{u} \cdot \hat{n}} \mathrm{d} A = 0 
\end{equation}
where $\mathbf{u}$ is the velocity field, $t$ is time, and $\rho$, $A$, and $V$ are the density, surface area, and total volume of the control volume, respectively. The vector $\mathbf{\hat{n}}$ is the outward unit normal on the CV surface. We begin with the region downstream of the flapper-valve, starting with the hammer and exhaust chambers. We then define the vent chamber, which provides the potential energy that is ultimately converted into the hammer kinetic energy delivered to the drill-bit.

\subsubsection{Hammer Chamber}

As shown in Fig.~\ref{fig:hammer_model}, the hammer chamber contains a moving CV that expands during the hammer strike and contracts when the flapper-valve is closed and the hammer is pushed back into position by the return spring. The CV is bounded by the internal walls of the chamber cylinder, the inlet that interfaces with the flapper-valve, and the hammer top (${\bf{u} \cdot \hat{n}}=0$). Assuming $\rho_h$ to be uniform over $V_h$, we expand the time derivative,
\begin{equation}
     \frac{\mathrm{d}}{\mathrm{d}t} \left( \rho_h \int_{V_h}  \mathrm{d} V \right) = \dot{\rho}_h V_h + \dot{V}_h \rho_h = \dot{m}_{v1} -\dot{m}_{leak}.
\end{equation}
Since $V_h$ is prescribed per Eq.~\ref{eq:Vh}, 
\begin{equation} \label{eq:Vdotpist}
    \dot{V}_h = A_h \dot{y},
\end{equation}
we can explicitly solve for the time evolution of hammer stagnation density as, 
\begin{equation} \label{eq:PistonDensityRate}
    \dot{\rho}_{0h} = \frac{\dot{m}_{v1} -\dot{m}_{\mathrm{leak}}}{V_h} - \left(\frac{\dot{V}_h }{V_h} \right)\rho_{0h}. 
\end{equation}
To define the relationship between thermodynamic variables, we assume the hammer expansion can largely be considered reversible adiabatic processes (i.e. isentropic, subscript~$s$).  
The validity of this assumption is relative to the rate of heat transfer between the hammer chamber and the environment during the expansion: for an adiabatic process to hold, it must be ``fast'' enough so that no significant amount of heat transfers into the chamber during the process. 
We can use isentropy to derive thermodynamic conditions between two states in the flow. For an ideal gas, 
\begin{equation} \label{eq:IdealGas}
    p = \rho R T, 
\end{equation}
where
\begin{equation}
    R = \frac{\bar{R}}{\bar{M}}
\end{equation}
is the specific gas constant, $\bar{R} = 8.314$ J mol$^{-1}$K$^{-1}$, $\bar{M}$ is the molar mass.
In the implemented reduced-order model, the hammer-chamber density state is taken directly as the chamber-average density, i.e. $\rho_h \equiv \rho_{0h}$. The hammer-chamber pressure is then closed through Eq.~\ref{eq:IdealGas} together with a chamber-average temperature state $T_h$.

\subsubsection*{Hammer-chamber energy balance (moving control volume)}

To evolve the hammer-chamber thermodynamic state, we apply a first-law control-volume (CV) balance to a
deforming CV that coincides with the hammer-chamber gas volume $V_h(t)$. The CV expands as the hammer
translates, such that the control surface velocity is $\mathbf{w}$ on the moving boundary. Denoting the
outward unit normal by $\hat{\mathbf{n}}$, and defining the specific total energy
$E \equiv e + \tfrac{1}{2}|\mathbf{u}|^2$, the energy balance for a moving CV can be written as
\begin{equation}
\frac{d}{dt}\int_{V_h(t)} \rho E dV
+
\oint_{\partial V_h(t)} \rho E (\mathbf{u}-\mathbf{w})\cdot \hat{\mathbf{n}} dA
=
\oint_{\partial V_h(t)} p \mathbf{u}\cdot \hat{\mathbf{n}} dA
-
\oint_{\partial V_h(t)} \mathbf{q}\cdot \hat{\mathbf{n}} dA
\label{eq:hammerEnergyMovingCV}
\end{equation}
where $\rho$ is the gas density, $p$ is pressure, $\mathbf{u}$ is the gas velocity, and $\mathbf{q}$ is the
conductive heat flux. Over a single hammer stroke the chamber walls are treated as adiabatic (i.e. $\oint \mathbf{q}\cdot \hat{\mathbf{n}} dA \approx 0$.

The only control surfaces that exchange mass with the CV are (i) the inlet surface at the valve (area $A_1$),
and (ii) a leakage surface (area $A_2$). All remaining boundaries satisfy
$(\mathbf{u}-\mathbf{w})\cdot \hat{\mathbf{n}} = 0$. Combining the advective energy flux and pressure-work
terms on each open surface yields an enthalpy-type transport term, giving
\begin{equation}
\frac{d}{dt}\int_{V_h(t)} \rho E dV
+
\dot{m}_{\ell}\left(E+\frac{p}{\rho}\right)
-
\dot{m}_{in}\left(E+\frac{p}{\rho}\right)
=
0
\label{eq:hammerEnergyPortForm}
\end{equation}
where $\dot{m}_{in}$ is the valve mass inflow rate into the chamber and $\dot{m}_{\ell}$ is the chamber
leakage mass outflow rate.

A well-mixed (uniform) closure is then adopted for the chamber volume, such that
$\int_{V_h}\rho E dV \approx \rho_h E_h V_h$, where $\rho_h(t)$ and $E_h(t)$ are chamber-average quantities.
Equation~\eqref{eq:hammerEnergyPortForm} becomes
\begin{equation}
\frac{d}{dt}\left(\rho_h E_h V_h\right)
+
\left(\dot{m}_{\ell}-\dot{m}_{in}\right)\left(E_h+\frac{p_h}{\rho_h}\right)
=
0.
\label{eq:hammerEnergyLumped}
\end{equation}
Expanding the time derivative and using the chamber mass balance,
\begin{equation}
\frac{d}{dt}\left(\rho_h V_h\right) = \dot{m}_{in}-\dot{m}_{\ell},
\label{eq:hammerMassBalance}
\end{equation}
cancels the terms proportional to $E_h$ and yields a compact evolution equation for the chamber-average
specific total energy,
\begin{equation}
\rho_h V_h \dot{E}_h
=
\left(\dot{m}_{in}-\dot{m}_{\ell}\right)\frac{p_h}{\rho_h}.
\label{eq:hammerEdot}
\end{equation}

For an ideal gas, the chamber-average specific total energy is modeled as
\begin{equation}
E_h = e_h + \frac{1}{2}v^2 = C_v T_h + \frac{1}{2}v^2
\label{eq:hammerEhDef}
\end{equation}
where $T_h(t)$ is the chamber-average temperature and $v(t)$ is an effective velocity scale associated with
the chamber kinetic-energy inventory. In the present implementation, this effective speed scale is closed by
$v \equiv |\dot{y}|$, using the hammer speed magnitude as a proxy for the unresolved chamber flow speed.
Substituting Eq.~\eqref{eq:hammerEhDef} into
Eq.~\eqref{eq:hammerEdot} and rearranging gives the chamber temperature ODE,
\begin{equation}
\rho_h C_v \dot{T}_h
=
\left(\frac{\dot{m}_{in}-\dot{m}_{\ell}}{V_h}\right)\frac{p_h}{\rho_h}
-
\rho_h \frac{d}{dt}\left(\frac{1}{2}v^2\right)
\label{eq:Th}
\end{equation}
The chamber pressure is then obtained from the ideal-gas relation,
\begin{equation}
p_h = \rho_h R T_h.
\label{eq:ppiston}
\end{equation}

The temperature variable $T_h$ obtained from Eq.~\ref{eq:Th} should be interpreted as an effective chamber-average thermodynamic state within the reduced-order model, rather than as a resolved local gas temperature. In particular, the formulation neglects spatial nonuniformity associated with valve expansion, inlet jet structure, and finite mixing time within $V_h$, and these approximations can produce exaggerated short-time temperature excursions during the initial filling transient. However, over the longer timescale relevant to the hammer cycle, the model predicts a net decrease in the effective chamber temperature, which is qualitatively consistent with the observed cooling tendency of the system.

For the exhaust-chamber closure and the vent-chamber timescale estimate below, we still use standard isentropic identities. Using temperature $T$, density $\rho$, and ratio of specific heats $\gamma$, the density ratio between two states can be written as~\cite{White2010},
\begin{equation} \label{eq:IsenDensRatio}
    \frac{ \rho_1}{\rho_2} = \frac{ V_2}{ V_1} = \left( \frac{T_1}{T_2} \right)^{\frac{1}{\gamma - 1}}.
\end{equation}
With the definition of the speed of sound $a$ for an ideal gas,
\begin{equation} \label{eq:speedofsoundIdealGas}
    a^2 = \left. \frac{\partial p}{ \partial \rho} \right|_s = \gamma R T,
\end{equation}
the Mach number using the local fluid velocity $U$ is
\begin{equation} \label{eq:MachNumbDef}
    M = \frac{U}{a}.
\end{equation}
and the stagnation-to-local temperature ratio becomes
\begin{equation} \label{eq:IsenTempRatio}
    \frac{T_0}{T} = 1 + \left( \frac{\gamma -1}{2} \right) M^2.
\end{equation}

To complete Eq.~\ref{eq:PistonDensityRate}, we must define the mass flow rates. Two sources of mass flow in and out of the hammer chamber exist: first is the leakage rate in the gap between the hammer and exhaust chambers ($\dot{m}_{leak}$), second is the flapper-valve mass flow rate ($\dot{m}_{v1}$).  
As depicted in Fig.~\ref{fig:hammer_model}, we approximate the flow field in the gap as laminar channel flow between two infinite parallel walls, where the gap between the cylinder and the piston head side walls $h$ is much less than the hammer height thickness $L_h$, $h << L_h$.  
Considering $\dot{y}$, the flow velocity profile over the gap dimension $r$ is \cite{schlichting2016boundary,White2010},
\begin{equation}
    u_{\mathrm{gap}} = \frac{r}{h} \dot{y} - \frac{h^2}{2 \mu} \left( \frac{\mathrm{d}p}{\mathrm{d}y} \right) \frac{r}{h} \left( 1 - \frac{r}{h} \right), 
\end{equation}
where $\mu$ is the fluid viscosity, $\mathrm{d}p/\mathrm{d}y$ is the pressure gradient through the gap, and 
\begin{equation} \label{eq:hgapsize}
    h = \frac{D_{ch} - D_h}{2},
\end{equation}
is the gap size.
Integrating it over $r$ to obtain the volumetric flow rate per unit span, 
\begin{equation*}
    \dot{q} = \int_{0}^h u_{\mathrm{gap}} \mathrm{d} r = \frac{1}{2} \dot{y} h - \left( \frac{\mathrm{d}p}{\mathrm{d}y} \right) \frac{h^3}{12 \mu}.
\end{equation*}
Approximating the gap span as $s \approx \pi D_h$, the pressure gradient $\mathrm{d}p/\mathrm{d}y \approx (p_{e} - p_h)/L_h$, and the fluid density as the density inside the piston, we have
\begin{equation}
    \label{eq:leakage}
    \dot{m}_{leak} = \rho \dot{q} s = \rho_{h} \left[\frac{1}{2} \dot{y} h + \left( \frac{p_h - p_{e}}{L_h }\right) \frac{h^3}{12 \mu} \right] \pi D_h.
\end{equation}
This expression is comparable to the conventional volumetric flow rate through a narrow gap for steady, fully developed Hagen-Poiseuille laminar flow with moving walls. This formulation does not account for gas compressibility, turbulent flow, or a choked limit in the mass flow rate, and assumes that the entire pressure difference $(p_e - p_h)$ is dissipated within the channel in the gap. Because $\mathrm{d}p/\mathrm{d}x \sim h^{-3}$ and $h \ll D_h$, this appears to be appropriate for the pressure levels in our system design.

The mass flow through the flapper-valve when open is approximated as the mass flow through an orifice~\cite[p. 394]{osti1993EPAHandbook}. For an un-choked system satisfying the pressure condition,
\begin{equation} \label{eq:chokedcondition}
    \frac{p_d}{p_u} > \left( \frac{2}{\gamma + 1 } \right)^ { \frac{\gamma}{\gamma-1} },
\end{equation}
the mass flow rate is defined as,
\begin{equation} \label{eq:massfloworificeunchoked}
    \dot{m}_{v} = \text{sign}\left( \Delta p \right) * C_D A_v \sqrt{ 2 \rho_u p_u \left( \frac{ \gamma }{\gamma - 1} \right) \left[ \left( \frac{p_d}{ p_u} \right)^{\frac{2}{\gamma}} - \left( \frac{p_d}{p_u} \right)^{ \frac{\gamma+1}{\gamma}} \right] },
\end{equation}
where $C_D$ is the discharge coefficient, $A_v$ is the valve flow area, and $\rho$ and $p$ are densities and pressures, respectively. In the case where the system is choked and condition \ref{eq:chokedcondition} is not satisfied, then
\begin{equation} \label{eq:massfloworificechoked}
    \dot{m}_{v} = \text{sign}\left( \Delta p \right) * C_D A_v \sqrt{ \gamma \rho_u p_u \left( \frac{2}{\gamma+1} \right)^{ \frac{\gamma + 1}{\gamma-1} } }.
\end{equation}
Because the flow can be bidirectional, the $u$ and $d$ subscripts correspond to the upstream and downstream conditions, defined as
\begin{equation} \label{eq:upstreamcondition}
    p_u > p_d.  
\end{equation}
To be consistent with the definition of Eq.~\ref{eq:PistonDensityRate}, the argument for $\text{sign}\left( \Delta p \right)$ for the flapper-valve is,
\begin{equation} \label{eq:valve1condition}
    \Delta p_{v1} = p_h - p_{v}.
\end{equation}
If $\Delta p_{v1} < 0$, then $p_u = p_{v}$, $\rho_u = \rho_{v}$, and $p_d = p_h$; otherwise, $p_u = p_{h}$, $\rho_u = \rho_{h}$, and $p_d = p_{v}$. Although it is unlikely that the latter occurs during normal operation, the model accounts for the possibility, especially since $A_{v1}$ is an input.

\subsubsection{Exhaust Chamber}
The dynamics of the exhaust chamber are similar to those of the hammer chamber, except that the flapper-valve is replaced by exhaust holes that are always open, and $\dot{m}_{leak}$ now represents flow \textit{into} the chamber.
From the geometric constraint, 
\begin{equation} \label{eq:Vdotexhaust}
    \dot{V}_e = -\dot{V}_h,
\end{equation}
the exhaust stagnation density time evolution becomes,
\begin{equation} \label{eq:ExhaustDensityRate}
    \dot{\rho}_{0e} = \frac{ - \dot{m}_{v2} + \dot{m}_{\mathrm{leak}}}{V_h} - \left(\frac{\dot{V}_e }{V_e} \right)\rho_{0e}. 
\end{equation}
We then obtain $T_{0e}$, $T_e$, $\rho_{e}$, and $p_e$ as before, 
\begin{equation} \label{eq:T0e}
    T_{0e} = \left( \frac{\rho_{0e}}{\rho_{atm}} \right)^{\gamma-1} T_{0atm},
\end{equation}
\begin{equation} \label{eq:Te}
    T_e = T_{0e} - \frac{\gamma - 1}{2} \left( \frac{\dot{y}^2}{\gamma R} \right),
\end{equation}
\begin{equation} \label{eq:rhoe}
    \rho_e = \left( \frac{T_e}{T_{0e}}\right)^{\frac{1}{\gamma-1}} \rho_{0e},
\end{equation}
\begin{equation} \label{eq:pe}
    p_e = \rho_e R T_e.
\end{equation}
We use Eqns.~\ref{eq:massfloworificeunchoked}-\ref{eq:massfloworificechoked} to define mass flow rate from the exhaust holes $\dot{m}_{v2}$, with parameters dictated by the geometry and flow of the holes themselves.  
Here, the relevant pressure differential is  
\begin{equation} \label{eq:valve2condition}
    \Delta p_{v2} = p_e - p_{atm}.
\end{equation}
When $\Delta p_{v2} > 0$, then $p_u = p_{e}$, $\rho_u = \rho_{e}$, and $p_d = p_{atm}$; otherwise, $p_u = p_{atm}$, $\rho_u = \rho_{atm}$, and $p_d = p_{e}$. The intent of the exhaust-flow-path design is to minimize flow resistance as much as possible, so that $p_e$ remains effectively at $p_{atm}$ during hammer-strike conditions, thereby maximizing the net force on the hammer.

\subsubsection{Vent Chamber}
The most upstream segment in the drill hammer section is the vent chamber, which is coupled to the surface pumping system via $\dot{m}_{in}$ and to the hammer chamber via the flapper-valve and $\dot{m}_{v1}$. 
Once the vent chamber is charged, it acts as the energy depot that delivers the potential energy stored in the fluid to the hammer as the fluid is discharged through the valve.

For the vent-chamber CV, bounded by the inner walls outlined in Fig.~\ref{fig:hammer_model}, the relevant mass flows are $\dot{m}_{in}$ and $\dot{m}_{v1}$. Noting that $V_v$ is constant, and assuming that the density is uniform within $V_v$, Eq.~\ref{eq:IntegralMassConservation} simplifies to
\begin{equation} \label{eq:rhodotPlenum}
    \dot{\rho}_v = \frac{\dot{m}_{in} - \dot{m}_{v1} } {V_v}.
\end{equation}
Assuming an isentropic process for the change in density and pressure in the vent chamber over the hammer strike period, 
\begin{equation} \label{eq:isentrope}
    p_v \rho_v^{-\gamma} = C,
\end{equation}
where $C$ is a constant, we can take the time derivative of Eq.~\ref{eq:isentrope}, solve for pressure, and combine with Eq.~\ref{eq:rhodotPlenum}, 
\begin{equation} \label{eq:dotpv}
    \dot{p}_v = \gamma \left( \frac{p_v}{\rho_v} \right) \dot{\rho}_v  = \gamma \left( \frac{\dot{m}_{in} - \dot{m}_{v1} } {V_v} \right) \frac{p_v}{\rho_v} .
\end{equation}
Although the vent-chamber volume $V_v$ is ideally the value designed and measured as part of the system, the effective volume available to do work is
\begin{equation} \label{eq:Vveff}
    V_{v\mathrm{eff}} = V_v \cdot \eta_v,
\end{equation}
where $\eta_v \leq 1$ is the volumetric efficiency of the vent chamber \cite{Choe2023Effect}.
Two major factors influence $\eta_v$ in our design. First, $V_v$ may be relatively small when compared to the timescale of the flow, especially while $\dot{m}_{v1}$ is choked. Specifically, the timescale associated with the pressure change within $V_v$ from Eq.~\ref{eq:dotpv} is
\begin{equation}
   t_{V_v} = \frac{\rho_v(t_0)V_v}{\dot{m}_{v1} - \dot{m}_{in}} \approx \frac{\rho_v(t_0)}{\rho^*} \left( \frac{V_v}{A_{v1}} \right)\left( \frac{1}{\sqrt{\gamma R T_v}} \right),
\end{equation}
where the * corresponds to $M=1$, and the approximation assumes $\dot{m}_{v1} \gg \dot{m}_{in}$. When we compare it to the timescale associated with a choked flapper-valve,
\begin{equation}
    t_{v1} = \frac{L_v}{\sqrt{\gamma R T_v}},
\end{equation}
where $L_v$ is a length scale of interest, we have
\begin{equation} \label{eq:VvTimeScale}
    \frac{\rho_v(t_0)}{\rho^*} \left( \frac{V_v}{A_{v1}} \right)\left( \frac{1}{\sqrt{\gamma R T_v}} \right) \gg \frac{L_v}{\sqrt{\gamma R T_v}} \rightarrow \frac{\rho_v(t_0)}{\rho^*} \left( \frac{V_v}{A_{v1}} \right) \gg L_v,
\end{equation}
in order for the vent chamber to behave as a stagnated or effectively infinite tank.
If condition \ref{eq:VvTimeScale} does not hold, that implies that non-negligible flow speeds may be established inside $V_v$. This leads to the second factor: as seen in Fig.~\ref{fig:hammer_model}, $V_v$ contains a considerable amount of internal flow geometry. If flow speeds inside $V_v$ are negligible, then the stagnation-pressure losses associated with flow past those internal geometries are also negligible. However, if internal flow speeds cannot be neglected, then $\eta_v$ begins to diverge from unity.
From Eqns.~\ref{eq:IsenDensRatio} and \ref{eq:IsenTempRatio}, setting $M=1$ for the choked conditions, $\gamma = 1.4$ for air, and letting $L_v \sim V_v/A_{ch}$, 
\begin{equation} \label{eq:VvTimeScaleVal}
    1.58 \left( \frac{V_v}{A_{v1}} \right) \gg \frac{V_v}{A_{ch}} \rightarrow 1.58 \gg \frac{A_{v1}}{A_{ch}}. 
\end{equation}
For our system, using parameters from Table~\ref{tab:parametersTable}, Eq.~\ref{eq:VvTimeScaleVal} simplifies to $12 \gg 1$. Given the tortuosity of the flow inside $V_v$, a reasonable range for $\eta_v$ is $[0.75, 1]$ \cite{Abd2017Effect,Paul2010Flow}. When $\eta_v < 1$, $V_v$ in Eqns.~\ref{eq:rhodotPlenum} and \ref{eq:dotpv} is replaced by $V_{v\mathrm{eff}}$ from Eq.~\ref{eq:Vveff}.

\subsection{Model Algorithm}
Algorithm~\ref{alg:PressureHammerEOM} illustrates how the equations of motion are coupled. It is implemented in MATLAB and uses the \texttt{ode113} solver, which is a variable-step, variable-order method well suited for non-stiff ODEs where high accuracy is required. It adapts the step size and order (1 to 13) to optimize efficiency, making it appropriate for smoothly varying thermodynamic and fluid-dynamics simulations.
The algorithm is the input function to the integrator, so the inputs are the state-vector component values $\bf{x}$ at the current time step, and the outputs are their corresponding derivatives $\bf{\dot{x}}$.
\begin{algorithm}[H]
\caption{Hammer and Thermodynamics Equations of Motion Input for Time Integrator} \label{alg:PressureHammerEOM}
\begin{algorithmic}[1]
\Ensure Inputs: $t$, ${\bf{x}} = [p_v, \rho_v, \rho_{0h}, \rho_{0e}, y, \dot{y}, T_h]$
\Comment{{\bf Output:} $\dot{p}_v$, $\dot{\rho}_v$, $\dot{\rho}_{0h}$, $\dot{\rho}_{0e}$, $\dot{y}$, $\ddot{y}$, $\dot{T}_h$}
\State Get physical parameters and thermodynamic properties: 
\renewcommand\labelitemi{{\boldmath$\cdot$}}
\begin{itemize}
    \item $g$, $V_v$, $V_{0h}$, $V_{0e}$, $A_h$, $R$, $\gamma$, $C_v$, $C_{v1}$, $C_{v2}$, $c_1$, $c_2$, $k$, $M_h$, $T_{0\text{atm}}$, $\rho_{\text{atm}}$, $p_{\text{atm}}$
\end{itemize}
\State Initialize: $p_v$, $\rho_v$, $\rho_{0h}$, $\rho_{0e}$, $y$, $\dot{y}$, $T_h$

\State Compute hammer chamber thermodynamics:
\renewcommand\labelitemi{{\boldmath$\cdot$}}
\begin{itemize}
    \item $\rho_h \gets \rho_{0h}$
    \item $p_h \gets$ Eq.~\ref{eq:ppiston}
\end{itemize}

\State Compute exhaust chamber thermodynamics:
\renewcommand\labelitemi{{\boldmath$\cdot$}}
\begin{itemize}
    \item $T_{0e}, T_e, \rho_e, p_e \gets$ Eq.~\ref{eq:T0e}, \ref{eq:Te}, \ref{eq:rhoe}, \ref{eq:pe}
\end{itemize}

\State Compute mass flow rates:
\renewcommand\labelitemi{{\boldmath$\cdot$}}
\begin{itemize}
    \item $\dot{m}_{v1} \gets$ Eqns.~\ref{eq:massfloworificeunchoked}-\ref{eq:massfloworificechoked} under condition \ref{eq:valve1condition}
    \item $\dot{m}_{\text{leak}} \gets$ Eq.~\ref{eq:leakage}
    \item $\dot{m}_{v2} \gets$ Eqns.~\ref{eq:massfloworificeunchoked}-\ref{eq:massfloworificechoked} under condition \ref{eq:valve2condition}
\end{itemize}

\State Update densities and pressures:
\renewcommand\labelitemi{{\boldmath$\cdot$}}
\begin{itemize}
    \item $\dot{\rho}_v, \dot{p}_v \gets$ Eqns.~\ref{eq:rhodotPlenum}, \ref{eq:dotpv}
    \item $\dot{\rho}_{0h} \gets$ Eq.~\ref{eq:PistonDensityRate}
    \item $\dot{\rho}_{0e} \gets$ Eq.~\ref{eq:ExhaustDensityRate}
\end{itemize}

\State Solve hammer dynamics:
\renewcommand\labelitemi{{\boldmath$\cdot$}}
\begin{itemize}
    \item $\ddot{y}, F_f, F_s \gets$ Eq.~\ref{eq:HammerFBDImplemented}, \ref{eq:Ff}, \ref{eq:Fsh}
\end{itemize}

\State Update hammer temperature:
\renewcommand\labelitemi{{\boldmath$\cdot$}}
\begin{itemize}
    \item $\dot{T}_h \gets$ Eq.~\ref{eq:Th}
\end{itemize}

\State Return: $ {\bf{\dot{x}}} = [\dot{p}_v, \dot{\rho}_v, \dot{\rho}_{0h}, \dot{\rho}_{0e}, \dot{y}, \ddot{y}, \dot{T}_h]$
\end{algorithmic}
\end{algorithm}

\subsection{Flapper Valve Opening Dynamics}
The dynamics of the flapper-valve as it opens and closes are not simulated as part of the model. This is primarily due to the complexity of the fluid-magneto-mechanical coupling that would define the detailed physics of the problem.  
As previously mentioned, a more advanced treatment would consider numerical simulations of at least the flow and the magnetic forces, and attempt to reduce its zeroth (or perhaps even first) order behavior into a lower order model implementable within our coupled framework.  
Instead, we include the valve opening and closing motion as a prescribed input to Algorithm~\ref{alg:PressureHammerEOM}, allowing it to be an arbitrary function of time.
It is nevertheless useful to explore a plausible hypothesis for its functional form. In that spirit, we can write Euler's equations at the valve-opening moment $t_0$, with assumptions that include: (1) the spatially averaged magnetic-force location is the same as the center of pressure, and is presumed to be at
\begin{equation} \label{eq:flapperleverarm}
    L_{l} = \frac{L_{v1}}{2},
\end{equation}
where $L_{v1} \approx D_{ch}$ is the diameter of the flapper; (2) the fluid force acting on the flapper surface is constant over time, and takes the form,
\begin{equation} \label{eq:flapperpressureforce}
    F_{fl} = A_{ch} \left( p_o - p_{atm} \right) = A_{ch} \Delta p;
\end{equation}
and (3) the magnetic force between the large magnet and the flapper takes the form,
\begin{equation} \label{eq:magforce}
    F_{m} = \frac{A_{ch} L_o^2 \Delta p \left( 1 - \varepsilon \right)}{\left( L_o + L_{v1} \sin \theta \right)^n},
\end{equation}
where $L_o$ is the control distance between the flapper seal surface and the large control magnet (i.e. it sets $p_o$), $\theta$ is the opening angle between the flapper seal surface and the flapper itself (i.e. $\theta = 0$ at $t_0$), and $\varepsilon$ is a small perturbation that allows the flapper to accelerate at $t_0$.
Eq.~\ref{eq:magforce} assumes the magnetic-force scaling in the local field follows $F_m \sim r^{-n}$, where $r$ is the distance between magnets, and, by design, that it cancels $F_{fl}$, modulo $\varepsilon$, at $\theta = 0$. Two dipole fields, between the flapper and the large magnet, are applicable after $r > r_{thr}$, where $n=4$ for axial displacement, but subsequent regions where $r < r_{thr}$ can yield $n < 4$ \cite{Vokoun2009Magnetostatic}. Although $r_{thr}$ can vary significantly as a function of magnet parameters, it is noted here as the far-field behavior.
Assumption (2) may be reasonable because the recirculation zone in the downstream wake of the flapper would likely remain near $p_{atm}$; furthermore, the valve may open fast enough relative to $\dot{p}_v$ to keep $p_v \approx p_o$ during the opening time. Assumption (1) also appears plausible given the valve geometry and magnet position in Fig.~\ref{fig:hammer_model} near $t_0$.
Hence, summing the moments about the flapper hinge, we can solve for the flapper angular acceleration, 
\begin{equation} \label{eq:angaccelFlapper}
    \ddot{\theta} = \frac{A_{ch} \Delta p L_{v1}}{2 I_{v1}} \left[ 1 - \frac{1 - \varepsilon}{\left( 1 + \frac{L_{v1}}{L_o} \sin \theta \right)^n} \right],
\end{equation}
where $I_{v1}$ is the flapper moment of inertia around its hinge.  Eq.~\ref{eq:angaccelFlapper} shows the scaling, 
\begin{equation}
    \ddot{\theta} \sim  \left[ 1 - \frac{1 - \varepsilon}{\left( 1 + \frac{L_{v1}}{L_o} \sin \theta \right)^n} \right],
\end{equation}
that produces plausible limiting behavior: at $t_0$, $\theta = 0$ and $\ddot{\theta} \sim \varepsilon$, at $t \gg t_0$, $\theta \gg 0$ yields $\ddot{\theta} \sim 1$ and $\theta \sim t^2$.  Taking the limit $L_o \rightarrow \infty$ makes the magnetic behavior weaker, such that we recover $\theta \sim \varepsilon t^2$, consistent with $\Delta p \rightarrow \varepsilon$ as $L_o \rightarrow \infty$ since $p_o$ is controlled by the same $r^{-n}$ magnetic force scaling. 
If the valve operates as intended, the magnetic-force term should drop off relatively quickly as $\theta$ increases, and we recover the functional form $\theta \sim t^2$.
If we let the time between $t_0$ to the fully open valve be $t_{open}$, we hypothesize that,
\begin{equation} \label{eq:opentimeCondition}
    t_{open} - t_0 \gg t_{close} - t_{open},
\end{equation}
where $t_{close}$ is the time when the flapper is back at its sealing surface. Three primary reasons make the conjecture in Eq.~\ref{eq:opentimeCondition} plausible. First, if the flapper collision with the mounting plate, once fully open, has a moderate restitution coefficient (steel impacting aluminum), then the return angular velocity will be near its maximum at $t_{open}$ immediately after the collision.
Second, it is plausible that at this point $p_v < p_o$ because enough time may have passed for $\dot{p}_v$ to become significant; this would reduce $F_{fl}$, which would otherwise decelerate the flapper on the return path.
Lastly, $F_m$ will now accelerate the flapper as $\theta \rightarrow 0$. Together, these effects imply that the average $\dot{\theta}$ on the return is likely significantly larger than the average $\dot{\theta}$ during opening, consistent with the condition in Eq.~\ref{eq:opentimeCondition}.

Hence, given the scaling from Eq.~\ref{eq:angaccelFlapper} and condition in Eq.~\ref{eq:opentimeCondition}, we postulate
\begin{align} \label{eq:Av1TimingEq}
A_{v1} = \frac{D_{v1}^2 \pi}{4} \left\{ \left( \frac{t}{t_{{open}}} \right)^2 \cdot \left[ \mathcal{H}(t) - \mathcal{H}(t - t_{{open}}) \right] + \right.  & \nonumber \\
\left. \frac{t - t_{{close}}}{t_{{open}} - t_{{close}}} \cdot \left[ \mathcal{H}(t - t_{{open}}) - \mathcal{H}(t - t_{{close}}) \right] \right\}, & 
\end{align}
where $\mathcal{H}$ is the heaviside function, and 
\begin{equation} \label{eq:Av1topen}
    t_{open} = 0.75 \cdot t_{close}.
\end{equation}
This functional form provides a closure for the system of equations, where the measured $t_{close}$ from experiments can be used as the sole input parameter to Eq.~\ref{eq:Av1TimingEq}.

\subsection{Parameter List and Ideal Operation Model Results} \label{sec:subParameterListModelRegOp}
To understand some of the canonical regimes and events of interest in the fluid-hammer dynamics, we solve the system of equations through Algorithm~\ref{alg:PressureHammerEOM} for the \textit{WiP} drill parameters and idealized friction and valve-opening conditions.
Table~\ref{tab:parametersTable} lists the input parameters and their sources relevant to the simulation. Values for $C_{v1}$ and $C_{v2}$ were obtained from the literature as flow-resistance quantities $K$, such that
\begin{equation} \label{eq:CVtoK}
    C_v = \left(\frac{1}{K} \right)^{\frac{1}{2}}.
\end{equation}
We approximate the behavior of the flapper-valve as that of a swing check valve with 0.5" diameter, yielding $K_{v1} = 5.1$ and $C_{v1} = 0.45$ from \citep[p.~385]{White2010}. For the exhaust holes, we approximate them as a single sharp-edged orifice exhausting into an infinite volume, producing $K_{v2} = 2.8$ and $C_{v2} = 0.6$ from \citep[p.~123]{Idelchik1996Handbook}. The orifice area is the summed total of all holes that exhaust to the environment, $A_{v2}$.
\textit{Measured} entries in the source column mean that values were obtained either through measurement of the hardware or from the CAD models, whereas \textit{calculated} entries are referenced to the relevant equations in the table. \textit{Controlled} means that the parameter is an input that can be varied, while \textit{calibrated} means that the value is derived either from experimental data or through iteration after observing model results.
The calibrated values listed that have minimal impact on drill performance based on $E$ are $\dot{m}_{in}$ and the wall-boundary values $k_{w}$ and $c_w$. The former can provide a small amount of additional energy to the hammer during its strike, but it is generally at least two orders of magnitude smaller than $\dot{m}_{v1}$ during the hammer-movement time. It is calibrated from the increase in $p_v$ after the flapper-valve has shut.
The wall-boundary values serve only to dissipate most of $E$ at the end of the strike while providing a stable numerical solution to the equations of motion. They alter the hammer \textit{return} dynamics after the hammer has struck the drill-bit. The only \textit{calibrated} value listed in Table~\ref{tab:parametersTable} that has a significant impact on $E$ is $c_2$. The calibration process is discussed in detail in Section~\ref{sec:LCExperimentsandModelComp}, where the final parameter value is given.

\begin{table}[H]
\centering
\resizebox{\textwidth}{!}{%
\begin{tabular}{l l l l l} 
\hline
\textbf{Symbol} & \textbf{Value} & \textbf{Unit} & \textbf{Source} & \textbf{Description} \\ \hline
$g$       & 9.81                    & m/s$^2$           & Standard      & Gravitational acceleration \\ 
$V_v$     & $7.25 \times 10^{-5}$   & m$^3$             & Measured CAD  & Vent chamber volume \\ 
$\eta_v$     & 1, (0.75)   & ND             & Controlled, (Literature \cite{Abd2017Effect,Paul2010Flow})  & Vent chamber volumetric efficiency \\ 
$D_{v1}$  & 0.01016                 & m                 & Measured      & Valve diameter \\ 
$t_{close}$  & expr. dependent      & s             & Measured      & Flapper valve close time \\ 
$A_{v1}$  & $8.11 \times 10^{-5}$, (function)   & m$^2$                 & Calculated, (Eq.~\ref{eq:Av1TimingEq})  & Flapper valve cross-sectional area \\ 
$C_{v1}$  & 0.45                     & ND               & Literature \protect{\citep[p.~385]{White2010}} & Flapper valve discharge coefficient \\ 
$\dot{m}_{\text{in}}$ & $7.14 \times 10^{-4}$ & kg/s     & Calibrated  & Mass flow rate into the vent chamber \\ 
$p_o$     & 100, (689476)           & psig, (Pa)         & Controlled & Target flapper-valve opening pressure \\ 
$D_{v2}$  & 0.01397                 & m                 & Measured & Exhaust hole diameter \\ 
$A_{v2}$  & $1.5332 \times 10^{-4}$ & m$^2$             & Calculated  & Exhaust cross-sectional area \\ 
$C_{v2}$  & 0.60                    & ND                & Literature \protect{\citep[p.~123]{Idelchik1996Handbook}} & Exhaust discharge coefficient \\ 
$c_1$     & 0                       & N                 & Controlled & Constant friction coefficient \\ 
$c_2$     & 0, (1.75)                & kg/s              & Controlled, (Calibrated) & Velocity-proportional Rayleigh friction coefficient \\ 
$V_{0e}$  & $3.967 \times 10^{-7}$  & m$^3$             & Measured CAD & Dead volume in the exhaust chamber \\ 
$V_{0h}$  & $1.131 \times 10^{-5}$  & m$^3$             & Measured CAD & Dead volume in the hammer chamber \\ 
$M_h$     & 0.143                   & kg                & Measured & Hammer mass \\ 
$k$       & 96                      & N/m               & Measured & Spring stiffness \\ 
$L_s$     & 0.127                   & m                 & Measured & Uncompressed spring length \\ 
$D_h$     & 0.02845                 & m                 & Measured & Hammer diameter \\ 
$D_{ch}$  & 0.02864                 & m                 & Measured & Hammer chamber diameter \\ 
$L_h$     & 0.030                   & m                 & Measured      & Hammer length \\ 
$L_{ch}$  & 0.1088                  & m                 & Measured CAD  & Hammer chamber length \\ 
$A_h$     & $6.359 \times 10^{-4}$  & m$^2$             & Eq.~\ref{eq:area}    & Area for the hammer \\ 
$t_{h}$   & model dependent, (expr. dependent)          & s                 & Calculated, (Measured)      & Hammer strike time \\ 
$y_0$     & 0, (expr. dependent)      & m               & Controlled, (Calculated) & Initial hammer position at $t_0$ \\ 
$h$       & $7.5 \times 10^{-5}$    & m                 & Eq.~\ref{eq:hgapsize}    & Gap around the hammer \\ 
$T_{0\text{atm}}$ & 300             & K                 & Standard      & Stagnation atmospheric temperature \\ 
Gas       & Air                     & ND                & Controlled    & Gas type  \\ 
$\gamma$  & 1.4                     & ND                & Standard      & Ratio of specific heats for gas \\ 
$R$       & 287.04                  & J/(kg$\cdot$K)    & Standard      & Specific gas constant \\ 
$p_{atm}$ & 14.7, (101352.9)        & psi, (Pa)         & Standard      & Atmospheric stagnation pressure \\ 
$\rho_{atm}$ & 1.1769               & kg/m$^3$          & Eq.~\ref{eq:IdealGas} & Atmospheric density \\ 
$k_{w}$  & $1 \times 10^6$          & N/m               & Calibrated    & End condition stiffness at the end of stroke \\ 
$c_{w}$  & 1000                     & N/(m/s)           & Calibrated    & End condition energy dissipation for velocity term \\  \hline
\end{tabular}}
\caption{List of parameters for WiP drill model and calculations. Values for $p_o$ are listed in different units; values for $c_2$ correspond to different model runs within this manuscript: $c_2 = 0$ in sec.~\ref{sec:subParameterListModelRegOp}, and $c_2 = 1.75$ in sec.~\ref{sec:LCExperimentsandModelComp}.}
\label{tab:parametersTable}
\end{table}

\subsubsection{Ideal Model Results} \label{sec:idealModelResults}
For simulations in this section, we set $c_2 = 0$ to remove any damping from the hammer-chamber walls, giving an ``ideal'' hammer strike. In the same spirit, the vent chamber is assumed to have no internal losses, such that $\eta_v = 1$, and the valve opening area is set as a step function at $t_0$, assuming the valve opens instantaneously to its full-bore area $A_{v1}$ and closes instantaneously at $t_{close} = 0.02$ s.
This analysis has two objectives: first, to provide a ``best possible'' value of $E$ given the current design and the uncertain parameters at hand; second, to provide signatures for key events when the drill is operating as intended, i.e. under normal operation.
Understanding these signatures allows us to identify abnormal drill operation and anomalous strikes.
This is one of the key points discussed in Section~\ref{sec:RockExperiments} when quantifying the energy states of strikes during abnormal operation.

\begin{figure}[H]
    \centering
    \begin{subfigure}[t]{0.475\textwidth}
        \centering 
        \includegraphics[trim=0cm .5cm 0cm .5cm, clip, width=1\linewidth]{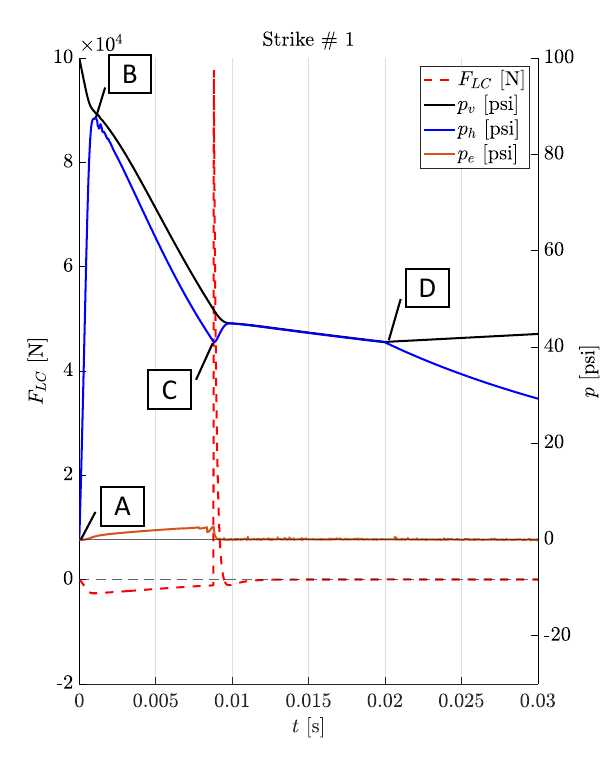}
        \caption{Model pressure and load-cell force results vs. time for a single strike.}
        \label{fig:ModelSingleStrikePandF}
    \end{subfigure}
    \begin{subfigure}[t]{0.48\textwidth}
        \centering
        \includegraphics[trim=0cm .5cm 0cm .5cm, clip, width=1\linewidth]{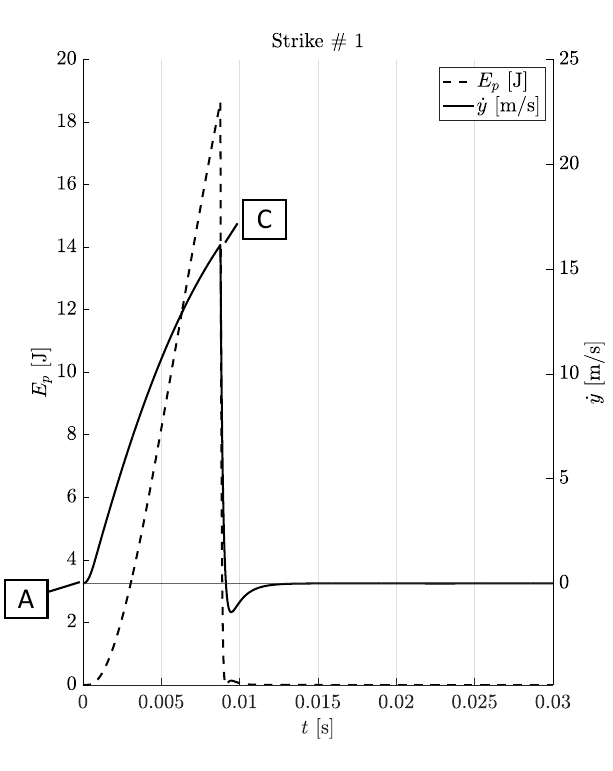}
        \caption{Model hammer velocity and kinetic energy results vs. time for a single strike.}
        \label{fig:ModelSingleStrikeEandV}
    \end{subfigure}
    \caption{Single strike modeled at $p_o = 100$ psi with key events labeled.}
    \label{fig:ModelStrikeExamples}
\end{figure}

\begin{figure}[htb!]
    \centering
    \begin{subfigure}[t]{0.48\textwidth}
        \centering 
        \includegraphics[trim=0cm .5cm 0cm .5cm, clip, width=1\linewidth]{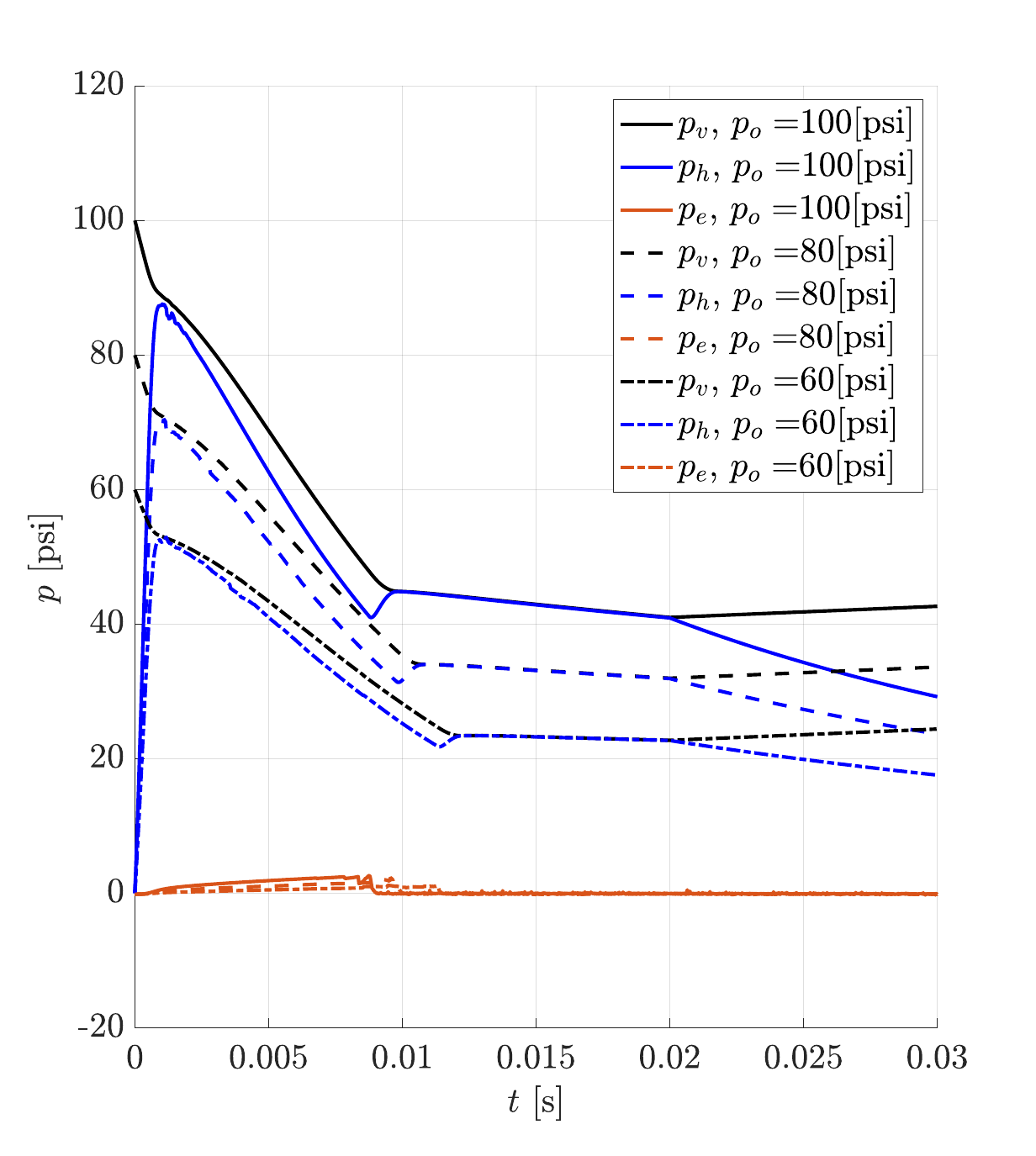}
        \caption{Model pressure results vs. time for strikes at varying flapper-valve opening threshold pressure.}
        \label{fig:ModelMultiple60-80-100P}
    \end{subfigure}
    \begin{subfigure}[t]{0.48\textwidth}
        \centering
        \includegraphics[trim=0cm .5cm 0cm .5cm, clip, width=1\linewidth]{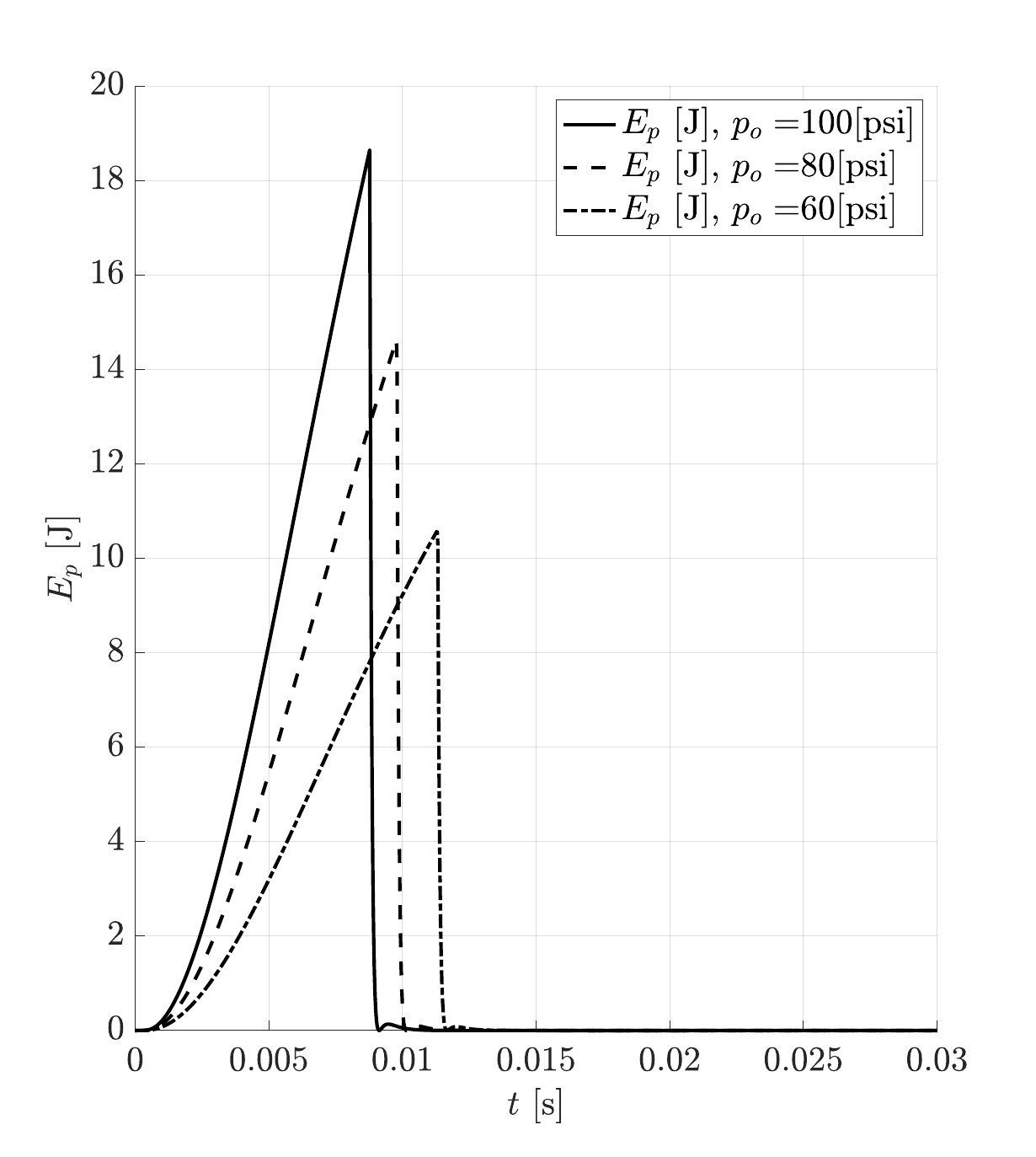}
        \caption{Model hammer kinetic energy results vs. time for strikes at varying flapper-valve opening threshold pressure.}
        \label{fig:ModelMultiple60-80-100E}
    \end{subfigure}
    \caption{Multiple strikes results modeled at varying  $p_o = [60,80,100]$ psi.}
    \label{fig:ModelMultiple60-80-100}
\end{figure}

\begin{figure}[H]
    \centering
    \begin{subfigure}[t]{0.475\textwidth}
        \centering 
        \includegraphics[trim=0cm 0cm 0cm 0cm, clip, width=1\linewidth]{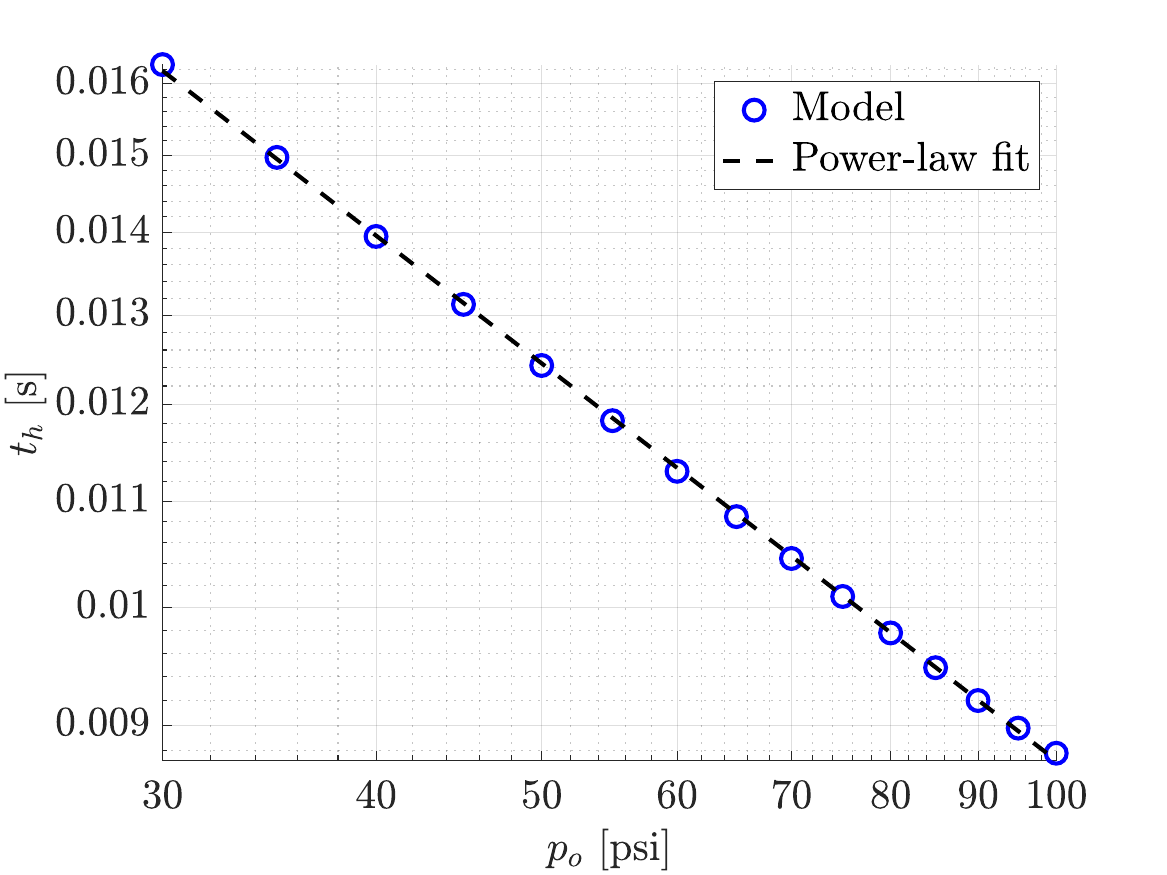}
        \caption{Ideal parameter model hammer strike time vs. flapper-valve opening threshold pressure in log-log scale.}
        \label{fig:ModelMultiple30-100th}
    \end{subfigure}
    \begin{subfigure}[t]{0.48\textwidth}
        \centering
        \includegraphics[trim=0cm 0cm 0cm 0cm, clip, width=1\linewidth]{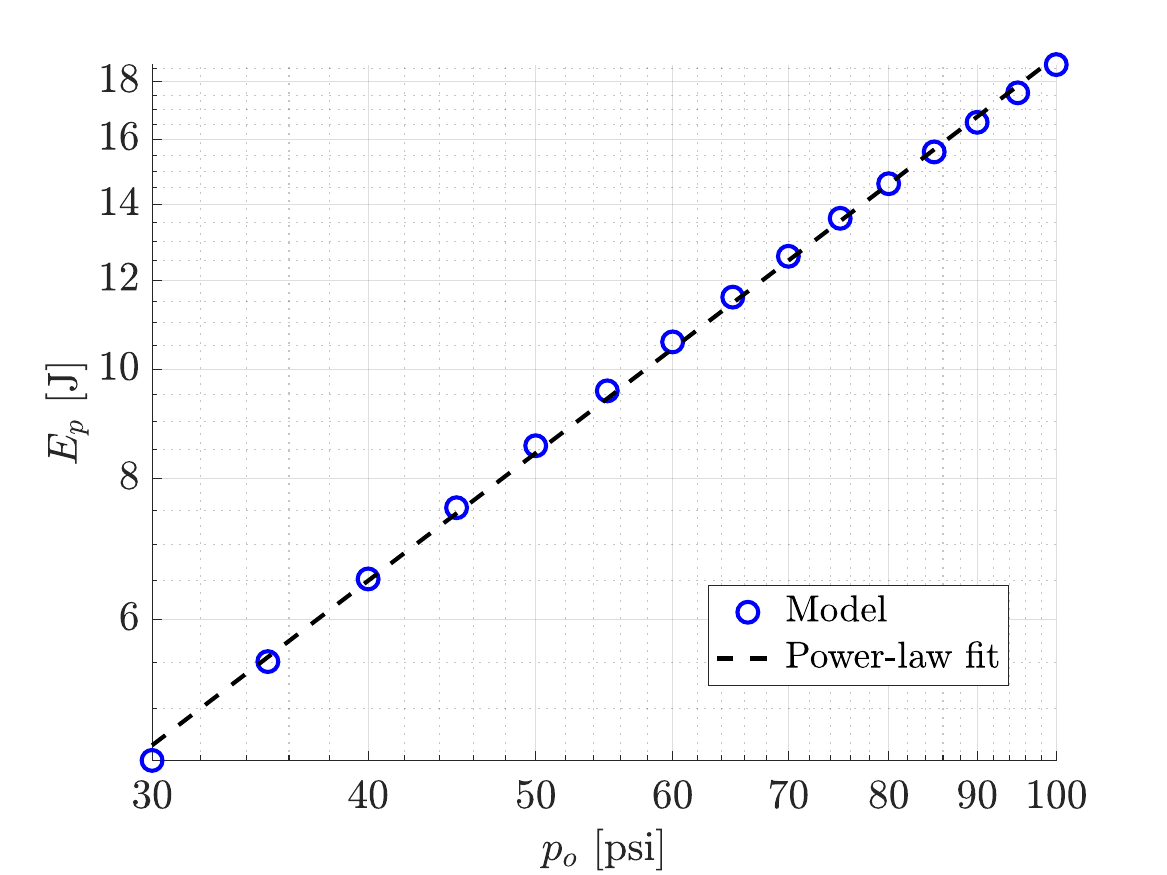}
        \caption{Ideal parameter model hammer strike kinetic energy time vs. flapper-valve opening threshold pressure in log-log scale.}
        \label{fig:ModelMultiple30-100E}
    \end{subfigure}
    \caption{Ideal parameter Model results vs. flapper-valve opening threshold pressure.}
    \label{fig:ModelMultiple30-100}
\end{figure}

A total of 15 simulations were run using the parameter values outlined in Table~\ref{tab:parametersTable}, while varying $p_o$ as the controlled input in increments of 5 psi from 30 to 100 psi. The first representative result, shown in Fig.~\ref{fig:ModelStrikeExamples}, is for $p_o = 100$ psi. Figure~\ref{fig:ModelSingleStrikePandF} shows $p_v$, $p_h$, and $p_e$ over time after the valve opens ($t_0 = 0$ s).
The hammer force,
    \begin{equation} \label{eq:FLCdef}
         F_{LC} = -F_h,
    \end{equation}
is emulated by an upward force on an imaginary load-cell at the base of the drill, $F_{LC}$, plotted in Fig.~\ref{fig:ModelSingleStrikePandF}. Figure~\ref{fig:ModelSingleStrikeEandV} shows the corresponding $\dot{y}$ and $E_p$ values. Here we refer to the hammer kinetic energy as $E_p$ because the model integrates the pressures to obtain accelerations, which is consistent with the notation used throughout the remainder of the text. The key events are labeled as follows:
\begin{enumerate}[label=\Alph*), font=\normalsize]  
    \item \textit{Flapper valve opens:} \label{evn:A_ValveOpens} At this point, $t_0$, $p_v$ begins to decay and $p_h$ begins to increase sharply as the gas in the vent chamber rushes in to fill the hammer chamber. As $p_h$ builds relative to $p_e$, the hammer begins to accelerate.
    Because $p_o \gg p_h$ at $t_0$, the downstream-to-upstream pressure ratio is initially below the threshold in Eq.~\ref{eq:chokedcondition}; the flow is therefore choked and the volumetric flow rate is approximately constant.
    
    \item \textit{Hammer chamber reaches peak pressure:} \label{evn:B_Peakph} At this point, $t_{\mathrm{pk}h}$, the hammer chamber has been largely filled and $p_h$ is near its peak value. This point is at or near the peak hammer acceleration, where the difference between $p_h$ and $p_e$ is greatest. The flow from the vent chamber into the hammer chamber is no longer choked and is significantly smaller than at $t_0$, and both pressures now begin to decay together while the accelerating hammer increases the hammer-chamber volume, $\dot{V}_h > 0$.
    During this phase, the difference between $p_v$ and $p_h$ is set by how much mass must flow from the vent chamber to the hammer chamber to keep up with $\dot{V}_h$.
    This difference is determined by the flapper-valve properties $C_{v1}$ and $A_{v1}$.
    
    \item \textit{Hammer strikes drill-bit:} \label{evn:C_HammerStrikes} Once the hammer strikes the drill-bit at $t_{h}$, $F_{LC}$ spikes as momentum is transferred from the hammer to the ``wall,'' and $\dot{V}_h \rightarrow 0$, so that $p_h \rightarrow p_v$ as mass in the vent chamber continues to flow into the hammer chamber. Once $p_h$ and $p_v$ have largely equalized, they decay together as a slow $\dot{m}_{leak}$ feeds the exhaust chamber from both the hammer and vent chambers. Although small, $p_e$ also reflects the effect of $\dot{V}_h \rightarrow 0$, as it quickly approaches $p_{atm}$ after the strike. Under normal operation, the distinct signature in the $p_h$ time series indicating that the hammer has struck the drill-bit is an abrupt change in slope, i.e. a peak in $\ddot{p}_h$, especially after the peak-pressure event \ref{evn:B_Peakph}. This can be seen in Fig.~\ref{fig:ModelSingleStrikeEandV} as the sharp deceleration spike in the hammer together with the drop in $\dot{y}$. As mentioned previously, the hammer dynamics after $t_h$ are determined by the wall parameters $k_w$ and $c_w$, which were chosen to dissipate the hammer energy effectively and provide a stable numerical solution.
    
    \item \textit{Flapper valve closes:} \label{evn:D_ValveCloses} The slow decrease in $p_v$ and $p_h$ through leakage into the exhaust chamber via $\dot{m}_{leak}$ continues until the flapper-valve closes at $t_{close}$ (for Fig.~\ref{fig:ModelStrikeExamples}, at 0.02 s). At that point, $p_v$ and $p_h$ diverge, with $\dot{m}_{in}$ now feeding only $V_v$ and increasing $p_v$, while $p_h$ continues to decay, albeit faster, since it no longer receives mass flow from $\dot{m}_{v1}$. Under normal operation, this distinct event occurs at the minimum of $p_v$, equivalent to when the vent chamber is most depleted of mass.
    The system then recharges $p_v$ until it once again reaches $p_o$, and the cycle repeats starting at \ref{evn:A_ValveOpens}.
\end{enumerate}
These key events are discussed further, in algorithmic form, in Section~\ref{sec:LCExperimentsandModelComp}, where experimental data are processed to extract $t_0$, $t_{\mathrm{pk}h}$, $t_h$, and $t_{close}$. For the ideal simulations in this section, where $t_{close} > t_{h}$, $t_h = \arg\max_t E_p(t)$ and $||E_p||_{\infty} = E(t_h)$.
As shown in Fig.~\ref{fig:ModelMultiple60-80-100}, the model results depict pressure and hammer kinetic-energy profiles for individual strikes at three different initial pressures, $p_o = 60$, 80, and 100 psi. Across this range, the overall shape of the curves remains consistent over time, suggesting no topological changes in the dynamic regime of the fluid-hammer system. As expected, increasing $p_o$ results in a higher peak hammer pressure, $||p_h||_{\infty}$, as shown in Fig.~\ref{fig:ModelMultiple60-80-100P}.
$p_e$ remains substantially lower than $p_h$ and has only a minimal effect on hammer acceleration, which is consistent with the exhaust-hole design objective of relieving pressure without substantially influencing hammer motion.
The hammer-strike time $t_h$ is inversely related to $p_o$, with lower $p_o$ values yielding reduced average hammer-chamber pressures, which translate to lower hammer acceleration and thus longer transit times.
Finally, the difference between $p_v$ and $p_h$ at $t_h$ also decreases as $p_o$ increases, most clearly when comparing the results for $p_o = 100$ and 60 psi in Fig.~\ref{fig:ModelMultiple60-80-100P}.
This trend is expected, because the pressure drop across the valve scales with $\dot{m}_{v1}$, which is proportionally related to $\dot{y}$ and is shown to increase between those cases in Fig.~\ref{fig:ModelMultiple60-80-100E}, while $E_p \sim \dot{y}^2$.
Fig.~\ref{fig:ModelMultiple30-100} shows $t_h$ and $||E_p||_{\infty}$ over multiple values of $p_o$ from 30 to 100 psi and captures these trends more explicitly. Fig.~\ref{fig:ModelMultiple30-100th} shows an inverse power-law scaling,
\begin{equation} \label{eq:thscalpo}
    t_h \sim p_o^{-0.51},
\end{equation}
with the data matching the fit well over the plotted results. Fig.~\ref{fig:ModelMultiple30-100E} also shows an almost linear scaling, but upon closer inspection the exponent is slightly above 1:
\begin{equation} \label{eq:Epscalpo}
    ||E_p||_{\infty} \sim p_o^{1.17}.
\end{equation}
This is seen approximately in the nearly constant incremental steps between adjacent simulations in Fig.~\ref{fig:ModelMultiple60-80-100E}, where $||E_p(80)||_{\infty}-||E_p(60)||_{\infty} \approx ||E_p(100)||_{\infty}-||E_p(80)||_{\infty}$. Eqns.~\ref{eq:thscalpo} and \ref{eq:Epscalpo}, although limited to these results, provide a baseline for the best performance achievable by the current drill hardware.
The instantaneous opening of the flapper-valve removes any delay in $\dot{m}_{v1}$ reaching its maximum, subsequently driving $\dot{p}_h$ to its maximum for a given $p_o$. This in turn maximizes the pressure force on the hammer, and therefore the hammer acceleration per Eq.~\ref{eq:HammerFBDImplemented}. Likewise, with $c_2 = 0$, no energy is lost to friction as the hammer accelerates toward the drill-bit. Under these ideal conditions, $t_h = [8.8, 9.8, 11.3] \times 10^{-3}$ s and $||E_p||_{\infty} = [18.6, 14.6, 10.6]$ J for $p_o = [100, 80, 60]$ psi.


\section{Percussive-Action Testbed} \label{sec:SetupPercussiveTB}
\label{sec:experiments}
\begin{figure}[b!]
    \centering
    \includegraphics[width=1\linewidth]{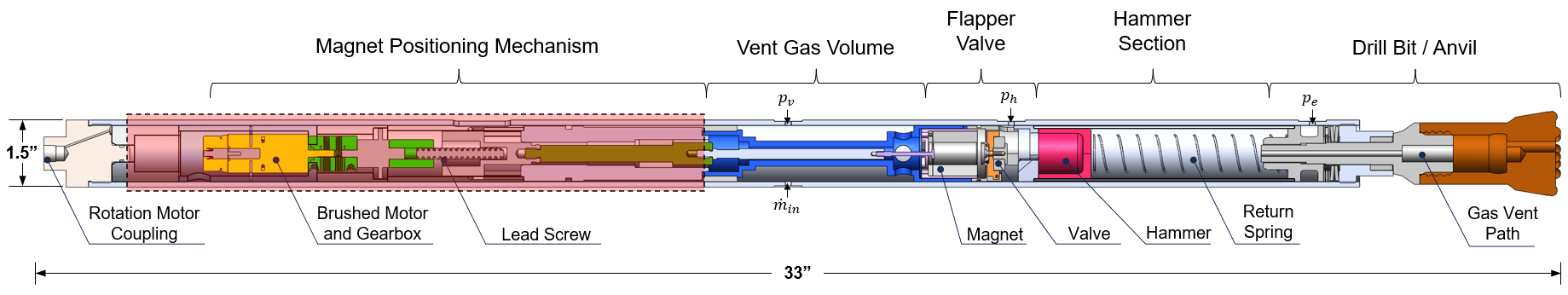}
    \caption{Cross-sectional view of the \textit{Percussive-Action Testbed}, a modified benchtop version of the \textit{WiP} BHA used to test the pneumatic percussion mechanism. The red shaded section highlights modified components.}
    \label{fig:Multi-Strike CAD}
\end{figure}

A set of experiments was conducted to quantify performance and characterize the pneumatic percussion mechanism on two rock simulants. Because these investigations focused on the percussive mechanism rather than full-system functionality, the \textit{WiP} BHA was modified into a benchtop version (see Fig.~\ref{fig:Multi-Strike CAD}) and tested at JPL under laboratory conditions. The outer BHA casing/shell, aside from added sensor and gas-inlet holes, and the forward components, including the bit, hammer, valve, and vent sections, remained unchanged and were reused in this benchtop configuration. In contrast, the rotation mechanism and liquid-CO$_2$ reciprocating section were replaced with a magnet-positioning unit. This unit includes a brushed motor and lead-screw mechanism that enables fine adjustment of the stand-off distance between the magnet and the flapper-valve, thereby allowing control of the target valve-opening pressure $p_o$ during testing. This pressure was a key operating parameter in these investigations because it governed the hammering conditions being evaluated. In addition, internal BHA features not directly required for percussion operation, namely the reciprocating section and integrated rotation mechanism, were removed for simplicity and to increase robustness during repeated percussion testing. During initial testing, however, it was determined that bit indexing was still required to assess percussive performance properly. Therefore, after removal of the integrated rotation mechanism, an external planetary-geared brushed motor was used to rotate the entire BHA in an oscillatory motion of $\pm 170^\circ$. A coupler was fabricated to link this external motor to the end of the BHA where the tether would normally connect.

Another change relative to the original \textit{WiP} drill design was that, in these experiments, the BHA was powered directly by a compressed-air line introduced into the vent chamber rather than by liquid CO$_2$ introduced through the tether attachment. This line provided adjustable pressure settings of up to 100 psi using an upstream regulator. A needle valve located at the air-line inlet controlled the mass flow into the vent chamber, $\dot{m}_{in}$. During testing, the vent chamber was first pressurized to the level set by the regulator, after which the valve magnet was gradually retracted until the valve opened at the desired pressure. The inlet needle valve was then adjusted to set the desired mass flow rate, enabling controlled drill strokes at a prescribed frequency. Thus, the magnet position controlled the starting pressure of each stroke, i.e. the valve-opening pressure, while the needle valve effectively determined the stroke rate, typically set to 0.5--2 Hz during testing. This range kept successive strokes sufficiently separated for data processing while still providing an acceptable rate of penetration. Because this BHA testbed was used primarily to investigate hammer action, it favored operational simplicity and adjustable magnet positioning, whereas future designs could use a fixed stand-off distance. Future builds intended for deeper drilling tests will instead include the substituted elements shown in Fig.~\ref{fig:Drill_Architecture}, namely the rotation mechanism and liquid reciprocating section, will be powered by liquid CO$_2$, and will incorporate an anchoring and anti-rotation device.

\begin{figure}[htb!]
    \centering
    \begin{subfigure}[t]{0.45\textwidth}
        \centering
        \includegraphics[height=3.5in]{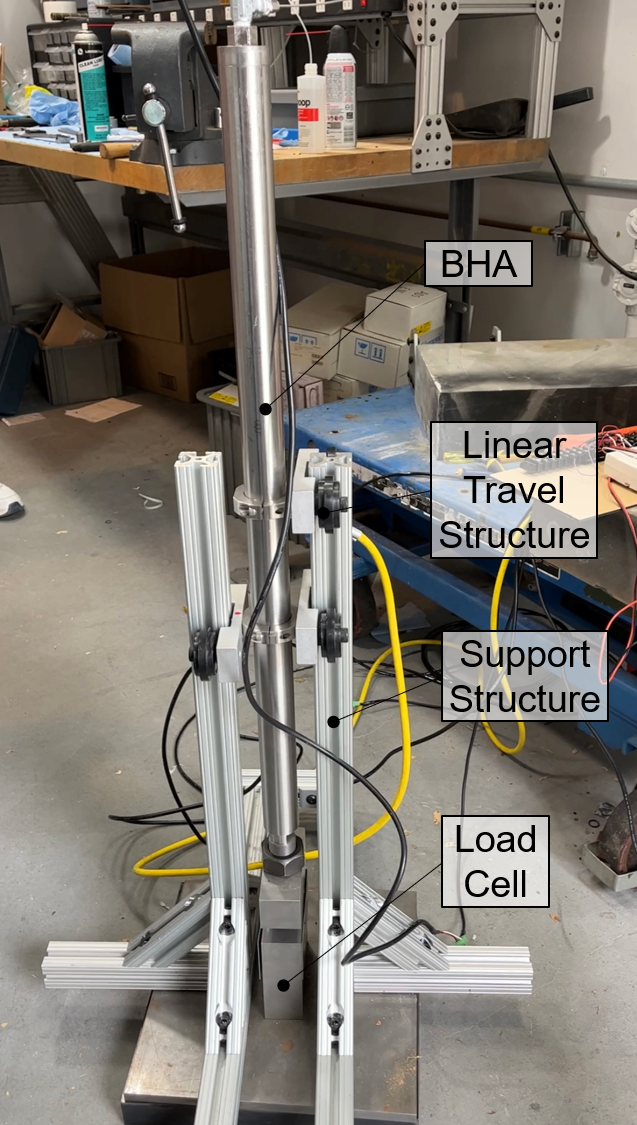}
        \caption{Load-cell characterization}
        \label{fig:setup_LC}
    \end{subfigure}
    \begin{subfigure}[t]{0.45\textwidth}
        \centering
        \includegraphics[height=3.5in]{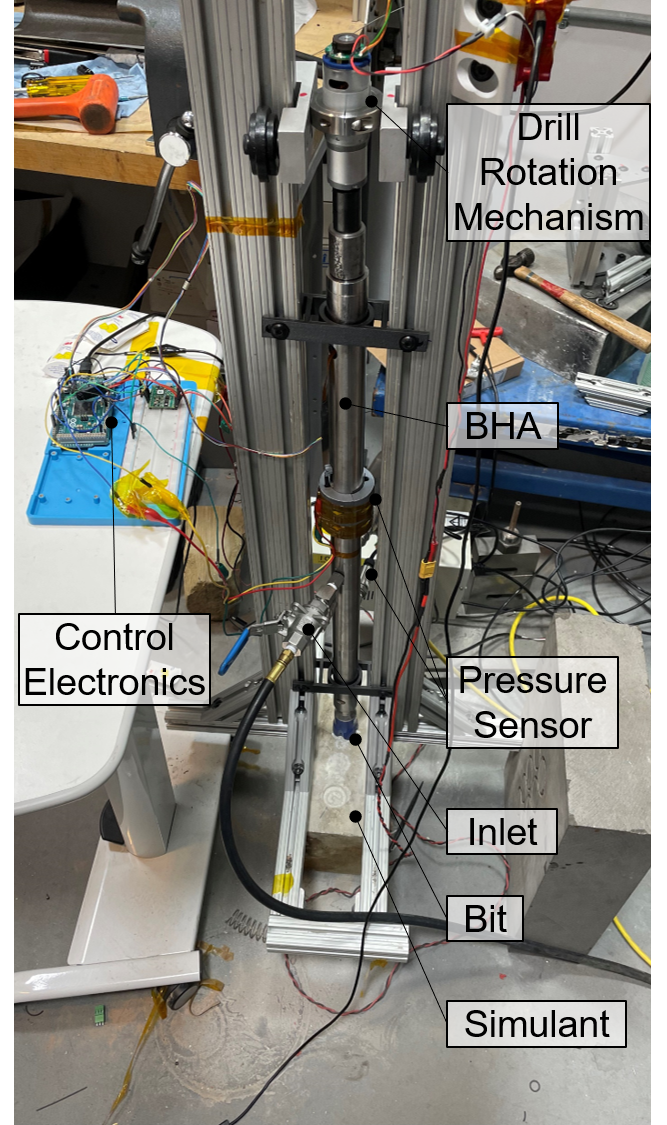}
        \caption{Simulant drill tests}
        \label{fig:setup_rock}
    \end{subfigure}
    \caption{Experimental setup of the borehole assembly (BHA)}
    \label{fig:setup}
\end{figure}

The final benchtop BHA has a total length of 33", a casing diameter of 1.5", and a mass of 3.8 kg. The experimental setup of the testbed with the vertically oriented BHA is shown in Fig.~\ref{fig:setup}. An aluminum frame supports the BHA with loose-fitting brackets that allow both rotation and vertical movement. The drill rotation mechanism engages the frame using guide wheels, allowing it to travel vertically with the drill as penetration progresses while remaining rotationally constrained. For data collection, the BHA was equipped with two pressure sensors to monitor and record the pressures within the vent and hammer chambers, i.e. $p_v$ and $p_h$; see Fig.~\ref{fig:Multi-Strike CAD} for their locations. The exhaust-chamber pressure was not measured during testing because of sensor limitations; however, it was measured during testbed commissioning and found to remain less than 1 psi above atmospheric pressure, as intended by design. The exhaust holes provide sufficient open area that little flow resistance develops within the exhaust chamber during a hammer strike. In subsequent calculations, $p_e = p_{atm}$ unless otherwise specified. All data were recorded from the start to the completion of each test, providing detailed documentation of drill performance under a range of operating conditions.

\begin{figure}[htb!]
    \centering
    \includegraphics[trim=45cm 10cm 5cm 10cm, clip,width=0.5\linewidth]{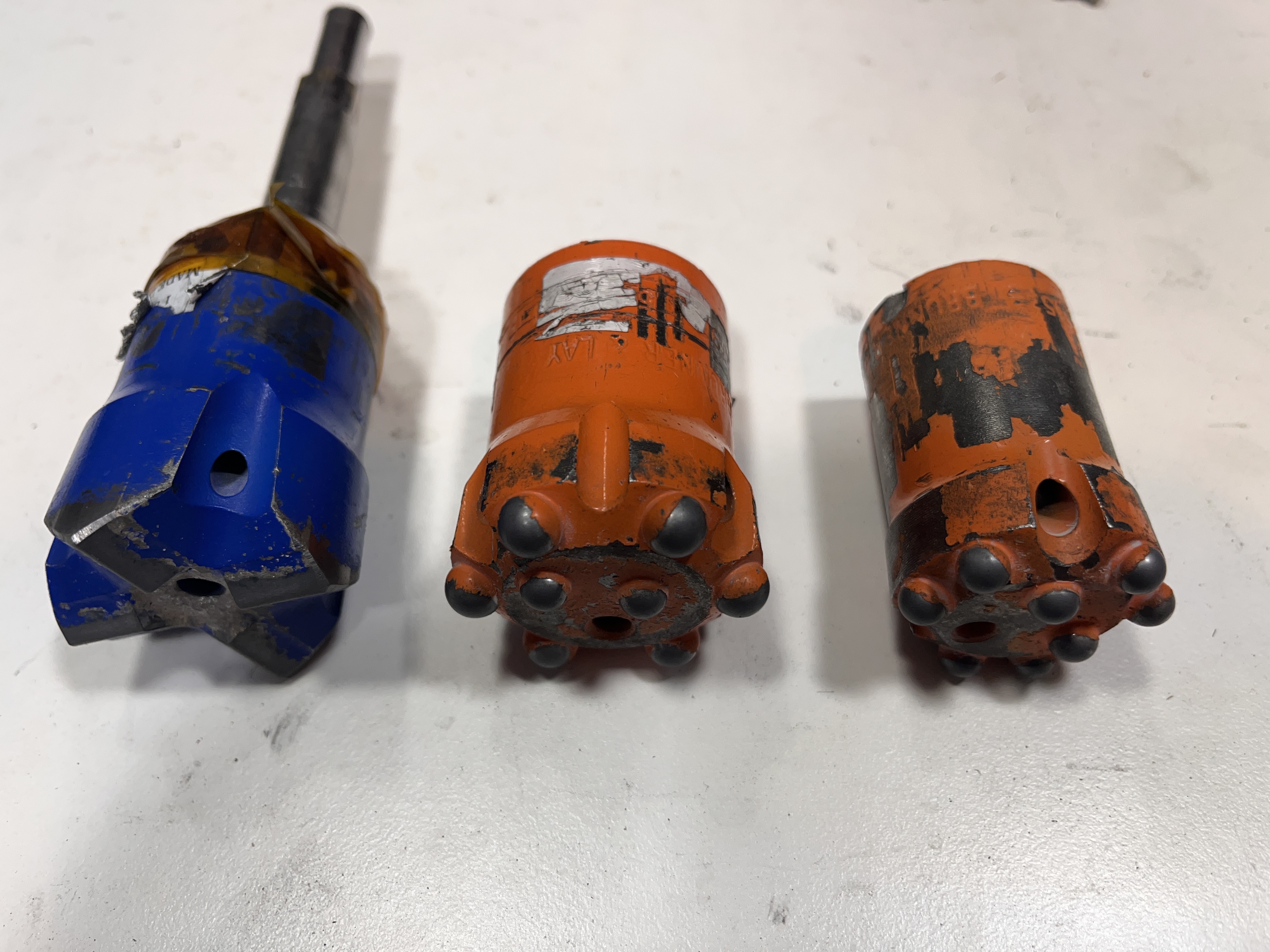}
    \caption{Investigated button-head drill-bits, 1-5/8" diameter (left) and 1-3/8" diameter (right)}
    \label{fig:bits}
\end{figure}

For initial system characterization and model validation, the percussion testbed was operated against a load-cell setup instead of a rock simulant. The bit was removed, and a shaft extension attached to the load-cell interfaced directly with the BHA through the drill-bit stem hole, allowing measurement of the hammer impact force that would otherwise be transferred to the bit (see Fig.~\ref{fig:setup_LC}). These tests, which primarily served model validation, did not require bit indexing. For simulant testing, two button-head drill-bits with outer diameters (ODs) of 1-3/8" and 1-5/8" were investigated (see Fig.~\ref{fig:bits}) on two simulant types, sandstone and Saddleback basalt. The drill-bits were obtained from Crowder Supply, with model numbers and descriptions listed in Tab.~\ref{tab:drillbits}. These commercially available modular bits interface directly with the drill-bit stem and provide a gas-vent path for built-up pressure inside the hammer section to clear pulverized cuttings. In addition to these exhaust ports, the testbed included extra exhaust ports at the base of the hammer section to preserve the low-flow-resistance condition that enabled $p_e = p_{atm}$. Each bit was tested in a new hole and placed on a flat surface of the simulant. The frame helps secure the simulant in place and prevents it from shifting during drilling impacts. Note that the smaller drill-bit is slightly smaller than the 1.5" drill-casing diameter, limiting the maximum hole depth to approximately the drill-bit length.
Three values of $p_o$ were explored during system characterization, while drill-performance evaluations on simulants were conducted at a fixed $p_o = 100$ psi to emulate the vent-chamber pressures expected for a likely Mars-like drill system. In all tests, the BHA was placed on either the load-cell or the simulant under its own weight, corresponding to a weight-on-bit (WOB) of 37 N. Table~\ref{tab:experiment_log} summarizes the experiments conducted, listing the test parameters and the total number of analyzed strikes, $N_{total}$.

\begin{table}[ht]
\centering
\resizebox{\textwidth}{!}{%
\begin{tabular}{c c c c l}
\hline
\textbf{Model \#} & \textbf{Bit Type} & \textbf{OD (in)} & \textbf{Measured OD (in)} & \textbf{Description} \\ \hline
BBH35 & Button Bit & 1-3/8 & 1.41 & Button-type tungsten carbide inserts. \\
BBH41 & Button Bit & 1-5/8 & 1.68 & Button-type tungsten carbide inserts.\\ \hline
\end{tabular}%
}
\caption{List of drill-bits used in rock simulant tests.}
\label{tab:drillbits}
\end{table}

Control of the testbed and data acquisition were managed through a custom LabView Virtual Instrument (VI), which allowed adjustment of the rotation rate and magnet position and recorded sensor data, i.e. pressure and load-cell measurements, at sampling frequencies up to 50 kHz. The shorter characterization tests using the load-cell setup were conducted at 50 kHz, whereas the much longer rock tests ($\approx$ 1 to 2 hours) were recorded at a lower frequency of 1 kHz to reduce file size and enable practical data processing.

\begin{table}[ht]
\centering
\resizebox{\textwidth}{!}{%
\begin{tabular}{l l c l l c c}
    \textbf{ID} & \textbf{Test Type} & \textbf{Target $p_o$ [psi]} & \textbf{Simulant} & \textbf{Drill Bit Type} & \textbf{Sampling Freq. [kHz]} & \textbf{$N_{total}$ \#} \\
    \hline
    1a  & System Char. & 100 & load-cell & N/A & 50 & 70 \\
    1b  & System Char. & 80 & load-cell & N/A & 50 & 103 \\
    1c  & System Char. & 60 & load-cell & N/A & 50 & 113 \\
    \hline
    2a  & Rock Test & 100 & Sandstone & Button-Head 1-3/8" Dia & 1 & 1926 \\
    2b  & Rock Test & 100 & Sandstone & Button-Head 1-5/8" Dia & 1 & 711 \\
    3a  & Rock Test & 100 & Saddleback Basalt & Button-Head 1-3/8" Dia & 1 & 2879 \\
    3b  & Rock Test & 100 & Saddleback Basalt & Button-Head 1-5/8" Dia & 1 & 1369 \\ \hline
\end{tabular}%
}
\caption{Test Log.}
\label{tab:experiment_log}
\end{table}

\section{Load-Cell Experimental Results and Model Comparison} \label{sec:LCExperimentsandModelComp} 
The first set of experiments aims to provide additional load-cell response data, $F_{LC}$, to complement the available pressure signals $p_v$ and $p_h$ in long-duration rock tests. In this case, $F_{LC}$ is no longer fictitious, as in the ideal-model discussion, and corresponds to the same definition as Eq.~\ref{eq:FLCdef}. This data is used to validate the relationship between the pressure signals, strike events, and the energy delivered during each strike. By incorporating load-cell data, we can compare the measured hammer acceleration and the integrated velocity and position over time with those calculated from $p_h$. Additionally, the load-cell provides a direct measure of hammer strike magnitude and timing, which can be cross-referenced with events inferred from the pressure time series.

Since the load-cell is not present during the long-duration rock tests, the strike-detection algorithms are first validated here to ensure that energy results based on $p_v$ and $p_h$ can be trusted in the subsequent tests. Table~\ref{tab:experiment_log} shows the two test types; here we focus on test IDs 1a--c, which vary the target flapper-valve opening pressure $p_o$ as the primary sweep parameter. These tests were conducted at a 50 kHz sampling rate to capture the impulse-force response on the load-cell.
In this section, we first discuss the algorithms developed for strike-data processing and then evaluate their performance. Table~\ref{tab:exprvariables_description} provides a list of calculated output variables, their units, and descriptions that are referenced throughout the experimental sections. The final objective is to compare these experimental results to the model predictions and demonstrate that the model developed in Sec.~\ref{sec:model} captures key trends in the system's physical behavior.  

\begin{table}[ht]
\centering
\begin{tabular}{l c p{10cm}}
\hline
\textbf{Symbol} & \textbf{Units} & \textbf{Description} \\ \hline
$p_{v}\left( t_0 \right)$ & psi & Initial Vent Pressure for all strikes\\
$E_{LC}$ & J    & Hammer kinetic energy from $F_{LC}$ data \\
$E_{p}$ & J     & Hammer kinetic energy from $p_h$ data \\
$\left.E_p \right|_{\Vert L_{int} \Vert_2}$ & J & Hammer kinetic energy at the end of the stroke for all strikes, where abnormal strikes are integrated to $\Vert L_{int} \Vert_2$ \\
$\left.E_p \right|_{L_{max}}$ & J & Hammer kinetic energy at the end of the stroke for all strikes, where abnormal strikes are integrated to $L_{max}$ \\
$t_{hp}$ & s & Hammer strike time from $p_h$ data for normal strikes \\
$t_{hLC}$ & s & Hammer strike time from $F_{LC}$ data for normal strikes\\
$t_{close}$ & s & Flapper valve close time for normal strikes \\
$L_{int}$ & m & Maximum hammer displacement for normal strikes\\
$\sum_{\text{norm}}  E_{p}$ & J & Total energy in experiment from normal strikes \\
$\sum_{\text{all}} \left.  E_{p} \right|_{\Vert L_{int} \Vert_2}$ & J & Total energy in experiment including abnormal strikes integrated with $\Vert L_{int} \Vert_2$ \\
$\sum_{\text{all}} \left.  E_{p} \right|_{L_{max}}$ & J & Total energy in experiment including abnormal strikes integrated with $L_{max}$ \\
$\left. \bar{E}_p \right|_{\Vert L_{int} \Vert_2}$ & MJ/m$^3$ & Mechanical specific energy for all strikes, with abnormal integrated as $\Vert L_{int} \Vert_2$ \\
$\left. \bar{E}_p \right|_{L_{max}}$ & MJ/m$^3$ & Mechanical specific energy for all strikes, with abnormal integrated as $L_{max}$ \\
$H_d$                               & m &                 Total drilled depth into simulant rock \\
$N_{total}$ & \# & Total number of strikes analyzed (sum of normal and abnormal strikes) \\
$N_{\text{norm}}$ & \# & Total number of normal strikes \\
$N_{\text{abn}}$ & \# & Total number of abnormal strikes \\ 
$N_{\text{missed}}$ & \# & Total number of strikes missed by algorithm \\ \hline
\end{tabular}
\caption{List of possible output values for different experiments in results tables Tab.~\ref{tab:experimental_results_mean}, Tab.~\ref{tab:experimental_results_min_max}, Tab.~\ref{tab:rockExperiment_results_mean}, and Tab.~\ref{tab:rockExperiment_results_variance}.}
\label{tab:exprvariables_description}
\end{table}

\subsection{Strike Analysis and Data Processing}
\begin{figure}[H]
    \centering
    \begin{subfigure}[t]{0.58\textwidth}
        \centering 
        \raisebox{0\height}{    
        \includegraphics[trim=2.25cm 0cm 2cm .5cm, clip, width=1\linewidth]{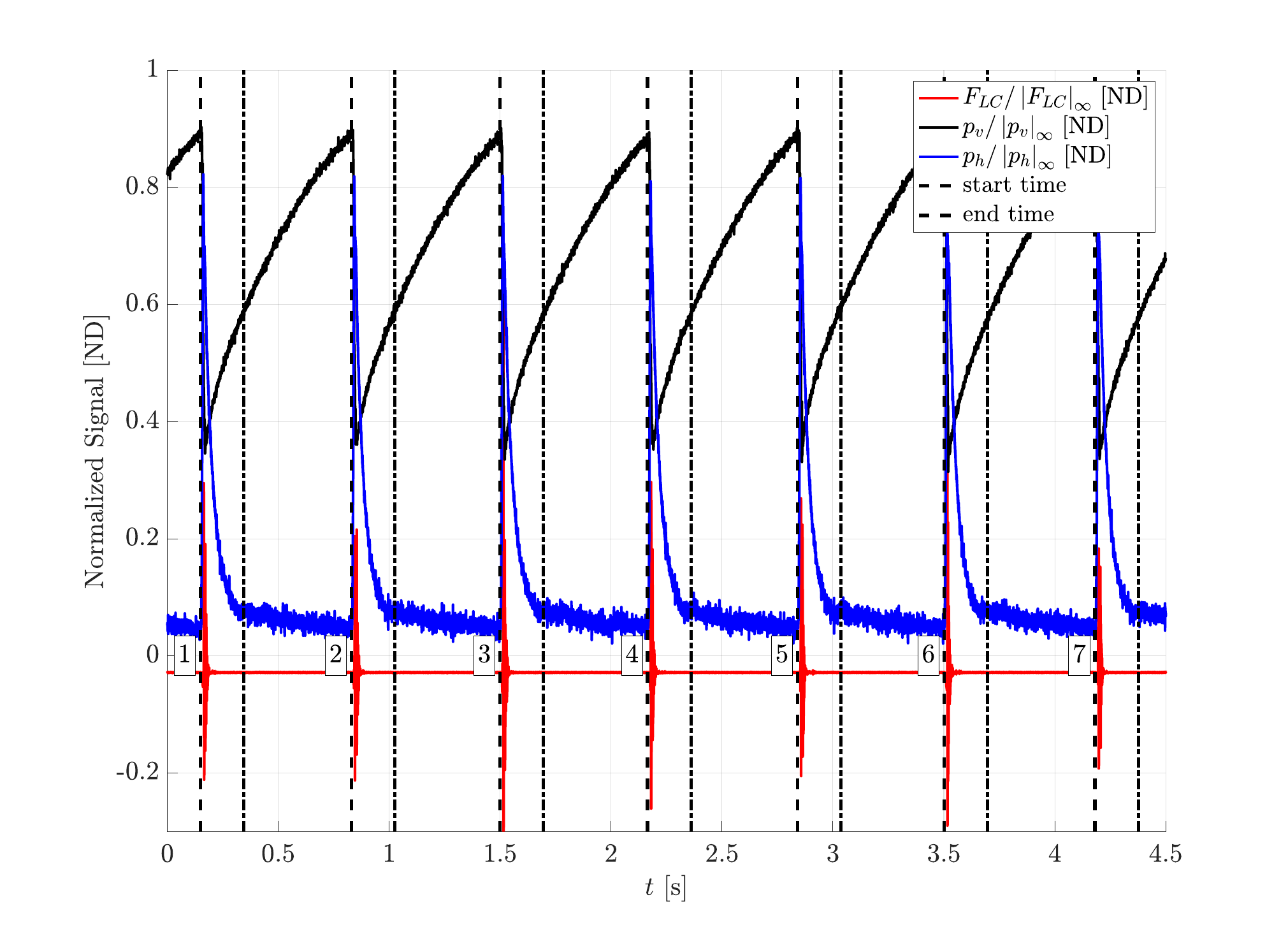}
        }
        \caption{Normal operation example of multiple strikes selected with start time, end time, strike count, and normalized measured signals.}
        \label{fig:multipleStrikeExample}
    \end{subfigure}
    \begin{subfigure}[t]{0.37\textwidth}
        \centering
        \includegraphics[width=1\linewidth]{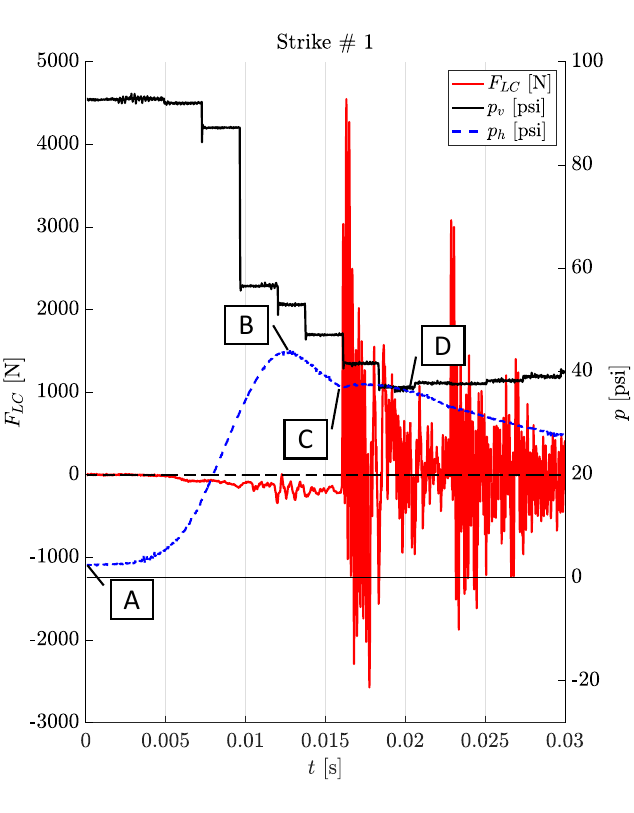}
        \caption{Normal operation example of single strike selected with dimensional measurements plotted, and key events labeled.}
        \label{fig:singleStrikeExample}
    \end{subfigure}
    \caption{Example of multiple and single strikes selected by algorithm \ref{alg:HammerStrikeDetection}.}
    \label{fig:StrikeExamples}
\end{figure}

The first challenge in data processing is to identify individual strikes from the pressure signals, specifically by finding $t_0$ in event \ref{evn:A_ValveOpens}. With the load-cell data available, we can develop an algorithm for extracting individual strikes using only $p_v$ and $p_h$, and then compare those results with the number of strikes measured by $F_{LC}$. The algorithm, shown in Appendix~\ref{sec:appendix1} as Algorithm~\ref{alg:HammerStrikeDetection}, compares the timing of the peak first derivative of $p_v$ with the subsequent peak value of $p_h$. If those events coincide within the tuned threshold parameters, then a strike is recorded. From Algorithm~\ref{alg:HammerStrikeDetection}, we let
\begin{equation}
    t_{se} = t_0+\Delta t_{\text{L}},
\end{equation}
be the end time of the strike event.  
The valve-opening time is calculated from the second derivative of $p_v$ as the datum and then shifted to capture the leading edge of the signal. The algorithm successfully captures all strikes logged by the load-cell that exceed a major peak of 500 N. To find $t_{\mathrm{pk}h}$ for event \ref{evn:B_Peakph}, when the hammer pressure reaches its maximum value, we find
\begin{equation} \label{eq:t_phfind}
    t_{\mathrm{pk}h} = \argmax_{t_0 < t < t_{se}}{p_h}.
\end{equation}
Next, to find the hammer strike time $t_h$ for event \ref{evn:C_HammerStrikes} under the normal operating regime, we identify the first dominant peak in $\ddot{p}_h$ after $t_{\mathrm{pk}h}$, such that
\begin{equation} \label{eq:t_hfind}
    t_{h} = \argmax_{t_{\mathrm{pk}h} < t < t_{se}}{\ddot{p}_h}.
\end{equation}
This is implemented using a moving-mean low-pass filter, applied both forward and backward to account for any phase lag, with a moving window of 0.0016 s. 
Lastly, we find the flapper-valve closing time $t_{close}$ corresponding to event \ref{evn:D_ValveCloses}. As described in Sec.~\ref{sec:subParameterListModelRegOp}, this is simply the time corresponding to the minimum $p_v$, where
\begin{equation} \label{eq:t_closefind}
    t_{close} = \argmin_{t_0 < t < t_{se}}{p_v}. 
\end{equation}
Figure~\ref{fig:multipleStrikeExample} shows the first seven strikes of test 1a with normalized measurements of $F_{LC}$, $p_v$, and $p_h$, and Fig.~\ref{fig:singleStrikeExample} zooms into the first strike with dimensional measurements for all recorded signals. Key events within a single strike are labeled and correspond to those in Fig.~\ref{fig:ModelStrikeExamples}.  
Although the sampling rate of the data acquisition card (DAQ) was set to 50 kHz, the pressure gauge measuring $p_v$ had a bandwidth of only 1 kHz. The stair-step response seen in Fig.~\ref{fig:singleStrikeExample} is therefore due to measurement oversampling.  
Different gauges were used for $p_v$ and $p_h$ primarily because the former had a digital readout that allowed the operator to confirm when the vent chamber was charged for safety purposes. Measurements of $F_{LC}$ and $p_h$ were able to maintain the full DAQ sampling rate. 

\subsubsection{Hammer Strike Time and Kinetic Energy Validation}
\begin{figure}[htb!]
    \centering
    \begin{subfigure}[t]{0.45\textwidth}
        \includegraphics[width=\textwidth,trim={0cm 0cm 1cm 0cm},clip]{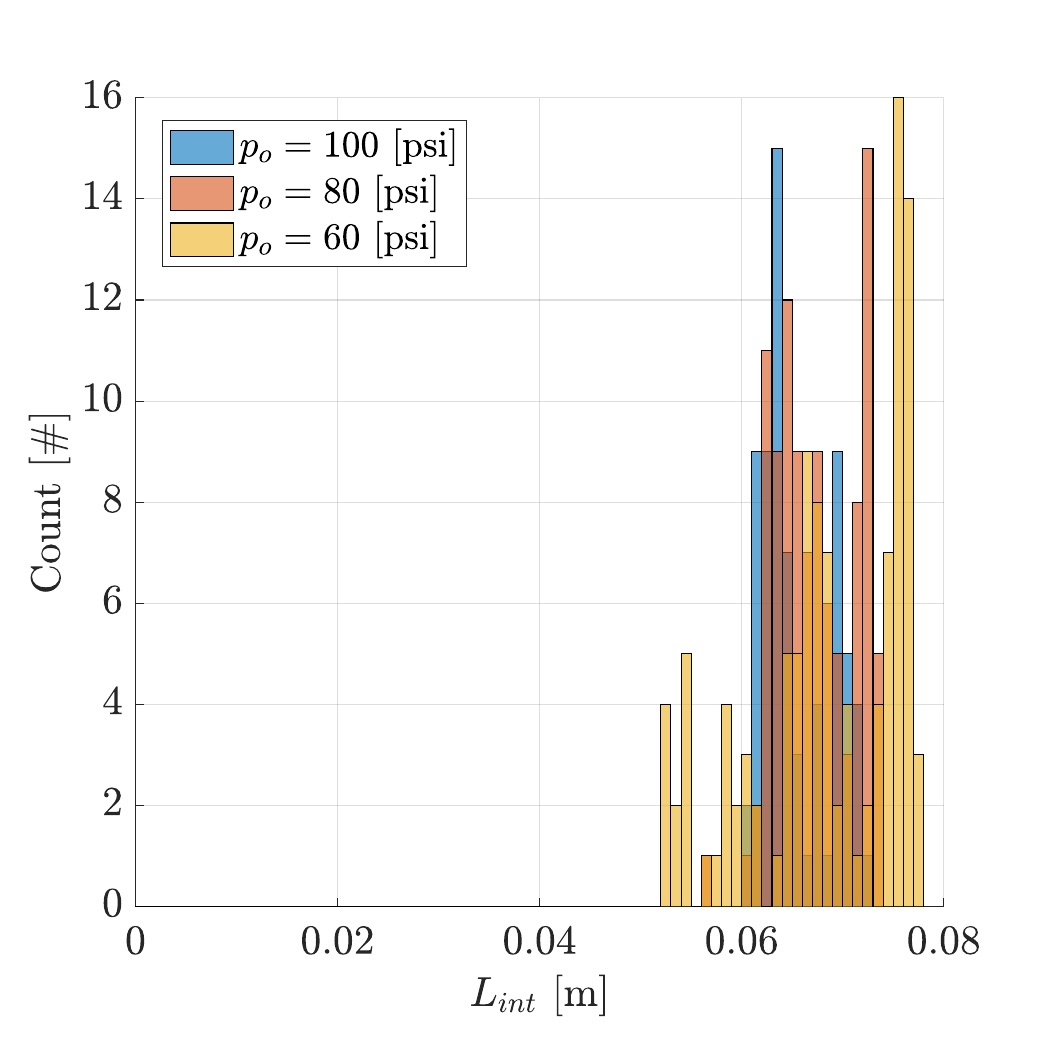}
        \caption[short]{$L_{int}$ distribution for $p_o$ of 100, 80, and 60 psi.}
        \label{fig:histLint_po100-80-60}
    \end{subfigure}
    \hfill
    \begin{subfigure}[t]{0.45\textwidth}
        \centering
        \includegraphics[width=\textwidth,trim={0cm 0cm 1cm 0cm},clip]{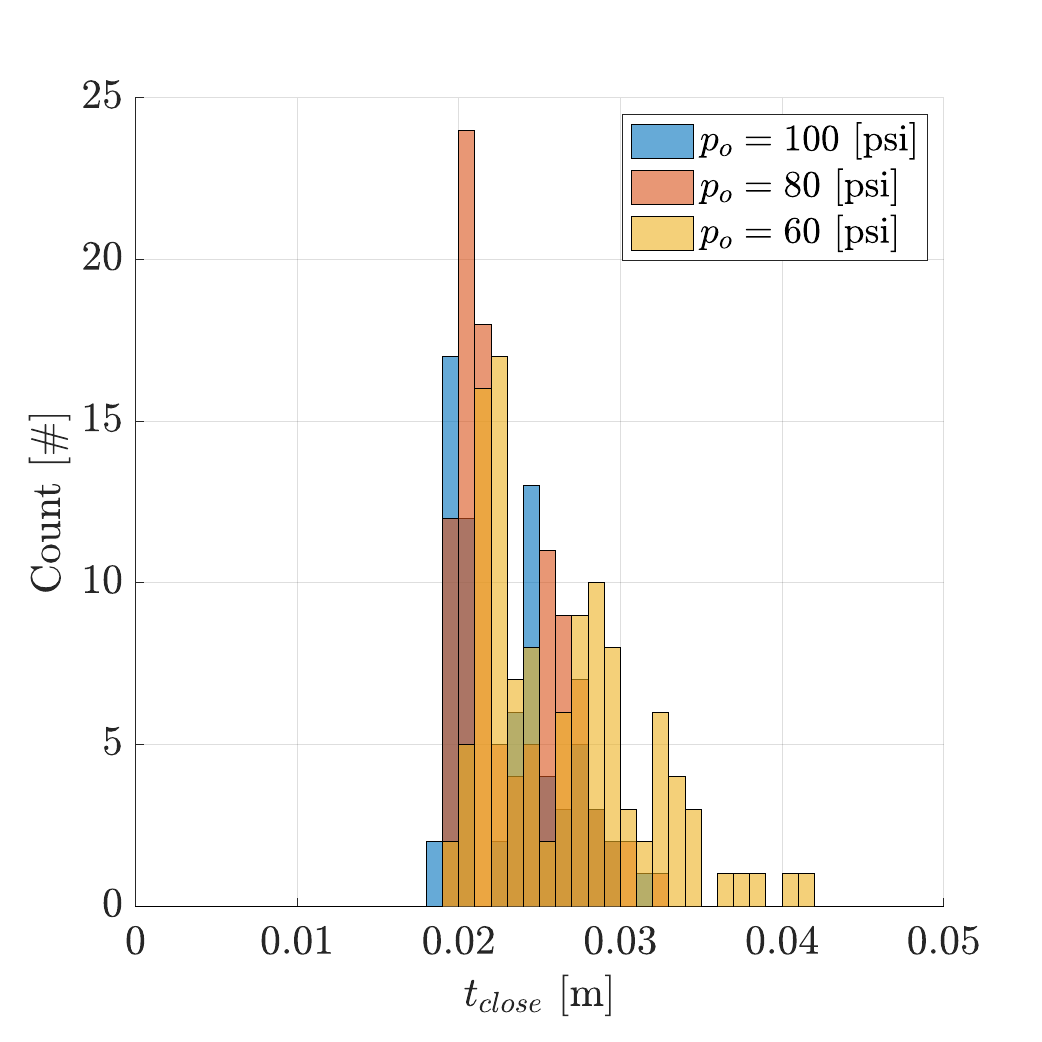}
        \caption[short]{$t_{close}$ distribution for $p_o$ of 100, 80, and 60 psi.}
        \label{fig:histtclose_po100-80-60}
    \end{subfigure}
    \hfill
    \begin{subfigure}[t]{0.45\textwidth}
        \centering
        \includegraphics[width=\textwidth,trim={0cm 0cm 1cm 0cm},clip]{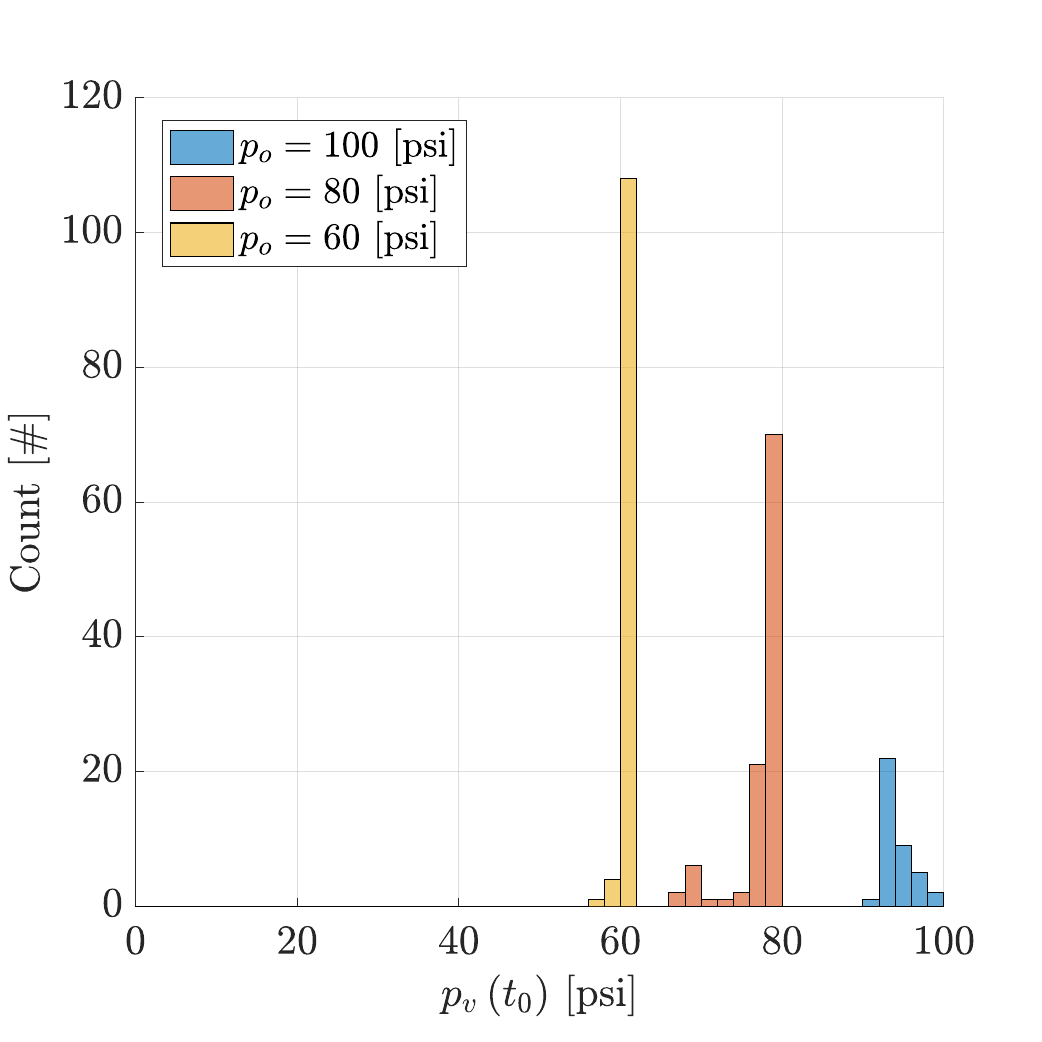}
        \caption[short]{$p_{v}(t_0)$ distribution for $p_o$ of 100, 80, and 60 psi.}
        \label{fig:histPvmax_po100-80-60}
    \end{subfigure}
    \hfill
     \begin{subfigure}[t]{0.45\textwidth}
        \centering
        \includegraphics[width=\textwidth,trim={0cm 0cm 1cm 0cm},clip]{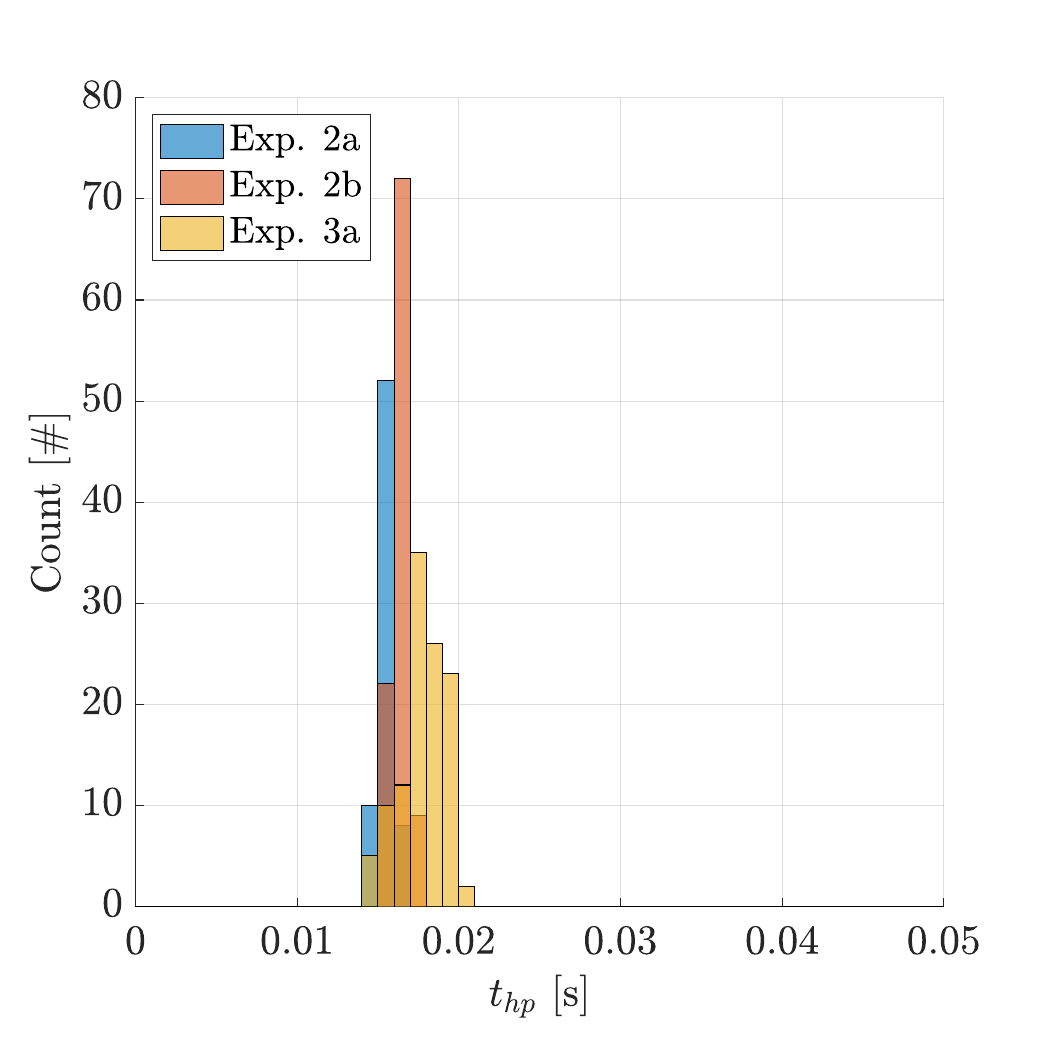}
        \caption[short]{$t_{hp}$ distribution for $p_o$ of 100, 80, and 60 psi.}
        \label{fig:histthp_po100-80-60}
    \end{subfigure}
   
    \caption{Relevant parameter distributions for comparison of the 100, 80, and 60 psi results from experiments 1a, 1b, and 1c in Table~\ref{tab:experiment_log}.}
    \label{fig:hist_po100-80-60}
\end{figure}

Since we do not directly measure the impact energy of each strike in our experiments, we rely on measuring other parameters that allow us to infer it. This map is defined in Eq.~\ref{eq:HammerKineticEnergy}, where $\dot{y}$ is obtained by solving the second-order differential equation in Eq.~\ref{eq:HammerFBDImplemented}.
Instead of solving for $\dot{p}_h$ and $\dot{p}_e$, defined in Sec.~\ref{sec:SubFlowThermoDynamics}, we augment Algorithm~\ref{alg:PressureHammerEOM} to incorporate the experimental $p_h$ and solve Eq.~\ref{eq:HammerFBDImplemented} directly for $\dot{y}$ and $y$. 
If we assumed that the hammer was initially flush with the flapper-valve mount at the start of each strike, i.e. $y_0 = 0$, we could solve for $y$ over time until it reached $L_{max}$, and the time $t_h$ would be when $y = L_{max}$. However, as discussed later in this section and shown in Fig.~\ref{fig:histLint_po100-80-60}, we cannot always assume that $y_0 = 0$ at the start of every strike; instead, we must treat it as an unknown parameter in the analysis.
The alternative method is to solve Eq.~\ref{eq:HammerFBDImplemented} for $y$ from an initial time $t_0$ to a predefined value $t_h$:
\begin{equation} \label{eq:ydotLC}
    \dot{y} = \int_0^{t_{h}} \left(\frac{F_h}{M_h} \right) \text{d}t,
\end{equation}
where $t_h$ must be known a priori. By integrating Eq.~\ref{eq:ydotLC} once more, we can obtain $y$ and determine the total displacement at the end of the strike:
\begin{equation} \label{eq:Lintdef}
    L_{int} = y(t_h).
\end{equation}
For consistency throughout, all $L_{int}$ shown are calculated using $t_h = t_{hp}$. To maintain consistency with the model and respect the physical constraints of the chamber and hammer geometries, the condition 
\begin{equation} \label{eq:LintLmaxcondition}
    L_{int} \leq L_{max}
\end{equation}
must be satisfied.
In experimental campaigns 1a--1c, we have two approaches for obtaining the right-hand side of Eq.~\ref{eq:ydotLC}. The first approach is based on load-cell measurements of acceleration, using Eq.~\ref{eq:FLCdef}, where $F_h$ is measured as the offset force on the load-cell during hammer acceleration. Knowing the mass of the hammer allows us to define the integrand in Eq.~\ref{eq:ydotLC} fully.
The second approach involves substituting the measured values of $p_h$, $p_e = p_{atm}$, and other system parameters into Eq.~\ref{eq:HammerFBDImplemented}, and solving the second-order ODE. While the second method is the only feasible option for all experimental campaigns, in this section, we validate it using the results from the first method, specifically by tuning $c_2$.
We first describe how to determine and validate $t_h$ from $F_{LC}$ and $p_h$ independently, and subsequently calculate and validate the impact energy from both datasets, ensuring consistency between them. Variables calculated from load-cell data carry the subscript $LC$, while those derived from $p_h$ carry the subscript $p$.

\begin{figure}[H]
    \centering
    \begin{subfigure}[t]{0.45\textwidth}
        \centering 
        \includegraphics[trim=0cm 0cm 0cm 0cm, clip, width=1\linewidth]{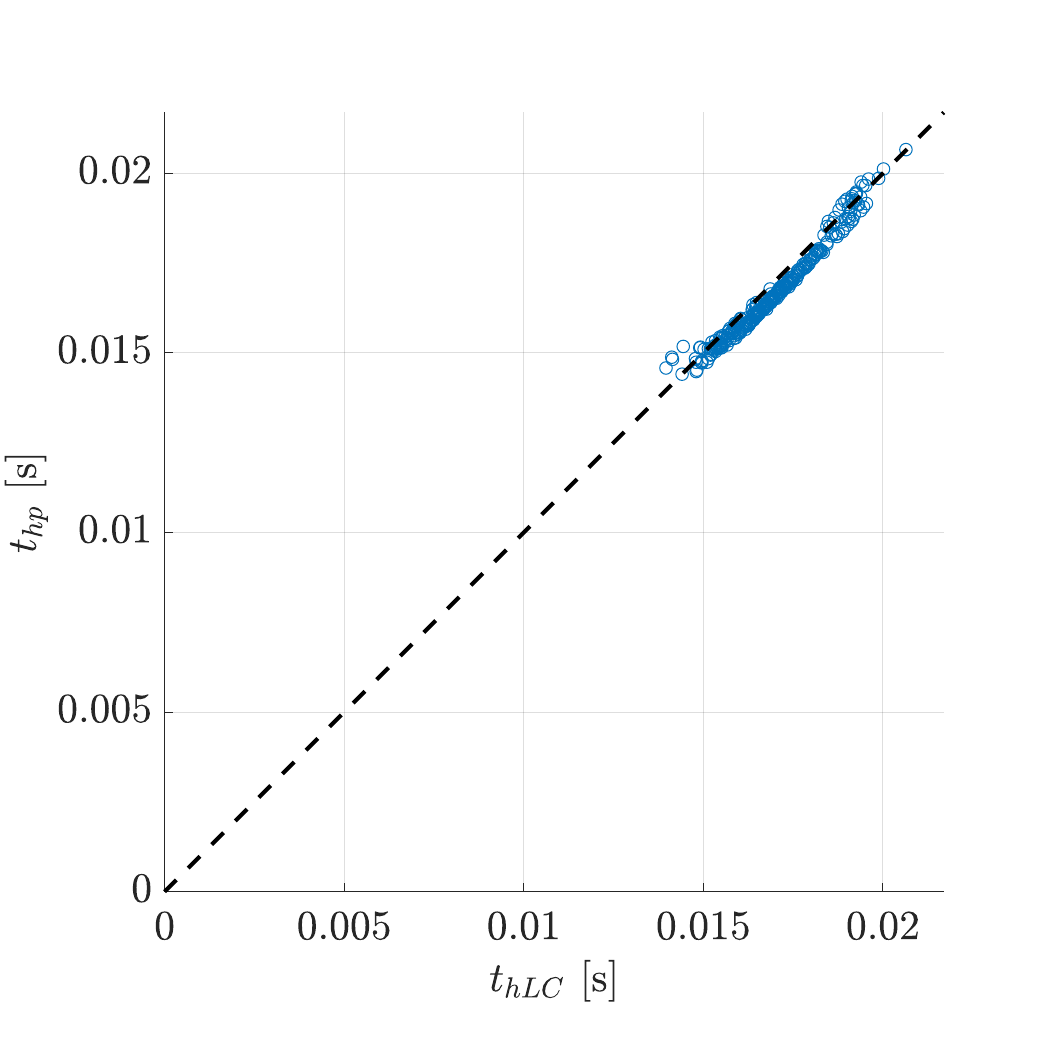}
        \caption{Comparison plot between $t_{hLC}$ and $t_{hp}$ .}
        \label{fig:CompthLCvsthP}
    \end{subfigure}
    \hfill
    \begin{subfigure}[t]{0.46\textwidth}
        \centering
        \includegraphics[trim=0cm 0cm 0cm 0cm, clip, width=1\linewidth]{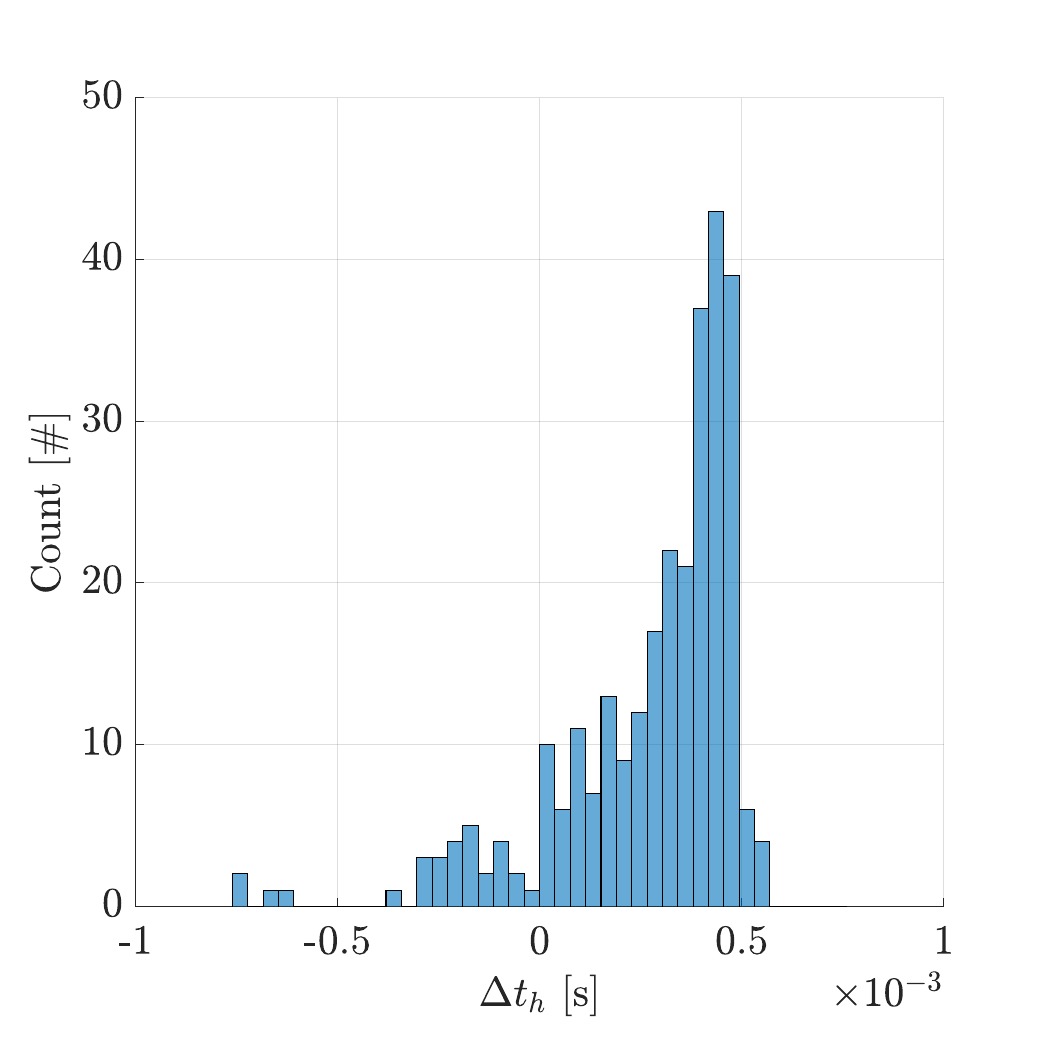}
        \caption{Histogram of error $\Delta t_{h} = t_{hLC} - t_{hp}$ .}
        \label{fig:histErrLCvsthP}
    \end{subfigure}
\caption{Aggregated data from experiments 1a, 1b, and 1c in Table~\ref{tab:experiment_log} and comparison of $t_h$ between the pressure- and load-cell-based methods for finding event \ref{evn:C_HammerStrikes}.}
    \label{fig:CompHistth}
\end{figure}

To determine the true hammer strike time $t_h$, we employ and compare two methods based on either $p_h$ or $F_{LC}$. 
As discussed for event \ref{evn:C_HammerStrikes}, Eq.~\ref{eq:t_hfind} is first used to find $t_h$ from $p_h$, and that result is denoted $t_{hp}$. 
The second method uses $F_{LC}$ and identifies the leading edge of the signal as it begins to rise immediately after the hammer strikes the load-cell. This algorithm is shown in Appendix~\ref{sec:appendix1} as Algorithm~\ref{alg:hammerStrikeTimeLC}, and its output is denoted $t_{hLC}$.
Since the load-cell provides a direct measurement of hammer contact, we treat it as the reference and verify that the pressure-based method matches it. Figure~\ref{fig:CompHistth} first shows this performance through the comparison plot in Fig.~\ref{fig:CompthLCvsthP}, where an exact match between the two methods lies on the unity-slope line.  
The data cluster tightly around that line, with $t_{hp}$ slightly underpredicting $t_{hLC}$. This can be seen in the error histogram in Fig.~\ref{fig:histErrLCvsthP}, where the mean is slightly above 0, but the distribution has a longer tail into the negative region than into the positive region. These results indicate that determining hammer strike time from $p_h$ using the implemented algorithm is valid, with a maximum deviation between the two measurements of $\left| \Delta t_h \right| < 8\times10^{-4}$ s and a mean deviation of $2.7\times 10^{-4}$ s. A summary of the results plotted in Fig.~\ref{fig:CompHistth} is given in Table~\ref{tab:ErrorStatistics}.   
\begin{table}[H]
\centering
\begin{tabular}{lcccc}
\hline
\textbf{Error Statistics} & \textbf{$\Delta t_h$ [s]} & \textbf{$\Delta \dot{y}$ [m/s]} & \textbf{$\Delta y$ [m]} & \textbf{$\Delta E$ [J]} \\ \hline
mean                 & 0.00027           & 0.043                     & -0.0026            & 0.198             \\ 
std                  & 0.00023           & 1.063                     & 0.0058             & 1.518             \\ 
min                  & -0.00076          & -2.233                    & -0.0158            & -3.053            \\ 
max                  & 0.00054           & 2.524                     & 0.0122             & 3.940             \\ \hline
\end{tabular}
\caption{Error statistics for 286 strikes combined in experiments 1a, 1b, and 1c in Table~\ref{tab:experiment_log}. The $\Delta(\cdot)$ values denote differences between load-cell and pressure values (subscripts $LC$ and $p$, respectively) at their corresponding hammer strike times.}
\label{tab:ErrorStatistics}
\end{table}
As expected from the simulation results in Fig.~\ref{fig:ModelSingleStrikePandF} during normal drill operation, the largest peak in $\ddot{p}_h$ after $p_h$ has passed through its maximum does correspond to the hammer striking the drill-bit.  
This method will be used in Sec.~\ref{sec:RockExperiments} to process the data set in which the simulant, rather than the load-cell, is struck by the hammer through the drill-bit.

With the hammer strike time measured, we have an end point for the integration of Eq.~\ref{eq:HammerFBDImplemented}. Experimental outputs provide values for $p_h(t)$, and we let $p_e = p_{atm}$, leaving $c_1$ and $c_2$ as free parameters. Since drill operation is vertical, there is no normal force between the hammer side walls and the cylinder, so $c_1 = 0$. The value of $c_2$ can then be calibrated by matching the integrated left- and right-hand sides of Eq.~\ref{eq:HammerFBDImplemented} and finding a best fit. Specifically, $F_{LC}$ from Eq.~\ref{eq:FLCdef} defines the integrand in Eq.~\ref{eq:ydotLC}, which, when integrated once in time, recovers $\dot{y}_{LC}(t)$.  

We can obtain $\dot{y}_p$ by solving the second-order ordinary differential equation in Eq.~\ref{eq:HammerFBDImplemented} for both $\dot{y}_p$ and ${y}_p$. This process is a reduced version of Algorithm~\ref{alg:PressureHammerEOM}, in which $p_h$ and $p_e$ are prescribed and only the hammer dynamics are solved. Since we cannot assume that the hammer begins at $y=0$ for every strike, the condition $y < L_{max}$ serves as a constraint to ensure that the solutions remain physical.  
We solve for $y_0$ iteratively using a shooting method with a relaxation parameter, as shown in Algorithm~\ref{alg:y0_iterative} in Appendix~\ref{sec:appendix1}.  
The algorithm provides $y_0$ as well as the $\dot{y}$, $y$ time-series, allowing us to evaluate the dynamics during each strike.   

\begin{figure}[H]
    \centering
    \begin{subfigure}[t]{0.45\textwidth}
        \centering 
        \includegraphics[trim=0cm 0cm 0cm 0cm, clip, width=1\linewidth]{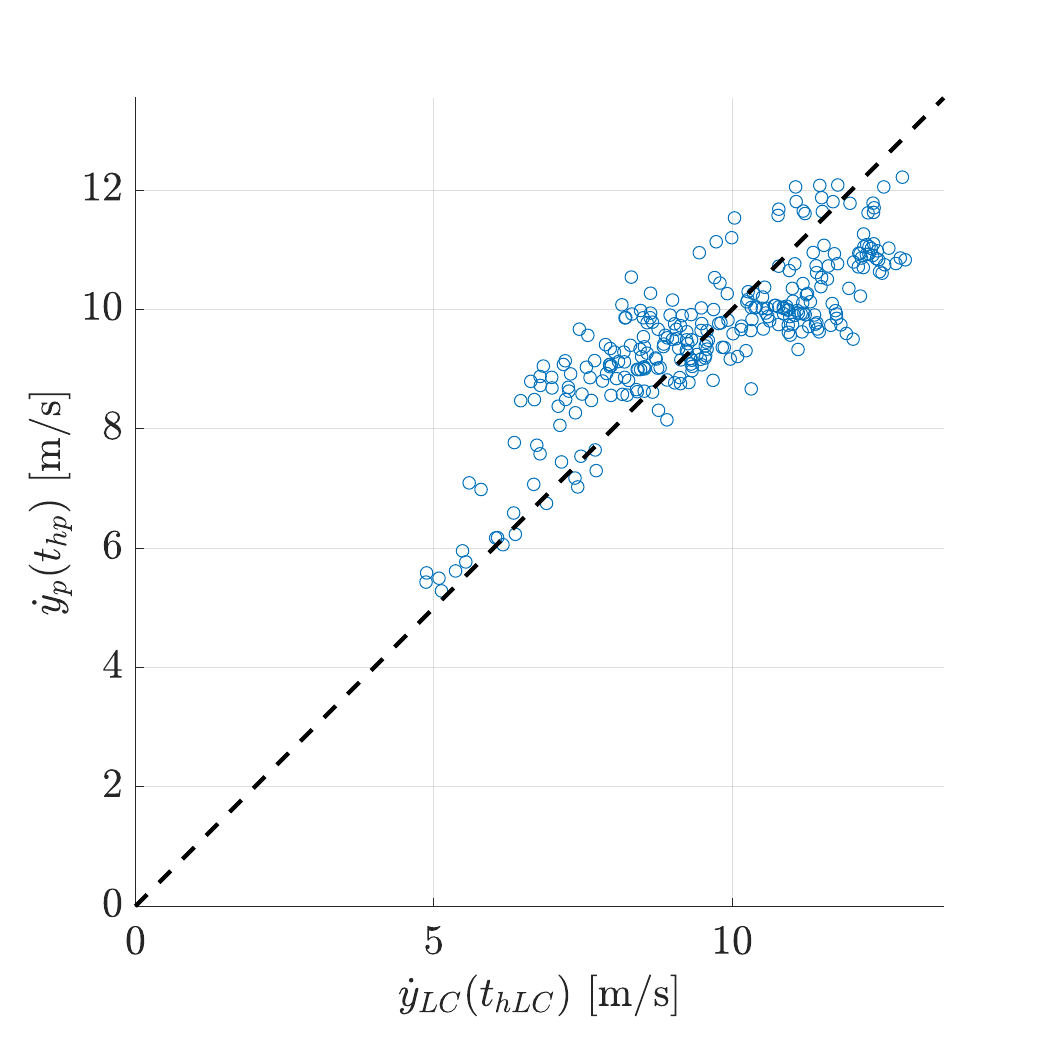}
        \caption{Comparison plot between $\dot{y}_{LC}$ and $\dot{y}_{p}$ at the hammer strike time.}
        \label{fig:Compydot}
    \end{subfigure}
    \hfill
    \begin{subfigure}[t]{0.45\textwidth}
        \centering
        \includegraphics[trim=0cm 0cm 0cm 0cm, clip, width=1\linewidth]{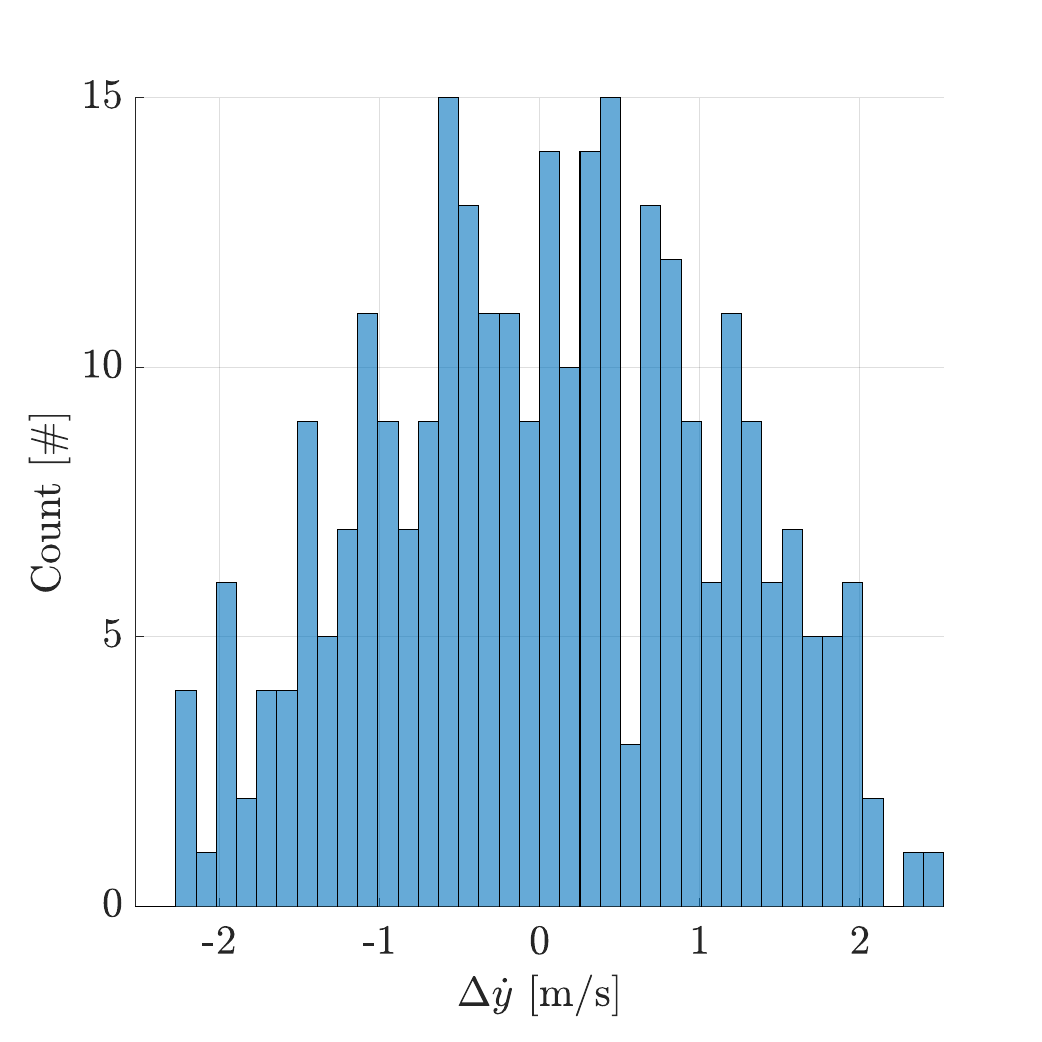}
        \caption{Histogram of error $\Delta \dot{y} = \dot{y}_{LC}(t_{hLC}) - \dot{y}_{p}(t_{hp})$ .}
        \label{fig:ErrHistydot}
    \end{subfigure}
\caption{Aggregated data from experiments 1a, 1b, and 1c in Table~\ref{tab:experiment_log} and comparison of $\dot{y}\left(t_h\right)$ obtained by integrating the hammer acceleration $\ddot{y}$ from pressure- and load-cell-based methods.}
    \label{fig:CompHistydot}
\end{figure}

\begin{figure}[H]
    \centering
    \begin{subfigure}[t]{0.45\textwidth}
        \centering 
        \includegraphics[trim=0cm 0cm 0cm 0cm, clip, width=1\linewidth]{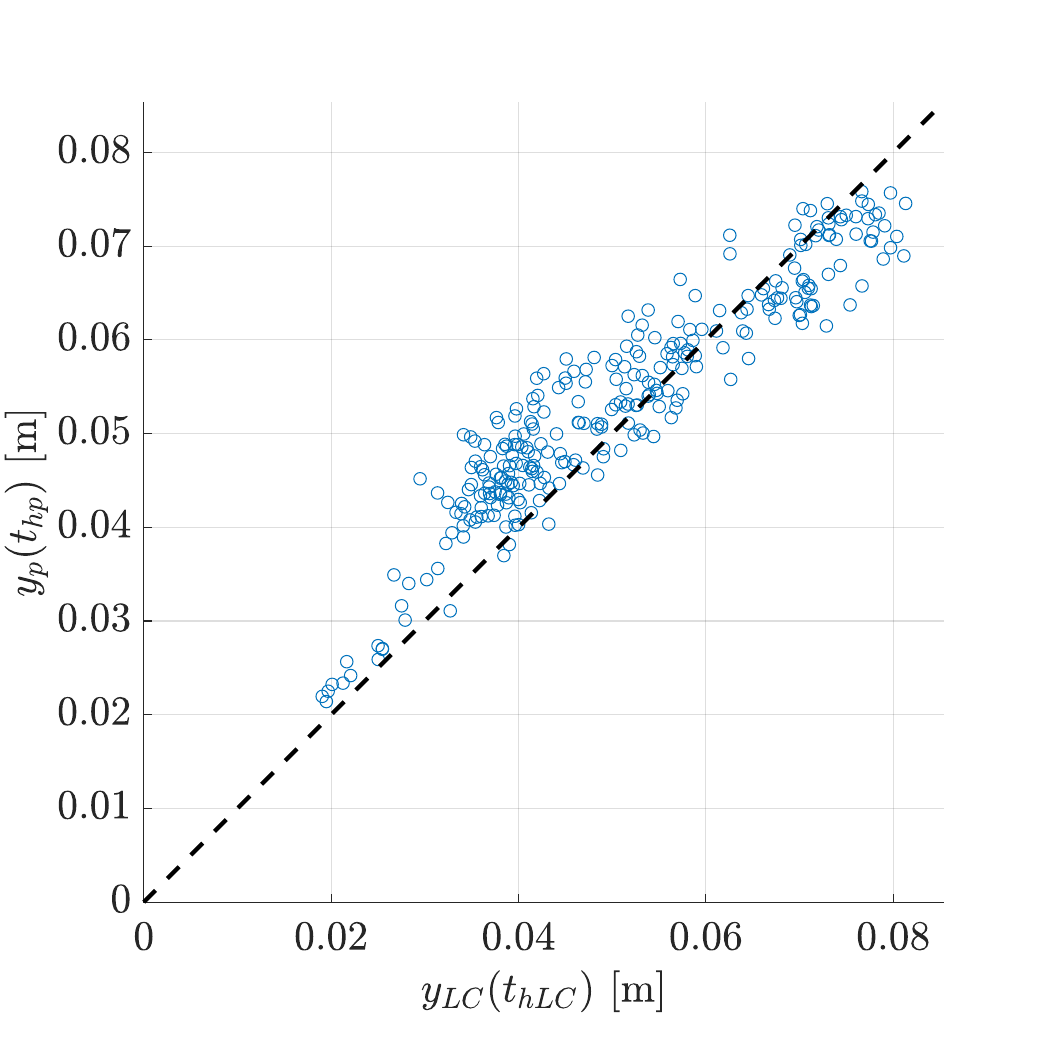}
        \caption{Comparison plot between $y_{LC}$ and $y_{p}$ at the hammer strike time.}
        \label{fig:Compy}
    \end{subfigure}
    \hfill
    \begin{subfigure}[t]{0.45\textwidth}
        \centering
        \includegraphics[trim=0cm 0cm 0cm 0cm, clip, width=1\linewidth]{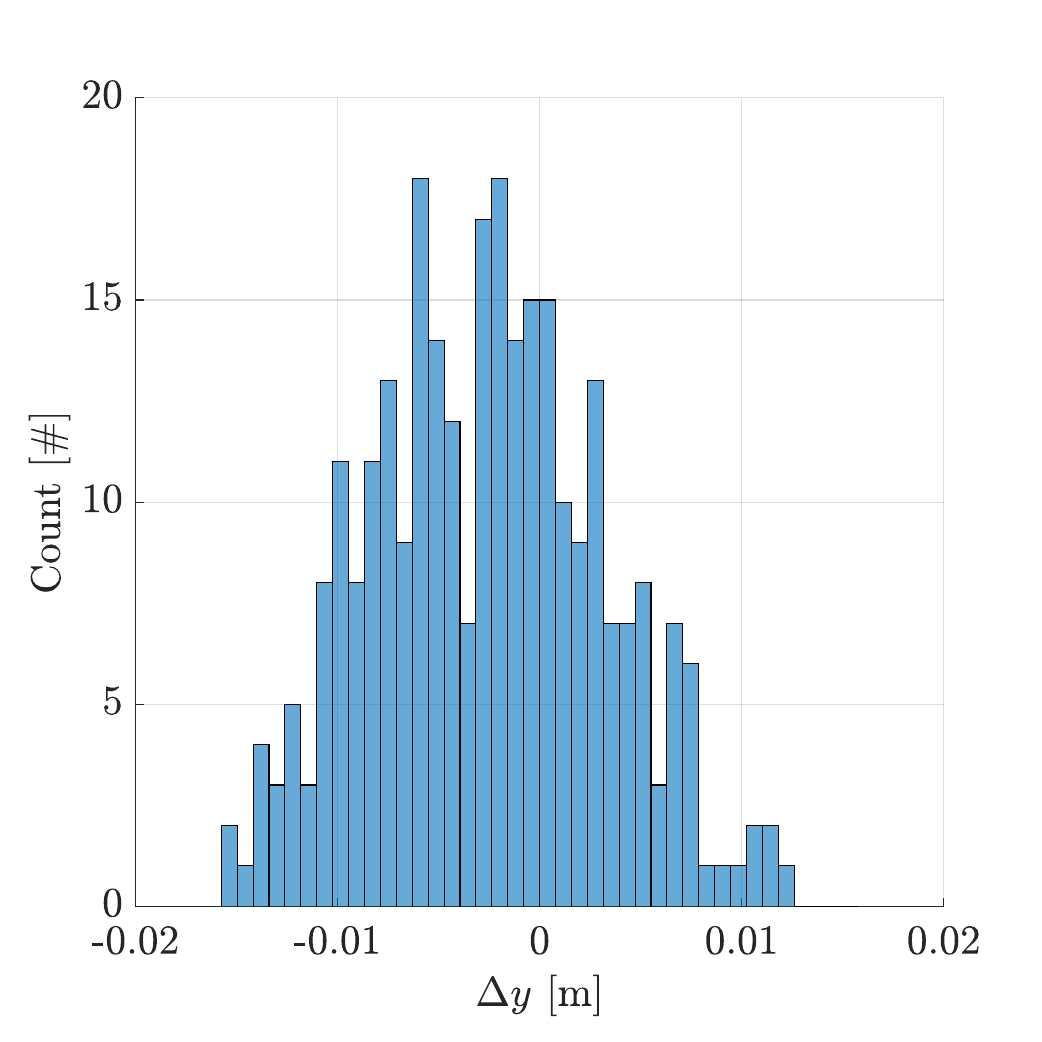}
        \caption{Histogram of error $\Delta y = y_{LC}(t_{hLC}) - y_{p}(t_{hp})$ .}
        \label{fig:ErrHisty}
    \end{subfigure}
\caption{Aggregated data from experiments 1a, 1b, and 1c in Table~\ref{tab:experiment_log} and comparison of $y\left(t_h \right)$ obtained by integrating hammer velocity $\dot{y}$ from pressure- and load-cell-based methods.}
    \label{fig:CompHisty}
\end{figure}

\begin{figure}[H]
    \centering
    \begin{subfigure}[t]{0.45\textwidth}
        \centering 
        \includegraphics[trim=0cm 0cm 0cm 0cm, clip, width=1\linewidth]{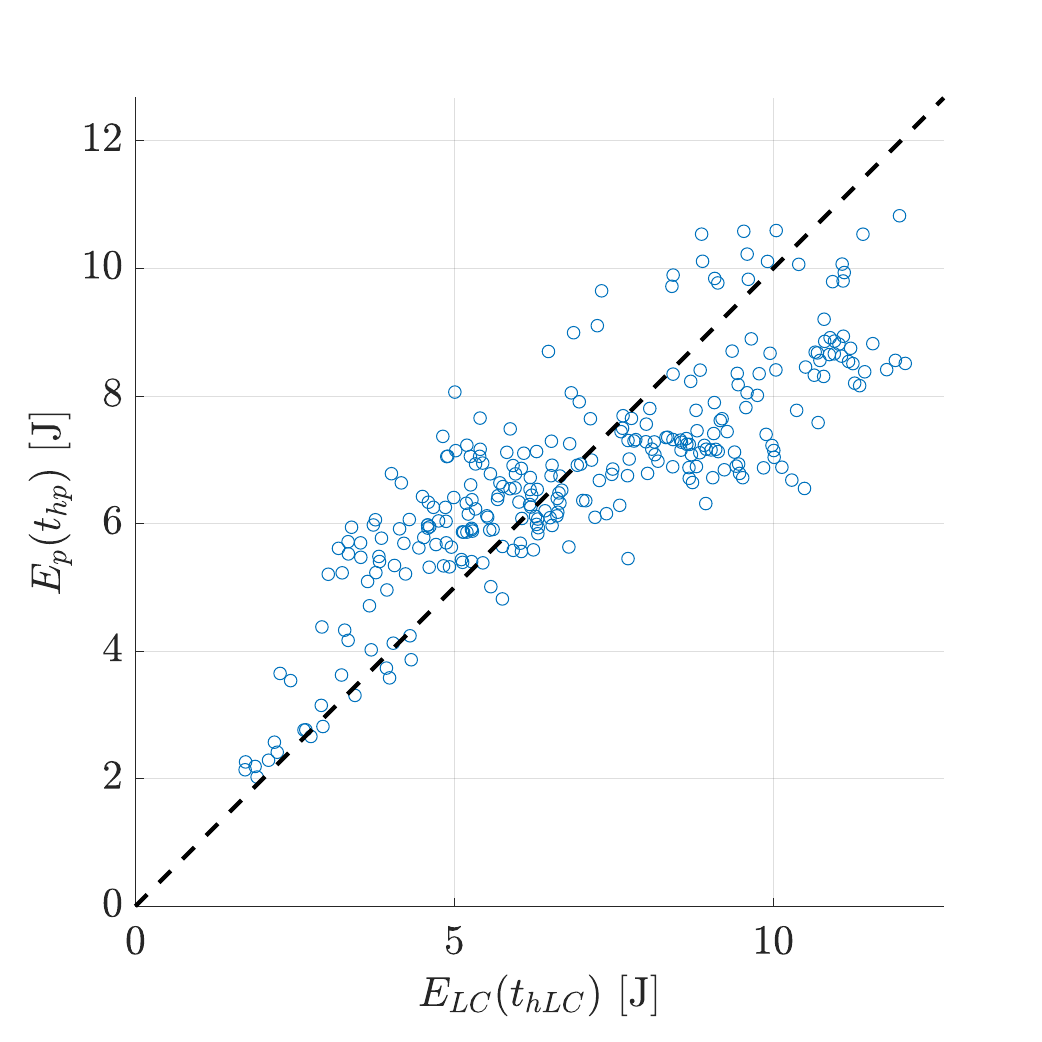}
        \caption{Comparison plot between $E_{LC}$ and $E_{p}$ at the hammer strike time.}
        \label{fig:CompE}
    \end{subfigure}
    \hfill
    \begin{subfigure}[t]{0.45\textwidth}
        \centering
        \includegraphics[trim=0cm 0cm 0cm 0cm, clip, width=1\linewidth]{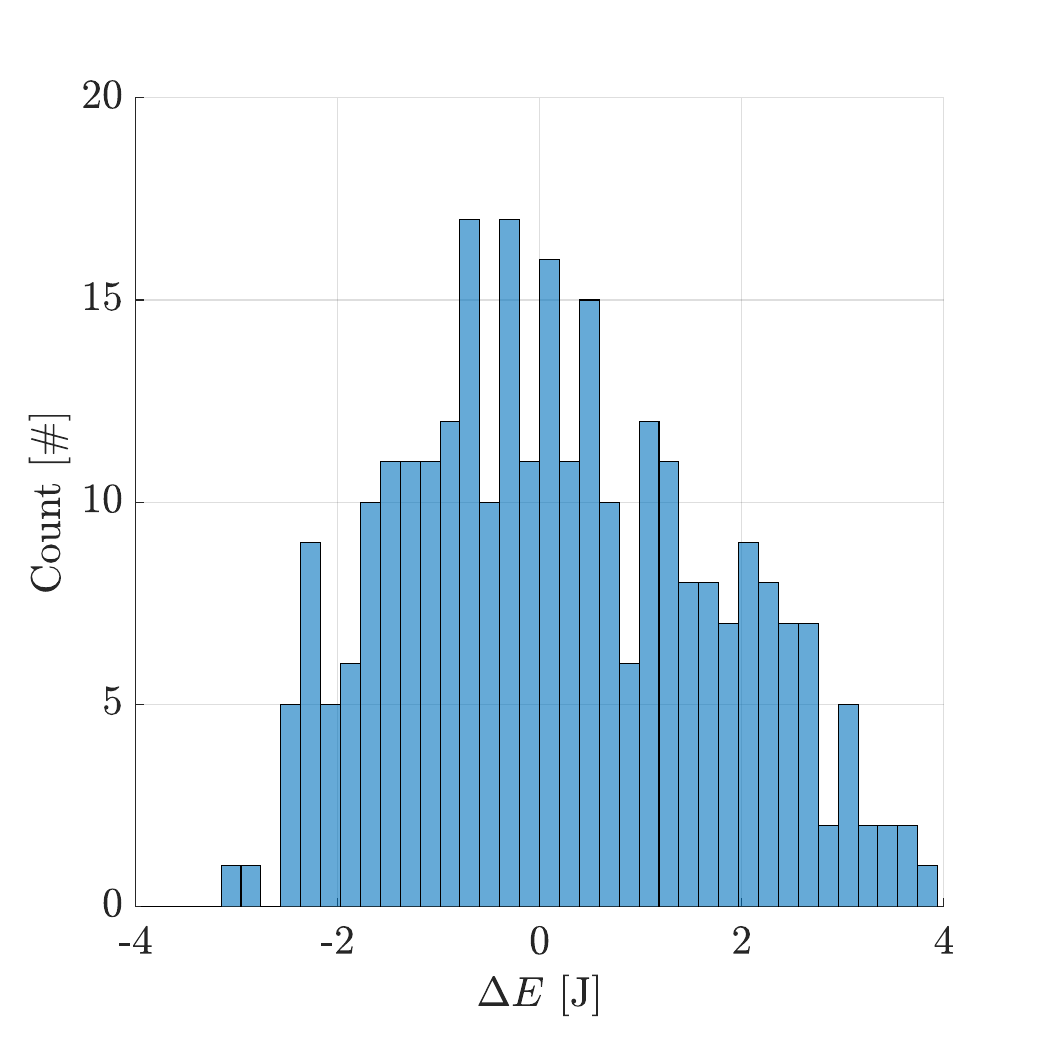}
        \caption{Histogram of error $\Delta E = E_{LC}(t_{hpLC}) - E_{p}(t_{hp})$ .}
        \label{fig:ErrHistE}
    \end{subfigure}
\caption{Aggregated data from experiments 1a, 1b, and 1c in Table~\ref{tab:experiment_log} and comparison of $E\left(t_h \right)$ for pressure- and load-cell-based methods.}
    \label{fig:CompHistE}
\end{figure}

We can then compare the two end-value velocities for different $c_2$, increasing it from 0 until $\Vert\Delta \dot{y} \Vert_2 \approx 0$. Figure~\ref{fig:CompHistydot} shows this comparison for $c_2 = 1.75$, obtained through this process. Fig.~\ref{fig:Compydot} shows the comparison between $\dot{y}_{LC}$ and $\dot{y}_p$, which largely clusters around the identity-slope line, with a normal-like distributed difference shown in Fig.~\ref{fig:ErrHistydot}. The average velocity error is near 0, the upper and lower bounds are on the order of $\pm$2 m/s, and these values are also listed in Table~\ref{tab:ErrorStatistics}.    
Figure~\ref{fig:CompHisty} shows the comparison and error-distribution histogram for $y_{LC}$ and $y_p$. Again, points are generally clustered around the unity-slope line in Fig.~\ref{fig:Compy}, but the mean $\Delta y$ is slightly shifted to the left: although the end velocities from the load-cell and pressure methods average to zero, the pressure method tends to slightly overpredict the $y$ time series. Results tabulated in Table~\ref{tab:ErrorStatistics} show a standard deviation between the two measurements of approximately 6 mm.  
Figure~\ref{fig:CompHistE} shows the comparison for our key metric, the hammer kinetic energy. Fig.~\ref{fig:CompE} displays a reasonable $E$ comparison plot, with data mostly tracking the unity-slope line, although with a larger variance than either $\dot{y}$ or $y$. This is mostly due to the mapping where $E \sim \dot{y}^2$, such that any variability in $\dot{y}$ is amplified when squared. This can be seen in the histogram of $\Delta E$ shown in Fig.~\ref{fig:ErrHistE}, with a mean around 0 and a standard deviation on the order of 1.5 J.  

\begin{figure}[H]
    \centering
    \begin{subfigure}[t]{0.32\textwidth}
        \includegraphics[width=\textwidth,trim={0cm 0cm 1cm 0cm},clip]{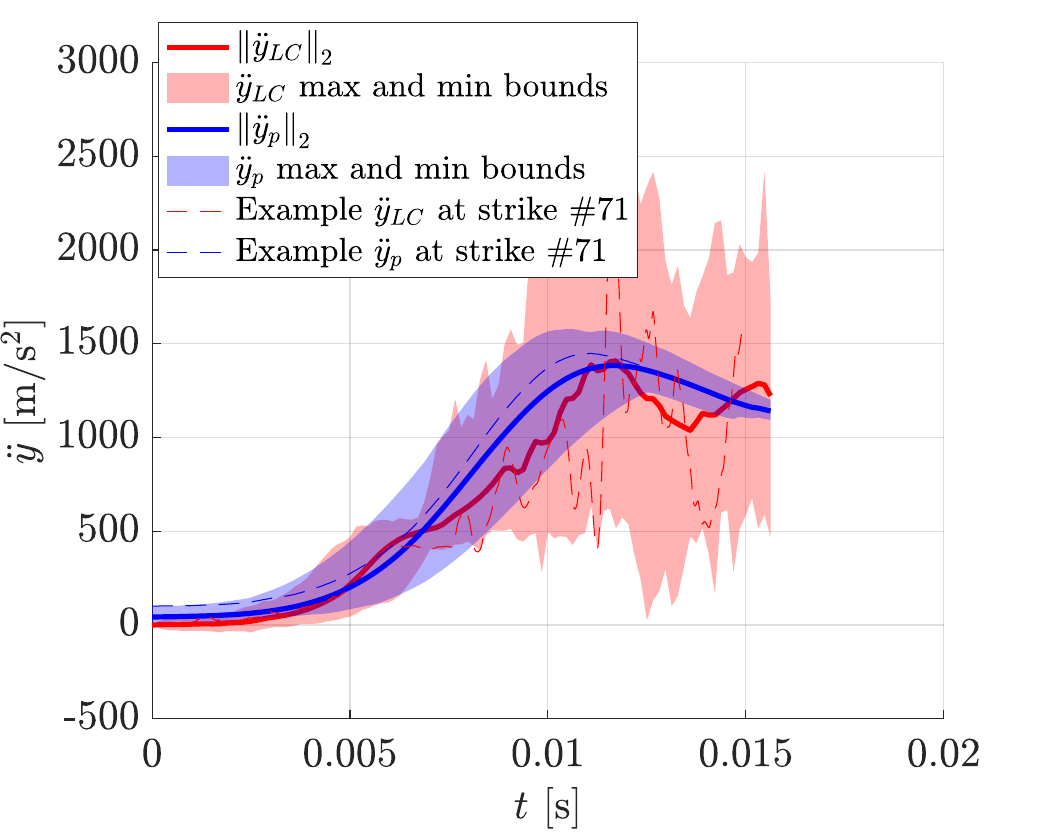}
        \caption[short]{$p_o = 100$ psi, and example strike \# 71.}
        \label{fig:yddotvstime_stats_po100}
    \end{subfigure}
    \hfill
    \begin{subfigure}[t]{0.32\textwidth}
        \centering
        \includegraphics[width=\textwidth,trim={0cm 0cm 1cm 0cm},clip]{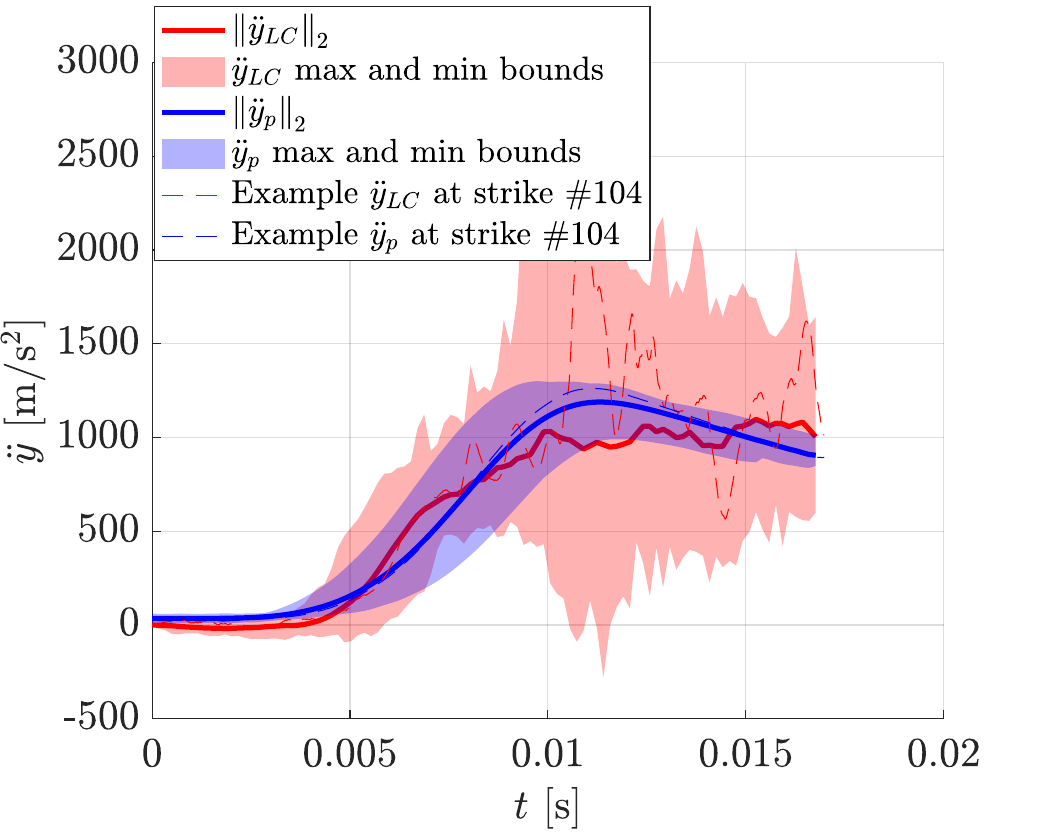}
        \caption[short]{$p_o = 80$ psi, and example strike \# 104.}
        \label{fig:yddotvstime_stats_po80}
    \end{subfigure}
    \hfill
    \begin{subfigure}[t]{0.32\textwidth}
        \centering
        \includegraphics[width=\textwidth,trim={0cm 0cm 1cm 0cm},clip]{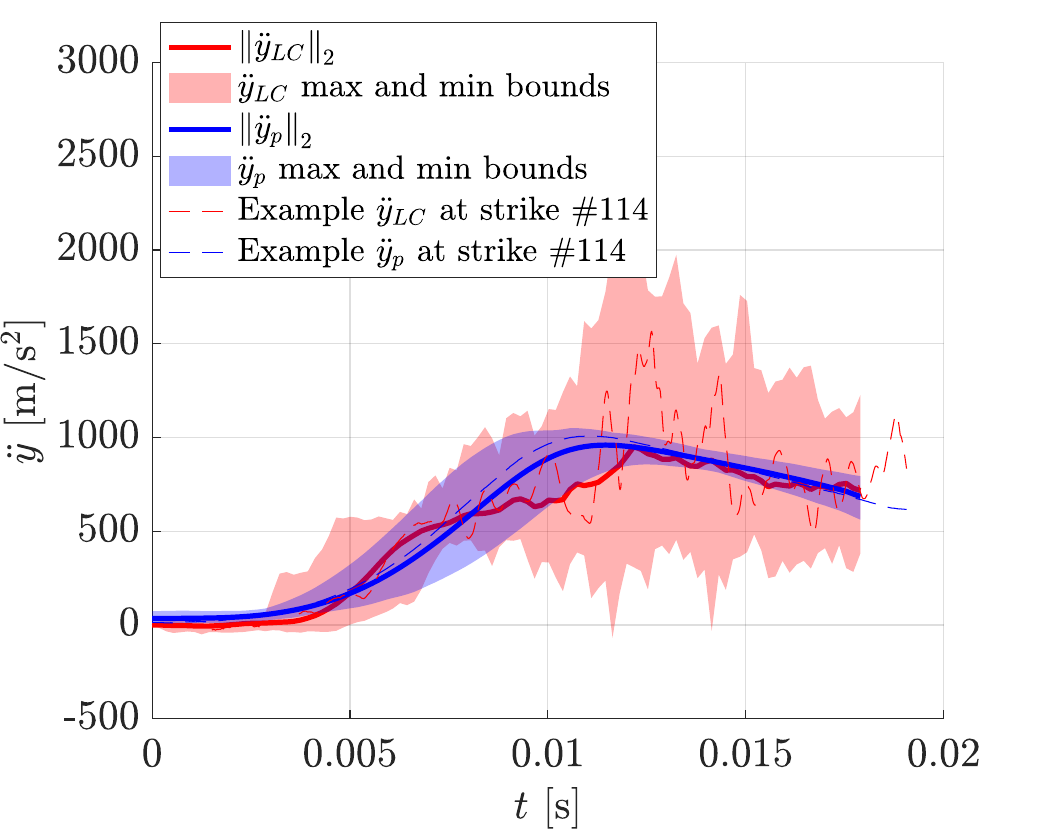}
        \caption[short]{$p_o = 60$ psi, and example strike \# 114.}
        \label{fig:yddotvstime_stats_po60}
    \end{subfigure}
   
    \caption{Relevant accelerations for the load-cell and pressure methods, with minimum and maximum bounds, and an individual strike example for experiments 1a, 1b, and 1c in Table~\ref{tab:experiment_log}.}
    \label{fig:yddotvstime_stats_po100-80-60}
\end{figure}

The normal-like distributions of Figs.~\ref{fig:ErrHistydot}, \ref{fig:ErrHisty}, and \ref{fig:ErrHistE} suggest that the errors likely do not stem from a biased set of unmodeled physics, but rather from randomness in the system. To understand this further, we examine the acceleration measurements, where $F_{LC}(t)$ is normalized by the well-defined, constant hammer mass $M_h$ and compared with the calculated acceleration from $p_h$.  
Figure~\ref{fig:yddotvstime_stats_po100-80-60} shows the $\ddot{y}$ min-max envelope, mean value, and an example strike over time for each $p_o$ in experiments 1a, 1b, and 1c. The min-max envelopes are built from strike data interpolated onto a common time series with a final time equal to the average $t_{hp}$ for each experiment. These values are listed in Table~\ref{tab:experimental_results_mean}, along with other mean scalar results for the load-cell experimental campaign.  
Although 50\% of strikes exceed the mean $t_{hp}$, as can be seen from the selected example strike in Fig.~\ref{fig:yddotvstime_stats_po60}, the plotted envelopes still capture the overall variability in both $F_{LC}$ and $p_h$.  
The most noticeable trend across all three plots is the consistently larger minimum-to-maximum variation of $\ddot{y}_{LC}$ when compared with $\ddot{y}_p$. Although the mean values from the two calculations follow one another well, the noise seen in the example strikes and in the min-max envelopes illustrates the root cause of the measurement errors reported in Table~\ref{tab:ErrorStatistics}.  
Yet, the mean and min-max envelope of $\ddot{y}_p$ lie reasonably well within the $\ddot{y}_{LC}$ values, with end conditions relatively close to one another when compared with the overall variance in $\ddot{y}$. The system dynamics act as a low-pass filter on the noisy $\ddot{y}$ measurements, reducing the total error after each integration step from $\ddot{y}$ to $\dot{y}$ and from $\dot{y}$ to $y$.   

Figure~\ref{fig:hist_po100-80-60} captures the summary of $L_{int}$, $t_{close}$, $p_v(t_0)$, and $t_{hp}$ for experiments 1a, 1b, and 1c. The complementary detailed statistics are reported in Tables~\ref{tab:experimental_results_mean} and \ref{tab:experimental_results_min_max}. General trends include a slight increase in the mean $L_{int}$ (Fig.~\ref{fig:histLint_po100-80-60}) as $p_o$ decreases, with a clear deficit between $L_{max}$ and $L_{int}$. Taking the geometry of the hammer chamber and the hammer height in Table~\ref{tab:parametersTable}, $L_{max} \approx 0.0788$ m, making the average strike approximately 1.2 cm less than the maximum possible displacement. Figure~\ref{fig:histLint_po100-80-60} suggests that lower $p_o$ strikes have a higher likelihood of being pushed against the flapper-valve mount and recovering the full chamber-length displacement. Figure~\ref{fig:histthp_po100-80-60} shows that $t_{hp}$ also has a tight distribution, especially when compared with that of $t_{close}$ (Fig.~\ref{fig:histtclose_po100-80-60}), although the variance of both $t_{close}$ and $t_{hp}$ appears to increase as $p_o$ decreases. 
The pressure at the start of a strike, $p_v(t_0)$, has a narrow distribution (see Fig.~\ref{fig:histPvmax_po100-80-60}) and is generally close to the target $p_o$ value. Although the trend is not as clear as for $L_{int}$, both $t_{close}$ and $t_{hp}$ also appear to increase as $p_o$ decreases. These three values are the main inputs from the experimental data set into the model results described next in Sec.~\ref{sec:ModelComp}.  

\begin{table}[!ht] 
    \centering
    \begin{tabular}{l c c c c}
    \hline
    \textbf{ID} & \textbf{1a} & \textbf{1b} & \textbf{1c} & \textbf{Units} \\ \hline
    $p_o$ & 100 & 80 & 60 & psi \\ 
    $p_{v}\left( t_0 \right)$ & 97.36 & 77.51 & 60.96 & psi \\ 
    $E_{p}(t_{hp})$ & 7.92 & 7.12 & 5.72 & J \\ 
    $E_{LC}(t_{hLC})$ & 7.47 & 7.56 & 6.10 & J \\ 
    $t_{hp}$ & 0.0155 & 0.0164 & 0.0178 & s \\ 
    $t_{hLC}$ & 0.0157 & 0.0168 & 0.0179 & s \\ 
    $t_{close}$ & 0.0230 & 0.0232 & 0.0264 & s \\ 
    $L_{int}$ & 0.0653 & 0.0673 & 0.0685 & m \\ 
    $N_{total}$ & 70 & 103 & 113 & \# \\ 
    $N_{\text{norm}}$ & 70 & 103 & 113 & \# \\ 
    $N_{\text{abn}}$ & 0 & 0 & 0 & \# \\ 
    $N_{\text{missed}}$ & 4 & 2 & 14 & \# \\ \hline
    \end{tabular}
    \caption{Mean values of experimental results 1a, 1b, and 1c in Table~\ref{tab:experiment_log}.}
    \label{tab:experimental_results_mean}
\end{table}

Another observable trend from Table~\ref{tab:experimental_results_mean} is that $t_{h}$ increases decisively as $p_o$ decreases, consistent with the ideal-model trend in Fig.~\ref{fig:ModelMultiple30-100th}.  
Mean energy values for $E_p$ and $E_{LC}$ are close to one another, although higher $p_o$ does not translate into a large increase in average strike energy. A much clearer trend is seen in the maximum and minimum values of $E_p$ in Table~\ref{tab:experimental_results_min_max}, where both increase monotonically with $p_o$. $E_{LC}$ presents a wider range, however.  
Finally, Table~\ref{tab:experimental_results_mean} reports the number of strikes missed during the analysis, which was as high as 11\% of the total for experiment 1c but significantly lower for 1a and 1b. These correspond to $F_{LC}$ time traces that were too noisy to analyze reliably, making it difficult for Algorithm~\ref{alg:hammerStrikeTimeLC} to select $t_{hLC}$. Even so, we were able to analyze most strikes and produce representative data for validating the pressure-based method, while also capturing trends that help identify drill performance.  

In summary, we showed that $t_h$ can be estimated within $1 \times 10^{-4}$ s of the load-cell strike signal, and that $E_p$ tracks well with $E_{LC}$, with a standard deviation of approximately 1.5 J, when $c_2 = 1.75$. The error driving the variance in $E$ can be traced back to noise in the load-cell measurements, while the $p_h$-derived hammer acceleration, velocity, and position present a much tighter min-max envelope that lies within that derived from $F_{LC}$. Overall, the pressure-based hammer dynamics appear to capture the true impact-energy levels well.  

\begin{table}[ht!]
    \centering
    \resizebox{\textwidth}{!}{%
    \begin{tabular}{l c c c c c c c c c}
    \hline
    \textbf{ID} & \multicolumn{3}{c}{\textbf{1a}} & \multicolumn{3}{c}{\textbf{1b}} & \multicolumn{3}{c}{\textbf{1c}} \\ 
    \textbf{$p_o$} & \multicolumn{3}{c}{100} & \multicolumn{3}{c}{80} & \multicolumn{3}{c}{60} \\ 
    \textbf{Symbol} & \textbf{Max} & \textbf{Min} & \textbf{STD} & \textbf{Max} & \textbf{Min} & \textbf{STD} & \textbf{Max} & \textbf{Min} & \textbf{STD} \\ \hline
    $p_{v}\left( t_0 \right)$ & 101.90 & 91.85 & 3.88 & 79.24 & 67.67 & 4.16 & 61.77 & 57.78 & 0.73 \\ 
    $E_{p}(t_{hp})$ & 10.83 & 5.63 & 1.56 & 8.94 & 3.58 & 1.09 & 7.82 & 2.03 & 1.53 \\
    $E_{LC}(t_{hLC})$ & 11.98 & 4.82 & 2.03 & 12.07 & 3.67 & 2.43 & 10.49 & 1.72 & 2.60 \\ 
    $t_{hp}$ & 0.0168 & 0.0145 & 0.00048 & 0.0175 & 0.0152 & 0.00052 & 0.0207 & 0.0144 & 0.00142 \\ 
    $t_{hLC}$ & 0.0169 & 0.0148 & 0.00046 & 0.0179 & 0.0154 & 0.00055 & 0.0206 & 0.0140 & 0.00144 \\ 
    $t_{close}$ & 0.0317 & 0.0187 & 0.00336 & 0.0330 & 0.0190 & 0.00419 & 0.0415 & 0.0198 & 0.00482 \\ 
    $L_{int}$ & 0.0720 & 0.0602 & 0.00348 & 0.0739 & 0.0569 & 0.00299 & 0.0781 & 0.0520 & 0.00748 \\ \hline
    \end{tabular}
    }
    \caption{Minimum, maximum, and standard deviation (STD) values of experimental results 1a, 1b, and 1c in Table~\ref{tab:experiment_log}.}
    \label{tab:experimental_results_min_max}
\end{table}

\subsection{Model Comparison}\label{sec:ModelComp}
Given the data set from experiments 1a, 1b, and 1c, with a calibrated friction coefficient $c_2 = 1.75$, we can now explore how these experimental results compare with predicted values from the full model derived in Sec.~\ref{sec:model}. Specifically, we solve the fluid-flow and thermodynamic equations together with the hammer dynamics using Algorithm~\ref{alg:PressureHammerEOM}.  
The model inputs are taken from Table~\ref{tab:experimental_results_mean} for $t_{close}$, $L_{int}$, and $p_v\left( t_0 \right)$ for the three $p_o$ values of 100, 80, and 60 psi. The two bounded parameters that have not been calibrated are the volumetric efficiency $\eta_v$, estimated at 0.75, and the ratio between $t_{open}$ and $t_{close}$ in Eq.~\ref{eq:opentimeCondition}; both are listed in Table~\ref{tab:parametersTable} with their sources.  
Both capture distinct phenomena that are worth exploring from a model-sensitivity standpoint: $\eta_v$ largely correlates with the internal flow geometry of the vent chamber, which can be made more efficient by streamlining the flow and increasing total volume relative to any narrow flow paths.  
The parameter $t_{open}$ directly relates to the flapper-valve design and its opening efficiency: the faster the flapper-valve opens, the greater the average force exerted on the hammer over the strike length.  
In Sec.~\ref{sec:idealModelResults}, both of these parameters were set to their equivalent lowest-dissipation values, with $\eta_v = 1$, $c_2 = 0$, and the flapper-valve assumed to open and close instantaneously at $t_0$ and $t_{close}$. Those results therefore mimic an ideal system with the maximum possible energy for the design.  
The model results in this section aim to replicate the true experimental drill-system behavior as closely as possible for model validation. Confidence that the model captures the dominant physics allows it to be used for design trade-off analysis when targeting a specific performance metric.  
\begin{figure}[H]
    \centering
    \begin{subfigure}[t]{0.32\textwidth}
        \includegraphics[width=\textwidth,trim={0cm 0cm 0cm 0cm},clip]{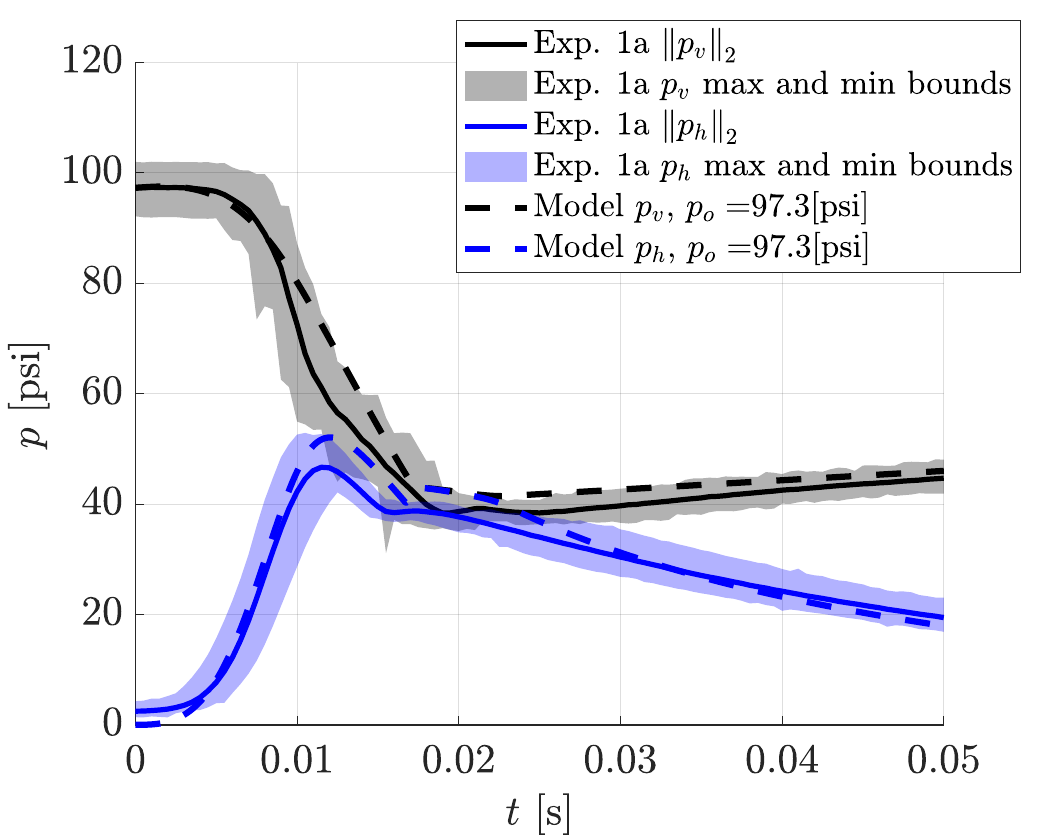}
        \caption[short]{Model and experimental $p_v$ and $p_h$ at $p_o = $ 100 psi}
        \label{fig:pvphvst-po100psi}
    \end{subfigure}
    \hfill
    \begin{subfigure}[t]{0.32\textwidth}
        \centering
        \includegraphics[width=\textwidth,trim={0cm 0cm 0cm 0cm},clip]{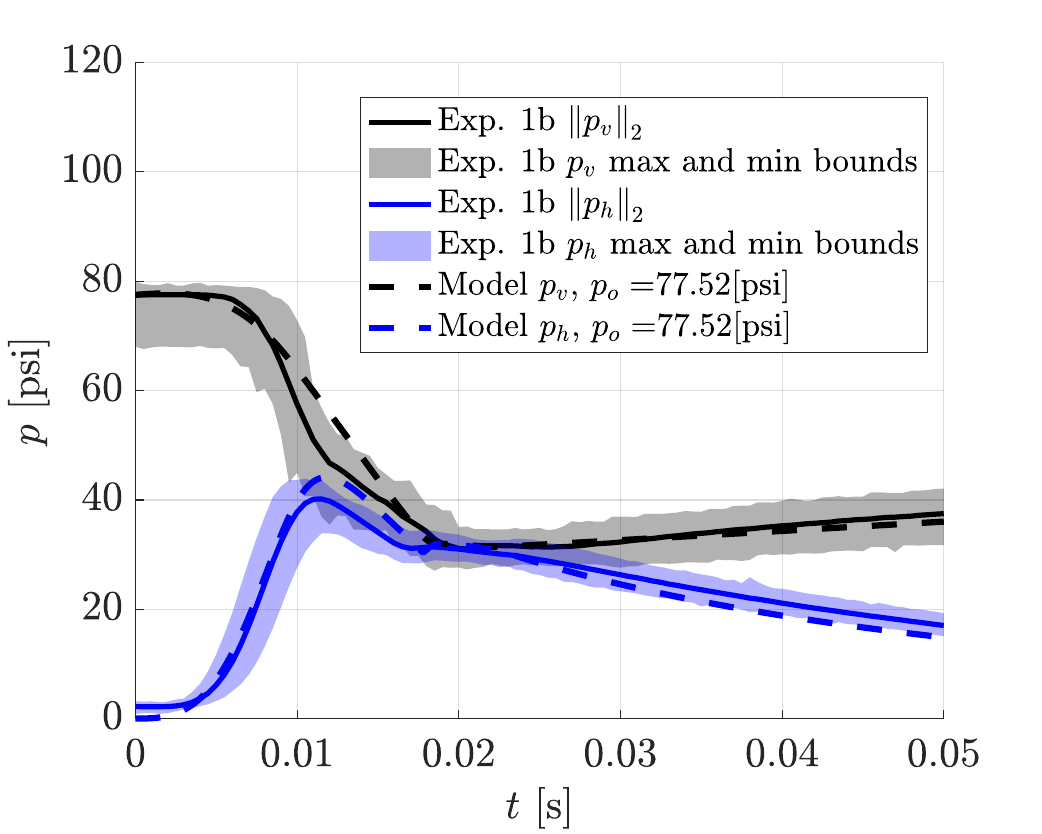}
        \caption[short]{Model and experimental $p_v$ and $p_h$ at $p_o = $ 80 psi.}
        \label{fig:pvphvst-po80psi}
    \end{subfigure}
    \hfill
    \begin{subfigure}[t]{0.32\textwidth}
        \centering
        \includegraphics[width=\textwidth,trim={0cm 0cm 0cm 0cm},clip]{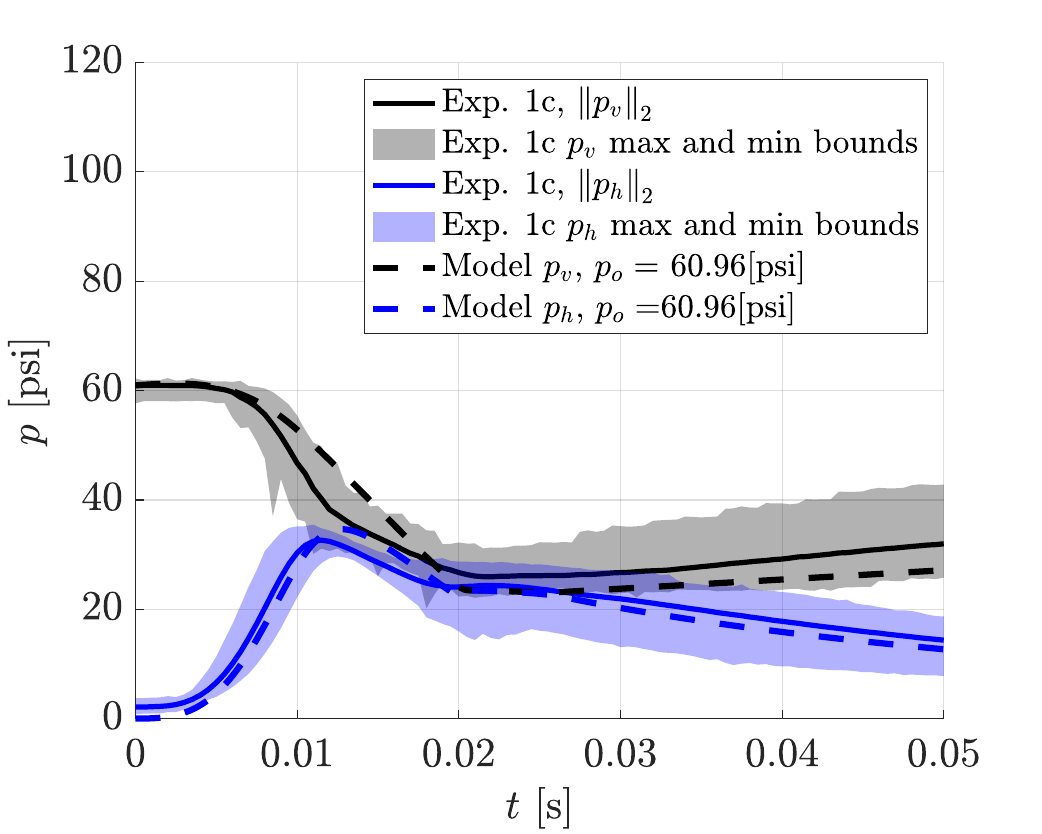}
        \caption[short]{Model and experimental $p_v$ and $p_h$ at $p_o = $ 60 psi}
        \label{fig:pvphvst-po60psi}
    \end{subfigure}
   
    \caption{Model and experimental $p_v$ and $p_h$ comparison for $p_o$ values of 100, 80, and 60 psi in experiments 1a, 1b, and 1c in Table~\ref{tab:experiment_log}.}
    \label{fig:pvphvst-po100-80-60psi}
\end{figure}

The first set of plots comparing the experimental vent and hammer pressures with those calculated by the model is shown in Fig.~\ref{fig:pvphvst-po100-80-60psi}. Similar to Fig.~\ref{fig:yddotvstime_stats_po100-80-60}, the min-max pressure envelope is derived by interpolating all strike data onto a common time mesh, here extending to 0.05 s. The plots show the min-max envelope, the mean value, and the model-calculated $p_h$ and $p_v$ curves for experiments 1a, 1b, and 1c using the inputs from Table~\ref{tab:experimental_results_mean}, $\eta_v = 0.75$, $c_2 = 0$, and the valve-area function in Eqs.~\ref{eq:Av1TimingEq} and \ref{eq:Av1topen}.   
Agreement between the mean and model-calculated curves is good, with the model-predicted pressures falling consistently within the experimental pressure envelopes. Figure~\ref{fig:pvphvst-po100psi} shows that the model slightly overestimates the peak magnitude of $p_h$ and predicts a slightly later impact time than the experiments. Yet the shape of the $p_h$ pressure rise relative to the $p_v$ decrease is captured reasonably well. The growth and decay rates of $p_v$ and $p_h$, respectively, are also reproduced well by the model, suggesting that the estimates for $\dot{m}_{in}$ and for the hammer-chamber to hammer-diameter gap $h$ are largely correct.  
\begin{table}[H]
\centering
\begin{tabular}{c c c c c}
\hline
\textbf{ID} & \textbf{$p_o$ [psi]} & \textbf{Model $t_h$ [s]} & \textbf{$\Vert t_{hp} \Vert_2$ [s]} & \textbf{$\Vert t_{hLC} \Vert_2$ [s]} \\ \hline
1a & 100 & 0.0167 & 0.0155 & 0.0157 \\
1b & 80  & 0.0177 & 0.0164 & 0.0168 \\
1c & 60  & 0.0199 & 0.0178 & 0.0179 \\ \hline
\end{tabular}%
\caption{Mean hammer impact-time comparison between the model and experiments 1a, 1b, and 1c from Table~\ref{tab:experiment_log}.}
\label{tab:hammerTimeModelComp}
\end{table}
The hammer impact time shifts later as $p_o$ decreases, a trend that is also captured well by the model in Figs.~\ref{fig:pvphvst-po80psi} and \ref{fig:pvphvst-po60psi}. Hammer strike-time values are listed in Table~\ref{tab:hammerTimeModelComp} and show that the model tends to overestimate them by $\sim 1 \times 10^{-3}$ s. The size of the experimental min-max envelope, which reflects experimental variance, also appears to increase as $p_o$ decreases, suggesting more erratic flapper-valve behavior as the control-magnet distance from the flapper increases.   

\begin{figure}[H]
    \centering
    \begin{subfigure}[t]{0.32\textwidth}
        \includegraphics[width=\textwidth,trim={0cm 0cm 0cm 0cm},clip]{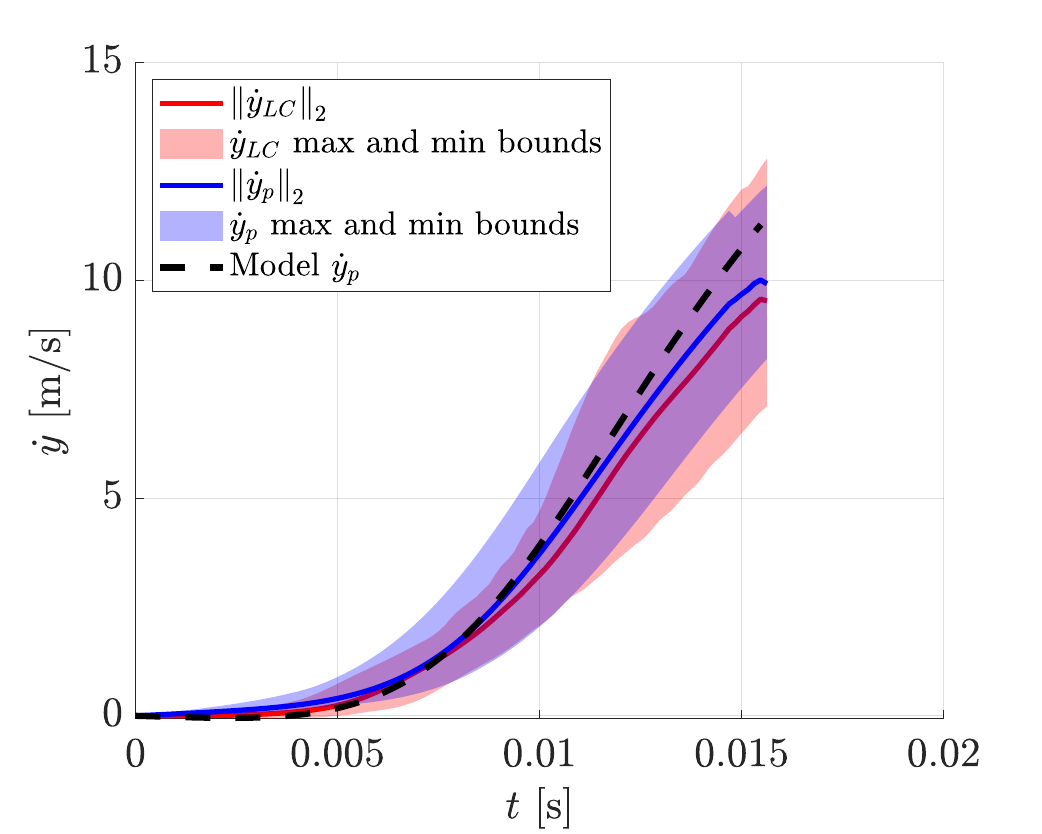}
        \caption[short]{Model and experimental $\dot{y}$ at $p_o = $ 100 psi}
        \label{fig:ydotvst-po100psi}
    \end{subfigure}
    \hfill
    \begin{subfigure}[t]{0.32\textwidth}
        \centering
        \includegraphics[width=\textwidth,trim={0cm 0cm 0cm 0cm},clip]{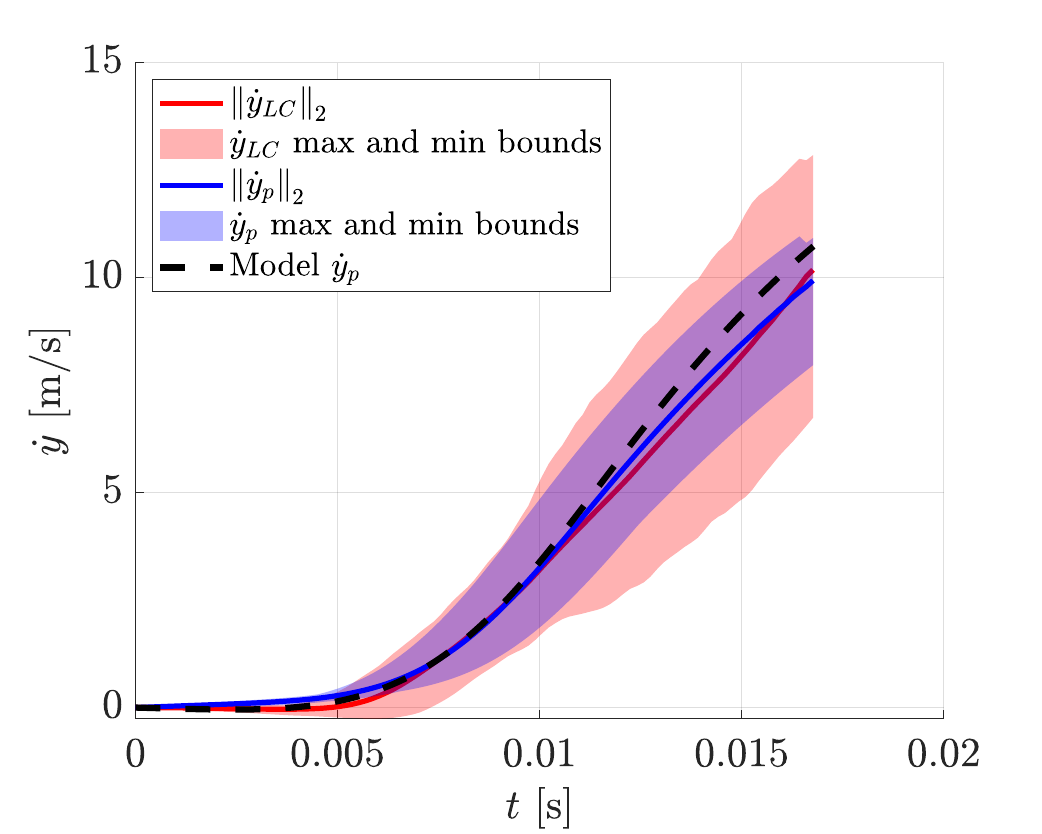}
        \caption[short]{Model and experimental $\dot{y}$ at $p_o = $ 80 psi.}
        \label{fig:ydotvst-po80psi}
    \end{subfigure}
    \hfill
    \begin{subfigure}[t]{0.32\textwidth}
        \centering
        \includegraphics[width=\textwidth,trim={0cm 0cm 0cm 0cm},clip]{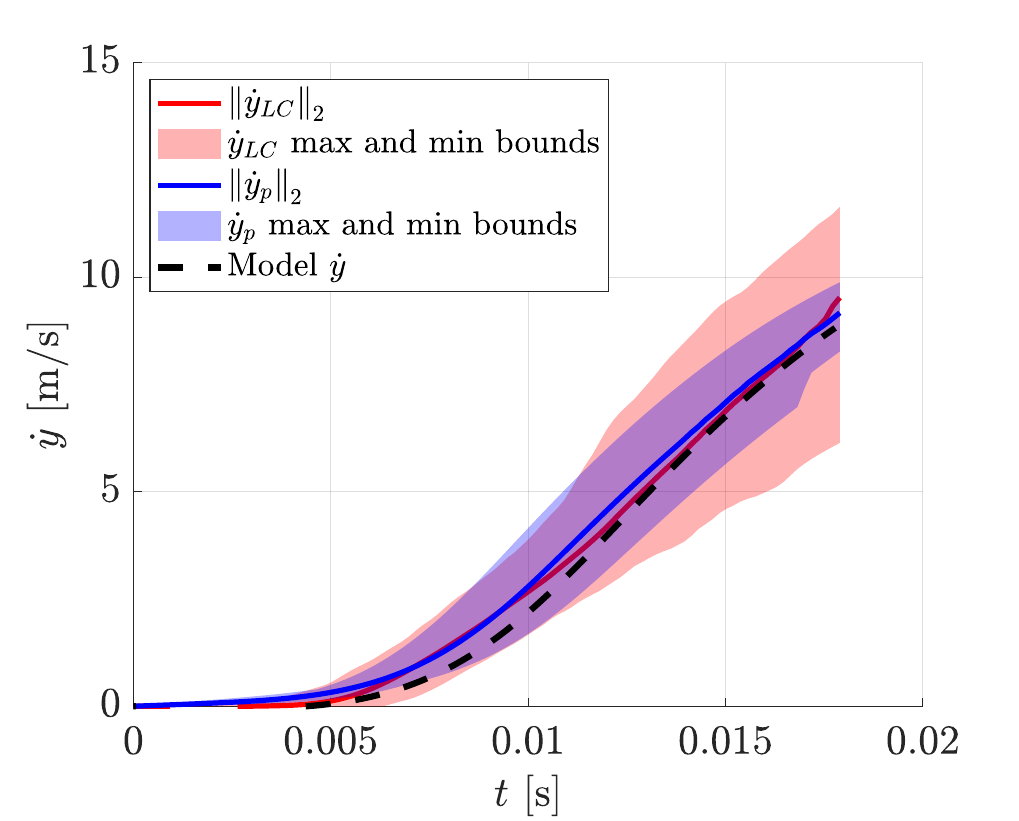}
        \caption[short]{Model and experimental $\dot{y}$ at $p_o = $ 60 psi}
        \label{fig:ydotvst-po60psi}
    \end{subfigure}
   
    \caption{Model and experimental $\dot{y}$ comparison for $p_o$ values of 100, 80, and 60 psi in experiments 1a, 1b, and 1c in Table~\ref{tab:experiment_log}.}
    \label{fig:ydotvst-po100-80-60psi}
\end{figure}
Figure~\ref{fig:ydotvst-po100-80-60psi} shows $\dot{y}$ comparisons in a manner similar to the $\ddot{y}$ results in Fig.~\ref{fig:yddotvstime_stats_po100-80-60}, with min-max envelopes for $\dot{y}_p$ and $\dot{y}_{LC}$ and their corresponding mean values. The example-strike data is now replaced by the model result for each $p_o$. These plots extend to the average hammer strike time for each condition, shown in Table~\ref{tab:experimental_results_mean}, which is also where the model result is truncated.  
Figure~\ref{fig:ydotvst-po100psi} shows a model curve that overpredicts both the load-cell- and pressure-integrated $\dot{y}$ means through most of the hammer-acceleration period, although the prediction still follows both reasonably well. This is consistent with the model $p_h$ curve in Fig.~\ref{fig:pvphvst-po100psi} and with the model overprediction of the peak $p_h$ value. This figure appears to be the only case where the load-cell variance is approximately the same as that of the pressure-based method, again reflecting the low-pass nature of the hammer dynamics as the response is integrated from $\ddot{y}$ to $\dot{y}$. Both the load-cell and pressure-method mean curves remain close to one another, with the model prediction falling within the experimental min-max envelope. Figure~\ref{fig:ydotvst-po80psi} shows a model prediction that is closer to the mean values than that in Fig.~\ref{fig:ydotvst-po100psi}, and the load-cell variance again exceeds that of the pressure-based method.  
Figure~\ref{fig:ydotvst-po60psi} shows results for the lowest experimental $p_o$ and perhaps the best agreement between the mean values of $\dot{y}$ and the model. Overall, the model compares well with the experiments, both quantitatively and in capturing the trends observed in $\dot{y}$.   

\begin{figure}[h]
    \centering
    \begin{subfigure}[t]{0.32\textwidth}
        \includegraphics[width=\textwidth,trim={0cm 0cm 0cm 0cm},clip]{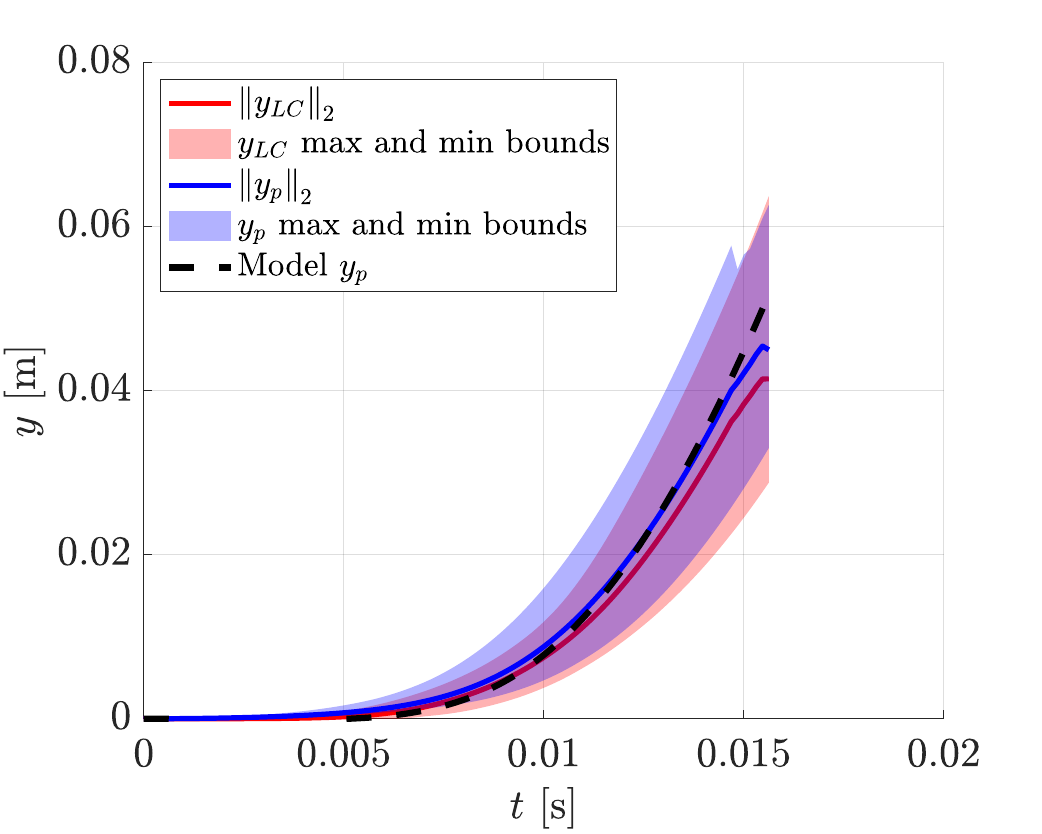}
        \caption[short]{Model and experimental ${y}$ at $p_o = $ 100 psi}
        \label{fig:yvst-po100psi}
    \end{subfigure}
    \hfill
    \begin{subfigure}[t]{0.32\textwidth}
        \centering
        \includegraphics[width=\textwidth,trim={0cm 0cm 0cm 0cm},clip]{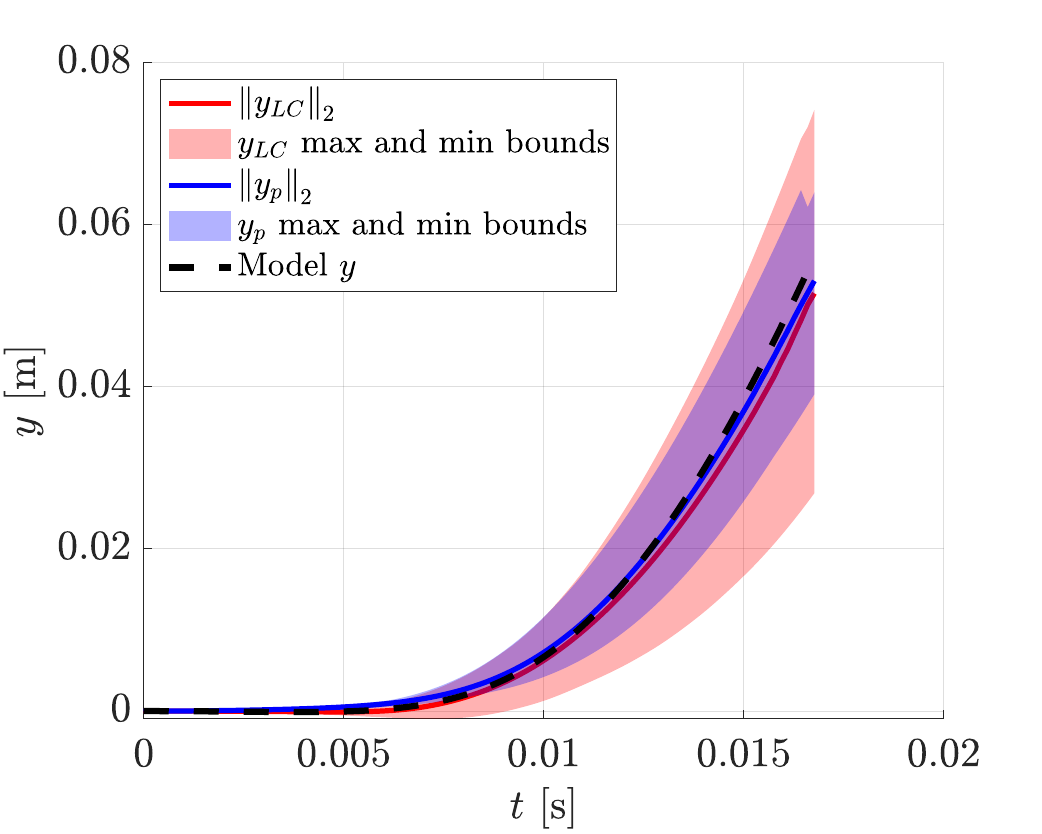}
        \caption[short]{Model and experimental ${y}$ at $p_o = $ 80 psi.}
        \label{fig:yvst-po80psi}
    \end{subfigure}
    \hfill
    \begin{subfigure}[t]{0.32\textwidth}
        \centering
        \includegraphics[width=\textwidth,trim={0cm 0cm 0cm 0cm},clip]{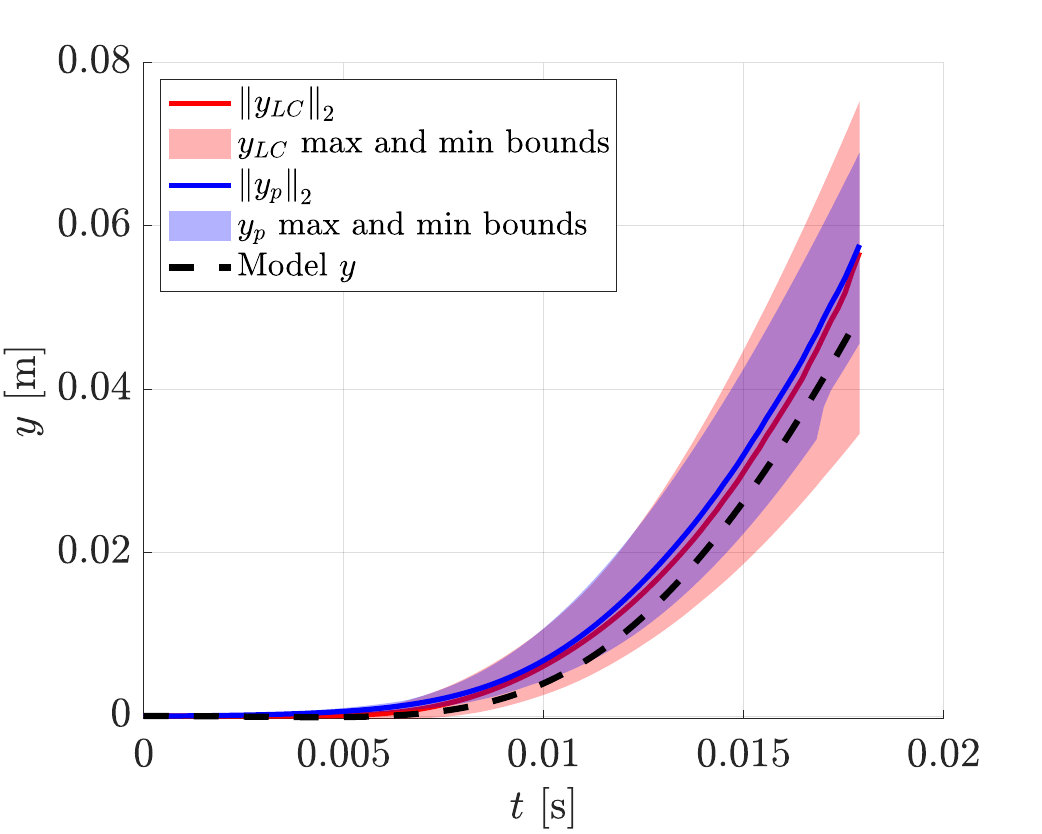}
        \caption[short]{Model and experimental ${y}$ at $p_o = $ 60 psi}
        \label{fig:yvst-po60psi}
    \end{subfigure}
   
    \caption{Model and experimental ${y}$ comparison for $p_o$ values of 100, 80, and 60 psi in experiments 1a, 1b, and 1c in Table~\ref{tab:experiment_log}.}
    \label{fig:yvst-po100-80-60psi}
\end{figure}
Figure~\ref{fig:yvst-po100-80-60psi} shows the same processed quantities as Fig.~\ref{fig:ydotvst-po100-80-60psi}, but for $y$. All min-max envelopes appear to have shrunk relative to those of $\dot{y}$, as expected given the low-pass nature of the dynamics. Interestingly, the modeled $y$ results in Figs.~\ref{fig:yvst-po100psi} and \ref{fig:yvst-po80psi} closely match the mean experimental $y$ curves, while Fig.~\ref{fig:yvst-po60psi} shows a slight underprediction of those same means. This can be attributed to the underprediction of the $\dot{y}$ curve early in the hammer-acceleration period at $p_o$ values of 100 and 80 psi, followed by overprediction later in the stroke, ending with a higher estimated $\dot{y}$ than both $\dot{y}_{LC}$ and $\dot{y}_{p}$.
This behavior appears to average out through the integration, such that the $y$ values match at the end of the stroke. This contrasts with the modeled $\dot{y}$ response in Fig.~\ref{fig:ydotvst-po60psi}, which matches both experimental values closely throughout the hammer stroke, but where a slight underprediction accumulates through the integral into a worse match for $y$ in Fig.~\ref{fig:yvst-po60psi}. Even so, the model appears to capture the hammer dynamics well when compared with the full experimental data set. 

\begin{figure}[H]
    \centering
    \begin{subfigure}[t]{0.32\textwidth}
        \includegraphics[width=\textwidth,trim={0cm 0cm 0cm 0cm},clip]{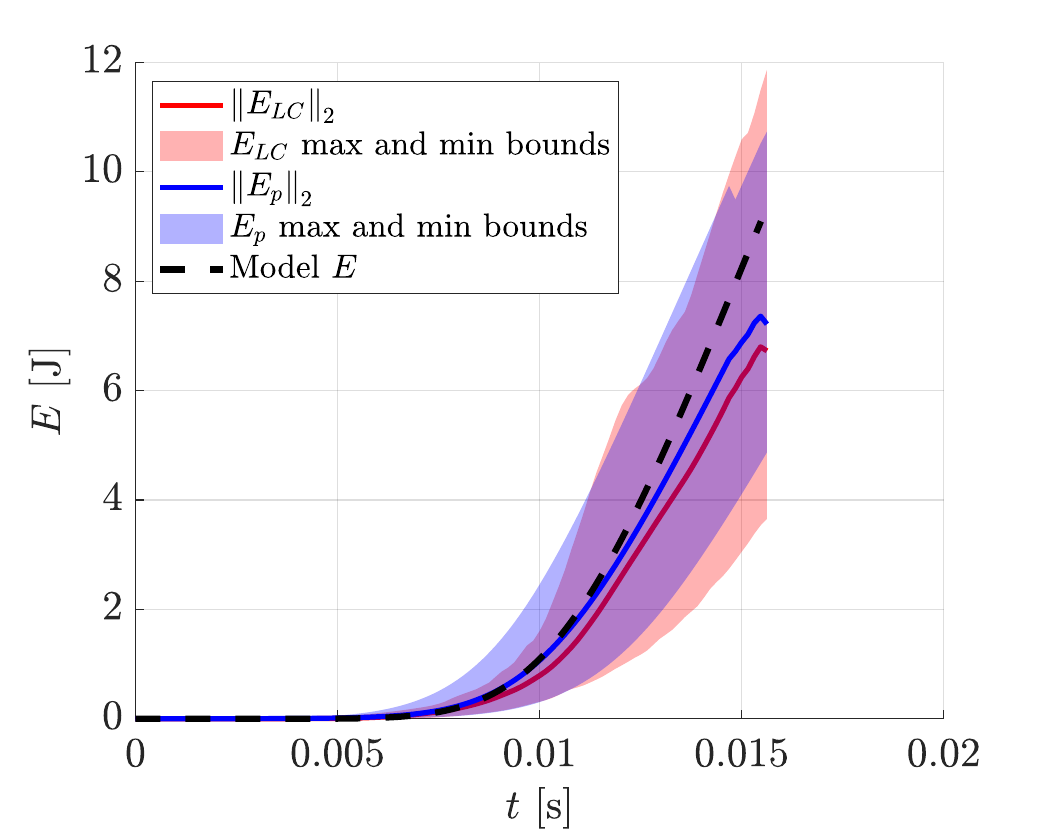}
        \caption[short]{Model and experimental ${E}$ at $p_o = $ 100 psi}
        \label{fig:Evst-po100psi}
    \end{subfigure}
    \hfill
    \begin{subfigure}[t]{0.32\textwidth}
        \centering
        \includegraphics[width=\textwidth,trim={0cm 0cm 0cm 0cm},clip]{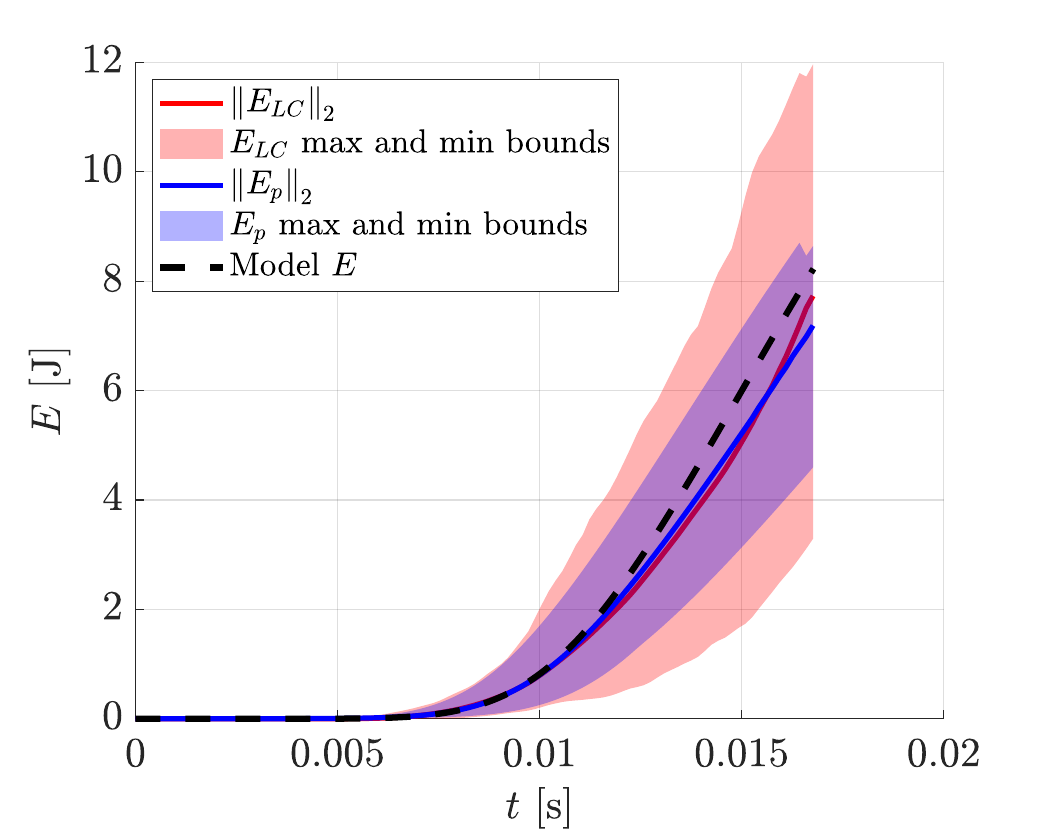}
        \caption[short]{Model and experimental ${E}$ at $p_o = $ 80 psi.}
        \label{fig:Evst-po80psi}
    \end{subfigure}
    \hfill
    \begin{subfigure}[t]{0.32\textwidth}
        \centering
        \includegraphics[width=\textwidth,trim={0cm 0cm 0cm 0cm},clip]{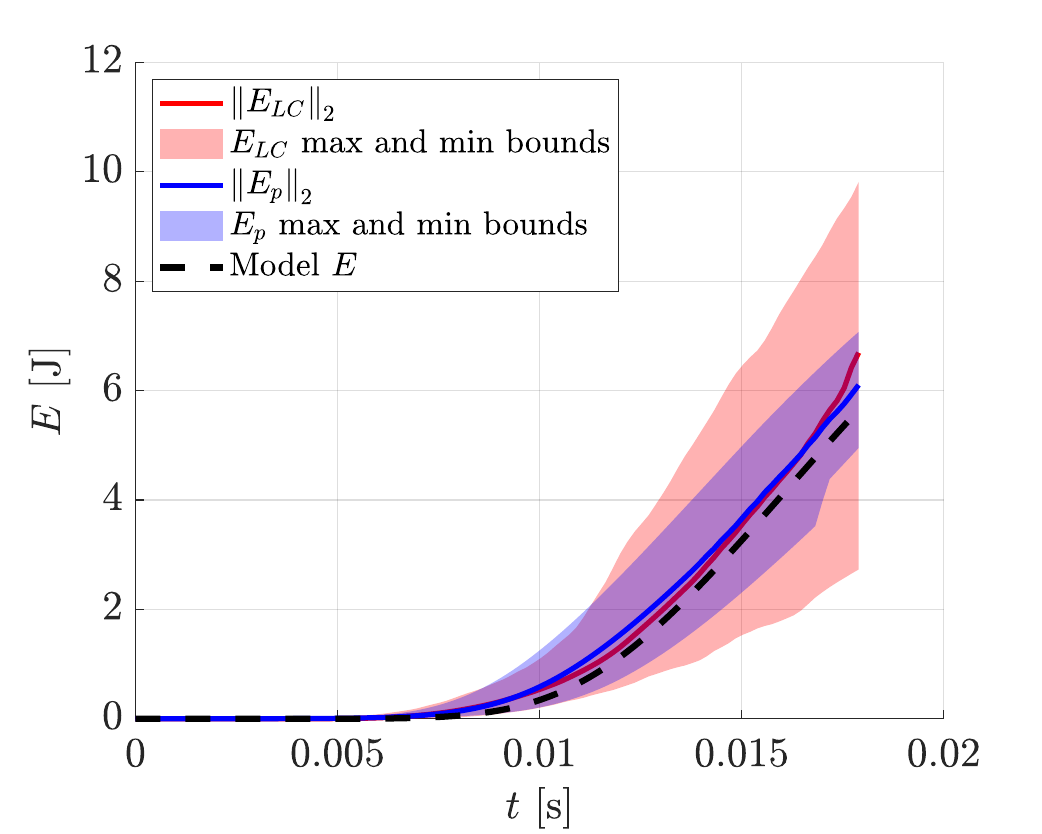}
        \caption[short]{Model and experimental ${E}$ at $p_o = $ 60 psi}
        \label{fig:Evst-po60psi}
    \end{subfigure}
   
    \caption{Model and experimental ${E}$ comparison for $p_o$ values of 100, 80, and 60 psi in experiments 1a, 1b, and 1c in Table~\ref{tab:experiment_log}.}
    \label{fig:Evst-po100-80-60psi}
\end{figure}
Figure~\ref{fig:Evst-po100-80-60psi} shows the same processed values as Figs.~\ref{fig:ydotvst-po100-80-60psi} and \ref{fig:yvst-po100-80-60psi}, but now for the main metric $E$. The trends are an amplified version of those in Fig.~\ref{fig:ydotvst-po100-80-60psi} because $E \sim \dot{y}^2$: the model overpredicts the experimental means in Fig.~\ref{fig:Evst-po100psi}, the estimate improves in Fig.~\ref{fig:Evst-po80psi}, and the agreement is closest in Fig.~\ref{fig:Evst-po60psi}. The end values at $t_{hp}$ still fall well within the min-max envelopes, again capturing both quantitative and qualitative features of the dynamics and the hammer impact-energy metric.
An interesting effect in Fig.~\ref{fig:Evst-po100psi} is that $E$ dips slightly at $t_{hp}$, similar to what occurs in Fig.~\ref{fig:ydotvst-po100psi}, yielding a slightly lower value. There are also small deviations in the final mean values for Figs.~\ref{fig:Evst-po80psi} and \ref{fig:Evst-po60psi}. This is likely a numerical artifact caused by the final common time-mesh point containing roughly 50\% fewer data points, which slightly skews the means. If the averaging limit is instead taken at the minimum $t_{hp}$ values from Table~\ref{tab:experimental_results_min_max}, the end-time anomalies disappear. The anomalous region is therefore confined to the small interval between the minimum and average values of $t_{hp}$.

\begin{table}[H]
\centering
\begin{tabular}{c c c c c}
\hline
\textbf{ID} & \textbf{$p_o$ [psi]} & \textbf{Model $E(\Vert t_{hp} \Vert_2)$ [J]} & \textbf{$\Vert E_p(\Vert t_{hp} \Vert_2) \Vert_2$ [J]} & \textbf{$\Vert E_{LC}(\Vert t_{hp} \Vert_2) \Vert_2$ [J]} \\ \hline
1a  & 100 & 9.10 & 7.22 & 6.73 \\
1b  & 80  & 8.23 & 7.19 & 7.73 \\
1c  & 60  & 5.56 & 6.10 & 6.70 \\ \hline
\end{tabular}%
    \caption{Mean hammer impact-energy comparison at the average experimental strike time $t_{hp}$ between the model and experiments 1a, 1b, and 1c from Table~\ref{tab:experiment_log}.}
\label{tab:mean_hammerenergy_mean_thp}
\end{table}
\begin{table}[H]
\centering
\resizebox{\textwidth}{!}{%
\begin{tabular}{c c c | c c | c c}
\hline
\textbf{ID} & \textbf{$p_o$ [psi]} & \textbf{Model $E(\Vert t_{hp} \Vert_2)$ [J]} & \textbf{$E_p(\Vert t_{hp} \Vert_2)$ max [J]} & \textbf{$E_p(\Vert t_{hp} \Vert_2)$ min [J]} & \textbf{$E_{LC}(\Vert t_{hp} \Vert_2)$ max [J]} & \textbf{$E_{LC}(\Vert t_{hp} \Vert_2)$ min [J]} \\ \hline
1a  & 100 & 9.10 & 10.7 & 4.87 & 11.9 & 3.66 \\
1b  & 80  & 8.23 & 8.64 & 4.60 & 12.0 & 3.29 \\
1c  & 60  & 5.56 & 7.07 & 4.95 & 9.82 & 2.73 \\ \hline
\end{tabular}%
}
    \caption{Minimum and maximum hammer impact-energy comparison at the average experimental strike time $t_{hp}$ between the model and experiments 1a, 1b, and 1c from Table~\ref{tab:experiment_log}.}
\label{tab:minmax_hammerenergy_mean_thp}
\end{table}
The final energy values at the mean $t_{hp}$ are listed in Table~\ref{tab:mean_hammerenergy_mean_thp}, with statistics in Table~\ref{tab:minmax_hammerenergy_mean_thp}. 
The true final energy values, using each method's calculated $t_h$, are listed in Table~\ref{tab:mean_hammerenergy_mean_th}.  
The trends seen in Fig.~\ref{fig:Evst-po100-80-60psi} are reflected in Tables~\ref{tab:mean_hammerenergy_mean_thp} and \ref{tab:minmax_hammerenergy_mean_thp}. The difference between the experimental values in Tables~\ref{tab:mean_hammerenergy_mean_thp} and \ref{tab:mean_hammerenergy_mean_th} suggests that hammer-acceleration profiles associated with the slower half of the measured $t_h$ values generally yield lower-than-expected performance in $E$. Experimental $E$ values for $p_o$ of 80 and 60 psi are particularly sensitive to this, with $\Vert E_p(\Vert t_{hp} \Vert_2) \Vert_2 - \Vert E_p(t_{hp}) \Vert_2 < 0$ and $\Vert E_{LC}\left(\Vert t_{hLC} \Vert_2 \right) \Vert_2 - \Vert E_{LC}(t_{hLC}) \Vert_2 < 0$, while the modeled energies gain 1 to 2 J between $\Vert t_{hp} \Vert_2$ and $t_h$.  
Exploring the root cause of this deviation is one of the future work items that could help explain how to make the impact energy more consistent at higher values. 
As noted, although model $E$ values tend to overpredict the experimental means, they fall within the bounds of both experimental methods at $\Vert t_{hp} \Vert_2$, and within the load-cell bounds for the final $t_h$ values shown in Table~\ref{tab:experimental_results_min_max}.    
\begin{table}[H]
\centering
\begin{tabular}{c c c c c}
\hline
\textbf{ID} & \textbf{$p_o$ [psi]} & \textbf{Model $E(t_h)$ [J]} & \textbf{$\Vert E_p(t_{hp}) \Vert_2$ [J]} & \textbf{$\Vert E_{LC}(t_{hLC}) \Vert_2$ [J]} \\ \hline
1a & 100 & 11.22 & 7.92 & 7.47 \\
1b & 80  & 9.31  & 7.12 & 7.56 \\
1c & 60  & 7.10  & 5.72 & 6.10 \\ \hline
\end{tabular}%
    \caption{Mean hammer impact-energy comparison at each method's calculated strike time between the model and experiments 1a, 1b, and 1c from Table~\ref{tab:experiment_log}.}
\label{tab:mean_hammerenergy_mean_th}
\end{table}
\begin{figure}[H]
    \centering
    \begin{subfigure}[t]{0.45\textwidth}
        \centering 
        \includegraphics[trim=0cm 0cm 0cm 0cm, clip, width=1\linewidth]{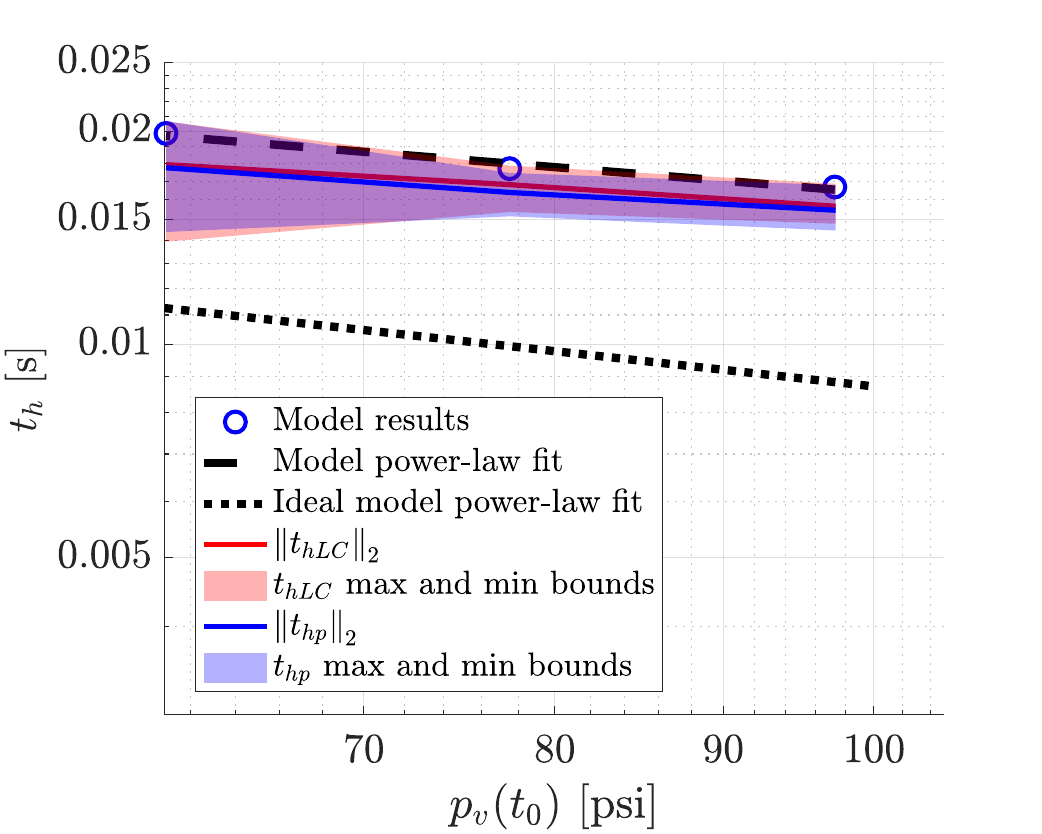}
        \caption{Experimental and model results of $t_{h}$ vs. mean values of $p_v(t_o)$ in log-log scale.}
        \label{fig:povsth_po100-80-60Comp}
    \end{subfigure}
    \hfill
    \begin{subfigure}[t]{0.45\textwidth}
        \centering
        \includegraphics[trim=0cm 0cm 0cm 0cm, clip, width=1\linewidth]{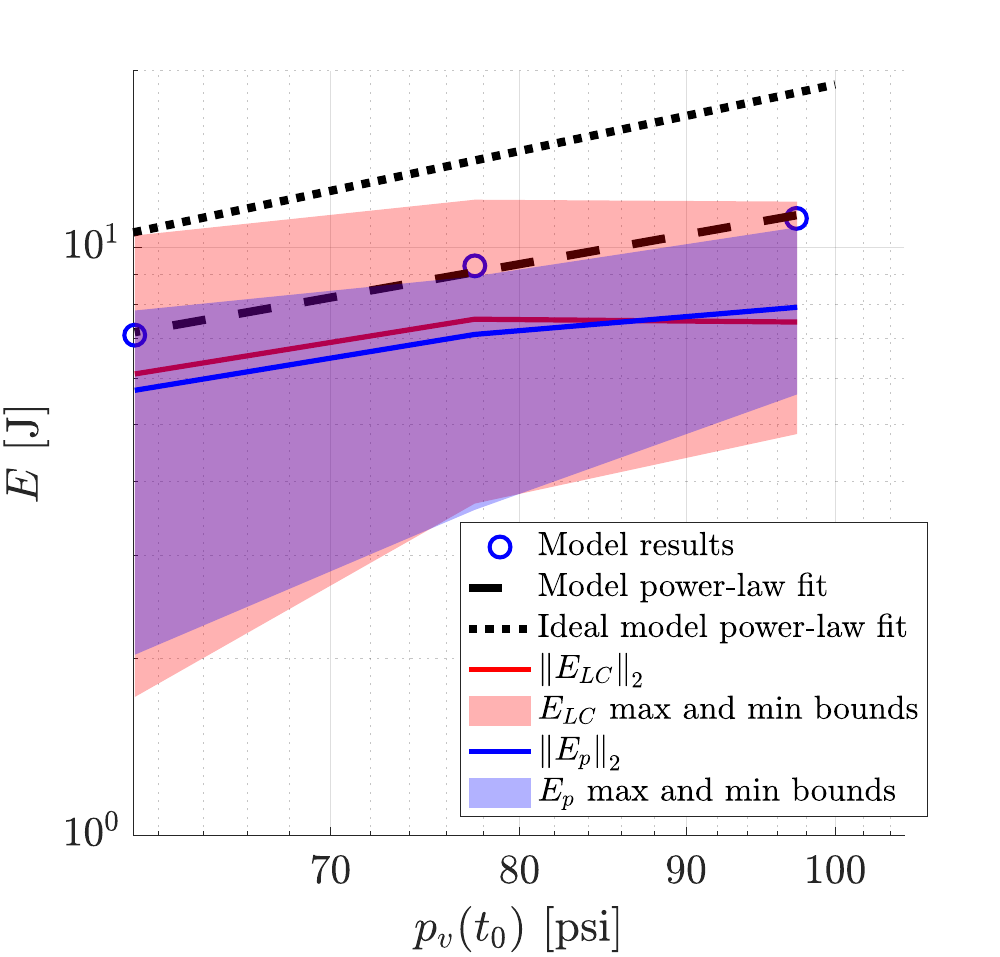}
        \caption{Experimental and model results of $E$ vs. mean values of $p_v(t_o)$ in log-log scale.}
        \label{fig:povsE_po100-80-60Comp}
    \end{subfigure}
    \caption{Modeled and experimental comparison of $t_h$ and $E$ vs. mean values of $p_v(t_o)$ for experiments 1a, 1b, and 1c in Table~\ref{tab:experiment_log}, including ideal-model results from Fig.~\ref{fig:ModelMultiple30-100}.}
    \label{fig:povsthE_po100-80-60Comp}
\end{figure}

Figure~\ref{fig:povsthE_po100-80-60Comp} shows the comparison of $t_{h}$ and $E$ as a function of $p_v\left(t_0 \right)$ for both model and experimental results, including model results produced with ideal parameters in Sec.~\ref{sec:idealModelResults}.  
Figure~\ref{fig:povsth_po100-80-60Comp} shows the trend discussed above: $t_h$ decreases with increasing $p_o \sim p_v\left(t_0 \right)$, and the size of the experimental min-max envelope also tends to decrease with increasing $p_o$.  
The calibrated model results in this section show a tendency to slightly overpredict $t_h$, but they still fall within the experimental min-max envelope.  
The ideal model, by contrast, is clearly separated, with considerably faster strikes than those observed experimentally.  
Figure~\ref{fig:povsE_po100-80-60Comp} shows the energy trends, where again the min-max envelope size decreases while $E$ increases with $p_o$.  
The exponent of the power-law coefficient is higher for the ideal model than for the calibrated-model results in this section, where $\alpha \approx 1$. The figure also shows that the calibrated model predictions fall within the load-cell measurement variation but effectively represent the upper bound of $E_p$.
The energy deficit between the experimental values and the ideal-model results is therefore clear. 
If design changes suggested by the calibrated model and experiments could be implemented such that (1) the flapper-valve opened \textit{faster}, so that the maximum $A_{v1}$ value was reached earlier in the hammer stroke; (2) losses inside the vent chamber were reduced to a negligible level; and (3) hammer side-wall friction was decreased, then the 100 psi case could approach 18 J rather than the calibrated-model prediction of 11 J, representing a 60\% increase.
Varying a number of other parameters from Table~\ref{tab:parametersTable} would also yield a better-performing drill. This is one of the main reasons for developing and validating the model presented in this manuscript. 

We conclude that the model is able to replicate the quantitative and qualitative trends of the dynamics well, and it typically predicts the upper bound of the energy level expected within an experiment. The experimentally reduced $E$ relative to the model expectation needs to be understood further, but it is likely caused by variations in the flapper-valve opening-area function $A_{v1}(t)$.  
As stated previously, the goal of the model is to capture the relevant physics and parametric trends that govern drill performance. In that regard, the formulation is largely successful and can be used as a design tool for future drill development.

\section{Rock Experimental Results} \label{sec:RockExperiments}

As described in Sec.~\ref{sec:SetupPercussiveTB}, a set of rock-simulant experiments followed the load-cell experiments, whose results are discussed in detail in Sec.~\ref{sec:LCExperimentsandModelComp}.  
The objective of the rock-simulant experiments, experimental sets 2 and 3 in Table~\ref{tab:experiment_log}, was to explore the performance of the \textit{WiP} drill hammer section on different Mars rock simulants using different drill-bits, and to provide both a baseline for comparison with other drill systems and a broader understanding of how drill-bit changes may affect drilling efficiency.
Two different Mars rock simulants, sandstone and Saddleback basalt, were tested using two drill-bits of the same tooth configuration but with outer diameters of 1-3/8'' and 1-5/8'' (see Fig.~\ref{fig:bits} and Table~\ref{tab:drillbits}). The simulants were selected to represent a relatively ``soft'' and ``hard'' rock, respectively. 
Both simulants were characterized through material tests in Sec.~\ref{sec:SimulantCharacterization} in conjunction with the drilling experiments. 
The performance of each rock-bit combination was evaluated using the methods developed and validated in Sec.~\ref{sec:LCExperimentsandModelComp} by measuring the hammer strike time $t_{hp}$ and calculating $\ddot{y}_p$, $\dot{y}_p$, $y_p$, and $E_p$ from the measured $p_h$.  
A distribution of $E_p$ was calculated for each experiment on a strike-by-strike basis and then summed to obtain total energies. The total hole volume was measured using a LIDAR scan ($\pm 1$ mm) to determine depth, while the drill-bit OD was measured with calipers (see Table~\ref{tab:drillbits}). The drill cross-sectional area was then multiplied by the depth to obtain the drilled volume. These values, when substituted into Eq.~\ref{eq:MSE}, give the mechanical specific energy metric (MSE), $\bar{E}$, for each combination.  
The MSE is a scalar measure of how much energy is required to remove a unit volume of material, with lower values indicating a more efficient system.  

\subsection{Simulant Characterization} \label{sec:SimulantCharacterization}

To better understand the mechanical behavior of the hard and soft Martian simulants, uniform Saddleback basalt (USB) and sandstone, respectively, under impact-drilling conditions, we conducted unconfined uniaxial compressive strength (UCS) tests on specimens extracted from the USB and sandstone samples. The specimens were cylindrical cores taken from the simulant rocks drilled in tests 2a--b and 3a--b in Table~\ref{tab:experiment_log}. 
Our objective was to provide a quantitative measure of rock strength for comparison with drilling performance between samples. Part of this study also investigated how material properties are affected by different strain rates, ranging from $1 \times 10^{-3} \, \text{s}^{-1}$ to $2 \times 10^{-2} \, \text{s}^{-1}$. To account for the intrinsic variability of these rock types, we repeated tests for each nominal condition and performed experiments on two batches of specimens collected and manufactured at different times \cite{schormair2006influence, kumar1968effect, lindholm1974dynamic, zhang2020dynamic}.

Despite the variability of the rock samples, consistent results were obtained for stiffness, strength, resilience, and strain energy density for each batch and strain-rate regime. Density values were $2655 \pm 86$ kg/m$^3$
for the USB batches, and $2024 \pm 24$ kg/m$^3$ for the sandstone batches, aligning with published values for standard test rocks such as Berea sandstone ($2110$ kg/m$^3$) and Dresser basalt \cite{zhang2020dynamic, malik2018strain, shangxin2020estimation, teale1965concept, bruno2005fundamental}. The results are summarized in Table~\ref{tab:UCSMaterial}.

\begin{table}[ht!]
\centering
\resizebox{\textwidth}{!}{%
\begin{tabular}{l l c c c c c}
\hline
\textbf{Simulant} & \textbf{Batch} & \textbf{Strain Rate (1/s)} & \textbf{Stiffness (GPa)} & \textbf{Strength (MPa)} & \textbf{Resilience (kJ/m$^3$)} & \textbf{Strain Energy Density (kJ/m$^3$)} \\ \hline
USB & A         & $1 \times 10^{-3}$ & $7.8$ & $77$ & $493$ & $493$ \\
USB & A         & $2 \times 10^{-2}$ & $107$ & $89$ & $206$ & $222$ \\
USB & B         & $1 \times 10^{-3}$ & $6.8$ & $65$ & $505$ & $605$ \\
USB & B         & $2 \times 10^{-2}$ & $214$ & $82$ & $24$ & $65$ \\  \hline
Sandstone & C   & $1 \times 10^{-3}$ & $3.69$ & $28$ & $103$ & $147$ \\
Sandstone & C   & $2 \times 10^{-2}$ & $4.11$ & $30$ & $104$ & $147$ \\
Sandstone & D   & $1 \times 10^{-3}$ & $3.90$ & $31$ & $127$ & $139$ \\
Sandstone & D   & $2 \times 10^{-2}$ & $3.19$ & $26$ & $105$ & $146$ \\ \hline
\end{tabular}
}
\caption{Mechanical properties of USB and sandstone under different strain rates.}
\label{tab:UCSMaterial}
\end{table}

Our findings indicate that material strength increased with strain rate, with a transition between $1 \times 10^{-3} \, \text{s}^{-1}$ and $2 \times 10^{-2} \, \text{s}^{-1}$, a phenomenon commonly referred to as strain-rate hardening. This effect occurs because faster loading rates allow less time for plastic deformation, resulting in higher stresses required for failure \cite{gong2019peak, lajtai1991effect, kumar1968effect, zhang2020dynamic}. The increase in strength was accompanied by a significant decrease in strain energy density, indicating increased embrittlement under faster loading conditions. Such findings have important implications for designing drilling equipment for Mars, where percussive drilling could be more energy efficient than rotary drilling in brittle rock types that become more susceptible to fracturing at higher strain rates \cite{schormair2006influence, shangxin2020estimation, kahraman2003dominant}.

These observations are consistent with previous studies on basaltic and other terrestrial rocks, which show that rock strength and stiffness generally increase with strain rate \cite{qi2009strain, gong2019peak, lajtai1991effect, kumar1968effect, lindholm1974dynamic, malik2018strain}. The values reported in Table~\ref{tab:UCSMaterial} are in agreement with those documented for basaltic rocks \cite{schultz1995limits, kumar1968effect, tandanand1975drillability}, albeit slightly lower, especially for stiffness. This discrepancy can be attributed to the natural variability in rock properties, even among samples of the same type.

The greater variability in strength at higher strain rates may be due to the activation of more cracks during deformation, rather than elastic or plastic deformation alone, as well as variability in the number and orientation of pre-existing cracks across samples \cite{lindholm1974dynamic, zhang2020dynamic}. These factors highlight the importance of collecting and analyzing multiple samples when characterizing rock mechanical properties.

\subsection{Normal vs. Abnormal Strike Classification and Integration Length}
Table~\ref{tab:experiment_log} outlines the load-cell and rock experiments carried out in all experimental campaigns, including the total number of strikes analyzed during each run. In Secs.~\ref{sec:idealModelResults} and \ref{sec:LCExperimentsandModelComp}, we introduced the distinction between a normal strike and a strike that deviates from the expected $p_h$ and $p_e$ trends. In normal operation, event~\ref{evn:A_ValveOpens} marks flapper-valve opening, followed sequentially by event~\ref{evn:B_Peakph}, the maximum in $p_h$; event~\ref{evn:C_HammerStrikes}, the hammer strike; and event~\ref{evn:D_ValveCloses}, closure of the flapper-valve and recharging of the vent chamber by $\dot{m}_{in}$.
When the algorithm cannot identify these events within a strike, or indicates that they occur out of order, the strike is classified as abnormal.
Strikes labeled as \textit{missed} are those flagged as possible hammer-acceleration events by Algorithm~\ref{alg:HammerStrikeDetection} but that could not be processed because they were either false positives or low-amplitude pressure fluctuations that should not have been classified as strikes.
Table~\ref{tab:rockExperiment_results_mean} lists the strike breakdown for each rock-simulant experiment, along with the other quantitative results. Missed strikes account for less than 10\% of the total possible strikes, i.e. $N_{total} + N_{missed}$.

\begin{figure}[H]
    \centering
    \begin{subfigure}[t]{0.47\textwidth}
        \centering 
        \includegraphics[width=1\linewidth]{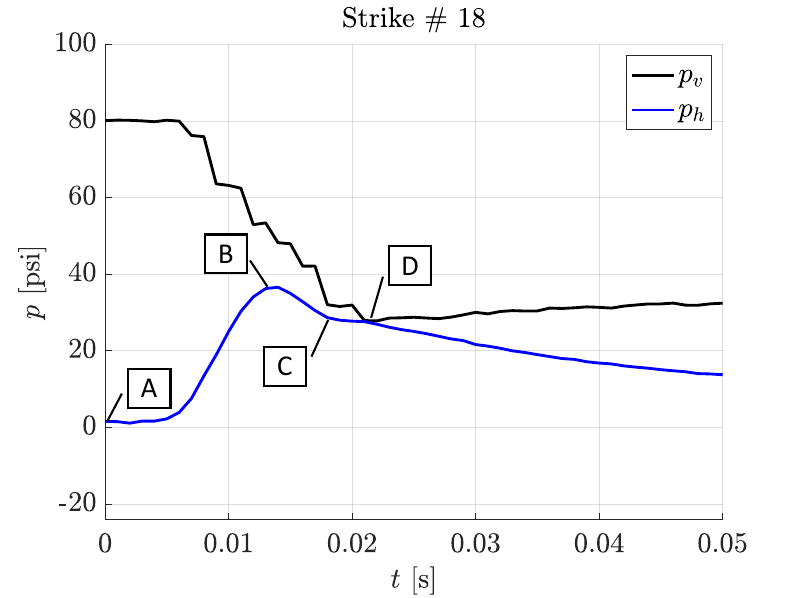}
        \caption[short]{Example of $p_v$ and $p_h$ time-series for a strike classified as \textit{normal} operation, with key events labeled.}
        \label{fig:ExprNormalStrikeExample}
    \end{subfigure}
    \begin{subfigure}[t]{0.47\textwidth}
        \centering
        \includegraphics[width=1\linewidth]{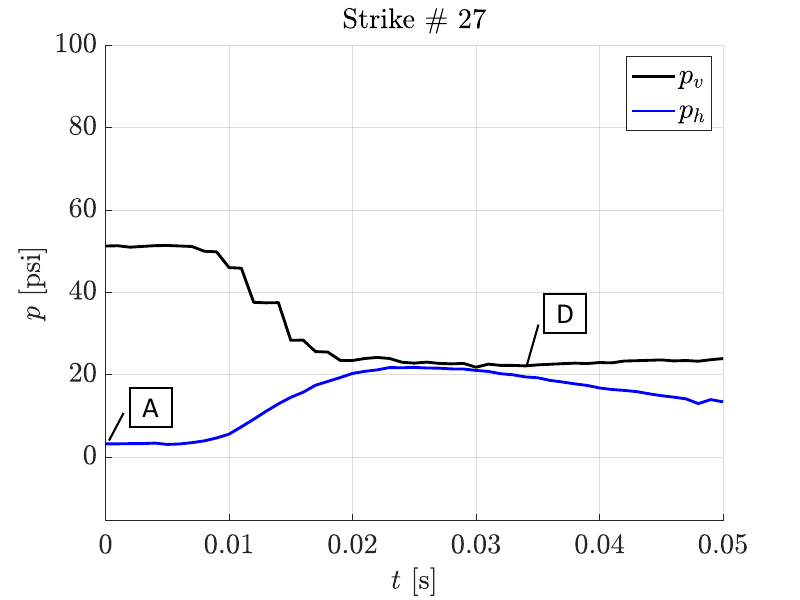}
        \caption[short]{Example of $p_v$ and $p_h$ time-series for a strike classified as \textit{abnormal} operation, with key events labeled.}
        \label{fig:ExprAbnormalStrikeExample}
    \end{subfigure}
    \caption{Single-strike examples for normal and abnormal classifications from experiment 2a in Table~\ref{tab:experiment_log}.}
    \label{fig:ExprNormalAbnormalStrikeExample}
\end{figure}

Figure~\ref{fig:ExprNormalAbnormalStrikeExample} shows example pressure traces from experiment 2a for both a normal strike (Fig.~\ref{fig:ExprNormalStrikeExample}) and an abnormal strike (Fig.~\ref{fig:ExprAbnormalStrikeExample}).
In the abnormal case, no clear $t_{hp}$ can be identified, nor is there a decisive peak in $p_h$ before the flapper-valve closes at event~\ref{evn:D_ValveCloses}. However, the vent-chamber gas clearly discharges into the hammer chamber, as indicated by the sharp drop in $p_v$ and the corresponding rise in $p_h$, even though the response differs from the expected normal-strike regime.
Review of the test videos confirms that these abnormal strikes still transfer momentum to the drill-bit, as evidenced by the audible impacts and the recoil of the drill after each event.
Because the goal is to account for the total energy delivered to the rock during each experiment, these strikes cannot be neglected. For experiments 2a, 2b, and 3b, abnormal strikes constitute more than 10\% but less than 50\% of the analyzed strikes, whereas in experiment 3a they account for roughly 80\%.

\begin{figure}[htb!]
    \centering
    \begin{subfigure}[t]{0.45\textwidth}
        \includegraphics[width=\textwidth,trim={0cm 0cm 0cm 0cm},clip]{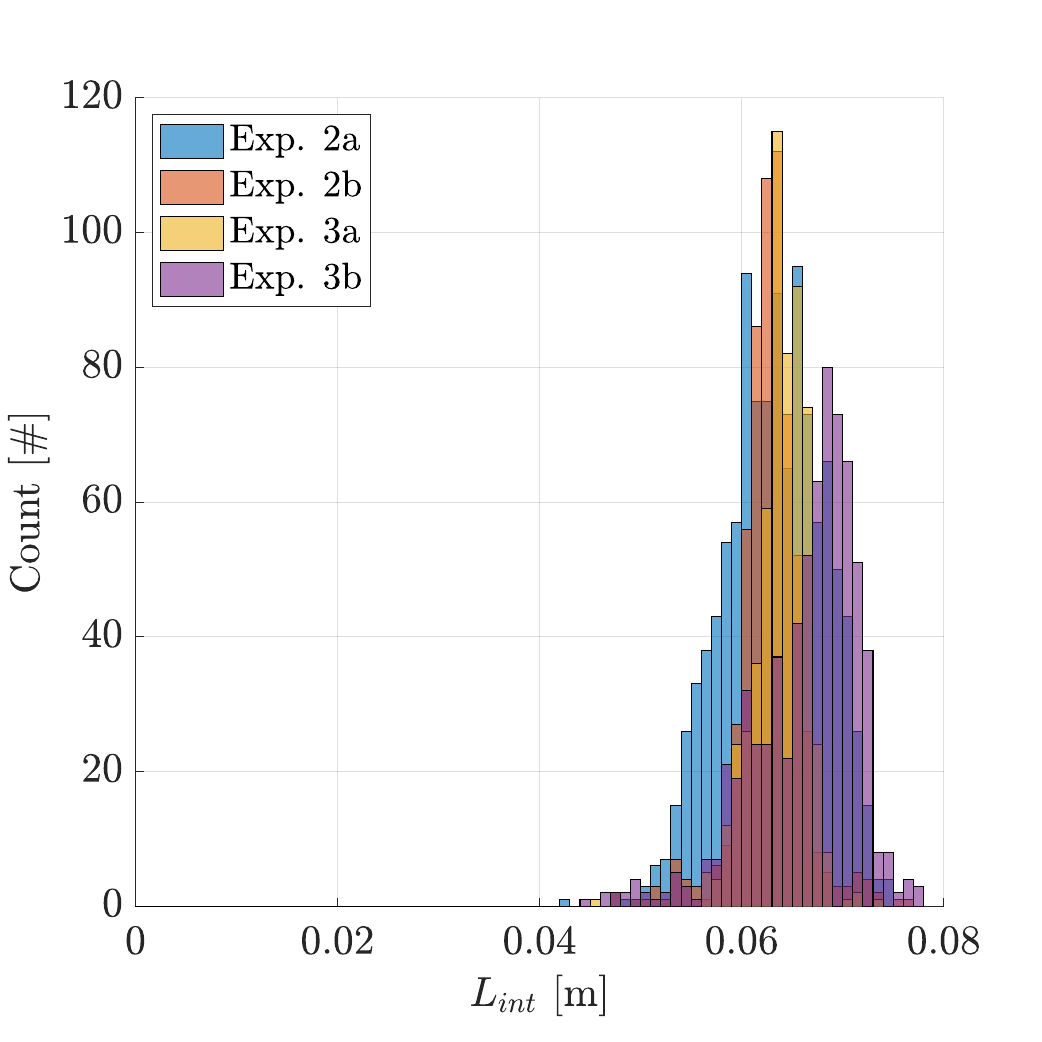}
        \caption[short]{$L_{int}$ distribution over all normal strikes analyzed.} 
        \label{fig:HistLint_Hole12567}
    \end{subfigure}
    \hfill
    \begin{subfigure}[t]{0.45\textwidth}
        \centering
        \includegraphics[width=\textwidth,trim={0cm 0cm 0cm 0cm},clip]{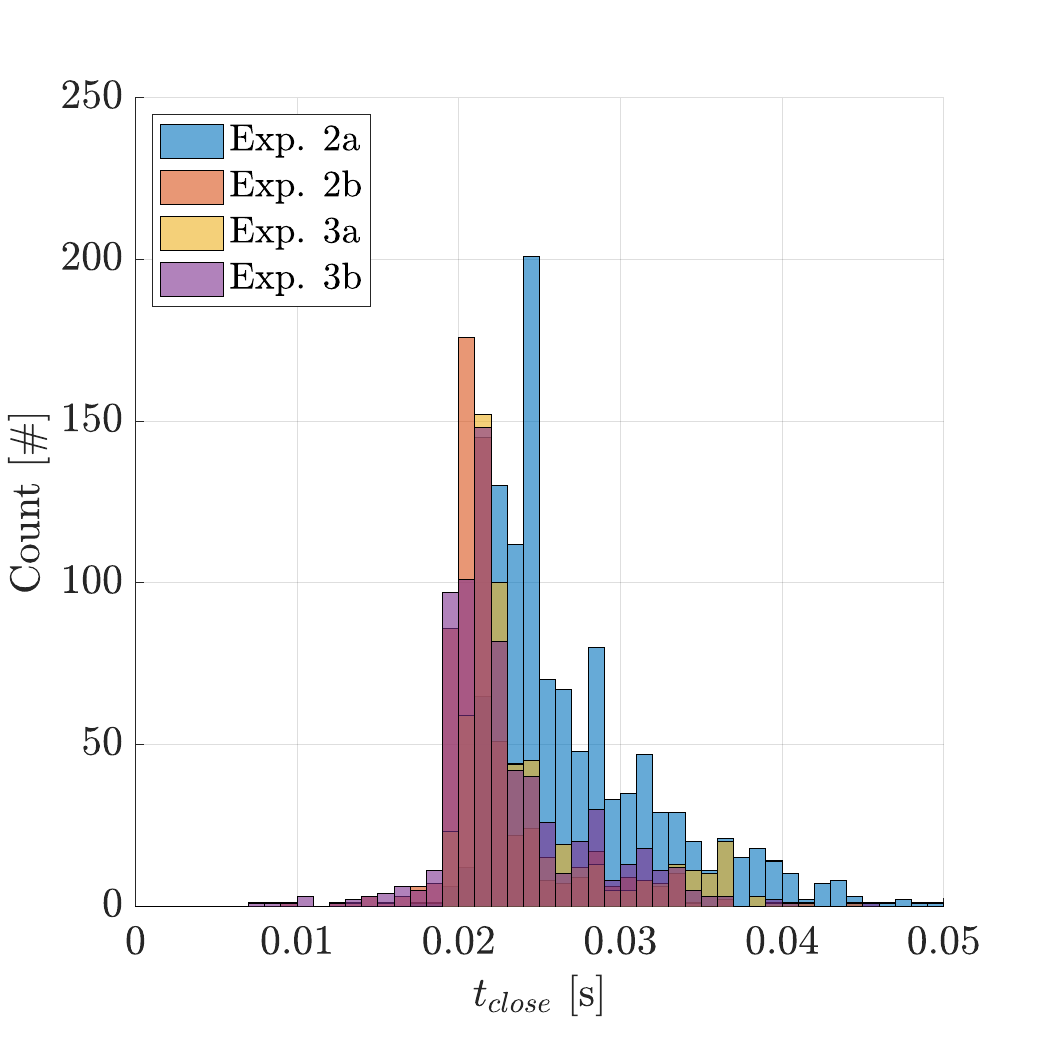}
        \caption[short]{$t_{close}$ distribution over all normal strikes analyzed.}
        \label{fig:Histtclose_Hole12567}
    \end{subfigure}
    \hfill
    \begin{subfigure}[t]{0.45\textwidth}
        \centering
        \includegraphics[width=\textwidth,trim={0cm 0cm 0cm 0cm},clip]{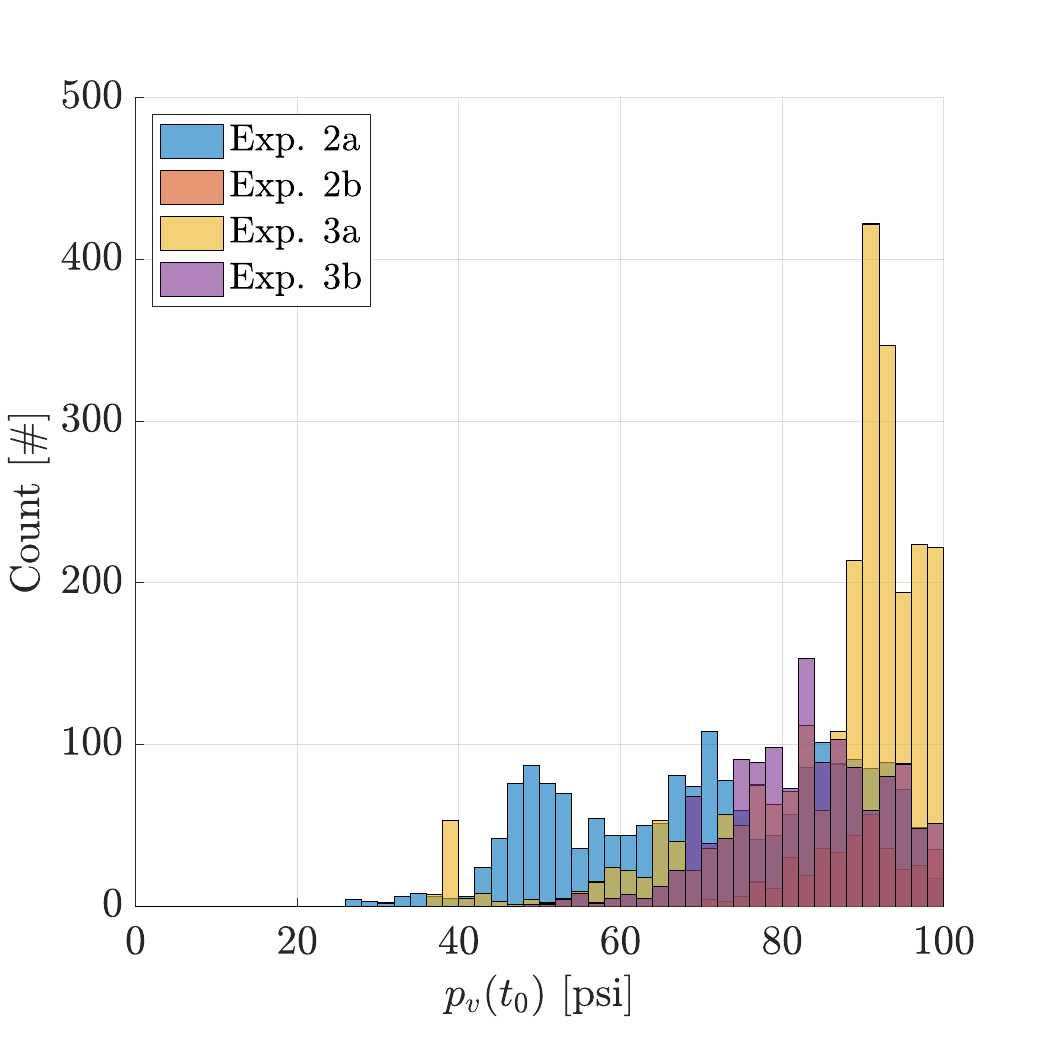}
        \caption[short]{$p_v(t_0)$ distribution over all normal and abnormal strikes analyzed.}
        \label{fig:Histpv0_Hole12567}
    \end{subfigure}
     \hfill
    \begin{subfigure}[t]{0.45\textwidth}
        \centering
        \includegraphics[width=\textwidth,trim={0cm 0cm 0cm 0cm},clip]{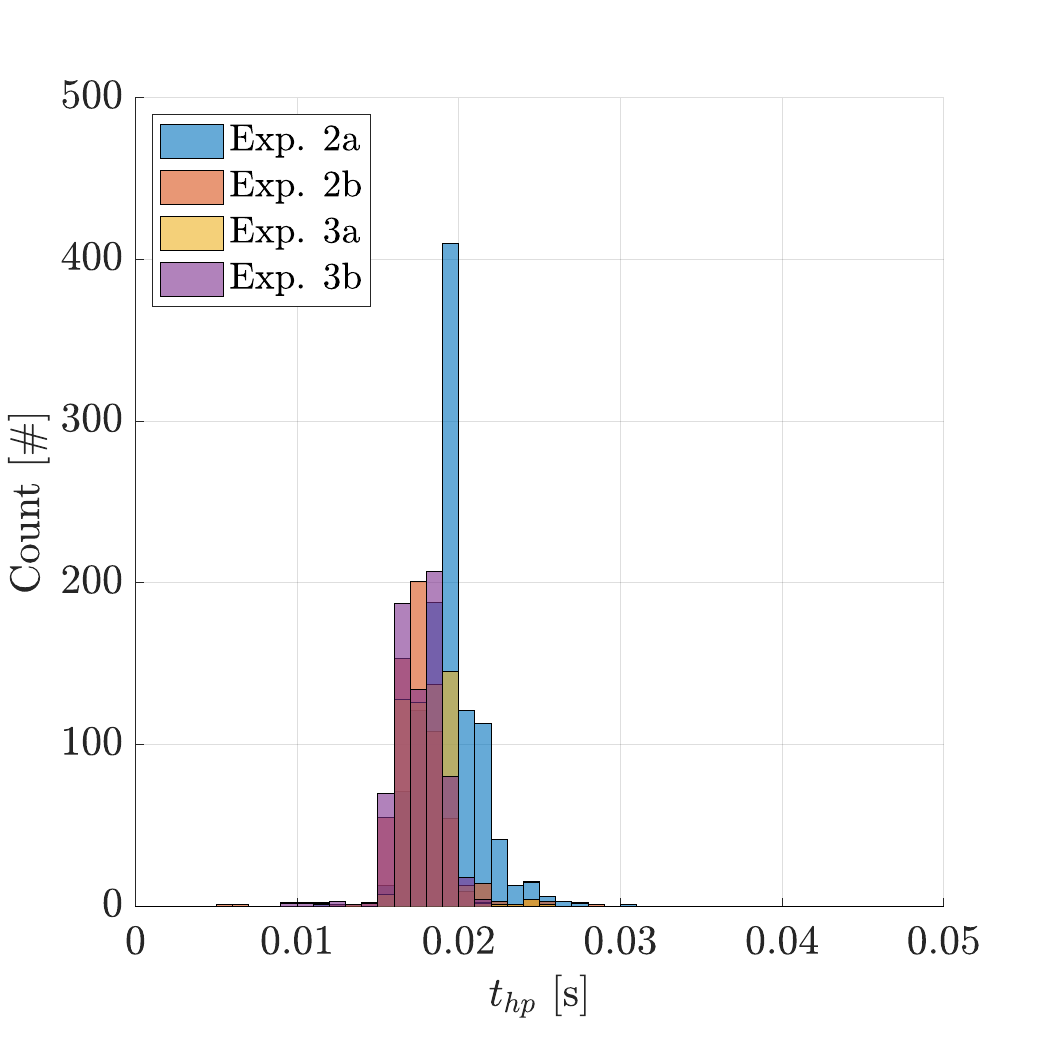}
        \caption[short]{$t_{hp}$ distribution over all normal strikes analyzed.}
        \label{fig:Histthp_Hole12567}
    \end{subfigure}
   
    \caption{Relevant parameter distributions for experiments 2a, 2b, 3a, and 3b in Table~\ref{tab:experiment_log}. Note that $L_{int}$ and $t_{close}$ are plotted for normal strikes only, while $p_v(t_0)$ includes both normal and abnormal strikes.}
    \label{fig:Hist_Hole12567}
\end{figure}

Although the root cause of the abnormal $p_h$ traces requires further investigation, a plausible explanation is variation in the opening-area function $A_{v1}$ caused by motion of the control magnet during an individual strike. Specifically, the interface between the component that retracts the magnet and the magnet holder is not perfectly constrained. Vibration during repeated 5--10 J impacts could shift the magnet vertically enough to change the magnetic force, which in turn would affect the valve opening pressure, the flapper-valve sealing force, and ultimately the time history of $A_{v1}$. A small post-closure leak between the vent and hammer chambers is one possible manifestation of this behavior.
This interpretation is partly supported by the broad distributions of $p_v\left(t_0 \right)$ in Fig.~\ref{fig:Histpv0_Hole12567}, especially when compared with Fig.~\ref{fig:histPvmax_po100-80-60} in Sec.~\ref{sec:LCExperimentsandModelComp}. Any substantial deviation of $p_v(t_0)$ from the nominal $p_o = 100$ psi implies that the flapper-valve opening pressure has shifted. In load-cell experiment 1a, the values cluster tightly near 100 psi, whereas in the rock-simulant experiments they span a much wider range and extend down to about 20 psi.

The primary issue with abnormal strikes is the absence of a measured $t_{hp}$, which is needed by Algorithm~\ref{alg:y0_iterative} to define the ODE solution and the integration length $L_{int}$ that sets the end of the strike and therefore the hammer kinetic energy. Nevertheless, two pieces of information can be used to bound or estimate the integration length. The first is the absolute upper bound,
\begin{equation} \label{eq:AbnormalStrikeLmaxCond}
    \left. L_{int} \right|_{abn}  = L_{max},
\end{equation}
where the subscripts \textit{norm} and \textit{abn} restrict the variable to strikes classified as normal or abnormal, respectively. Equation~\ref{eq:AbnormalStrikeLmaxCond} sets $L_{int}$ to its maximum allowed value under the constraint in Eq.~\ref{eq:LintLmaxcondition}. Assuming that the hammer traverses $L_{max}$ during an abnormal strike will likely overestimate the strike energy, because it allows the hammer to accelerate longer than if $\left. L_{int} \right|_{abn} < L_{max}$.
The second plausible assumption is that $L_{int}$ does not vary substantially between normal and abnormal strikes, so that the abnormal-strike integration length can be estimated from the average value of the normal strikes,
\begin{equation} \label{eq:AbnormalStrikeLintaveCond}
    \left. L_{int} \right|_{abn}  \approx \left. \Vert L_{int} \Vert_2 \right|_{norm}.
\end{equation}
Equation~\ref{eq:AbnormalStrikeLintaveCond} therefore provides a best estimate of the most likely strike energy given the absence of a measured $t_{hp}$.
Accordingly, we report energy estimates for experiments 2 and 3 using both Eqs.~\ref{eq:AbnormalStrikeLmaxCond} and \ref{eq:AbnormalStrikeLintaveCond}, providing an upper bound and a likely estimate for the total energy in each run. The MSE obtained from Eq.~\ref{eq:AbnormalStrikeLmaxCond} is therefore the most conservative, i.e. least efficient, estimate for the subset of abnormal strikes.
The symbols and definitions used in the experimental-results tables are again summarized in Table~\ref{tab:exprvariables_description}.

\subsection{Results}
\begin{figure}[htb!]
    \centering
    \begin{subfigure}[t]{0.45\textwidth}
        \centering 
        \includegraphics[trim=0cm 0cm 0cm 0cm, clip, width=1\linewidth]{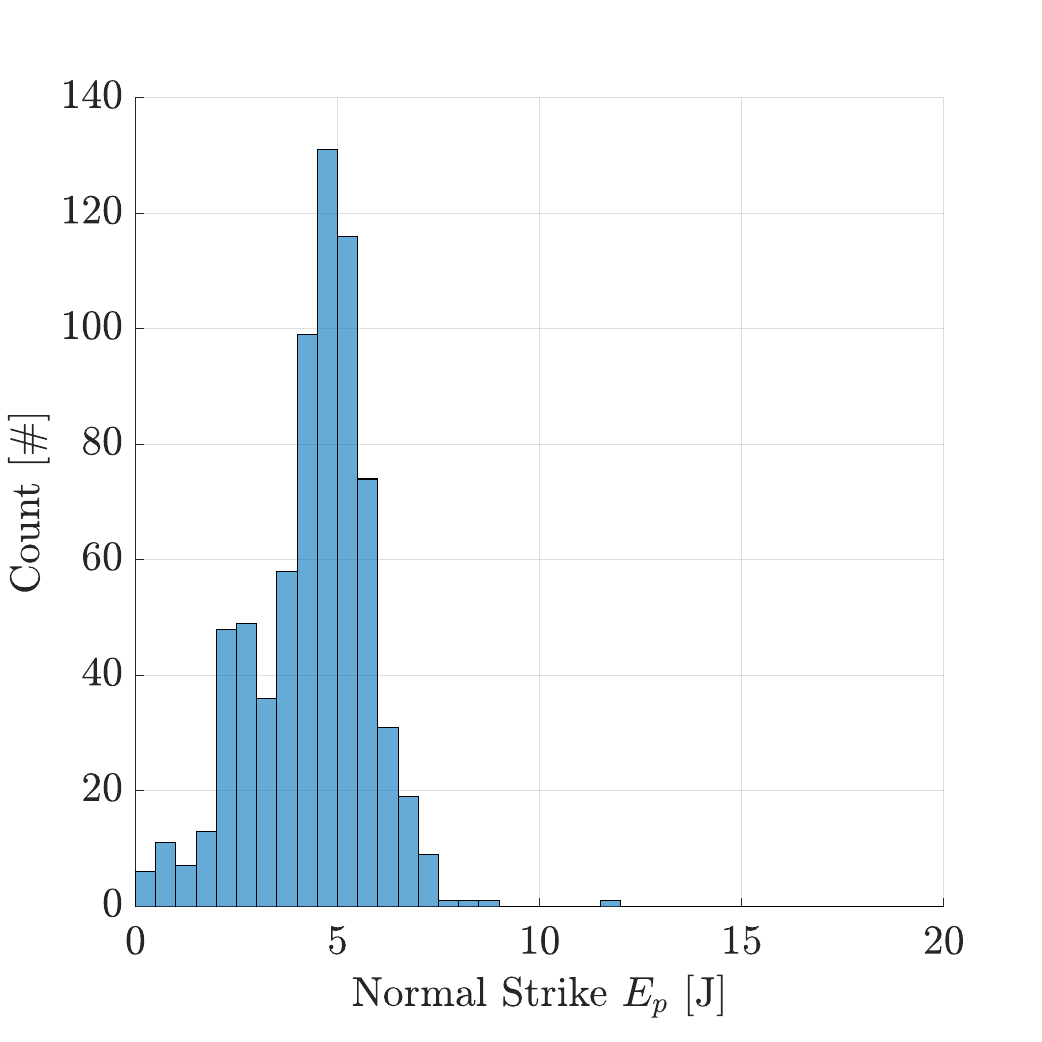}
        \caption{$E_p$ distribution at $t_{hp}$ classified under normal operation.}
        \label{fig:HistEpNormal_Hole6}
    \end{subfigure}
    \hfill
    \begin{subfigure}[t]{0.45\textwidth}
        \centering
        \includegraphics[trim=0cm 0cm 0cm 0cm, clip, width=1\linewidth]{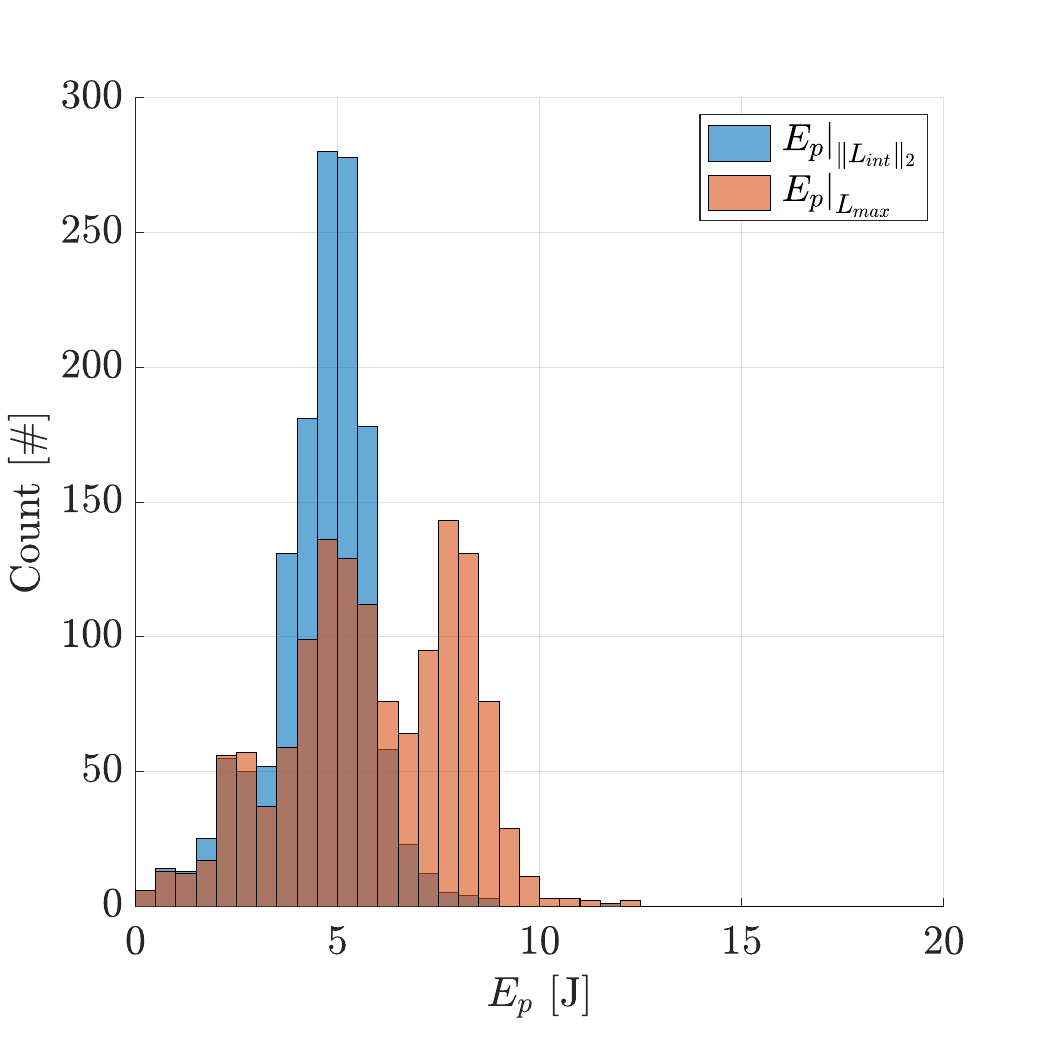}
        \caption{$E_p$ distribution at the end of all strikes, with those classified under abnormal operation either integrated to $\Vert L_{int} \Vert_2$ of normal strikes or $L_{max}$.}
        \label{fig:HistEpAbnormal_Hole6}
    \end{subfigure}
    \caption{Hammer kinetic-energy profiles for normal and all-strike classifications for experiment 3b in Table~\ref{tab:experiment_log}.}
    \label{fig:HistEp_Hole6}
\end{figure}

We begin by examining the remaining plots in Fig.~\ref{fig:Hist_Hole12567}, which show the distributions of $L_{int}$ (Fig.~\ref{fig:HistLint_Hole12567}) and $t_{close}$ (Fig.~\ref{fig:Histtclose_Hole12567}) for normal strikes only.
Neither drill-bit selection nor rock simulant is expected to directly change the energy generated by the hammer section itself. Any variance observed in the scalar parameters in Fig.~\ref{fig:Hist_Hole12567} must therefore arise upstream of the bit. Because all experiments in sets 2 and 3 were run at $p_o = 100$ psi, one would otherwise expect hammer-section performance similar to that in experiment 1a.
Even when only normal strikes are considered, the distributions of these output variables are much broader than those in Fig.~\ref{fig:hist_po100-80-60}. Yet the mean values, listed in Table~\ref{tab:rockExperiment_results_mean}, still remain reasonably close to those of experiment 1a in Table~\ref{tab:experimental_results_mean}.
As discussed in connection with Fig.~\ref{fig:povsthE_po100-80-60Comp}, even modest variation in $t_{hp}$ indicates valve behavior that underdelivers relative to the model expectation for $E$. Figure~\ref{fig:Histthp_Hole12567} shows a much broader distribution of $t_{hp}$ than the 1a distribution in Fig.~\ref{fig:histthp_po100-80-60}, so a broad distribution of $E_p$ is expected as well. Overall, Fig.~\ref{fig:Hist_Hole12567} shows that the hammer performance during the rock experiments is substantially more variable than in load-cell experiment 1a and in the calibrated model results.
Understanding the source of this input-parameter variability is likely key to improving drill performance and producing consistently high-energy strikes.
\begin{figure}[htb!]
    \centering
    \begin{subfigure}[t]{0.45\textwidth}
        \centering 
        \includegraphics[trim=0cm 0cm 0cm 0cm, clip, width=1\linewidth]{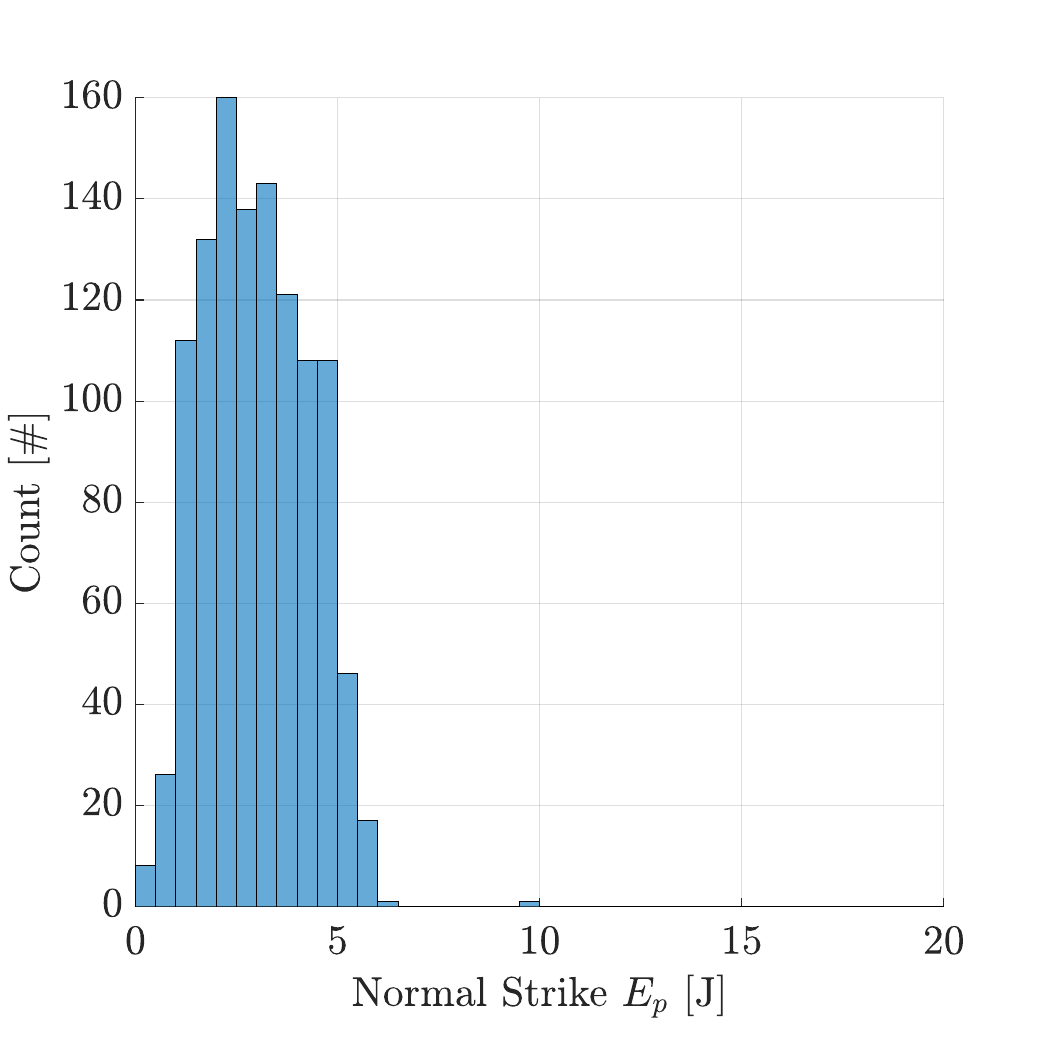}
        \caption{$E_p$ distribution at $t_{hp}$ classified under normal operation.}
        \label{fig:HistEpNormal_Hole1n2}
    \end{subfigure}
    \hfill
    \begin{subfigure}[t]{0.45\textwidth}
        \centering
        \includegraphics[trim=0cm 0cm 0cm 0cm, clip, width=1\linewidth]{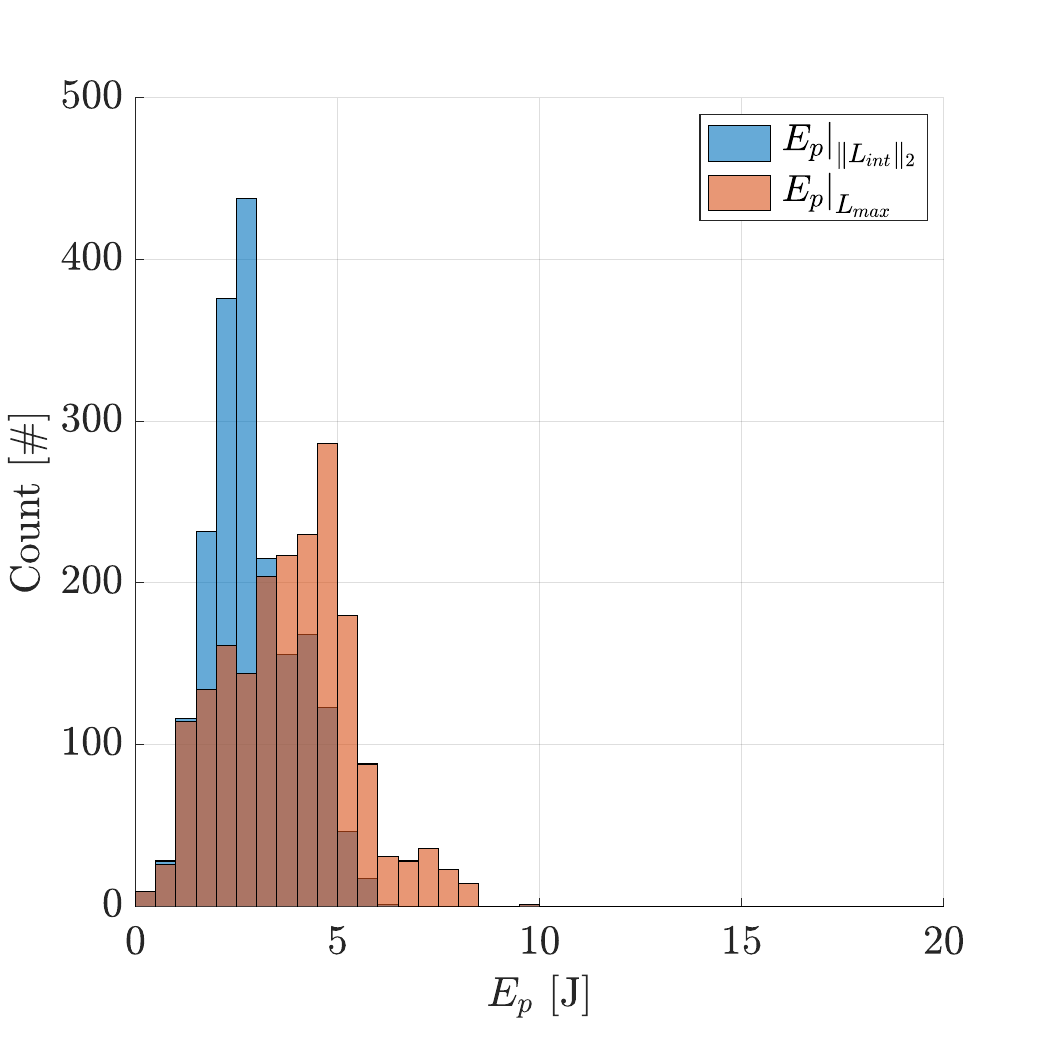}
        \caption{$E_p$ distribution at the end of all strikes, with those classified under abnormal operation either integrated to $\Vert L_{int} \Vert_2$ or $L_{max}$.}
        \label{fig:HistEpAbnormal_Hole1n2}
    \end{subfigure}
    \caption{Hammer kinetic-energy profiles for normal and all-strike classifications for experiment 2a in Table~\ref{tab:experiment_log}.}
    \label{fig:HistEp_Hole1n2}
\end{figure}
Next, a series of figures shows the distribution of $E_p$ for each rock-simulant experiment, 2a, 2b, 3a, and 3b, in Table~\ref{tab:experiment_log}. Each figure contains two plots: one for normal-strike $E_p$, and a second showing the total $E_p$ distribution for both normal and abnormal strikes, first assuming a maximum strike length (Eq.~\ref{eq:AbnormalStrikeLmaxCond}) and then assuming a strike length equal to the average of the normal strikes (Eq.~\ref{eq:AbnormalStrikeLintaveCond}). Figure~\ref{fig:HistEp_Hole1n2} shows the $E_p$ values for experiment 2a, where the normal-strike distribution centers around 3--4 J per strike (see Fig.~\ref{fig:HistEpNormal_Hole1n2}). Figure~\ref{fig:HistEpAbnormal_Hole1n2} then shows how that distribution changes once the two different $\left. L_{int} \right|_{abn}$ assumptions are applied: the conservative estimate based on Eq.~\ref{eq:LintLmaxcondition} shifts the center visibly to the right, closer to 5 J, and adds a series of strikes near 9 J; in contrast, assuming that the average $L_{int}$ remains the same between normal and abnormal classifications yields a similar but amplified version of the normal distribution in Fig.~\ref{fig:HistEpNormal_Hole1n2}. The mean results are listed in Table~\ref{tab:rockExperiment_results_mean}, with statistics in Table~\ref{tab:rockExperiment_results_variance}. 
\begin{figure}[H]
    \centering
    \begin{subfigure}[t]{0.45\textwidth}
        \centering 
        \includegraphics[trim=0cm 0cm 0cm 0cm, clip, width=1\linewidth]{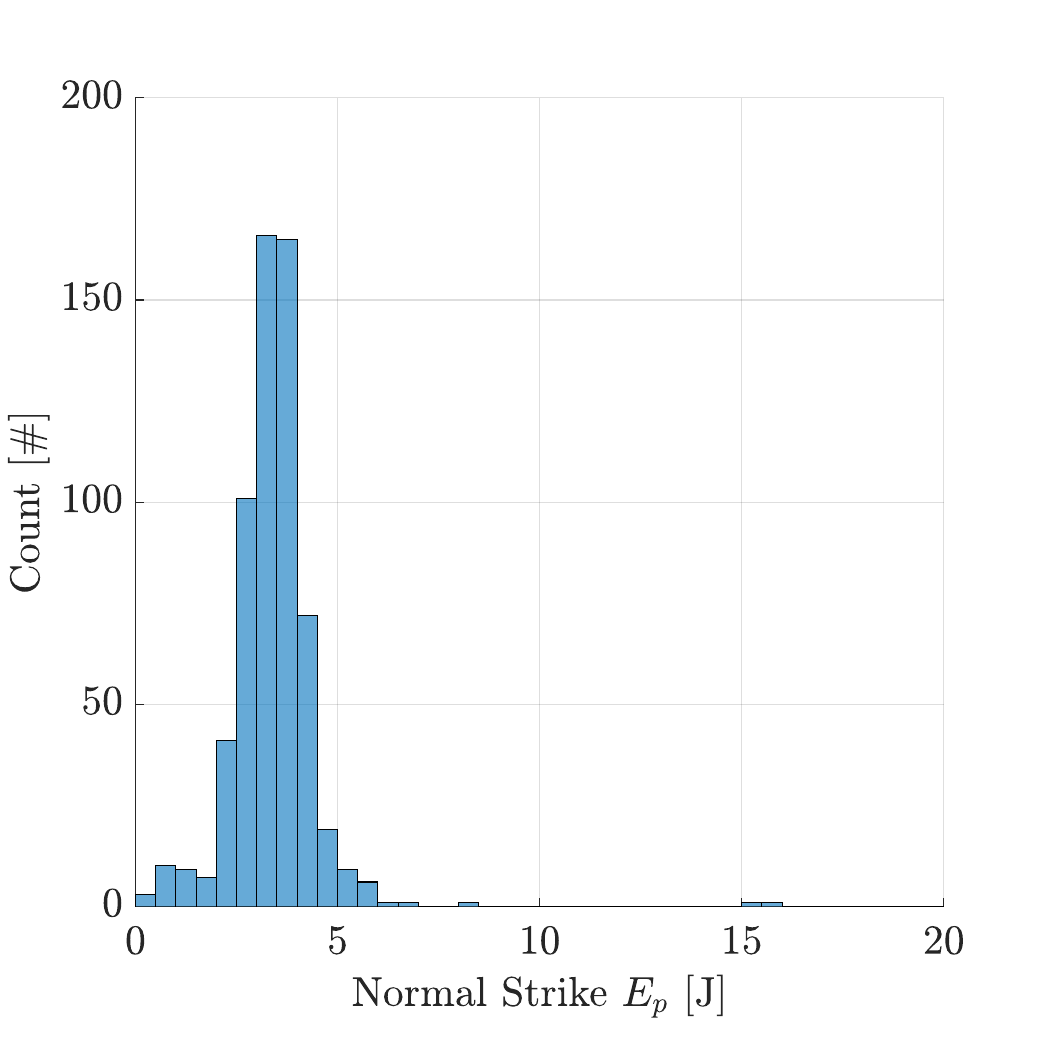}
        \caption{$E_p$ distribution at $t_{hp}$ classified under normal operation.}
        \label{fig:HistEpNormal_Hole7}
    \end{subfigure}
    \hfill
    \begin{subfigure}[t]{0.45\textwidth}
        \centering
        \includegraphics[trim=0cm 0cm 0cm 0cm, clip, width=1\linewidth]{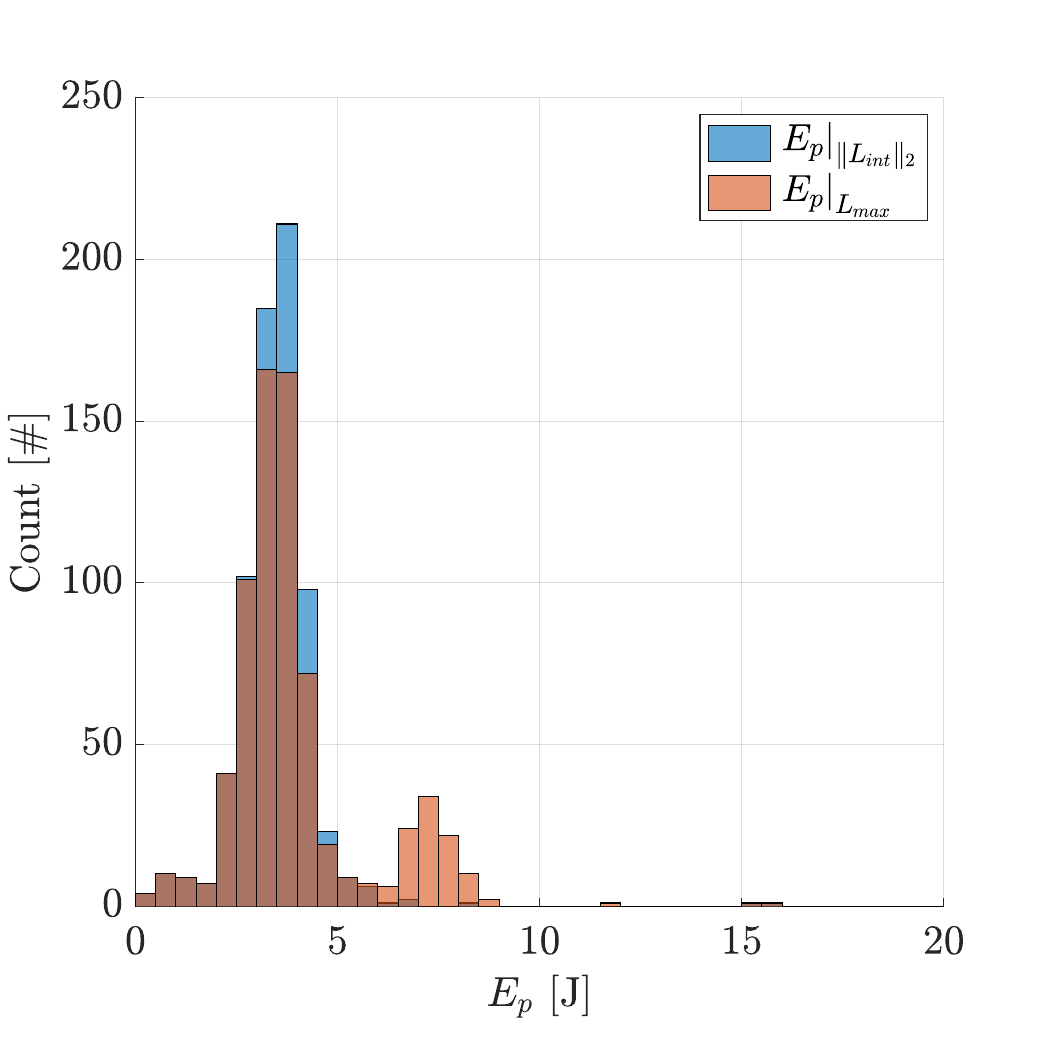}
        \caption{$E_p$ distribution at the end of all strikes, with those classified under abnormal operation either integrated to $\Vert L_{int} \Vert_2$ of normal strikes or $L_{max}$.}
        \label{fig:HistEpAbnormal_Hole7}
    \end{subfigure}
    \caption{Hammer kinetic-energy profiles for normal and all-strike classifications for experiment 2b in Table~\ref{tab:experiment_log}.}
    \label{fig:HistEp_Hole7}
\end{figure}
Figure~\ref{fig:HistEp_Hole7} shows a slightly higher mean for experiment 2b than for Fig.~\ref{fig:HistEp_Hole1n2}. The majority of strikes in 2b are classified as normal operation, so in comparing Fig.~\ref{fig:HistEpNormal_Hole7} with Fig.~\ref{fig:HistEpAbnormal_Hole7}, the assumptions for abnormal-strike $L_{int}$ affect only approximately 100 strikes out of a distribution of about 700. Once again, when the abnormal integrated length is taken as $L_{max}$, Fig.~\ref{fig:HistEpAbnormal_Hole7} shows an additional peak at around 6.5 J, while the average-$L_{int}$ assumption moves those added strikes onto the normal-strike distribution from Fig.~\ref{fig:HistEpNormal_Hole7}. From the values tabulated in Table~\ref{tab:rockExperiment_results_mean}, we see an increase in the mean $E_p$ energy from 2a to 2b.  
\begin{figure}[H]
    \centering
    \begin{subfigure}[t]{0.45\textwidth}
        \centering 
        \includegraphics[trim=0cm 0cm 0cm 0cm, clip, width=1\linewidth]{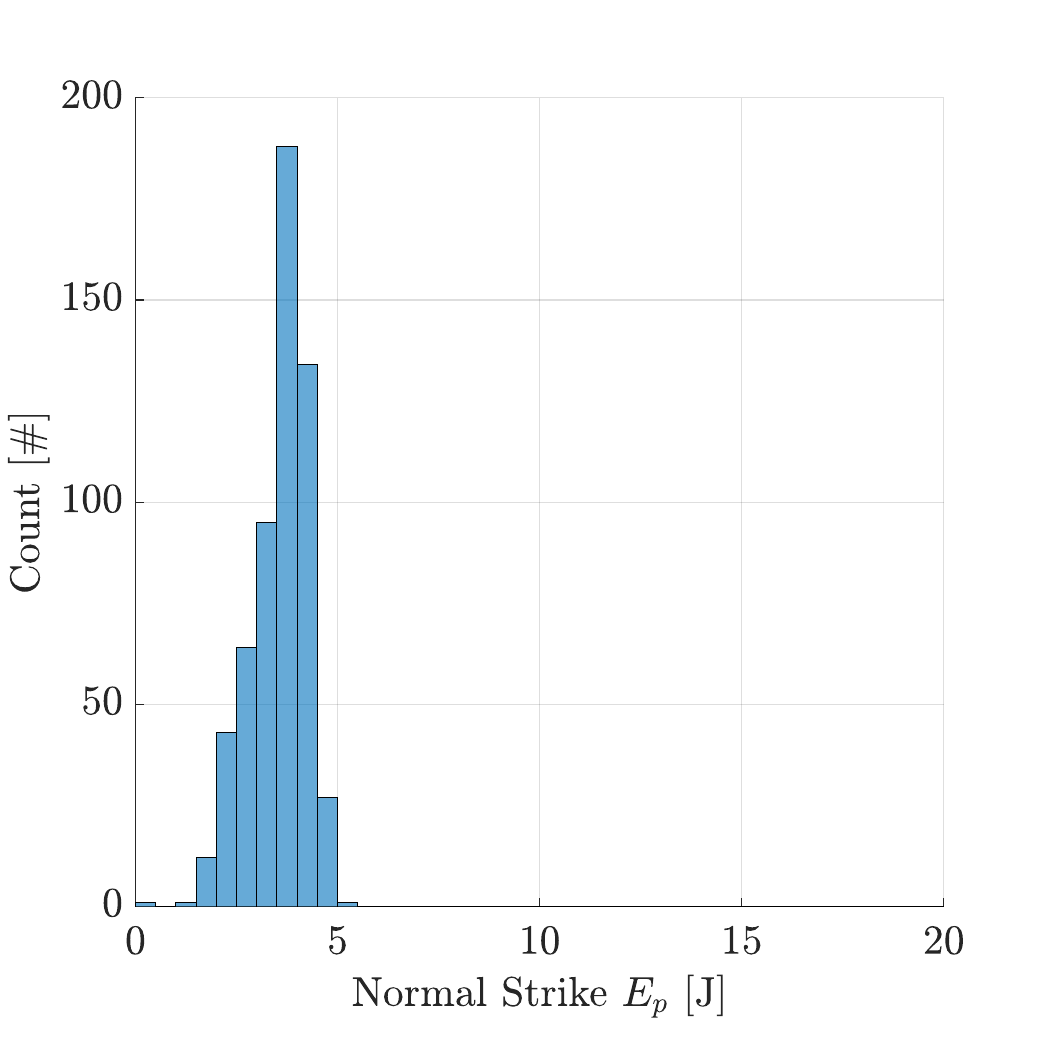}
        \caption{$E_p$ distribution at $t_{hp}$ classified under normal operation.}
        \label{fig:HistEpNormal_Hole5}
    \end{subfigure}
    \hfill
    \begin{subfigure}[t]{0.45\textwidth}
        \centering
        \includegraphics[trim=0cm 0cm 0cm 0cm, clip, width=1\linewidth]{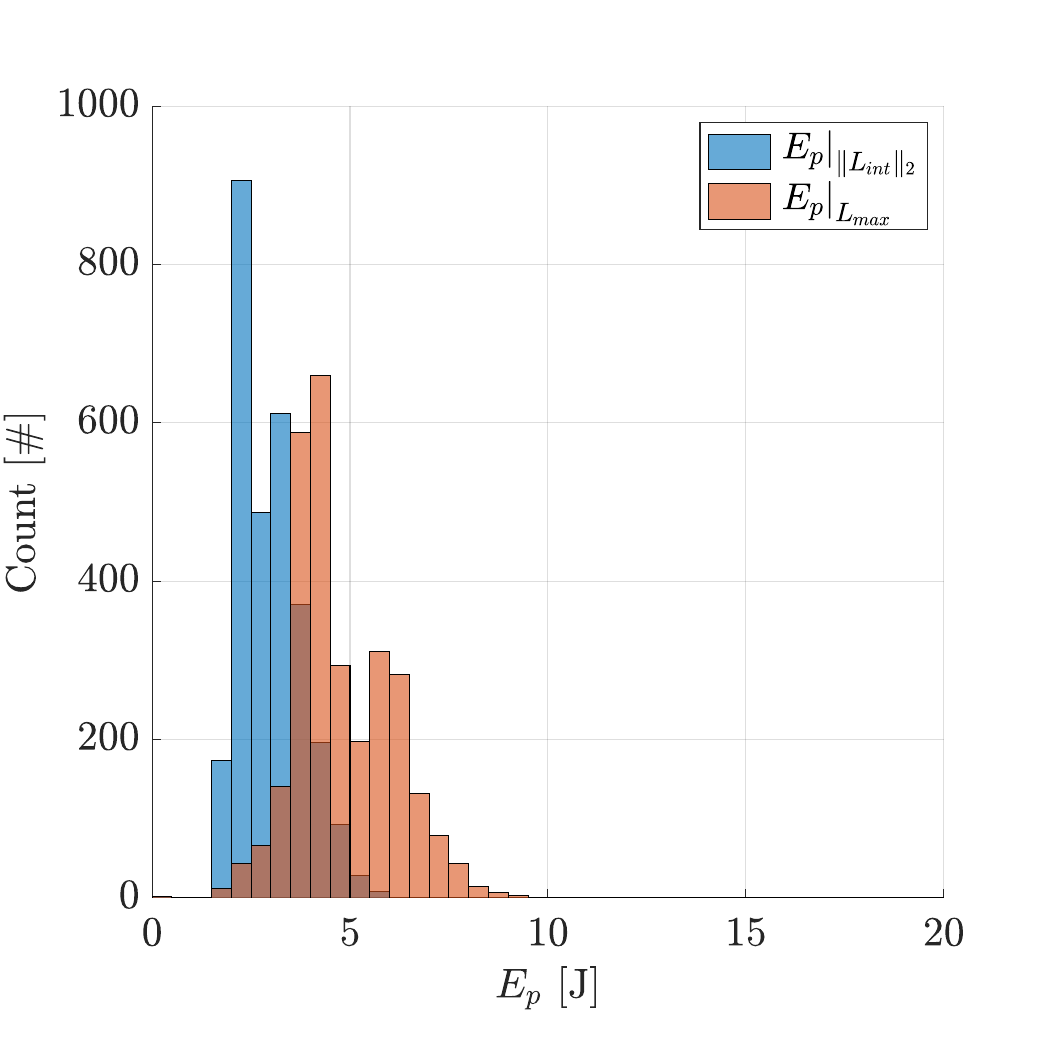}
        \caption{$E_p$ distribution at the end of all strikes, with those classified under abnormal operation either integrated to $\Vert L_{int} \Vert_2$ of normal strikes or $L_{max}$.}
        \label{fig:HistEpAbnormal_Hole5}
    \end{subfigure}
    \caption{Hammer kinetic-energy profiles for normal and all-strike classifications for experiment 3a in Table~\ref{tab:experiment_log}.}
    \label{fig:HistEp_Hole5}
\end{figure}
Figure~\ref{fig:HistEpNormal_Hole5}, from experiment 3a, shows a slight improvement in the mean of the normal-strike distribution compared with 2a and 2b. However, because 80\% of strikes are classified as abnormal, the results in Fig.~\ref{fig:HistEpAbnormal_Hole5} dominate the performance averages, with a larger difference between the means of the two abnormal-strike integrated-length assumptions than in the previous two experiments (see Table~\ref{tab:rockExperiment_results_mean}).   
Experiment 3b, shown in Fig.~\ref{fig:HistEp_Hole6}, appears to provide the best drilling performance of all the rock-simulant runs, with the normal-strike mean in Fig.~\ref{fig:HistEpNormal_Hole6} near 4.5 J, compared with less than 4 J in experiments 2a, 2b, and 3a. The conservative abnormal integration-length estimate based on $L_{max}$ pushes the average to almost 6 J, producing the bimodal distribution shown in Fig.~\ref{fig:HistEpAbnormal_Hole6}, though this remains below the $\approx 7.5$--8 J level of experiment 1a. 
Overall, the performance captured by the $E_p$ distributions in Figs.~\ref{fig:HistEp_Hole1n2}--\ref{fig:HistEp_Hole6} is between 35 and 75\% of that in experiment 1a. This was expected from the variance seen in the scalar outputs in Fig.~\ref{fig:Hist_Hole12567}, even before computing the impact energy for each strike.  
By bounding the upper performance of the abnormal-strike energy and computing the normal-strike energy directly, we can still produce reasonable estimates of the MSE for each simulant-bit combination.  
\begin{figure}[htb!]
    \centering
        \includegraphics[trim=0cm 0cm 0cm 0cm, clip, width=1\linewidth]{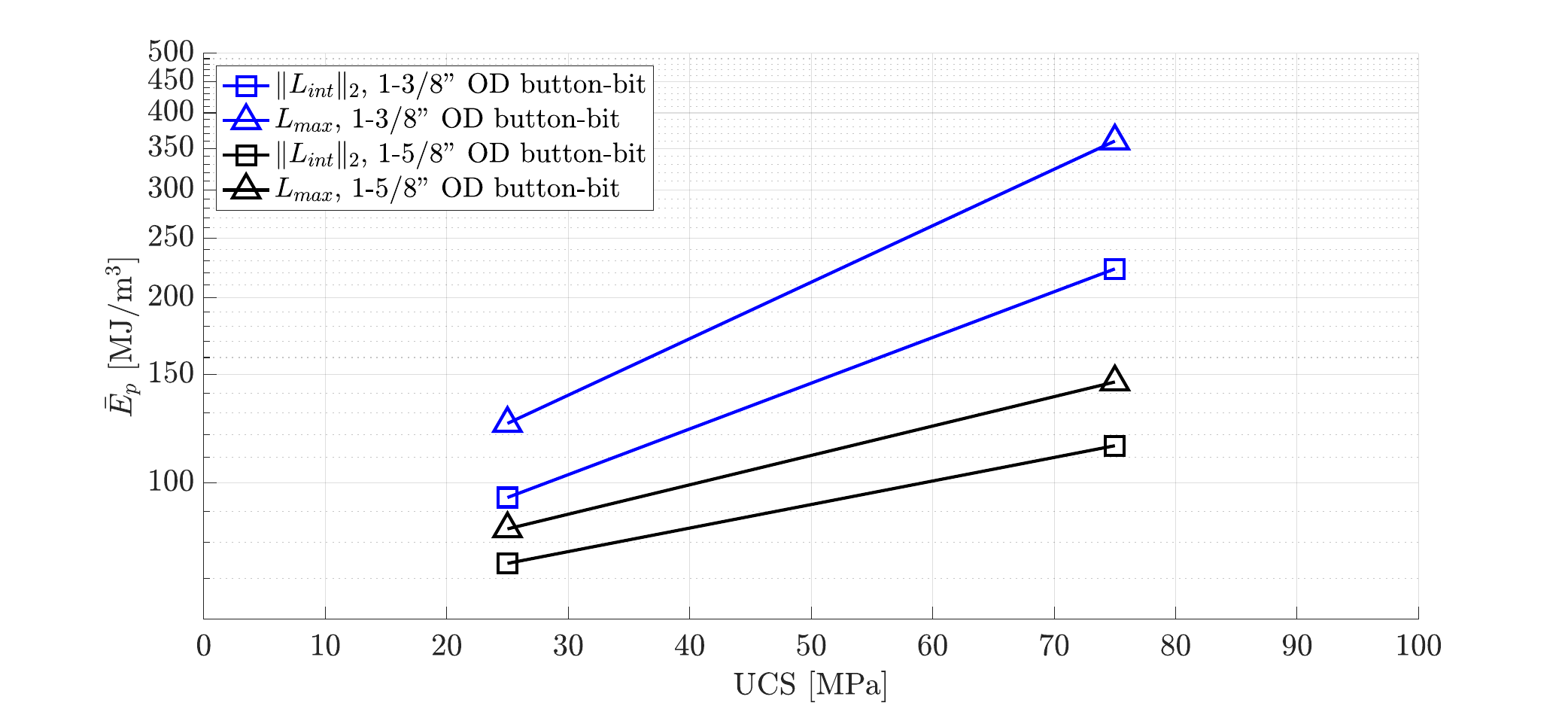}
        \caption{$\bar{E}$ over UCS strength of different rocks for experiments 2a, 2b, 3a, and 3b.}
        \label{fig:MSEvsUCS_all}
\end{figure}
Also tabulated in Table~\ref{tab:rockExperiment_results_mean} and plotted in Fig.~\ref{fig:MSEvsUCS_all} are the MSE values $\bar{E}_p$ for the two abnormal integrated-length assumptions, namely $\left. \bar{E}_p \right|_{L_{max}}$ in Eq.~\ref{eq:AbnormalStrikeLmaxCond} and $\left. \bar{E}_p \right|_{\Vert L_{int} \Vert_2}$ in Eq.~\ref{eq:AbnormalStrikeLintaveCond}. The UCS values used are the average values for USB and sandstone noted in Sec.~\ref{sec:SimulantCharacterization}.
We find a consistent trend between bit choice and rock compressive strength: for both the small and large button bits, drilling through sandstone is on average about twice as easy as drilling through Saddleback basalt. Interestingly, increasing the bit diameter also improved drilling efficiency: the larger-diameter bit is approximately 1.25--1.4 times more efficient than the smaller bit in sandstone, and approximately 2--2.5 times more efficient in Saddleback basalt.   
\begin{table}[htb!]
\centering
\begin{tabular}{l c c c c c}
\hline
 & \multicolumn{4}{c}{\textbf{ID}} & \\
\textbf{Symbol} & \textbf{2a} & \textbf{2b} & \textbf{3a} & \textbf{3b} & \textbf{Units} \\
\hline
$p_o$ & 100 & 100 & 100 & 100 & psi \\
$p_v(t_0)$ & 71.6 & 95.5 & 95.5 & 87.5 & psi \\
$E_p$ & 3.01 & 3.40 & 3.55 & 4.39 & J \\
$\left.E_p\right|_{\Vert L_{int} \Vert_2}$ & 2.89 & 3.46 & 2.96 & 4.60 & J \\
$\left.E_p\right|_{L_{max}}$ & 3.82 & 3.93 & 4.78 & 5.86 & J \\
$t_{hp}$ & 0.0194 & 0.0175 & 0.0181 & 0.0174 & s \\
$t_{close}$ & 0.0288 & 0.0224 & 0.0244 & 0.0231 & s \\
$L_{int}$ & 0.0632 & 0.0629 & 0.0639 & 0.0665 & m \\
$\sum_{\text{norm}} E_p$ & 3380 & 2080 & 2000 & 3120 & J \\
$\sum_{\text{all}} \left. E_p \right|_{\Vert L_{int} \Vert_2}$ & 5560 & 2460 & 8520 & 6290 & J \\
$\sum_{\text{all}} \left. E_p \right|_{L_{max}}$ & 7360 & 2790 & 13700 & 8010 & J \\
$\left. \bar{E}_p \right|_{\Vert L_{int} \Vert_2}$ & 94.7 & 74.0 & 223 & 115 & MJ/m$^3$ \\
$\left. \bar{E}_p \right|_{L_{max}}$ & 125 & 84.2 & 360 & 146 & MJ/m$^3$ \\
$H_d$ & 0.0583 & 0.0232 & 0.0380 & 0.0383 & m \\
$N_{total}$ & 1926 & 711 & 2879 & 1369 & \# \\
$N_{\text{norm}}$ & 1121 & 613 & 566 & 711 & \# \\
$N_{\text{abn}}$ & 805 & 98 & 2310 & 658 & \# \\
$N_{\text{missed}}$ & 51 & 3 & 237 & 136 & \# \\
\hline
\end{tabular}
\caption{Mean values of rock-simulant experimental results, as described in Table~\ref{tab:exprvariables_description}, for experiments 2a, 2b, 3a, and 3b in Table~\ref{tab:experiment_log}.}
\label{tab:rockExperiment_results_mean}
\end{table}
\begin{table}[htb!]
\centering
\resizebox{\textwidth}{!}{%
\begin{tabular}{lccc|ccc|ccc|ccc}
\hline
\textbf{ID} & \multicolumn{3}{c|}{\textbf{2a}} & \multicolumn{3}{c|}{\textbf{2b}} & \multicolumn{3}{c|}{\textbf{3a}} & \multicolumn{3}{c}{\textbf{3b}} \\ \hline
\textbf{Symbol} & \textbf{Max} & \textbf{Min} & \textbf{STD} & \textbf{Max} & \textbf{Min} & \textbf{STD} & \textbf{Max} & \textbf{Min} & \textbf{STD} & \textbf{Max} & \textbf{Min} & \textbf{STD} \\ \hline
$p_{v}\left( t_0 \right)$ & 102 & 27 & 17 & 102 & 71 & 7.62 & 103 & 65 & 6.74 & 102 & 49 & 8.74 \\
$E_p$ & 10 & 0.0052 & 1.28 & 16 & $1.6 \times 10^{-7}$ & 1.11 & 5.02 & 0.086 & 0.71 & 12 & 0.099 & 1.43 \\
$\left.E_p \right|_{\Vert L_{int} \Vert_2}$ & 10 & $1.2 \times 10^{-10}$ & 1.09 & 16 & $1.6 \times 10^{-7}$ & 1.06 & 6.51 & 0.0012 & 0.80 & 12 & 0.099 & 1.28 \\
$\left.E_p \right|_{L_{max}}$ & 10 & $1.3 \times 10^{-10}$ & 1.58 & 16 & $1.6 \times 10^{-7}$ & 1.72 & 9.51 & 0.0003 & 1.25 & 12 & 0.099 & 2.14 \\
$t_{hp}$ & 0.030 & 0.011 & 0.0018 & 0.068 & 0.0060 & 0.0026 & 0.026 & 0.013 & 0.0012 & 0.021 & 0.009 & 0.0013 \\
$t_{close}$ & 0.20 & 0.017 & 0.015 & 0.060 & 0.009 & 0.0039 & 0.041 & 0.019 & 0.0046 & 0.046 & 0.008 & 0.0044 \\
$L_{int}$ & 0.078 & 0.043 & 0.0049 & 0.078 & 0.047 & 0.0034 & 0.073 & 0.045 & 0.0025 & 0.078 & 0.045 & 0.0052 \\ \hline
\end{tabular}
}
\caption{Minimum, maximum, and standard deviation values of rock-simulant experimental results, as described in Table~\ref{tab:exprvariables_description}, for experiments 2a, 2b, 3a, and 3b in Table~\ref{tab:experiment_log}.}
\label{tab:rockExperiment_results_variance}
\end{table}
The general trend of MSE relative to UCS follows the expected increase as drilling proceeds into stronger rock, consistent with the discussion in Sec.~\ref{sec:introduction}. The efficiency improvement obtained by increasing drill-bit size suggests that the button-insert configuration and the available percussion energy are better matched to the larger-OD drill-bit, and implies that exploring other drill-bit geometries may yield a non-negligible change in drilling performance.   

In comparison with other drilling systems from Table~\ref{tab:review}, our system appears to perform better, while remaining in the same general energy range, than systems baselined against similar rock strengths. For example, the Auto-Gopher I and II performance in gypsum and limestone with UCS values of 40 and 45 MPa, respectively, yields MSE values of 273--652 MJ/m$^3$. The Honeybee Icebreaker drill in 40 MPa ice-cemented ground and 45 MPa limestone shows MSE values of 420--733 MJ/m$^3$. The RedWater drill in $-60^{\circ}$C ice, for which UCS was not reported, shows MSE results around 63 MJ/m$^3$, which is closest to our sandstone range of 74--125 MJ/m$^3$, although typical UCS values for hard ice are still considerably lower than those of sandstone \citep{dundas2023large, kahraman2003dominant, abubakar2018penetration}.    
Although a full system comparison must also account for upstream compression efficiency, a strict comparison of the percussive mechanism suggests that \textit{WiP} drill values may be more than twice as efficient as other similar drill systems. This remains a rough literature comparison, and a more detailed analysis is needed to understand the trade space among drill-bit geometry, percussion impact energy, and overall system design, but the present study suggests that this is a promising avenue to pursue.

\section{Representative Mars drilling case}

\begin{figure}[htb!]
    \centering
    \begin{subfigure}[t]{0.48\textwidth}
        \centering 
        \includegraphics[trim=0cm .5cm 0cm .5cm, clip, width=1\linewidth]{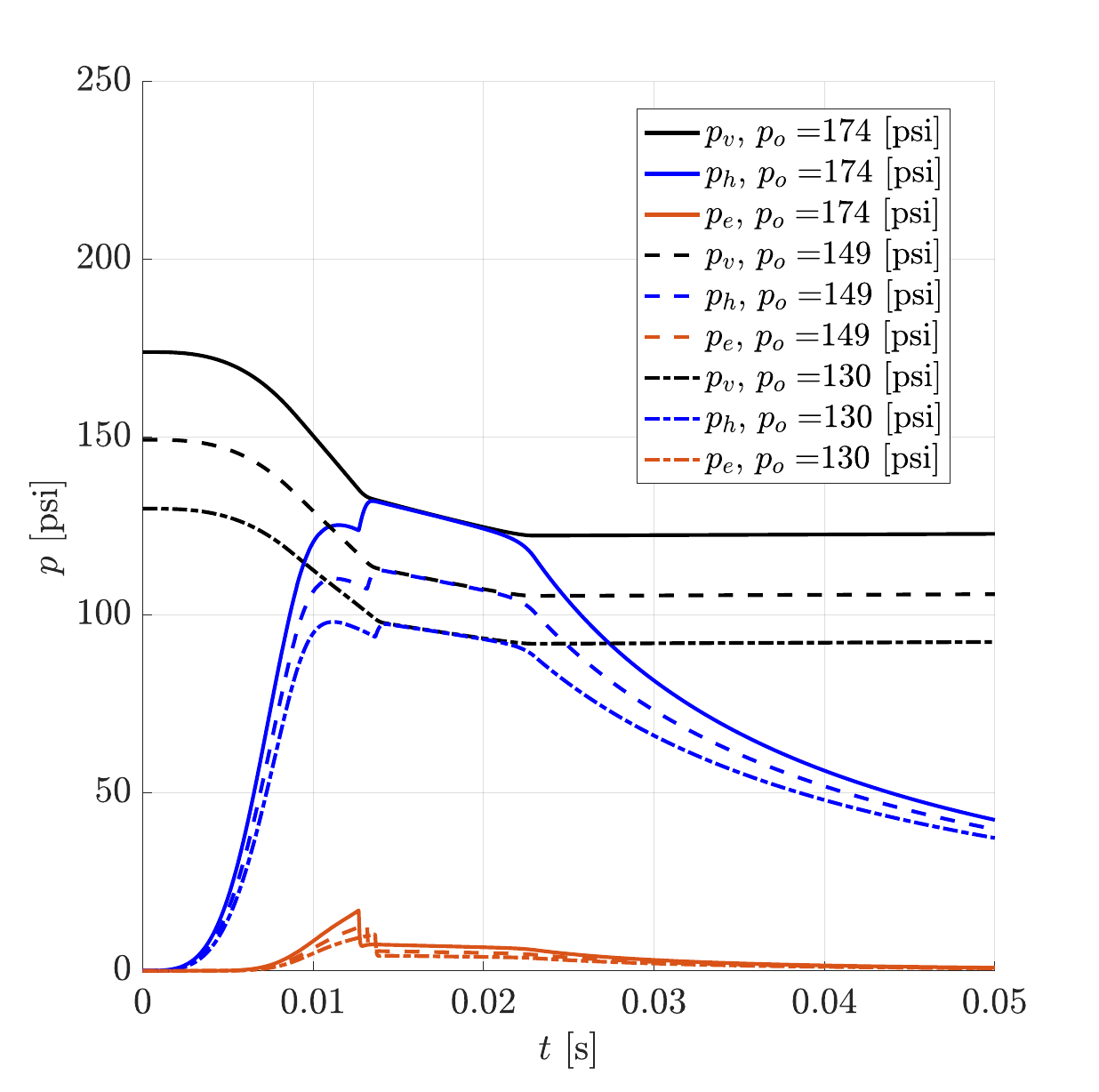}
        \caption{Model pressure results vs. time for strikes at varying flapper-valve opening-threshold pressure.}
        \label{fig:ModelMultipleMars60-80-100P}
    \end{subfigure}
    \begin{subfigure}[t]{0.48\textwidth}
        \centering
        \includegraphics[trim=0cm .5cm 0cm .5cm, clip, width=1\linewidth]{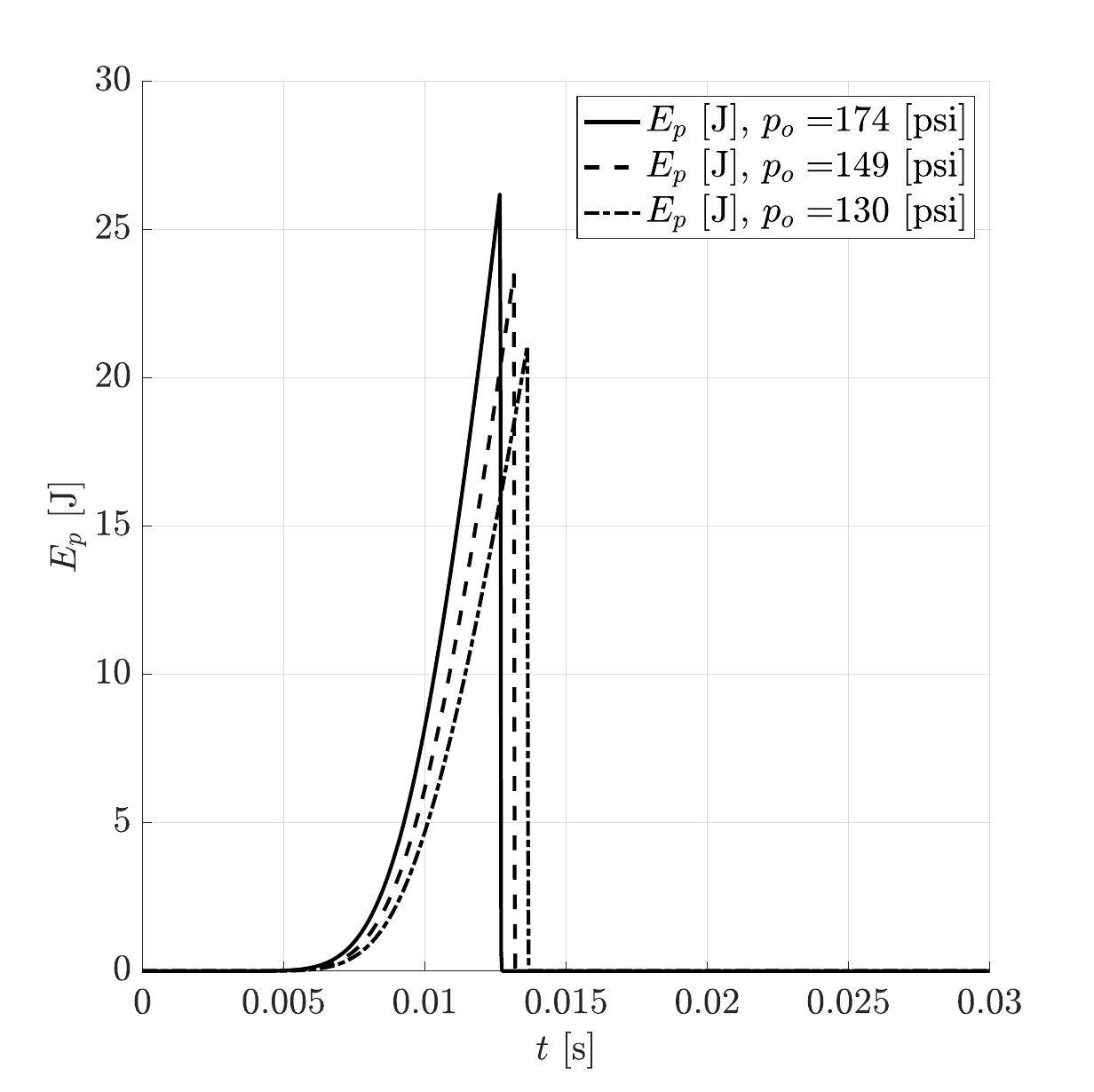}
        \caption{Model hammer kinetic-energy results vs. time for strikes at varying flapper-valve opening-threshold pressure.}
        \label{fig:ModelMultipleMars60-80-100E}
    \end{subfigure}
    \caption{Multiple-strike results modeled at varying $p_o = [60, 80, 100]$ psi under representative Mars operating conditions.}
    \label{fig:ModelMultipleMars60-80-100}
\end{figure}

As mentioned in Sec.~\ref{sec:introduction}, the pneumatic drill architecture outlined here could enable deep drilling on Mars by using the Martian atmosphere as the primary working fluid for both actuation and cuttings removal. In light of this potential, the results in this section present drill performance for a representative set of Mars operating conditions.  
Specifically, the drill geometry is represented through the model parameters in Table~\ref{tab:parametersTable}, and the coupled fluid-structural-thermodynamic model is solved for CO$_2$ with chamber temperature $T_{v0} = T_{0\mathrm{atm}} = 300$ K. 
Figure~\ref{fig:ModelMultipleMars60-80-100} shows the resulting pressure and energy histories for three vent-chamber pressures, $p_{0v} = \left[ 174, \ 149, \ 130 \right]$ psi, chosen to remain clearly above the CO$_2$ triple-point pressure of approximately 75 psi.  
Consistent with previous results, increasing $p_{v0}$ increases hammer impact energy, with $E_p = \left[ 26.2, \ 23.5, \ 21.1 \right]$ J, respectively. Hammer travel times also decrease as $p_{v0}$ increases, while remaining near $t_{hp} \approx 0.012$ s for all three simulated cases.  

These results provide a basis for estimating ROP on Mars for a future system concept using the MSE results from Table~\ref{tab:rockExperiment_results_mean}. Taking the conservative value of $\left. \bar{E}_p \right|_{L_{\mathrm{max}}}$ for sandstone with the 1-5/8" drill-bit, 84.2 MJ/m$^3$, the predicted penetration per impact is 0.23, 0.21, and 0.19 mm/impact. For Saddleback basalt, using 146 MJ/m$^3$, the corresponding values are 0.13, 0.12, and 0.11 mm/impact. 
At the relatively slow multi-strike testbed operating frequency of 0.5 Hz, the drill could therefore penetrate approximately 19 to 42 cm per hour in these two simulant materials. Because that frequency reflects the present benchtop testbed rather than an optimized flight implementation, these values should be interpreted as conservative first-order estimates rather than upper-bound performance predictions.

Although additional drill-design iterations remain future work, this representative Mars case already supports the suitability of the \textit{WiP} concept for deep Martian drilling. The system combines impact energies of roughly 21--26 J per strike with a wireline architecture that avoids long drill strings, uses atmospheric CO$_2$ as the working fluid for both actuation and cuttings transport, and remains compatible with the low-power mission assumptions outlined earlier in the manuscript. Taken together, these results indicate that the concept is not only mechanically viable under representative Mars conditions, but also well matched to the mass, power, and operational constraints of a robotic Mars lander mission.

\section{Discussion}
\label{sec:conclusion}
This work supports the feasibility of a wireline pneumatic (WiP) drill as a candidate architecture for deep Martian subsurface access. The central premise is that a downhole assembly powered by compressed atmospheric CO$_2$ can combine percussion, limited bit indexing, and cuttings transport without requiring a heavy drill string or substantial downhole electrical actuation. That combination is particularly attractive for Mars, where low available WOB, strict power limits, and the need to penetrate beyond the shallow regolith layer all favor a compact downhole system that derives as much function as possible from the working fluid itself.

The prototype and model results together show that the proposed actuation concept is mechanically credible. The magnetic flapper-valve successfully converts a quasi-steady fluid supply into repeated hammer strikes, and the reduced-order model captures the dominant coupling among chamber pressure evolution, valve timing, hammer acceleration, and delivered impact energy. The comparison between the calibrated model and the load-cell experiments is especially important because it establishes more than a qualitative match in waveform shape. It shows that the pressure-based method can recover strike timing and impact-energy trends with enough fidelity to support design trades, even when direct force measurements are unavailable. In that sense, the model should be viewed as a design and interpretation tool rather than as a high-fidelity predictor of every detail of valve and hammer motion.

That distinction is important for how the present results should be interpreted. This manuscript does not claim that the current reduced-order model closes every uncertainty relevant to flight-system prediction, nor that the benchtop hardware fully represents a deployable Mars drill. What it does show is that the dominant first-order relationships among opening pressure, valve behavior, strike timing, hammer travel, and impact energy are sufficiently captured to guide design decisions. For a planetary drilling technology that is still at the prototype stage, that is a meaningful step: it converts the system from a largely conceptual mechanism into one whose major sensitivities can be measured, modeled, and intentionally traded.

At the same time, the experiments make clear where the present prototype falls short of its own idealized potential. The gap between the ideal-model predictions and the calibrated experimental results indicates that the dominant losses are not fundamental to pneumatic percussion on Mars, but instead arise from the current implementation, particularly the flapper-valve opening behavior, vent-side flow losses, and hammer friction. The measured spread in strike timing, integration length, and impact energy, especially at lower opening pressures, is consistent with a mechanism whose effective opening-area function varies from strike to strike. That variability matters because it broadens the delivered-energy distribution and lowers average performance even when peak performance remains promising. One of the strongest outcomes of this study is therefore not only that the drill works, but that the main shortfalls now appear identifiable and, in principle, engineerable.

The rock-simulant results clarify how that performance translates into drilling effectiveness. The measured MSE values increase with rock strength as expected, and the larger drill-bit consistently performs better than the smaller one for the same button layout and percussion system. This suggests that the tested button-insert geometry is better matched to the larger diameter and to the available impact energy, rather than indicating a simple monotonic dependence on bit size alone. The UCS characterization also supports the broader use case for percussion on Mars: as loading rate increases, the tested materials exhibit behavior consistent with greater brittleness, which reinforces the case for a percussion-dominant drilling mode in competent rock. Taken together, these results suggest that future gains are likely to come from co-optimizing hammer energy, valve timing, and drill-bit geometry rather than from modifying any one of those in isolation.

The abnormal-strike behavior observed in the rock experiments is an important caveat. Those strikes could not be ignored because they still transferred momentum to the drill-bit, but they also complicated energy reconstruction by obscuring the pressure features used to identify $t_{hp}$. The bounded treatment adopted here is therefore reasonable for first-order performance estimation, but it also highlights a broader design requirement: a planetary drill intended for autonomous operation must deliver not only high mean energy, but also repeatable strike mechanics that remain interpretable from its onboard telemetry. The likely link between abnormal behavior and motion of the control magnet or flapper-valve sealing state points to the valve subsystem as the highest-leverage target for the next hardware iteration.

This also helps clarify where the remaining technical risk now resides. The present results do not suggest that the core challenge is whether a pneumatically driven downhole hammer can fracture rock under Mars-relevant constraints; rather, they suggest that the remaining challenge is achieving stable, repeatable, and efficient operation across many thousands of strikes while preserving the advantages of the architecture. In other words, the main open problem has shifted from basic feasibility to engineering robustness. That is a substantially more tractable development problem, and it is one that can be attacked systematically through valve redesign, tighter mechanical tolerancing, improved vent-path geometry, and integrated end-to-end testing.

From a mission perspective, the representative Mars case strengthens the argument that this concept is worth pursuing. When the experimentally informed MSE values are combined with the Mars operating model, the resulting penetration estimates remain meaningful even when evaluated conservatively and at the relatively slow testbed operating frequency. More importantly, the architecture aligns with the system-level constraints that motivate it in the first place: wireline deployment limits structural mass growth with depth, atmospheric CO$_2$ can serve as the working fluid, and the required operating regime appears compatible with the low-power envelope of a robotic lander mission. A full mission assessment must still include atmospheric compression, thermal control, anchoring, and long-duration cuttings removal, none of which is closed by the present experiments alone. Nevertheless, the current results show that the core percussive mechanism is viable, that its performance can be modeled and interpreted with useful accuracy, and that its drilling efficiency is sufficiently promising to justify continued development toward an integrated Mars-capable system.

Viewed in that broader context, the main contribution of this study is not simply the demonstration of a pneumatic hammer, nor only a favorable comparison in MSE against selected prior systems. Rather, it is the combination of architecture, experimentally anchored modeling, and simulant drilling data into a single framework for evolving the concept toward a mission-relevant drill. The paper therefore advances the WiP concept from an interesting idea to an engineering basis for subsequent development. A next-generation prototype that incorporates a more stable flapper-valve mechanism, anchoring, and integrated cuttings-removal testing would be a natural next step, and the results presented here provide a defensible starting point for that progression.


\section*{Funding Acknowledgement}
The research was carried out at the Jet Propulsion Laboratory, California Institute of Technology, under a contract with the National Aeronautics and Space Administration (80NM0018D0004).
\newpage 

\bibliography{bibliography}

@inproceedings{van2005development,
  title={Development of the ESA ExoMars rover},
  author={Van Winnendael, M and Baglioni, P and Vago, J},
  booktitle={Proc. 8th int. symp. artif. intell., robot. automat. Space},
  pages={5--8},
  year={2005},
  organization={Citeseer Munich}
}

@inproceedings{badescu2006ultrasonic,
  title={Ultrasonic/sonic Gopher for subsurface ice and brine sampling: analysis and fabrication challenges and testing results},
  author={Badescu, M and Sherrit, S and Olorunsola, A and Aldrich, J and Bao, X and Bar-Cohen, Y and Chang, Z and Doran, PT and Fritsen, CH and Kenig, F and others},
  booktitle={Smart Structures and Materials 2006: Industrial and Commercial Applications of Smart Structures Technologies},
  volume={6171},
  pages={48--56},
  year={2006},
  organization={SPIE}
}

@incollection{palmowski2022redwater,
  title={RedWater: Extraction of Water from Mars’ Ice Deposits},
  author={Palmowski, Joseph and Zacny, Kris and Mellerowicz, Boleslaw and Vogel, Brian and Bocklund, Andrew and Stolov, Leo and Yen, Bernice and Sabahi, Dara and Ware, Lilly and Faris, David and others},
  booktitle={Earth and Space 2022},
  pages={355--362},
  year={2022}
}

@article{magnani2011testing,
  title={Testing of ExoMars EM drill tool in mars analogous materials},
  author={Magnani, P and Re, E and Fumagalli, A and Senese, S and Ori, GG and Gily, A and Baglioni, P},
  journal={Proceedings Advanced Space Technologies for Robotics and Automation (ASTRA). Noordwijk, The Netherlands: European Space Agency},
  year={2011}
}

@inproceedings{zacny2011testing,
  title={Testing of a 1 meter Mars Icebreaker drill in a 3.5 meter vacuum chamber and in an Antarctic Mars analog site},
  author={Paulsen, Gale and Zacny, Kris  and Mckay, Chris and Glass, Brian and Szczesiak, Mateusz and Craft, Jack and Santoro, Chris and Shasho, Jeff and Davila, Alfonso and Marinova, Margarita and others},
  booktitle={AIAA space 2011 conference \& exposition},
  pages={7236},
  year={2011}
}

@inproceedings{badescu2013auto,
  title={Auto-Gopher: a wireline deep sampler driven by piezoelectric percussive actuator and EM rotary motor},
  author={Badescu, Mircea and Ressa, Aaron and Lee, Hyeong Jae and Bar-Cohen, Yoseph and Sherrit, Stewart and Zacny, Kris and Paulsen, Gale L and Beegle, Luther and Bao, Xiaoqi},
  booktitle={Sensors and Smart Structures Technologies for Civil, Mechanical, and Aerospace Systems 2013},
  volume={8692},
  pages={782--789},
  year={2013},
  organization={SPIE}
}

@inproceedings{barcohen2016autoASCE,
author = {Bar-Cohen, Yoseph and Zacny, Kris and Badescu, Mircea and Lee, Hyeong Jae and Sherrit, Stewart and Bao, Xiaoqi and Paulsen, Gale L. and Beegle, Luther},
publisher = {Pasadena, CA: Jet Propulsion Laboratory, National Aeronautics and Space Administration, 2016},
title = {{The Auto-Gopher: a wireline rotary-percussive deep sampler}},
booktitle={ASCE Earth and Space 2016 Conference, Orlando, Florida, April 11-15, 2016},
year = {2016},
version = {V2},
doi = {2014/45800},
url = {https://hdl.handle.net/2014/45800}
}

@article{bar2017auto,
  title={Auto-Gopher-2-Wireline deep sampler driven by percussive piezoelectric actuator and rotary EM motors},
  author={Bar-Cohen, Yoseph and Zacny, Kris and Badescu, Mircea and Lee, Hyeong Jae and Sherrit, Stewart and Bao, Xiao Qi and Freeman, David and Paulsen, Gale L and Beegle, Luther},
  journal={Advances in science and technology},
  volume={100},
  pages={207--212},
  year={2017},
  publisher={Trans Tech Publ}
}

@inproceedings{badescu2019auto,
  title={Auto-Gopher-II: an autonomous wireline rotary-hammer ultrasonic drill test results},
  author={Badescu, Mircea and Bar-Cohen, Yoseph and Sherrit, Stewart and Bao, Xiaoqi and Lee, Hyeong Jae and Jackson, Shannon and Metz, Brandon and Valles, Zachary C and Zacny, Kris and Mellerowicz, Boleslaw and others},
  booktitle={Sensors and Smart Structures Technologies for Civil, Mechanical, and Aerospace Systems 2019},
  volume={10970},
  pages={271--280},
  year={2019},
  organization={SPIE}
}

@article{zacny2013reaching,
  title={Reaching 1 m deep on Mars: the Icebreaker drill},
  author={Zacny, K and Paulsen, G and McKay, CP and Glass, B and Dav{\'e}, A and Davila, AF and Marinova, M and Mellerowicz, B and Heldmann, J and Stoker, C and others},
  journal={Astrobiology},
  volume={13},
  number={12},
  pages={1166--1198},
  year={2013},
  publisher={Mary Ann Liebert, Inc. 140 Huguenot Street, 3rd Floor New Rochelle, NY 10801 USA}
}

@article{wippermann2020penetration,
  title={Penetration and performance testing of the HP$^3$ Mole for the InSight Mars mission},
  author={Wippermann, Torben and Hudson, Troy L and Spohn, Tilman and Witte, Lars and Scharringhausen, Marco and Tsakyridis, Georgios and Fittock, Mark and Kr{\"o}mer, Olaf and Hense, Simon and Grott, Matthias and others},
  journal={Planetary and Space Science},
  volume={181},
  pages={104780},
  year={2020},
  publisher={Elsevier}
}

@article{shenhar2005final,
author = {Shenhar, Joram and Tucker, Matthew and Farrand, Jennifer},
title = {Final Report: Phase II 
Autonomous Tethered Corer for Deep Drilling 
Contract: NNC04CB04C},
institution = {Raytheon UTD},
journal = {},
year = {2005}
}

@article{shenhar2006autnomous,
author = {Shenhar, Joram and Barfield, T. and Dolgin, B. and Farrand, J. and Goldstein, B. and Hill, J. and Tucker, M. and Wallace, R. and Lee, S.},
title = {Autonomous Tethered Corer (ATC) for Deep Drilling - Basalt Field Test, Idaho Falls, ID},
institution = {Raytheon UTD},
journal = {},
year = {2006}
}

@article{guerrero2008final,
author = {Jose Guerrero},
title = {Final Report: Hardware Base Technology Task 
Modular Planetary Drill System (MPDS),   
JPL Contract No. 1264531},
institution = {ATK Space},
journal = {},
year = {2008}
}

@inproceedings{george2012planetary,
  title={Planetary drilling and resources at the Moon and Mars},
  author={George, Jeffrey A},
  booktitle={Pioneer Natural Resources Geoscience, Engineering and Drilling Technology Conference},
  number={JSC-CN-27285},
  year={2012}
}

@article{magnani2004deep,
  title={Deep drill (DeeDri) for Mars application},
  author={Magnani, PG and Re, E and Ylikorpi, T and Cherubini, G and Olivieri, A},
  journal={Planetary and Space Science},
  volume={52},
  number={1-3},
  pages={79--82},
  year={2004},
  publisher={Elsevier}
}

@incollection{paulsen2006robotic,
  title={Robotic drill systems for planetary exploration},
  author={Paulsen, Gale and Zacny, Kris and Chu, Phil and Mumm, Erik and Davis, Kiel and Frader-Thompson, Seth and Petrich, Kyle and Glaser, David and Bartlett, Paul and Cannon, Howard and others},
  booktitle={Space 2006},
  pages={7512},
  year={2006}
}

@article{cannon2007marte,
  title={MARTE: technology development and lessons learned from a Mars drilling mission simulation},
  author={Cannon, Howard N and Stoker, Carol R and Dunagan, Stephen E and Davis, Kiel and G{\'o}mez-Elvira, Javier and Glass, Brian J and Lemke, Lawrence G and Miller, David and Bonaccorsi, Rosalba and Branson, Mark and others},
  journal={Journal of Field Robotics},
  volume={24},
  number={10},
  pages={877--905},
  year={2007},
  publisher={Wiley Online Library}
}

@article{stoker20082005,
  title={The 2005 MARTE robotic drilling experiment in R{\'\i}o Tinto, Spain: objectives, approach, and results of a simulated mission to search for life in the martian subsurface},
  author={Stoker, Carol R and Cannon, Howard N and Dunagan, Stephen E and Lemke, Lawrence G and Glass, Brian J and Miller, David and Gomez-Elvira, Javier and Davis, Kiel and Zavaleta, Jhony and Winterholler, Alois and others},
  journal={Astrobiology},
  volume={8},
  number={5},
  pages={921--945},
  year={2008},
  publisher={Mary Ann Liebert, Inc. 140 Huguenot Street, 3rd Floor New Rochelle, NY 10801~…}
}

@inbook{zacny2009extraterrestrial,
author = {Zacny, Kris and Bar-Cohen, Yoseph and Davis, Kiel and Coste, Pierre and Paulsen, Gale and Sherrit, Stewart and George, Jeffrey and Derkowski, Brian and Gorevan, Steve and Boucher, Dale and Guerrero, Jose and Kubota, Takashi and Thomson, Bradley J. and Stanley, Scott and Thomas, Peter and Lan, Nicholas and McKay, Christopher and Onstot, Tullis C. and Stoker, Carol and Glass, Brian and Wakabayashi, Sachiko and Whyte, Lyle and Visentin, Gianfranco and Re, Edoardo and Richter, Lutz and Badescu, Mircea and Bao, Xiaoqi and Fincher, Roger and Hoshino, Takeshi and Magnani, Piergiovanni and Menon, Carlo},
publisher = {John Wiley \& Sons, Ltd},
isbn = {9783527626625},
title = {Extraterrestrial Drilling and Excavation},
booktitle = {Drilling in Extreme Environments},
chapter = {6},
pages = {347-557},
doi = {https://doi.org/10.1002/9783527626625.ch6},
url = {https://onlinelibrary.wiley.com/doi/abs/10.1002/9783527626625.ch6},
eprint = {https://onlinelibrary.wiley.com/doi/pdf/10.1002/9783527626625.ch6},
year = {2009}
}

@article{stoker2023mission,
  title={A mission simulating the search for life on Mars with automated drilling, sample handling, and life detection instruments performed in the hyperarid core of the Atacama Desert, Chile},
  author={Stoker, Carol R and Glass, Brian J and Stucky, Thomas R and Dave, Arwen I and Kobayashi, Linda T and Quinn, Richard C and Moreno-Paz, Mercedes and S{\'a}nchez-Garc{\'\i}a, Laura and Mora, Maria F and Kehl, Florian and others},
  journal={Astrobiology},
  volume={23},
  number={12},
  pages={1284--1302},
  year={2023},
  publisher={Mary Ann Liebert, Inc., publishers 140 Huguenot Street, 3rd Floor New~…}
}

@inproceedings{glass2024trident,
  title={TRIDENT Drill Validation at Mars and Lunar Analog Field Sites},
  author={Glass, B and Stoker, C and Battah, H and Boelter, S and Fortuin, C and King, I and Stevenson, T and Stucky, T},
  booktitle={55th Lunar and Planetary Science Conference (LPSC)},
  year={2024}
}

@techreport{bruno2005fundamental,
  title={Fundamental research on percussion drilling: improved rock mechanics analysis, advanced simulation technology, and full-scale laboratory investigations},
  author={Bruno, Michael S},
  year={2005},
  institution={Terralog Technologies Inc.}
}

@book{hossain2015fundamentals,
  title={Fundamentals of sustainable drilling engineering},
  author={Hossain, M Enamul and Al-Majed, Abdulaziz Abdullah},
  year={2015},
  publisher={John Wiley \& Sons},
  pages={17-72}
}

@inproceedings{han2005dynamically,
  title={Dynamically modelling rock failure in percussion drilling},
  author={Han, Gang and Bruno, Mike and Dusseault, Maurice B},
  booktitle={ARMA US Rock Mechanics/Geomechanics Symposium},
  pages={ARMA--05},
  year={2005},
  organization={ARMA}
}

@inproceedings{teale1965concept,
  title={The concept of specific energy in rock drilling},
  author={Teale, Robert},
  booktitle={International journal of rock mechanics and mining sciences \& geomechanics abstracts},
  volume={2},
  number={1},
  pages={57--73},
  year={1965},
  organization={Elsevier}
}

@article{sliwa2012drilling,
  title={Drilling bits in percussive-rotary drilling technology (down the hole DTH)},
  author={{\'S}liwa, Tomasz and {\'S}nie{\.z}ek, Pawe{\l}},
  journal={AGH Drilling, Oil, Gas},
  volume={29},
  number={4},
  pages={453--462},
  year={2012},
  publisher={Akademia G{\'o}rniczo-Hutnicza im. Stanis{\l}awa Staszica w Krakowie. Wydawnictwo AGH}
}

@book{tandanand1975drillability,
  title={Drillability determination: a drillability index for percussion drills},
  author={Tandanand, Sathit and Unger, Harold F},
  volume={8073},
  year={1975},
  publisher={US Department of the Interior, Bureau of Mines}
}

@article{zhang2022energy,
  title={Energy requirement for rock breakage in laboratory experiments and engineering operations: A review},
  author={Zhang, Zong-Xian and Ouchterlony, Finn},
  journal={Rock mechanics and rock engineering},
  volume={55},
  number={2},
  pages={629--667},
  year={2022},
  publisher={Springer}
}

@article{oparin2022evaluation,
  title={Evaluation of the energy efficiency of rotary percussive drilling using dimensionless energy index},
  author={Oparin, VN and Karpov, VN and Timonin, VV and Konurin, AI},
  journal={Journal of Rock Mechanics and Geotechnical Engineering},
  volume={14},
  number={5},
  pages={1486--1500},
  year={2022},
  publisher={Elsevier}
}

@article{li2000analysis,
  title={Analysis of impact hammer rebound to estimate rock drillability},
  author={Li, X and Rupert, G and Summers, David A and Santi, P and Liu, D},
  journal={Rock Mechanics and Rock Engineering},
  volume={33},
  pages={1--13},
  year={2000},
  publisher={Springer}
}

@article{zacny2008drilling,
  title={Drilling systems for extraterrestrial subsurface exploration},
  author={Zacny, Kris and Bar-Cohen, Y and Brennan, M and Briggs, G and Cooper, G and Davis, K and Dolgin, B and Glaser, D and Glass, B and Gorevan, S and others},
  journal={Astrobiology},
  volume={8},
  number={3},
  pages={665--706},
  year={2008},
  publisher={Mary Ann Liebert, Inc. 140 Huguenot Street, 3rd Floor New Rochelle, NY 10801~…}
}

@misc{knez2021review,
  title={A review of different aspects of off-earth drilling. Energies, 14 (21), 7351},
  author={Knez, D and Khalilidermani, M},
  year={2021}
}

@article{khalilidermani2022survey,
  title={A Survey of application of mechanical specific energy in petroleum and space drilling},
  author={Khalilidermani, Mitra and Knez, Dariusz},
  journal={Energies},
  volume={15},
  number={9},
  pages={3162},
  year={2022},
  publisher={MDPI}
}

@book{horne2015drilling,
  title={Drilling on mars--mathematical model for rotary-ultrasonic core drilling of brittle materials},
  author={Horne, Mera Fayez},
  year={2015},
  publisher={University of California, Berkeley}
}

@inproceedings{tosi2024deep,
  title={Deep Access Subsurface Extraction \& Retrieval (DASER)},
  author={Tosi, Luis Phillipe C and Sherrill, Kristopher and Howe, A Scott and Perl, Scott M and Veismann, Marcel and Gori, Marcello and Parker, Ceth W and King, Isabel},
  booktitle={2024 IEEE Aerospace Conference},
  pages={1--14},
  year={2024},
  organization={IEEE}
}

@inproceedings{paulsen2016planetary,
  title={Planetary Deep Drill for Mars, Europa, and Enceladus},
  author={Paulsen, G and Shara, M and Zacny, K and Mellerowicz, B and Spring, J and Ridilla, A and Sharpe, R and Bowsher, J and Hoisington, N and Abrashkin, J},
  booktitle={47th Annual Lunar and Planetary Science Conference},
  number={1903},
  pages={1077},
  year={2016}
}

@inproceedings{bar2012deep,
  title={Deep drilling and sampling via the wireline auto-gopher driven by piezoelectric percussive actuator and EM rotary motor},
  author={Bar-Cohen, Yoseph and Badescu, Mircea and Sherrit, Stewart and Zacny, Kris and Paulsen, Gale L and Beegle, Luther and Bao, Xiaoqi},
  booktitle={Sensors and smart structures technologies for civil, mechanical, and aerospace systems 2012},
  volume={8345},
  pages={590--597},
  year={2012},
  organization={SPIE}
}

@article{zacny2005strategies,
  title={Strategies for drilling on Mars},
  author={Zacny, KA and Cooper, GA},
  journal={American Geophysical},
  year={2005}
}

@incollection{zacny2009drilling,
  title={Drilling and excavation for construction and in-situ resource utilization},
  author={Zacny, Kris and Bar-Cohen, Yoseph},
  booktitle={Mars: prospective energy and material resources},
  pages={431--459},
  year={2009},
  publisher={Springer}
}

@article{glass2020future,
  title={Future space drilling and sample acquisition: A collaborative industry-government workshop},
  author={Glass, BJ and New, M and Voytek, MA},
  journal={Icarus},
  volume={338},
  pages={113378},
  year={2020},
  publisher={Elsevier}
}

@article{edwards2020deep,
  title={Deep Trek: Mission Concepts for Exploring Subsurface Habitability \& Life on Mars: A Window into Subsurface Life in the Solar System},
  author={Edwards, Charles D and Stamenkovi{\'c}, Vlada and Boston, Penelope J and Lynch, Kennda L and Rivera-Valent{\'\i}n, Edgard G and others},
  year={2020}
}

@book{aadnoy2022petroleum,
  title={Petroleum rock mechanics: drilling operations and well design},
  author={Aadnoy, Bernt S and Looyeh, Reza},
  year={2022},
volume={2nd Edition},
  publisher={Gulf professional publishing},
 pages={172-174}
}

@article{westergaard1940plastic,
  title={Plastic state of stress around a deep well},
  author={Westergaard, Harold Malcolm},
  year={1940}
}

@book{hossain2018drilling,
  title={Drilling Engineering Problems and Solutions: A Field Guide for Engineers and Students},
  author={Hossain, M Enamul and Islam, Muhammad Rafiqul},
  year={2018},
  publisher={John Wiley \& Sons}
}

@article{miles1949stresses,
  title={Stresses around a deep well},
  author={Miles, AJ and Topping, AD},
  journal={Transactions of the AIME},
  volume={179},
  number={01},
  pages={186--191},
  year={1949},
  publisher={SPE}
}

@article{liu2018rock,
  title={The rock breaking mechanism analysis of rotary percussive cutting by single PDC cutter},
  author={Liu, Weiji and Zhu, Xiaohua and Li, Bo},
  journal={Arabian Journal of Geosciences},
  volume={11},
  pages={1--11},
  year={2018},
  publisher={Springer}
}

@book{White2010,
  title={Fluid mechanics},
  author={White, Frank M},
  year={2010},
  publisher={Boston: McGraw-Hill Higher Education}
}

@book{schlichting2016boundary,
  title={Boundary-layer theory},
  author={Schlichting, Hermann and Gersten, Klaus},
  year={2016},
  publisher={springer}
}

@article{osti1993EPAHandbook,
title = {Handbook of chemical hazard analysis procedures},
author = {{U.S. Environmental Protection Agency}},
abstractNote = {The fact that hazardous materials pose a threat to public safety and the environment is of vital concern to industry and all levels of government, particularly in the aftermath of the tragedy in Bhophal, India, that took over 2000 lives and injured tens of thousands of others in the course of a few hours. The primary purpose of the handbook and its associated computer program is to provide emergency planning personnel with the resources necessary to undertake comprehensive evaluations of potentially hazardous facilities and activities within their respective jurisdictions and thereby formulate a basis for their planning efforts.},
doi = {},
url = {https://www.osti.gov/biblio/6645955}, journal = {},
number = {},
volume = {},
place = {United States},
year = {1993},
month = {1}
}

@book{Idelchik1996Handbook,
  title={Handbook of Hydraulic Resistance},
  author={Idelchik, I. E.},
  year={1996},
  edition={3rd},
  publisher={Begell House},
  address={New York, NY},
  isbn={978-1567001545}
}

@article{Vokoun2009Magnetostatic,
  title={Magnetostatic interactions and forces between cylindrical permanent magnets},
  author={Vokoun, David and Beleggia, Marco and Heller, Lud{\v{e}}k and {\v{S}}ittner, Petr},
  journal={Journal of magnetism and Magnetic Materials},
  volume={321},
  number={22},
  pages={3758--3763},
  year={2009},
  publisher={Elsevier}
}

@article{Choe2023Effect,
  title={Effect of the geometrical shapes of the helical-spiral shroud intake valve on swirl generation in cylinder of diesel engine},
  author={Choe, Sung Gong and Choe, Tong Ho and Ho, In Chol and Mun, Myong Hak and Kim, Il Jun and Ri, Jong Hun and Jong, Ryong Uhn and Kim, Yong Chol},
  journal={Results in Engineering},
  volume={18},
  pages={101132},
  year={2023},
  publisher={Elsevier}
}

@article{Abd2017Effect,
  title={Effect of shroud and orientation angles of inlet valve on flow characteristic through helical--spiral inlet port in diesel engine},
  author={Abd El-Sabor Mohamed, A and Abo-Elfadl, Saleh and Nassib, Abd El-Moneim M},
  journal={Journal of Engineering for Gas Turbines and Power},
  volume={139},
  number={10},
  pages={102802},
  year={2017},
  publisher={American Society of Mechanical Engineers}
}

@article{Paul2010Flow,
  title={Flow field development in a direct injection diesel engine with different manifolds},
  author={Paul, Benny and Ganesan, V},
  journal={International Journal of Engineering, Science and Technology},
  volume={2},
  number={1},
  pages={80--91},
  year={2010}
}

@article{qi2009strain,
  title={Strain-rate effects on the strength and fragmentation size of rocks},
  author={Qi, C and Wang, M and Qian, Q},
  journal={International Journal of Impact Engineering},
  volume={36},
  number={12},
  pages={1355--1364},
  year={2009},
  publisher={Elsevier}
}

@article{gong2019peak,
  title={A peak-strength strain energy storage index for rock burst proneness of rock materials},
  author={Gong, F and Yan, J and Li, X and Luo, S},
  journal={International Journal of Rock Mechanics and Mining Science},
  volume={117},
  pages={76--89},
  year={2019},
  publisher={Elsevier}
}

@article{schultz1995limits,
  title={Limits on strength and deformation properties of jointed basaltic rock masses},
  author={Schultz, RA},
  journal={Rock Mechanics and Rock Engineering},
  volume={28},
  number={1},
  pages={1--15},
  year={1995},
  publisher={Springer}
}

@article{lajtai1991effect,
  title={The effect of strain rate on rock strength},
  author={Lajtai, EZ and Duncan, EJS and Carter, BJ},
  journal={Rock Mechanics and Rock Engineering},
  volume={24},
  number={2},
  pages={99--109},
  year={1991},
  publisher={Springer}
}

@article{kumar1968effect,
  title={The effect of stress rate and temperature on the strength of basalt and granite},
  author={Kumar, A},
  journal={Geophysics Research Letters},
  volume={95},
  pages={2757--2772},
  year={1968},
  publisher={American Geophysical Union}
}

@article{lindholm1974dynamic,
  title={The dynamic strength and fracture properties of Dresser basalt},
  author={Lindholm, US and Yeakley, LM and Nagy, A},
  journal={International Journal of Rock Mechanics and Mining Science},
  volume={11},
  number={10},
  pages={339--349},
  year={1974},
  publisher={Elsevier}
}

@article{zhang2020dynamic,
  title={Dynamic compressive properties of Kalgoorlie basalt rock},
  author={Zhang, X and Chiu, YW and Hao, H and Hsieh, A and Dight, P and Liu, K},
  journal={International Journal of Rock Mechanics and Mining Science},
  volume={134},
  pages={104492},
  year={2020},
  publisher={Elsevier}
}

@article{malik2018strain,
  title={Strain rate effect on the mechanical behavior of basalt: Observations from static and dynamic tests},
  author={Malik, A and Chakraborty, T and Rao, KS},
  journal={Thin-Walled Structures},
  volume={157},
  pages={107049},
  year={2018},
  publisher={Elsevier}
}

@inproceedings{schormair2006influence,
  title={The influence of anisotropy on hard rock drilling and cutting},
  author={Schormair, N and Thuro, K and Plinninger, RJ},
  booktitle={10th Congress of the International Association for Engineering Geology and the Environment (IAEG 2006)},
  year={2006},
  address={Nottingham, United Kingdom},
  publisher={Geological Society of London}
}

@article{shangxin2020estimation,
  title={Estimation of optimal drilling efficiency and rock strength by using controllable drilling parameters in rotary non-percussive drilling},
  author={Shangxin, F and Yujie, W and Guolai, Z and Yufei, Z and Shanyong, W and Ruilang, C and Enshang, X},
  journal={Journal of Petroleum Science and Engineering},
  volume={186},
  pages={106704},
  year={2020},
  publisher={Elsevier}
}

@article{dundas2023large,
  title = {A Large New Crater Exposes the Limits of Water Ice on Mars},
  author = {Dundas, Colin M. and Mellon, Michael T. and Miljkovi{\'c}, Katarina and Collins, Gareth S. and others},
  journal = {Geophysical Research Letters},
  volume = {50},
  number = {2},
  pages = {e2022GL100747},
  year = {2023},
  doi = {10.1029/2022GL100747},
  note = {Cites the compressive strength of intact ice at approximately 10 MPa}
}

@article{kahraman2003dominant,
  title = {Dominant rock properties affecting the penetration rate of percussive drills},
  author = {Kahraman, S. and Bilgin, N. and Feridunoglu, C.},
  journal = {International Journal of Rock Mechanics and Mining Sciences},
  volume = {40},
  number = {5},
  pages = {711--723},
  year = {2003},
  doi = {10.1016/S1365-1609(03)00063-9},
  note = {Reports UCS values for various sandstones ranging from 20 MPa to 149 MPa}
}

@article{abubakar2018penetration,
  title = {Penetration Rate and Specific Energy Prediction of Rotary-Percussive Drills: Contributions of Rock Properties},
  author = {Abu Bakar, M. Z. and Qureshi, M. A.},
  journal = {Journal of Mining Science},
  volume = {54},
  number = {2},
  pages = {233--245},
  year = {2018},
  doi = {10.1134/S106273911802363X},
  note = {Characterizes Murree Sandstone with a UCS of 125 MPa}
}

\newpage
\appendix
\section{Algorithm Descriptions}
This appendix provides a list and details of key algorithms used for data processing in this manuscript.

\subsection{Individual Strike Search Algorithm} \label{sec:appendix1}

Algorithm~\ref{alg:HammerStrikeDetection} outlines the steps used to find strike events. It references threshold values in Table~\ref{tab:PressureStrikeDetection} that were tuned so that strike events detected using the algorithm were equivalent to all strikes where the peak load exceeded 500 N. All data presented in this manuscript were verified visually to ensure the quality of the processed results. Algorithm~\ref{alg:hammerStrikeTimeLC} finds the leading edge of the load-cell force signal and outputs the hammer strike time for that signal as $t_{hLC}$. It uses Algorithm~\ref{alg:firstDerivative} to calculate derivatives using a 5th-order finite-difference method. Algorithm~\ref{alg:y0_iterative} is a shooting method that solves for the hammer dynamics in the ODE Eq.~\ref{eq:HammerFBDImplemented} by iterating the initial hammer position $y_0$. It uses the built-in MATLAB ODE solver \texttt{ode45}, an adaptive 4th- and 5th-order Runge-Kutta method with variable time step to solve non-stiff ordinary differential equations efficiently.

\begin{algorithm}[H]
\caption{Hammer Strike Detection Based on Pressure Signals} \label{alg:HammerStrikeDetection}
\begin{algorithmic}[1]
\Ensure Inputs: $t$, $p_h(t)$, $p_v(t)$
\Comment{{\bf Output:} $t_0$, $t_{se}$}
\State \textbf{Pre-process signals:}
\renewcommand\labelitemi{{\boldmath$\cdot$}}
\begin{itemize}
    \item $p_h(t)$: Hammer pressure signal
    \item $p_v(t)$: Vent pressure signal
    \item Compute $\dot{p}_v(t)$: First derivative of vent pressure, normalized 
    \item Compute $\ddot{p}_v(t)$: Second derivative of vent pressure, normalized 
\end{itemize}

\State \textbf{Find local maxima using \texttt{findpeaks}:}
\renewcommand\labelitemi{{\boldmath$\cdot$}}
\begin{itemize}
    \item Find peak locations in $p_h(t)$: $\texttt{findpeaks}(p_h)$ s.t. MinPeakProminence $\gets$ mt1thr, MinPeakWidth $\gets$ wthr
    \item Find peak locations in $\dot{p}_v(t)$: $\texttt{findpeaks}(\dot{p}_v)$ s.t. MinPeakProminence $\gets$ mt2thr
    \item Find peak locations in $\ddot{p}_v(t)$: $\texttt{findpeaks}(\ddot{p}_v)$ s.t. MinPeakProminence $\gets$ mt3thr
\end{itemize}

\State \textbf{For each detected peak in hammer pressure $p_h(t)$, find the closest peak in the first derivative of vent pressure $\dot{p}_v(t)$:}
\renewcommand\labelitemi{{\boldmath$\cdot$}}
\begin{itemize}
    \item Loop through all peaks in $p_h(t)$
    \item For each peak in $p_h(t)$:
    \begin{itemize}
        \item Compute time differences between the current peak in $p_h(t)$ and all earlier peaks in $\dot{p}_v(t)$
        \item Find the closest peak in $\dot{p}_v(t)$ that occurs before the current peak in $p_h(t)$: 
        \[
        t_{\mathrm{min}} = \displaystyle\argmin_{\Delta t > 0} \left( t_{\mathrm{strike}, p_h} - t_{\mathrm{peak}, \dot{p}_v} \right)
        \]
        \item If the time difference $\Delta t_{\mathrm{min}}$ is within a suitable range, record this as a valid strike event
    \end{itemize}
\end{itemize}

\State \textbf{Calculate time range for each strike event:}
\renewcommand\labelitemi{{\boldmath$\cdot$}}
\begin{itemize}
    \item For each valid strike event:
    \begin{itemize}
        \item Calculate the start time of the strike event by finding the closest peak in $\ddot{p}_v(t)$ (second derivative of vent pressure) to the left of the peak in $\dot{p}_v(t)$:
        \[
        t_{0} = \displaystyle\argmin_{\Delta t > 0} \left( t_{\mathrm{peak}, \dot{p}_v} - t_{\mathrm{peak}, \ddot{p}_v} \right)
        \]
        \item Assign this as the strike start time
        \item Calculate the strike duration using a preset time length:
        \[
        t_{se} = t_0+\Delta t_{\text{L}},
        \]
    \end{itemize}
\end{itemize}

\State Return: $t_0$, $t_{se}$
\end{algorithmic}
\end{algorithm}

\begin{algorithm}[H]
\caption{Hammer Strike Time Detection from Load Cell Data}\label{alg:hammerStrikeTimeLC}
\begin{algorithmic}[1]
\Ensure Inputs: $t$, $F_{LC}$
\Comment{{\bf Output:} $t_{hLC}$}

\State \textbf{Compute derivatives:}
\renewcommand\labelitemi{{\boldmath$\cdot$}}
\begin{itemize}
    \item $\dot{F}_{LC} \gets DF(F_{LC}, dt)$ \Comment{First derivative of load-cell force}
    \item $\ddot{\bar{F}}_{LC}  \gets \text{DF}(\dot{F}_{LC}, dt) / \left\Vert \left|DF(\dot{F}_{LC}, dt)\right| \right\Vert_{\infty}$ \Comment{Normalized 2nd derivative}
\end{itemize}
\State \textbf{Determine search window of interest:}
\renewcommand\labelitemi{{\boldmath$\cdot$}}
\begin{itemize}
    \item $t_{\mathrm{LCmax}} \gets $Find largest peak location in $F_{LC}$: $\texttt{findpeaks}(F_{LC})$ 
\end{itemize}
\State \textbf{Locate and assign strike time:}
\[
t_{hLC} = \min \{ t \mid \ddot{\bar{F}}_{LC}(t) > F_{\mathrm{thr}}, \; t_0 < t < t_{\mathrm{LCmax}} \}
\]
\State \textbf{Return:} $t_{hLC}$
\end{algorithmic}
\end{algorithm}

\begin{algorithm}[H]
\caption{First Derivative Calculation using 5th-Order Finite Difference}\label{alg:firstDerivative}
\begin{algorithmic}[1]
\Ensure Inputs: $f$, $h$
\Comment{{\bf Output:} $df$}
\Function{df}{$f$, $h$}
    \State $n \gets \text{length}(f)$
    \State \textbf{Initialize} $df \gets \text{zeros}(1, n)$
    \For{$i = 3$ to $n-2$}
        \State $dfout(i) \gets \frac{f(i-2) - 8 \cdot f(i-1) + 8 \cdot f(i+1) - f(i+2)}{12 \cdot h}$
    \EndFor
    \State \textbf{Handle boundary points:}
    \State $df(1) \gets \frac{-25 \cdot f(1) + 48 \cdot f(2) - 36 \cdot f(3) + 16 \cdot f(4) - 3 \cdot f(5)}{12 \cdot h}$
    \State $df(2) \gets \frac{-3 \cdot f(1) + 10 \cdot f(2) - 18 \cdot f(3) + 6 \cdot f(4) - f(5)}{12 \cdot h}$
    \State $df(n-1) \gets \frac{3 \cdot f(n) - 10 \cdot f(n-1) + 18 \cdot f(n-2) - 6 \cdot f(n-3) + f(n-4)}{12 \cdot h}$
    \State $df(n) \gets \frac{25 \cdot f(n) - 48 \cdot f(n-1) + 36 \cdot f(n-2) - 16 \cdot f(n-3) + 3 \cdot f(n-4)}{12 \cdot h}$
    \State \textbf{Return} $df$
\EndFunction
\end{algorithmic}
\end{algorithm}

\begin{table}[h!]
\centering
\small
\begin{tabular}{c p{7.2cm} c c}
\hline
\textbf{Variable} & \textbf{Description} & \textbf{Value} & \textbf{Units} \\
\hline
wthr & Number of samples for width threshold & 20 & \# \\
mt1thr & Amplitude threshold for hammer pressure ($p_h$) & 4 & psi \\
mt2thr & Amplitude of normalized peaks for first derivative of vent pressure ($\dot{p}_v$) & 0.12 & ND \\
mt3thr & Amplitude of normalized peaks for second derivative of vent pressure ($\ddot{p}_v$) & 0.1 & ND \\
tthr & Closeness of peaks between hammer pressure and vent pressure derivatives & 0.045 & s \\
$\Delta t_{\text{L}}$ & Maximum length of strike when event occur & 0.18 & s \\
$F_{\mathrm{thr}}$ & Threshold for normalized second derivative  & 0.05 & ND \\
\hline
\end{tabular}
\caption{Threshold values used in Algorithms~\ref{alg:HammerStrikeDetection} and \ref{alg:hammerStrikeTimeLC}.} \label{tab:PressureStrikeDetection}
\end{table}

\begin{algorithm}[H]
\caption{Iterative ODE Solution with Relaxation}
\label{alg:y0_iterative}
\begin{algorithmic}[1]
\Ensure Inputs: $p_h(t)$, $p_e(t)$, $M_h$, $c_1$, $c_2$, $k$, $A_h$, $L_s$, $L_{ch}$, $L_h$, $g$
\Comment{{\bf Output:} $y_0, \ \vec{y}_{\text{init}}, \ \vec{\Delta e}, \ t_X, \ X$}
\State \textbf{Initialize:}
\State $i_c \gets 0$ \Comment{Iteration counter}
\State $\Delta e \gets \infty$ \Comment{Initial error value}
\State $\alpha_1 \gets 0.95$ \Comment{Relaxation factor}
\State $\vec{\Delta e} \gets$ array of NaN values of size $i_{c,\text{max}}$
\State $\vec{y}_{\text{init}} \gets$ array of NaN values of size $i_{c,\text{max}}$
\State \textbf{Set initial conditions:} $\vec{X}_{\text{init}} \gets [y_0, v_0]$

\While{$\Delta e > \Delta e_{\text{thr}}$ \textbf{and} $i_c < i_{c,\text{max}}$}
    \State $f_{\text{ode}}(t, \vec{y}) \gets $ Eq.~\ref{eq:HammerFBDImplemented}, with inputs: $p_h(t)$, $p_e(t)$, $M_h$, $c_1$, $c_2$, $k$, $A_h$, $L_s$, $L_{ch}$, $L_h$, $g$
    \State Solve ODE using \texttt{ode45}: $[t_X, X] \gets$ \texttt{ode45}$(f_{\text{ode}}, t_{\text{strike}}, \vec{X}_{\text{init}})$
    \State Extract position and velocity: $\vec{y}_{h} \gets X(:, 1)$, $\vec{v}_{h} \gets X(:, 2)$
    
    \State \textbf{Update initial guess:}
    \State $y_0^{\text{new}} \gets (L_{ch} - L_h) - \vec{y}_{h}(\text{end})$
    
    \State \textbf{Calculate error:}
    \State $\Delta e \gets |y_0 - y_0^{\text{new}}|$
    
    \State \textbf{Update iteration variables:}
    \State $i_c \gets i_c + 1$
    \State $\vec{\Delta e}(i_c) \gets \Delta e$
    \State $\vec{y}_{0} \gets \alpha_1 y_0^{\text{new}} + (1 - \alpha_1) y_0$
    \State $\vec{X}_{\text{init}} \gets [y_0, v_0]$
    \State $\vec{y}_{\text{init}}(i_c) \gets y_0$
\EndWhile
\State \textbf{Return} $y_0, \ \vec{y}_{\text{init}}, \ \vec{\Delta e, t_X, \ X}$

\end{algorithmic}
\end{algorithm}

\end{document}